\documentclass[reqno,centertags,11pt]{amsart}
\usepackage{amsmath,amsthm,amscd,amssymb} \usepackage{latexsym}
\usepackage{graphicx}
\usepackage{pstricks,multido,graphicx,psfrag}

 \newcommand{\N}{{\mathbb{N}}}
 \newcommand{\R}{{\mathbb{R}}}
 \newcommand{\C}{{\mathbb{C}}}
 
 \newcommand{\Z}{{\mathbb{Z}}}

 \newcommand{\beq}{\begin{equation}}
\newcommand{\eeq}{\end{equation}}
\newcommand{\bdm}{\begin{displaymath}}
\newcommand{\edm}{\end{displaymath}} \newcommand{\ba}{\begin{align}}
\newcommand{\ea}{\end{align}} \newcommand{\bpf}{\begin{proof}}
\newcommand{\epf}{\end{proof}}

\newcommand{\la}{\langle} \newcommand{\ra}{\rangle}

 \newcommand{\vphi}{\varphi}

 \allowdisplaybreaks
\newcommand{\mathcalD}{\mathcal{D}}
\newcommand{\mathcalE}{\mathcal{E}}

\newcommand{\mathcalK}{\mathcal{K}}

\newcommand{\mathcalO}{\mathcal{O}}

\newcommand{\mathcalR}{\mathcal{R}}

\newtheorem{thm}{Theorem} 
\newtheorem{lem}[thm]{Lemma} \newtheorem{rem}{Remark}
\newtheorem{cor}[thm]{Corollary}

\theoremstyle{definition} 
\newtheorem{definition}[thm]{Definition}

 \newtheorem{remark}[thm]{Remark}

\newcounter{theoremi}[thm] 
\newcommand{\itemthm}{\refstepcounter{theoremi}
{\rm(\roman{theoremi})}{~}}

\numberwithin{thm}{section} \numberwithin{equation}{section}

\begin{document}

\title[Spectral properties of a limit-periodic Schr\"{o}dinger
operator] {Spectral properties of  a limit-periodic Schr\"{o}dinger
operator in dimension two.} \author[Y.~Karpeshina and
Y.-R.~Lee]{Yulia Karpeshina and Young-Ran~Lee}
% Yulia's address
\address{Department of Mathematics, University of Alabama at
Birmingham,
1300 University Boulevard, Birmingham, AL 35294.}%
\email{karpeshi@math.uab.edu}%
%Young-Ran's address
\address{Department of Mathematics, Sogang University, Shinsu-dong 1,
    Mapo-gu, Seoul, 121-742, South Korea.}%
\email{younglee@sogang.ac.kr}

\thanks{Supported in part by NSF grant DMS-0800949 (Y.K.) and the National
Research Foundation of Korea (NRF)-grant 2010-0015575 (Y.-R.L.).}
%\keywords{Non-linear Schr\"odinger equation, decay of eigenfunctions}
%\subjclass[2000]{35B20, 35B40, 35P30 }
\date{\today}
%\thanks{version 14, file Schrodinger14.tex}

\maketitle  \begin{abstract}  We study Schr\"{o}dinger operator
$H=-\Delta+V(x)$ in  dimension two, $V(x)$ being a limit-periodic
potential. We prove that the spectrum of $H$ contains a semiaxis and
there is a family of generalized eigenfunctions at every point of
this semiaxis with the following properties. First, the
eigenfunctions are close to plane waves $e^{i\langle \vec k,\vec
x\rangle }$ at the high energy region. Second, the isoenergetic
curves in the space of momenta $\vec k$ corresponding to these
eigenfunctions have a form of slightly distorted circles with  holes
(Cantor type structure). Third, the spectrum corresponding to the
eigenfunctions (the semiaxis) is absolutely continuous.
\end{abstract}

\section{Introduction.}\label{section1}

We study  the operator
    \begin{equation}
    H=-\Delta+V(x) \label{limper}
    \end{equation}
    in two dimensions, $V(x)$ being a limit-periodic potential:
    \begin{equation}V(x)=\sum _{r=1}^{\infty}V_r(x),\label{V}
    \end{equation}
   where $\{V_r\}_{r=1}^{\infty}$ is a family of periodic potentials
with doubling periods and decreasing $L_{\infty}$-norms. Namely,
$V_r$ has orthogonal periods $2^{r-1}\vec{d_1},\ 2^{r-1}\vec{d _2}$
and $\|V_r\|_{\infty}<\hat Cexp(-2^{\eta r})$ for some $\eta>\eta
_0>0$. Without loss of generality we assume that $\hat C=1$, $\vec
d_1=(d_1,0)$, $\vec d_2=(0,d_2)$ and $\int _{Q_r}V_r(x)dx=0$, $Q_r$
being the elementary cell of periods corresponding to $V_r(x)$. We
assume that all $V_r(x)$ are trigonometric polynomials with the
lengths growing at most linearly with  period. Namely, there exists
a positive number $R_0 <\infty$, such that each potential admits
Fourier representation: \bdm
 V_r(x)=\sum _{q \in  \Z^2\setminus \{0\},\ 2^{-r+1}|
 q|<R_0} v_{r,q}
  \exp i \la 2^{-r+1}\tilde q,x \ra ,\ \ \ \tilde q=2\pi
  \left(\frac{q_1}{d_1}, \frac{q_2}{d_2}\right),
\edm $\langle \cdot ,\cdot\rangle $ being the canonical dot product
in $\R^2$.

The one-dimensional analog of (\ref{limper}), (\ref{V}) is already
thoroughly investigated. It is proven in \cite{1}--\cite{9} that the
spectrum of the operator $H_1u=-u''+Vu$ is generically a Cantor type
set. It has a positive Lebesgue measure \cite{1,8}. The spectrum is
absolutely continuous \cite{1,3}, \cite{6}-\cite{7}. Generalized
eigenfunctions can be represented in the form of $e^{ikx}u(x)$,
$u(x)$ being limit-periodic \cite{6,8,9}. The case of a
complex-valued potential is studied in \cite{10}. Integrated density
of states is investigated in \cite{11}-\cite{14}. Spectral
properties of Schr\"odinger operators in $l^2(\Z)$ with
limit-periodic potentials are recently investigated in \cite{DG}. It
regards such potentials as generated by continuous sampling along
the orbits of a minimal translation of a Cantor group. It is shown
that the spectrum is a Cantor set of positive Lebesgue measure and
purely absolutely continuous for a dense set of sampling functions,
and it is a Cantor set of zero Lebesgue measure and purely singular
continuous for a dense $G_{\delta }$ set of sampling functions.
Properties of eigenfunctions of discrete multidimensional
limit-periodic Schr\"odinger operator are studied in \cite{15}. As
to the continuum multidimensional case, it is proven  in \cite{14}
that the integrated density of states for (\ref{limper}) is the
limit of densities of states for periodic operators.

    We concentrate here on properties of the spectrum and
eigenfunctions of (\ref{limper}), (\ref{V}) in the high energy
region. We prove the following results for the two-dimensional case.
    \begin{enumerate}
    \item The spectrum of the operator (\ref{limper}), (\ref{V})
    contains a semiaxis. A proof of an analogous result by different means
can be found
    in the  paper \cite{20}. In \cite{20}, the authors consider the
    operator $H=(-\Delta
    )^l+V$, $8l>d+3$, $d\neq 1(\mbox{mod}4)$, $d$ being the dimension of the space.
     This obviously includes our case $l=1$, $d=2$.
    However,
    there is an additional rather strong restriction on the potential
    $V(x)$ in \cite{20}, which we don't have here: in \cite{20}
all the lattices of periods $Q_r$ of periodic potentials
      $V_r$ need to contain a nonzero vector $\gamma$ in common,
      i.e., V(x) is periodic in a direction $\gamma $.

    \item There are generalized eigenfunctions $\Psi_{\infty }(\vec k, \vec
    x)$,
    corresponding to the semiaxis, which are    close to plane waves:
    for every $\vec k $ in an extensive subset $\mathcal{G} _{\infty }$ of
$\R^2$, there is
    a solution $\Psi_{\infty }(\vec k, \vec x)$ of the  equation
    $H\Psi _{\infty }=\lambda _{\infty }\Psi _{\infty }$ which can be
described by
    the formula:
    \begin{equation}
    \Psi_{\infty }(\vec k, \vec x)
    =e^{i\langle \vec k, \vec x \rangle}\left(1+u_{\infty}(\vec k, \vec
    x)\right), \label{aplane}
    \end{equation}
    \begin{equation}
    \|u_{\infty}\|=_{|\vec k| \to \infty}O(|\vec k|^{-\gamma _1}),\ \ \
\gamma _1>0,
    \label{aplane1}
    \end{equation}
    where $u_{\infty}(\vec k, \vec x)$ is a limit-periodic
    function:
    \begin{equation}
    u_{\infty}(\vec k, \vec x)=\sum_{r=1}^{\infty}  u_r(\vec k, \vec
    x),\label{aplane2}
    \end{equation}
    $u_r(\vec k, \vec x)$ being periodic with periods $2^{r-1} \vec{d_1},\
2^{r-1} \vec{d_2}$.
    The  eigenvalue $\lambda _{\infty }(\vec k)$ corresponding to
    $\Psi_{\infty }(\vec k, \vec x)$ is close to $|\vec k|^{2}$:
    \begin{equation}
    \lambda _{\infty }(\vec k)=_{|\vec k| \to \infty}|\vec k|^{2}+
    O(|\vec k|^{-\gamma _2}),\ \ \ \gamma _2>0. \label{16a}
    \end{equation}
     The ``non-resonant" set $\mathcal{G} _{\infty }$ of the
       vectors $\vec k$, for which (\ref{aplane}) -- (\ref{16a}) hold, is
       an extensive
       Cantor type set: $\mathcal{G} _{\infty }=\cap _{n=1}^{\infty }\mathcal{G}
_n$,
       where $\{\mathcal{G} _n\}_{n=1}^{\infty}$ is a decreasing sequence of
sets in $\R^2$. Each $\mathcal{G} _n$ has a finite number of holes
in each bounded
       region. More and more holes appears when $n$ increases,
       however
       holes added at each step are of smaller and smaller size.
       The set $\mathcal{G} _{\infty }$ satisfies the estimate:
       \begin{equation} \frac{\left|\left(\mathcal{G} _{\infty
}\cap
        \bf B_R\right)\right|}{\left|\bf B_R\right|}=_{R\to \infty }
1+O(R^{-\gamma _3}),\quad \gamma_3 >0,
         \label{full}
       \end{equation}
       where $\bf B_R$ is the disk of radius $R$ centered at the
       origin, $|\cdot |$ is the Lebesgue measure in $\R^2$.

       \item The set $\mathcal{D}_{\infty}(\lambda)$,
defined as a level (isoenergetic) set for $\lambda _{\infty }(\vec
k)$, $$ {\mathcal D} _{\infty}(\lambda)=\left\{ \vec k \in
\mathcal{G} _{\infty } :\lambda _{\infty }(\vec k)=\lambda
\right\},$$ is proven to be a slightly distorted circle with the
infinite number of holes. It can be described by  the formula:
\begin{equation} {\mathcal D}_{\infty}(\lambda)=\{\vec k:\vec
k=\varkappa _{\infty}(\lambda, \vec{\nu})\vec{\nu},
    \ \vec{\nu} \in {\mathcal B}_{\infty}(\lambda)\}, \label{D}
    \end{equation}
where ${\mathcal B}_{\infty }(\lambda )$ is a subset of the unit
circle $S_1$. The set ${\mathcal B}_{\infty }(\lambda )$ can be
interpreted as the set of possible directions of propagation for the
almost plane waves (\ref{aplane}).  The set ${\mathcal B}_{\infty
}(\lambda )$ has a Cantor type structure and an asymptotically full
measure on $S_1$ as $\lambda \to \infty $:  \begin{equation}
L\left({\mathcal B}_{\infty}(\lambda )\right)=_{\lambda \to \infty
}2\pi +O\left(\lambda^{-\gamma _3/2}\right), \label{B}
\end{equation} here and below $L(\cdot)$ is the length of a curve.
The value $\varkappa _{\infty }(\lambda ,\vec \nu )$ in (\ref{D}) is
the ``radius" of ${\mathcal D}_{\infty}(\lambda)$ in a direction
$\vec \nu $. The function $\varkappa _{\infty }(\lambda ,\vec \nu
)-\lambda^{1/2}$ describes the deviation of ${\mathcal
D}_{\infty}(\lambda)$ from the perfect circle of the radius
$\lambda^{1/2}$. It is proven that the deviation is asymptotically
small: \begin{equation} \varkappa _{\infty }(\lambda ,\vec \nu
)=_{\lambda \to \infty} \lambda^{1/2}+O\left(\lambda^{-\gamma _4
}\right), \ \ \ \gamma _4>0. \label{h} \end{equation}

\item Absolute continuity of the branch of the
    spectrum (the semiaxis) corresponding to $\Psi_{\infty }(\vec k, \vec
x)$ is proven. \end{enumerate}

       To prove the results listed above we  develop a modification of the
Kolmogorov-Arnold-Moser (KAM) method. This paper is inspired by
\cite{21,22,24}, where the method is used for periodic problems. In
\cite{21} KAM method is applied to classical Hamiltonian systems. In
\cite{22,24} the technique developed in \cite{21}  is applied to
semiclassical approximation for
         multidimensional periodic Schr\"{o}dinger operators at high
         energies.

         We consider a sequence of operators
    $$ H_0=-\Delta , \ \ \ \ \ \
H^{(n)}=H_0+\sum_{r=1}^{M_n} V_r,\ \ \ n\geq 1, \ M_n \to \infty
\mbox{ as } n \to \infty .$$ Obviously, $\|H-H^{(n)}\|\to 0$ as
$n\to \infty $, where $\|\cdot \|$ is the norm in the class of
bounded operators. Clearly,
\begin{equation}H^{(n)}=H^{(n-1)}+W_n, \ \
W_n=\sum_{r=M_{n-1}+1}^{M_n} V_r. \label{W_n} \end{equation} We
consider each operator $H^{(n)}$, $n\geq 1$, as a perturbation of
the previous operator $H^{(n-1)}$. Every operator $H^{(n)}$ is
periodic, however the periods go to infinity as $n \to \infty$. We
show that there is a $\lambda_*$, $\lambda_*=\lambda_*(V)$, such
that the semiaxis $[\lambda _*, \infty )$ is contained in the
spectra of {\bf all} operators $H^{(n)}$. For every operator
$H^{(n)}$ there is a set of eigenfunctions (corresponding to the
semiaxis) being close to plane waves:
    for every $\vec k $ in an extensive subset $\mathcal{G} _n$ of $\R^2$, there
is
    a solution $\Psi_{n}(\vec k, \vec x)$ of the differential equation
    $H^{(n)}\Psi _n=\lambda _n\Psi _n$, which can be described by
    the formula:
    \begin{equation}
    \Psi_n (\vec k, \vec x)
    =e^{i\langle \vec k, \vec x \rangle}\left(1+\tilde u_{n}(\vec k, \vec
    x)\right),\ \ \
    %%%%%%\label{n}
  %%%%%%   \end{equation}
%%%%%%\begin{equation}
\|\tilde u_{n}\|_{L_{\infty }(\R^2)}\underset{|\vec k| \to
\infty}{=}O(|\vec k|^{-\gamma _1}),\ \ \ \gamma _1>0, \label{na}
\end{equation}
    where $\tilde u_{n}(\vec k, \vec x)$ has periods $2^{M_n-1}\vec d_1,
    2^{M_n-1}\vec d_2$.\footnote{Obviously, $\tilde u_{n}(\vec k, \vec x)$
    is simply related to functions $u_{r}(\vec k, \vec x)$
used in (\ref{aplane2}): $\tilde u_{n}(\vec k, \vec x)=\sum
_{r=M_{n-1}+1}^{M_n} u_{r}(\vec k, \vec x)$. }
    The corresponding eigenvalue $\lambda ^{(n) }(\vec k)$ is close to $|\vec
k|^{2}$:
    \begin{equation} \lambda ^{(n)}(\vec k)=_{|\vec k| \to \infty}|\vec k|^{2}+
    O\left(|\vec k|^{-\gamma _2}\right),\ \ \ \gamma _2>0.
    \label{back}
    \end{equation}
     The non-resonant set $\mathcal{G} _{n}$
       for which (\ref{back}) holds,
        is proven to be
       extensive in $\R^2$:
              \begin{equation} \frac{\left|\mathcal{G} _{n
}\cap
        \bf B_R \right|}{\left|\bf B_R\right|}=_{R\to \infty }
1+O(R^{-\gamma _3}).
         \label{16b}
       \end{equation}
        Estimates (\ref{na}) -- (\ref{16b}) are uniform in $n$.
The set ${\mathcal D}_{n}(\lambda)$ is defined as the level
(isoenergetic) set for non-resonant eigenvalue $\lambda ^{(n) }(\vec
k)$: $$ {\mathcal D} _{n}(\lambda)=\left\{ \vec k \in \mathcal{G}
_n:\lambda ^{(n)}(\vec k)=\lambda \right\}.$$ This set is proven to
be a slightly distorted circle with a finite number of holes (Fig.
\ref{F:1}, \ref{F:2}). The set ${\mathcal D} _{n}(\lambda)$ can be
described by the formula:
\begin{figure}
\begin{minipage}[t]{7cm}
\includegraphics[totalheight=.25\textheight]{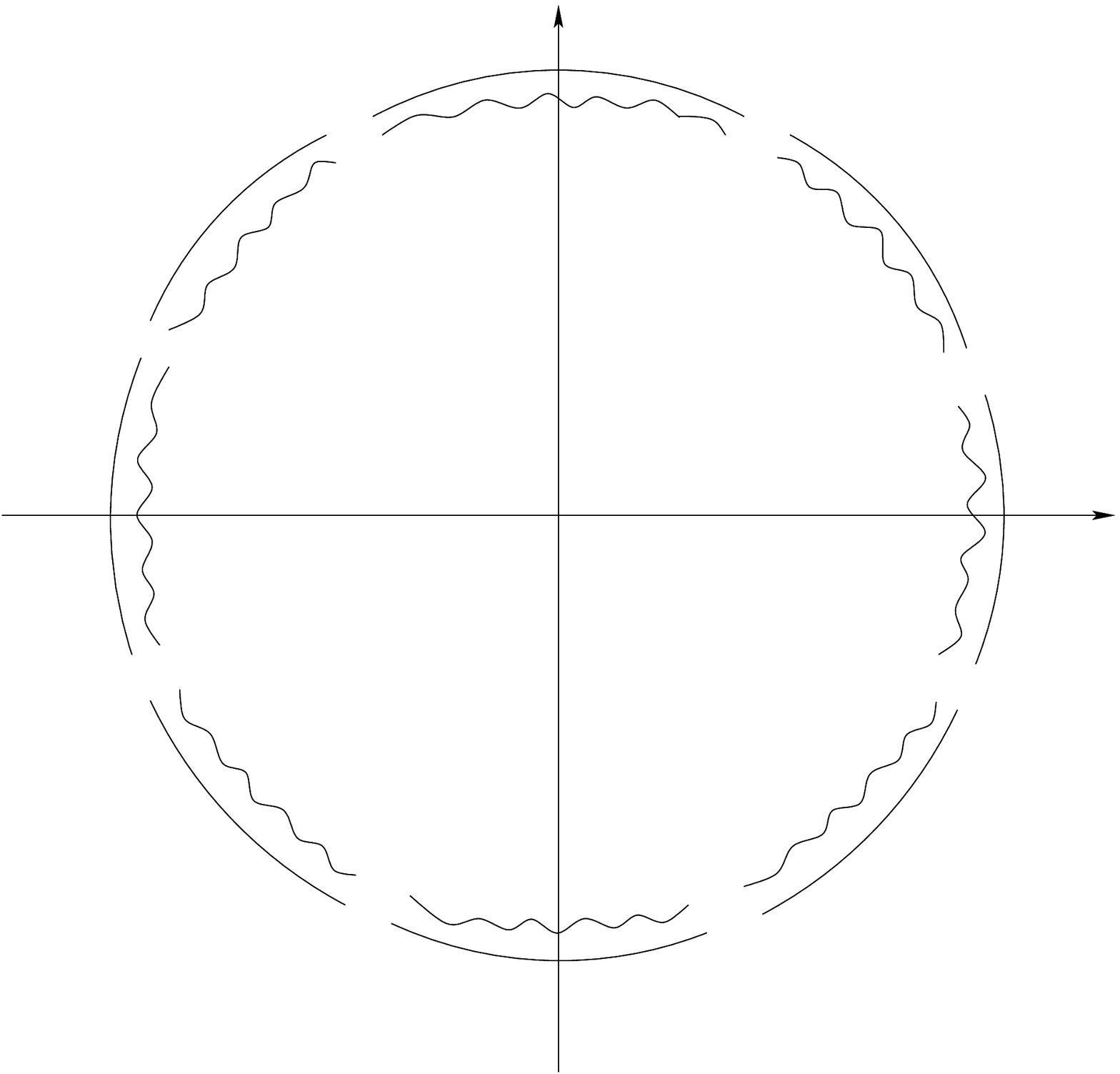}
\caption{Distorted  circle with holes,
$\mathcal{D}_1(\lambda)$.}\label{F:1} \end{minipage} \hfill
\begin{minipage}[t]{7cm}
\includegraphics[totalheight=.25\textheight]{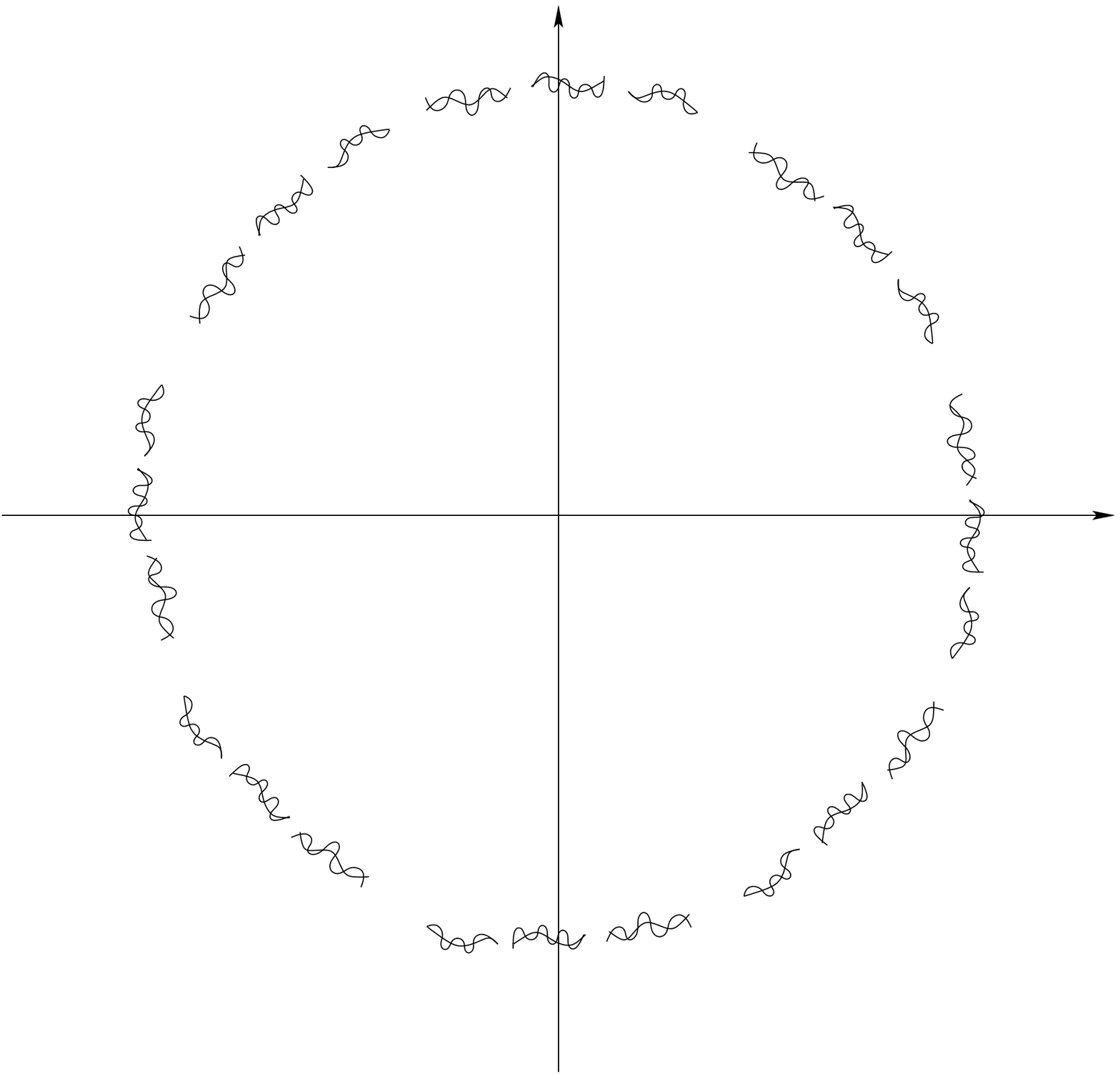}
\caption{Distorted circle with holes,
$\mathcal{D}_2(\lambda)$.}\label{F:2} \end{minipage} \hfill
\end{figure}

\begin{equation} {\mathcal D}_{n}(\lambda)=\{\vec k:\vec k=
    \varkappa_{n}(\lambda, \vec{\nu})\vec{\nu},
    \ \vec{\nu} \in {\mathcal B}_{n}(\lambda)\}, \label{Dn}
    \end{equation}
where ${\mathcal B}_{n}(\lambda )$ is a subset  of the unit circle
$S_1$. The set ${\mathcal B}_{n}(\lambda )$ can be interpreted as
the set of possible directions of propagation for  almost plane
waves (\ref{na}). It is shown that $\{{\mathcal
B}_n(\lambda)\}_{n=1}^{\infty }$  is a decreasing sequence of sets,
since on each step more and more directions are excluded.  Each
${\mathcal B}_{n}(\lambda )$ has an asymptotically full measure on
$S_1$ as $\lambda \to \infty $: \begin{equation} L\left({\mathcal
B}_{n}(\lambda )\right)=_{\lambda \to \infty }2\pi
+O\left(\lambda^{-\gamma _3 /2}\right), \label{Bn} \end{equation}
the estimate being uniform in $n$. The set ${\mathcal B}_{n}$ has
only a finite number of holes, however their number is growing with
$n$. More and more holes of a smaller and smaller size are added at
each step. The value $\varkappa_{n}(\lambda ,\vec \nu
)-\lambda^{1/2}$ gives the deviation of ${\mathcal D}_{n}(\lambda)$
from the perfect circle of the radius $\lambda^{1/2}$  in the
direction $\vec \nu $. It is proven that the deviation is
asymptotically small uniformly in $n$: \begin{equation}
\varkappa_{n}(\lambda ,\vec \nu)
=\lambda^{1/2}+O\left(\lambda^{-\gamma _4 }\right),\ \ \ \
\frac{\partial \varkappa_{n}(\lambda ,\vec \nu)}{\partial \varphi
}=O\left(\lambda^{-\gamma _5 }\right), \ \ \ \gamma _4, \gamma _5>0,
\label{hn} \end{equation} $\varphi $ being an angle variable $\vec
\nu =(\cos \varphi ,\sin \varphi )$.

On each step  more and more points are excluded from the
non-resonant sets $\mathcal{G} _n$, thus $\{ \mathcal{G} _n
\}_{n=1}^{\infty }$ is a decreasing sequence of sets. The set
$\mathcal{G} _\infty $ is defined as the limit set: $\mathcal{G}
_\infty=\cap _{n=1}^{\infty }\mathcal{G} _n $. It has the infinite
number of holes at each bounded region, but nevertheless satisfies
the relation (\ref{full}). For every $\vec k \in \mathcal{G} _\infty
$ and every $n$, there is a generalized eigenfunction of $H^{(n)}$
of the type  (\ref{na}). It is proven that  the sequence of $\Psi
_n(\vec k, \vec x)$ has a limit in $L_{\infty }(\R^2)$ as $n\to
\infty$, when $\vec k \in \mathcal{G} _\infty $. The function $\Psi
_{\infty }(\vec k, \vec x) =\lim _{n\to \infty }\Psi _n(\vec k, \vec
x)$ is a generalized eigenfunction of $H$. It can be written in the
form (\ref{aplane}) -- (\ref{aplane2}). Naturally, the corresponding
eigenvalue $\lambda _{\infty }(\vec k) $ is the limit of $\lambda
^{(n)}(\vec k )$ as $n \to \infty $.

We consider the limit ${\mathcal B}_{\infty}(\lambda)$ of ${\mathcal
B}_n(\lambda)$:
    $${\mathcal B}_{\infty}(\lambda)=\bigcap_{n=1}^{\infty} {\mathcal
B}_n(\lambda),\ \ \ {\mathcal B}_n\subset {\mathcal B}_{n-1}.$$
    This set has a Cantor type structure on the unit circle.
    It is proven that ${\mathcal B}_{\infty}(\lambda)$ has an asymptotically
    full measure on the unit circle (see (\ref{B})).
    We prove
    that the sequence $\varkappa _n(\lambda ,\vec \nu )$, $n=1,2,... ,$,
describing the
     isoenergetic curves $\mathcal{D}_n$,  quickly converges as $n\to \infty$.
Hence, ${\mathcal D}_{\infty}(\lambda)$ can be described as the
limit of ${\mathcal D}_n(\lambda)$ in the sense (\ref{D}), where
$\varkappa _{\infty}(\lambda, \vec{\nu})=\lim _{n \to \infty}
\varkappa _n(\lambda, \vec{\nu})$ for every $\vec{\nu} \in {\mathcal
B}_{\infty}(\lambda)$. It is shown that the derivatives of the
functions $\varkappa _n(\lambda, \vec{\nu})$ (with respect to the
angle variable $\varphi $ on the unit circle) have a limit as $n\to
\infty $ for every $\vec{\nu} \in {\mathcal B}_{\infty}(\lambda)$.
We denote this limit by $\frac{\partial \varkappa_{\infty}(\lambda
,\vec \nu)}{\partial \varphi }$. Using (\ref{hn}) we  prove that
    \begin{equation}\frac{\partial \varkappa_{\infty}(\lambda ,\vec \nu)}{\partial
\varphi }=O\left(\lambda^{-\gamma _5 }\right).\label{Dec9a}
\end{equation} Thus, the limit curve ${\mathcal
D}_{\infty}(\lambda)$ has a tangent vector in spite of its Cantor
type structure, the tangent vector being the limit of the
corresponding tangent vectors for ${\mathcal D}_n(\lambda)$ as $n\to
\infty $.  The curve ${\mathcal D}_{\infty}(\lambda)$ looks as
  a slightly distorted circle with
the infinite number of holes.

Absolute continuity of the branch of the spectrum  $[\lambda
_*(V),\infty )$, corresponding to the functions $\Psi _{\infty
}(\vec k, \vec x)$, $\vec k\in \mathcal{G} _{\infty }$, follows from
 continuity properties of  level curves
$\mathcal{D}_{\infty }(\lambda )$ with respect to $\lambda $, and
from convergence of  spectral projections corresponding to $\Psi
_{n}(\vec k, \vec x)$, $\vec k\in \mathcal{G} _{\infty }$, to
spectral projections of $H$ in the strong sense and uniformly in
$\lambda $, $\lambda >\lambda _* $.

  The  limit-periodic operator $H=(-\Delta )^l+V$, $l\geq 6
         $, $d=2$ is considered in \cite{KL,cpde}. The results proved in \cite{KL,cpde} are
         analogous to 1-4 on pages \pageref{aplane}, \pageref{h}. The main difficulty of the case $l=1$ comparing with
          $l\geq 6
         $ is  in  starting the recurrent procedure.  Step 1 here is really cumbersome comparing with the case
         $l\geq 6$. This technical  difficulty is related to the fact that perturbation theory for a periodic operator
         $(-\Delta )^l +V_{per}$ is much simpler for $l>1$ then for $l=1$, since Bloch eigenvalues are well-spaced for
         $l>1$ at high energies.

         Further steps of approximation procedure are  similar to those in \cite{KL,cpde}
         up to some technical modifications.
The main technical difficulty to overcome is construction of
   non-resonance sets
$\mathcal{B} _n(\lambda)$ for every sufficiently large $\lambda $,
$\lambda > \lambda_* (V)$, the last bound being uniform in $n$. The
set $\mathcal{B} _n(\lambda)$ is obtained  by deleting a ``resonant"
part from $\mathcal{B}_{n-1}(\lambda)$. Definition of $\mathcal{B}
_{n-1}\setminus \mathcal{B}_{n}$, naturally, includes Bloch
eigenvalues of $H^{(n-1)}$. To describe $\mathcal{B} _{n-1}\setminus
\mathcal{B}_{n}$ one needs
  both non-resonant
eigenvalues  (\ref{back}) and resonant eigenvalues. No
 suitable  formulae are known for resonant eigenvalues. Absence of
formulae causes difficulties in estimating the size of $\mathcal{B}
_n\setminus \mathcal{B}_{n-1}$. Let us describe  shortly how we
treat this problem in the second and higher steps of approximation.
Indeed, on $n$-th step  of approximation we start with the operator
$H^{(n-1)}$. It has elementary cell of periods $Q_{n-1}$. The
corresponding Bloch decomposition is denoted by
$H^{(n-1)}(t^{(n-1)})$, quasimomentum $t^{(n-1)}$ belonging to the
dual elementary cell $K_{n-1}$.  We take $t^{(n-1)}(\varphi )$ being
equal to $\vec \varkappa _{n-1}(\varphi )$ modulo $K_{n-1}$, where
$\vec \varkappa _{n-1}(\varphi )$ describes
 $\mathcal{D}_{n-1}
       (\lambda)$ (Fig. 1,2), $\vec \varkappa _{n-1}(\varphi)=\varkappa _{n-1} (\lambda
, \vec \nu
       ) \vec \nu$, $\vec \nu =(\cos \varphi ,\sin \varphi )$.
 When $\vec
\varkappa _{n-1}(\varphi )$ is a point on
$\mathcal{D}_{n-1}(\lambda)$, vector $\vec \nu =(\cos \varphi ,\sin
\varphi )$ belongs to $\mathcal{B}_{n-1}(\lambda)$. The operator
$H^{(n-1)}(t^{(n-1)})$ has a simple eigenvalue equal to $\lambda $,
formulas \eqref{na}--\eqref{hn} with $n-1$ instead of $n$  being
valid. All other eigenvalues are separated from $\lambda $ by the
distance much greater than $\|W_{n-1}\|$, see \eqref{W_n}. It is
convenient to denote such $ H^{(n-1)}(t^{(n-1)})$  by $
H^{(n-1)}(\vec \varkappa _{n-1}(\varphi ))$. Next, perturbation
$W_n$ of operator $H^{(n-1)}$ has  bigger periods  than $H^{(n-1)}$.
We assign these bigger periods to $H^{(n-1)}$. The corresponding
Bloch decomposition we denote by $\tilde H^{(n-1)}(t^{(n)})$, where
$t^{(n)}$ is quasimomentum in the dual elementary cell $K_n$.
According to Bloch's theory, for any $t^{(n)}$ the spectrum of
$\tilde H^{(n-1)}(t^{(n)})$ is the union of the spectra of
$H^{(n-1)}(t^{(n)}+\vec b)$, where $\vec b$ belongs to the lattice
$P^{(n)}$ generated by $K_n$ in $K_{n-1}$, see Fig.4, page
\pageref{F:6}. We take $t^{(n)}(\varphi )$ being equal to $\vec
\varkappa _{n-1}(\varphi )$ modulo $K_{n}$. This means that
$t^{(n-1)}(\varphi )$ and $t^{(n)}(\varphi )$ satisfy the relation
$t^{(n-1)}(\varphi )=t^{(n)}(\varphi )+\vec b_*$ for some $\vec
b_*\in P^{(n)}$ where $t^{(n-1)}(\varphi )$ is $\vec \varkappa
_{n-1}(\varphi )$ modulo $K_{n-1}$. It is convenient to denote such
operator $\tilde H^{(n-1)}(t^{(n)})$  by $\tilde H^{(n-1)}(\vec
\varkappa _{n-1}(\varphi ))$. Obviously, $\tilde H^{(n-1)}(\vec
\varkappa _{n-1}(\varphi ))$ has an eigenvalue equal to $\lambda $,
since $H^{(n-1)}(\vec \varkappa _{n-1}(\varphi )+\vec b_*)$ does,
all other eigenvalues of $H^{(n-1)}(\vec \varkappa _{n-1}(\varphi
)+\vec b_*)$ being separated from $\lambda $.  We say that $\varphi
$ is resonant if $H^{(n-1)}\left(\vec \varkappa _{n-1}(\varphi
)+\vec b\right)$ has an eigenvalue close to $\lambda $ for some
$\vec b\neq \vec b_*$. Note that it happens if and only if the
operator \begin{equation}\label{det-S} I+S_n(\varphi
)=\Bigl(H^{(n-1)}\bigl(\vec \varkappa _{n-1}(\varphi )
       +\vec b\bigr)-\lambda-\epsilon \Bigr) \Bigl(H_{0}\bigl(\vec \varkappa
_{n-1}(\varphi )
       +\vec b\bigr)+\lambda\Bigr)^{-1}\end{equation}
       has an eigenvalue equal to zero for some small $\epsilon $ and $\vec b\neq \vec b_*$,
       the operator $H_0$ being the free operator with the same
       periods as $H^{(n-1)}$.
       Assume for a moment we consider a polyharmonic operator
       $H=(-\Delta )^{l}+V$, $l>1$. Then, $S_n$ is in the trace class
       and the determinant of $I+S_n$ exists.
       The resonant set  $\mathcal{B}_{n-1}\setminus \mathcal{B}_{n}$ can
be described (in terms of  $\varphi $) as the set of solutions of
the equation $Det \bigl(I+S_n(\varphi )\bigr)=0$. To obtain
$\mathcal{B}_{n-1}\setminus
       \mathcal{B}_{n}$ we take all values of $\epsilon $ in a small interval
       and values of $\vec b$ in a finite set, $\vec b\neq 0$.
       To estimate the size of $\mathcal{B}_{n-1}\setminus
       \mathcal{B}_{n}$ we introduce a complex angle variable $\varphi
       $, i.e., we extend our considerations to
        a complex neighborhood $\varPhi _0$ of $[0,2\pi
       )$. We show that the    determinant \eqref{det-S} is an  analytic function of
        $\varphi $ in $\varPhi _0$, and, by this,
         reduce the
        problem of estimating the size of $\mathcal{B}_{n-1}\setminus
       \mathcal{B}_{n}$
         to
        a problem in complex analysis. We use Rouche's theorem  to count
          zeros of the determinants and to investigate how far the zeros
move when
         $\varepsilon $ changes or $W_n$ is added to $H^{(n-1)}$. It enables us to estimate the size
         of the zero set of the determinants, and, hence, the size of
         the non-resonance set $\varPhi _n\subset \varPhi _0$,  which is
defined as a
         non-zero set for the determinants.
          Proving that the non-resonance set $\varPhi _n$
         is sufficiently large, we
          obtain estimates
          (\ref{Bn}) for
         $\mathcal{B}_n$, the set  $\mathcal{B}_n$ being defined by the real part of
         $\varPhi _n$.

          To obtain $\varPhi _n$ we delete  from $\varPhi _0$ more and more
holes of smaller and
         smaller radii at each step. Thus, the non-resonance set $\varPhi
         _n\subset \varPhi _0$ has a structure of Swiss
         Cheese (Fig. \ref{F:7}, \ref{F:8}), pages \pageref{pageF:7},
         \pageref{pageF:8}.
 Deleting  resonance set from $\varPhi _0$ at each
         step of the recurrent procedure we call  a ``Swiss
         Cheese Method" .  The essential
         difference of our method from those applied  in similar
         situations before (see e.g. \cite{21}--\cite{B3}) is that
we construct a
         non-resonance set not only in the whole space of a parameter
         ($\vec k\in \R^2$ here), but also on   all isoenergetic curves
         ${\mathcal D}_n(\lambda )$ in
         the space of parameter, corresponding to sufficiently large $\lambda $. Estimates for the
         size of non-resonance sets on a curve require more subtle
         technical considerations (``Swiss
         Cheese construction") than those sufficient for
         description of a non-resonant set in the whole space of
         the parameter.

         When $l=1$ (the present case) the determinant of
$I+S_n(\varphi )$ (see (\ref{det-S})) does not exist, since
$S_n(\varphi )$ is not from a trace class. We approximate
$S_n(\varphi )$ by finite dimensional operators $S_n^{(N)}$ and
consider the solutions of $Det \bigl(I+S_n^{(N)}(\varphi )\bigr)=0$.
The accumulation points of these solutions as $N\to \infty $ we call
the solutions of $``Det" \bigl(I+S_n(\varphi )\bigr)=0$. Swiss
Cheese method is applied here with such a modification.

          The
         requirement for super exponential decay of $\|V_r\|$ as
         $r\to \infty $ is essential, since it is needed to
         ensure convergence of the recurrent procedure. At every step we use  the upper bounds on $\|V_r\|$ to prove perturbation
         formulae for Bloch eigenvalues and eigenfunctions when $\lambda >\lambda_*(V)$, $\lambda_*$ being the same for all steps.
        It is not  important  that  potentials $V_r$ have doubling
         periods, in the sense that
         the periods of the type $q^{r-1}\vec d_1,\ q^{r-1}\vec d_2, q \in
         \N,$ can be treated in the same way as the doubling.

         The plan of the paper is the following. Section 2 (page \pageref{sec:first})
          is devoted to the first step of the recurrent
         procedure.  Sections 3 (page \pageref{chapt4}),  4 (page \pageref{chapt5}) and
          5 (page \pageref{Section 5})   describe second,
         third and $n$-th steps of the recurrent procedure, respectively. Discussion of
         convergence of the procedure and proofs of the results
         1 -- 3,
         listed above, are in Section 6 (page \pageref{chapt7}). Absolute continuity is proved in Section 7
         (page \pageref{chapt8}). Proofs of geometric lemmas
         and appendices
         are in Sections 8 (page \pageref{sec:geometric-1}) and 9 (page \pageref{appendices}), respectively.

         Further, we denote by $c, C$ absolute constants, by
         $C(\|V\|)$ a value depending only on $\|V\|$, etc.

%%%%%%%%%%%%%%%%%%%%%%%%%%%%%%%%%%%%%%%%%%%%%%%%
\section{The first approximation} \label{sec:first}
%%%%%%%%%%%%%%%%%%%%%%%%%%%%%%%%%%%%%%%%%%%%%%%%

We fix  $0<s_1< 10^{-4}$. Let $k>0$  and be large enough so that
$k^{s_1}>500R_0,d_1,d_2, d_1^{-1},d_2^{-1}$. We define the first
operator $H^{(1)}$ by
 \beq\label{E:H_1}
 H^{(1)} := -\Delta +W_1, \quad W_1 := \sum_{r=1}^{M_1} V_r, \quad 2^{M_1} \approx
 k^{s_1}.\ \ \footnote{We write $a(k)\approx b(k)$
when the inequalities $\frac{1}{2}b(k) \leq a(k) \leq 2b(k) $ hold.}
 \eeq
Obviously, operator $H^{(1)}$ has a periodic potential. We denote
its periods by $(a_1,0),\ (0,a_2)$, $a_1=2^{M_1-1}d_1$,
$a_2=2^{M_1-1}d_2$.  We employ Bloch theory (see e.g. \cite{RS}) for
this  operator, i.e., consider a family of operators $H^{(1)}(t)$,
$t\in K_1$, where $K_1$ is the elementary cell of the dual lattice,
$K_1=[0,2\pi a_1^{-1})\times [0,2\pi a_2^{-1})$. Vector $t$ is
called quasi-momentum. Each operator $H^{(1)}(t)$ acts in
$L_2(Q_1)$, $Q_1=[0,a_1]\times [0,a_2]$. The operator $H^{(1)}(t)$
is defined by \eqref{E:H_1} and quasi-periodic boundary conditions:
 \beq
 \begin{gathered}
    u(a_1,x_2)= \exp(it_1a_1)u(0,x_2),\quad
    u(x_1,a_2)= \exp(it_2a_2)u(x_1,0), \label{quasi}\\
    u_{x_1}(a_1,x_2)=\exp(it_1a_1)u_{x_1}(0,x_2),\quad
    u_{x_2}(x_1,a_2)=\exp(it_2a_2)u_{x_2}(x_1,0).
 \end{gathered}
 \eeq
Each operator $H^{(1)}(t)$, $t\in K_1$, has a discrete bounded below
spectrum $\Lambda ^{(1)}(t)$,
 \bdm
 \Lambda ^{(1)}(t)=\bigcup _{n=1}^{\infty } \{\lambda_n^{(1)}(t)\},\quad
 \lambda^{(1)} _n(t) \to \infty  \text{ as } n\to \infty .
 \edm
The spectrum $\Lambda ^{(1)}$ of the original operator $H^{(1)}$ is
the union of the spectra of the operators $H^{(1)}(t)$ over all $t
\in K_1$: $\Lambda ^{(1)}=\cup _{t\in K_1}\Lambda ^{(1)}(t)$. The
functions $\lambda_n^{(1)}(t)$ are continuous, so  $\Lambda ^{(1)}$
has a band structure. Extending all the eigenfunctions of
$H^{(1)}(t)$ for all $t \in K_1$ by the quasi-periodic boundary
conditions to $\R^2$ yields a complete system of generalized
eigenfunctions of $H^{(1)}$.

Let $H_0^{(1)}$ be the operator \eqref{E:H_1} corresponding to
$W_1=0$. We consider that it has periods $a_1, a_2$ and that
operators $H_0^{(1)}(t)$ are defined in  $L_2(Q_1)$. The
eigenfunctions of the operator $H_0^{(1)}(t)$  are plane waves
satisfying \eqref{quasi}. They are naturally indexed by points in
$\Z^2$:
 \bdm
 \Psi _j^0(t,x)=|Q_1|^{-1/2}\exp i \la \vec p_j(t),x \ra,\quad
 |Q_1|=a_1a_2,
 \edm
 where here and
below
 $
 \vec p_j (t) := (2\pi j_1/a_1+t_1, 2\pi j_2/a_2+t_2).$
 The eigenvalue corresponding to $ \Psi _j^0(t,x)$ is equal to
 $p_j^2(t)$,
$p_j(t) := |\vec p_j (t)|.$

\begin{figure} %%%%%%%%\centering

\begin{pspicture}(0,0)(6.5,6.5)
\psset{unit=.9in}

\def\4row{
\pscircle(0.5,0.5){0.5} \qdisk(0.5,0.5){1.5pt}
\psline[linewidth=0.5pt](-0.2,0.5)(1.2,0.5)
\psline[linewidth=0.5pt](0.5,-0.2)(0.5,1.2)}

\multips(0,0)(0.4,0.0){4}{\4row}

\multido{\n=0+0.4}{4} {\multips(\n,0)(0.0,0.4){4}{\4row}}

\psline[linewidth=2pt](0.9,0.9)(1.3,0.9)
\psline[linewidth=2pt](0.9,0.9)(0.9,1.3)
\psline[linewidth=2pt](0.9,1.3)(1.3,1.3)
\psline[linewidth=2pt](1.3,0.9)(1.3,1.3)

\end{pspicture}

\hspace{5mm}

\caption{The isoenergetic surface $S_0(\lambda)$ of the free
operator $H_0^{(1)}$}\label{F:3}
\end{figure}
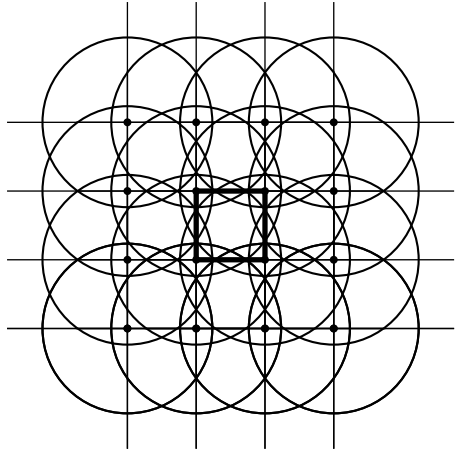

\begin{figure}
 \centering
\psfrag{Phi_2}{$\varPhi_2$}
\includegraphics[totalheight=.2\textheight]{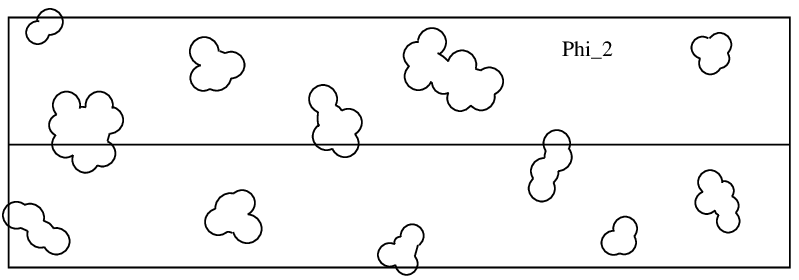}
\caption{The set $\varPhi
   _2$.}\label{F:7}
\end{figure}

Next, we introduce an isoenergetic surface \footnote{``surface'' is
a traditional term. In our case, it is a curve.} $S_0(\lambda)$ of
the free operator $H_0^{(1)}$. A point $t\in K_1$ belongs to
$S_0(\lambda)$ if and only if $H_0^{(1)}(t)$ has an eigenvalue equal
to $\lambda$, i.e., there exists  $j \in \Z^2$ such that
$p_j^{2}(t)=\lambda$. This surface can be obtained as follows: the
circle of radius $k=\sqrt \lambda$ centered at the origin is divided
into pieces by the dual lattice $\{\vec p_q(0)\}_{q \in \Z^2}$, and
then all pieces are translated in a parallel manner into the cell
$K_1$ of the dual lattice. We also can  get $S_0(\lambda)$ by
drawing sufficiently many circles of radii $k$ centered at the dual
lattice $\{\vec p_q(0)\}_{q \in \Z^2}$ and by looking at the figure
in the cell $K_1$. As the result of either of these two procedures,
we obtain a circle of radius $k$ ``packed into the bag $K_1$,'' as
is shown in the Fig.~\ref{F:3}. Note that each piece of $S_0(\lambda
)$ can be described by an equation $p_j^{2}(t)=\lambda$ for a fixed
$j$. If $t\in S_0(\lambda )$, then $j$ can be uniquely defined from
the last equation, unless $t$ is a  point of  self-intersection of
the isoenergetic surface. A point $t$ is a self-intersection of
$S_0(\lambda )$ if and only if $p_q^{2}(t)=p_j^{2}(t)=\lambda $ for
at least one pair of indices $q,\ j$, $q\neq j$.

Note that any vector $\vec \varkappa $ in $\R ^2$ can be uniquely
represented in the form $\vec \varkappa =\vec{p}_j(t)$, where $j\in
\Z ^2$ and $t\in K_1$. Let $\mathcal K_1$ be the parallel shift into
$K_1$: $$ {\mathcal K_1}:\R ^2\to K_1,\quad {\mathcal
K_1}\vec{p}_j(t)=t. $$ Suppose $\Omega \subset \R ^2$. In order to
obtain ${\mathcal K}_1\Omega$, it is necessary to partition $\Omega
$ by the lattice with nodes at the points $\vec{p}_q(0)$, $q\in \Z
^2$ and to shift all parts in a parallel manner into a single cell.
It is obvious that $ \left|{\mathcal K_1}\Omega \right|\leq
\left|\Omega \right|$ for any $\Omega $. If $\Omega $ is a smooth
curve, then %%%%%\begin{equation}\label{2.51*}
$ L({\mathcal K_1}\Omega )\leq L(\Omega).$ For any pair of sets
$\Omega _1$ and $\Omega _2$, ${\mathcal K_1}\bigl(\Omega _1\cup
\Omega _2\bigr)={\mathcal K_1}\Omega _1\cup{\mathcal K_1}\Omega _2$.
Obviously, ${\mathcal K_1}S_k=S_0(\lambda )$ and $L
\bigl(S_0(\lambda )\bigr)=L (S_k )=2\pi k$, $k=\sqrt \lambda $,
$S_k$ being the circle of radius $k$ centered at the origin.

The operator $H^{(1)}(t)$, $t\in K_1$, has the following matrix
representation in the basis of plane waves $\Psi _j^0(t,x),\ j \in
\Z^2$:
 \bdm
 H^{(1)}(t)_{mq}=p_m^{2}(t)\delta _{mq}+w_{m-q},\quad   m,q \in \Z^2.
 \edm
Here and below, $\delta_{mq}$ is the Kronecker symbol, $w_{m-q}$ are
Fourier coefficients of $W_1$: $$w_q=\frac{1}{|Q_1|}\int
_{Q_1}W_1(x)\exp -i\langle \vec p_q,x \rangle dx,\ \ \ \overline{
w}_q=w_{-q}. $$ Using assumptions on potentials $V_r$, we easily
obtain: \begin{equation}\label{w_q} w_0=0,\ \ \ \ |w_q|\leq
\|W_1\|,\ \ \ w_q=0,\ \mbox{when } p_q>R_0. \end{equation}
 The matrix
$H^{(1)}(t)_{mq}$ describes an operator in $l_2(\Z^2)$ unitary
equivalent to $H^{(1)}(t)$ in $L_2(Q_1)$. From now on,  we denote
the operator in $l_2(\Z^2)$ also by  $H^{(1)}(t)$. Since the
canonical basis in $l_2(\Z^2)$ does not depend on $t$ and all
dependence on $t$ is in the matrix, the matrix $H^{(1)}(t)_{mq}$
and, hence, the operator $H^{(1)}(t):l_2(\Z^2)\to l_2(\Z^2)$ can be
analytically extended in $t$ from $K_1$ to $\C^2$. From now on, when
we refer to $H^{(1)}(t)$ for $t\in \C^2$, we mean the operator in
$l_2(\Z^2)$.

%%%%%%%%%%%%%%%%%%%%%%%%%%%%%%%%%%%%%%%%%%%%%%%%%%%%%%%%%%%%%%%%%%%%%%%%%%%%%%%%%%%%%%%%%%
\subsection{Perturbation theory} We formulate the main results of
the perturbation theory considering $H^{(1)}(t)$ as a perturbed
operator of the free operator $H^{(1)}_0$, i.e., we construct the
perturbation series for eigenvalues and spectral projections for $t$
in a small neighborhood of a non-resonant set $\chi_1$.
 \begin{lem}[Geometric Lemma]\label{Lem:Geometric-1}
 For arbitrarily small positive $\delta <s_1$ and $\beta:\ 4s_1< 2\beta \leq  1-15s_1-8\delta$
  and sufficiently large $\lambda$, $\lambda >\lambda_0(\beta , s_1,\delta )$, there exists a non-resonant set
  $\chi_1(\lambda, \beta , s_1, \delta)\subset S_0(\lambda)$
  satisfying

  \itemthm For any $t \in \chi_1(\lambda, \beta ,s_1, \delta)$, there exists a unique $j \in \Z^2$ such that
  $p_j(t)=k$,
  $k:=\sqrt{\lambda }$. The following inequalities hold:
       \beq\label{InE:Geometric-1-1}
       4\min _{\substack{i, i+q\neq j\\ 0< p_q <k^{s_1}}}
       |p_i^{2}(t)-p_j^{2}(t)||p_{i+q}^{2}(t)-p_j^{2}(t)|
       >k^{2\beta},
       \eeq
       \beq\label{InE:Geometric-1-2}
       2\min _{0< p_q <k^{s_1}}
       |p_{j+q}^{2}(t)-p_j^{2}(t)|
       >k^{1-3s_1-\delta}.
       \eeq

  \itemthm For any $t$ in the complex $(k^{2\beta-2-s_1-2\delta })$-neighborhood of the non-resonance set in $\C^2$, there exists a unique $j\in \Z ^2$ such that
       \beq\label{InE:Geometric-1-nbh}
       |p_j^2(t)-k^2| < 3k^{2\beta-1-s_1-2\delta }.
       \eeq
      Estimates \eqref{InE:Geometric-1-1} and \eqref{InE:Geometric-1-2} hold.

  \itemthm The non-resonance set $\chi_1(\lambda ,\beta , s_1, \delta)$ has an asymptotically full measure on $S_0(\lambda)$ in the sense of
       \beq\label{E:Geometric-1-full}
       \frac{L\left(S_0(\lambda )\setminus \chi _1(\lambda,\beta ,s_1, \delta)\right)}
       {L\left(S_0(\lambda)\right)}\underset{\lambda\to \infty }{=}O(k^{-\delta}).
       \eeq
 \end{lem}

 \begin{cor}\label{cor:geometric-1}
 If $t$ belongs to the complex $(k^{2\beta-2-s_1-2\delta })$-neighborhood of the non-resonant set
 $\chi _1(\lambda , \beta, s_1,\delta )$ in $\C^2$, then for any $z\in \C$ lying on the circle
  \beq\label{E:circle-1}
  C_1=\{ z: |z-k^{2}| =k^{2\beta -1-s_1-\delta }\}
  \eeq
 and for any $i$ and $q$ in $ \Z^2$ with $0<p_q<k^{s_1}$
 \beq \label{c-1}
 2|p_i^{2}(t)-z| \geq k^{2\beta-1-s_1-\delta }
 \eeq
  \beq\label{InE:circle-1}
  16| p_i^{2}(t)-z| | p_{i+q}^{2}(t)-z| > k^{2\beta -4s_1-2\delta}
  \eeq
 \end{cor}

The set $\chi _1(\lambda , \beta, s_1, \delta
 )$ is defined below. The lemma and the corollary are proven in Section
\ref{sec:geometric-1}. Section \ref{5.1.1} is introductory.
Properties (i), (ii)  and Corollary \ref{cor:geometric-1} are proven
in Section \ref{S:7.1.1} (Corollary \ref{cor:t}). Section
\ref{S:7.1.2} is devoted to the proof of (iii).
 Here we just note  the following. An analog of the
lemma and corollary is proven earlier in \cite{19}. However, in
 this work we need more detailed description of $\chi _1(\lambda , \beta, s_1,\delta
 )$ than that in \cite{19}. Moreover, we need a complex analog of  $\chi _1(\lambda , \beta, s_1, \delta
 )$.

 We define  $\chi _1(\lambda , \beta, s_1, \delta
 )$ by the formula:
 \begin{equation}
  \chi _1(\lambda , \beta, s_1,\delta
 )=\mathcalK_1\mathcalD_0(\lambda)_{nonres}, \label{chi _1}
 \end{equation}
 where $\mathcalD_0(\lambda)_{nonres}$ is  a subset of the sphere
 $S_k$,
 \begin{equation}
 \mathcalD_0(\lambda)_{nonres}=\{k\vec \nu, \ \vec \nu =(\cos
 \varphi , \sin \varphi ),\ \varphi \in \Theta _1\}, \label{D_0}
 \end{equation}
 the set $\Theta _1$ being defined as the real part of a set $\varPhi _1\subset \C$:
 \begin{equation}
 \Theta _1=\varPhi _1\cap [0,2\pi ),\label{Theta_1}
 \end{equation}
 \begin{equation}
 \varPhi _1=\varPhi _0\setminus \mathcalO^{(1)}, \label{Phi_1}
 \end{equation}
 where $\varPhi _0$ is the complex $k^{-\delta }$ neighborhood of
 $[0,2\pi)$,
 \begin{equation}
 \varPhi _0=\{\varphi \in \C: \Re \varphi \in [0,2\pi), |\Im \varphi
 |<k^{-\delta }\}, \label{Phi_0}
 \end{equation}
 $\mathcalO^{(1)}$ is the union of discs $\mathcalO^{\pm}_m$:
 \begin{equation}
\mathcalO^{(1)}=\cup _{m\in \Z^2,\ 0<p_m<4k,\ \pm}\
\mathcalO^{\pm}_m. \label{O^1} \end{equation} Discs
$\mathcalO^{\pm}_m$ are centered at the points $\varphi _m^{\pm}$,
which are solutions in $\C$ (two for each $m$) of the equations:
\begin{equation} |\vec k (\varphi )+\vec p_m|^2_*=k^2, \ \ \ \vec k
(\varphi ) =k(\cos \varphi ,\sin \varphi ), \label{eqm}
\end{equation} here and below $|\vec x|_*^2=x_1^2+x_2^2$ for any
$\vec x=(x_1,x_2)$, $x_1,x_2\in \C$. The radii $r_m$ of the discs
are chosen to ensure the properties of the set $\chi _1(\lambda,
\beta, s_1,\delta
 )$ to be valid. They are given by the formula (\ref{r_m}).
\label{pageF:7}

 Let $E_j(t)$ be the spectral projection of the free operator corresponding to the eigenvalue $p_j^{2}(t)$, $(E_j)_{rm}=\delta
_{jr}\delta _{jm} $. In the $(k^{2\beta-2-s_1-2\delta
})$-neighborhood of $\chi _1(\lambda, \beta ,s_1,\delta )$, we
define functions $g_r^{(1)}(k,t)$ and operator-valued functions
$G_r^{(1)}(k,t)$, $r=1, 2, \ldots$ as follows.
 \begin{gather}
 g_r^{(1)}(k,t)=\frac{(-1)^r}{2\pi ir}\mbox{Tr}
 \oint_{C_1}((H_0^{(1)}(t)-z)^{-1}W_1)^rdz,\label{E:g_r}\\
 G_r^{(1)}(k,t)=\frac{(-1)^{r+1}}{2\pi i}
 \oint_{C_1}((H_0^{(1)}(t)-z)^{-1}W_1)^r (H_0^{(1)}(t)-z)^{-1}dz. \label{E:G_r}
 \end{gather}
 To find $g_r^{(1)}(k,t)$ and $G_r^{(1)}(k,t)$, it is
necessary to compute the residues of a rational function of a simple
structure, whose numerator does not depend on $z$, while the
denominator is a product of factors of the type $(p_i^{2}(t)-z)$.
For all $t$ in the non-resonance set  the integrand has a single
pole within $C_1$ at the point $z=k^{2}=p_j^{2}(t)$. By computing
the residue at this point, we obtain explicit expressions for
$g_r^{(1)}(k,t)$ and $G_r^{(1)}(k,t)$. For example,
$g_1^{(1)}(k,t)=0$, \begin{equation}
 \begin{aligned}
    g_2^{(1)}(k,t)&=\sum _{q\in \Z^2,q\neq 0}| w_q| ^2(p_j^{2}(t)-
        p_{j+q}^{2}(t))^{-1} \\
    &=-\sum _{q\in \Z^2,q\neq 0}\frac{| w_q| ^2p_q^2(0)}{(p_j^{2}(t)-
    p_{j+q}^{2}(t))(p_j^{2}(t)-p_{j-q}^{2}(t))},\label{2.14}
 \end{aligned}
\end{equation} \begin{equation}\label{2.15}
G_1^{(1)}(k,t)_{rm}=\frac{w_{j-m}}{p_j^{2}(t)-p_m^{2}(t)}\delta
_{rj}+ \frac{w_{r-j}}{p_j^{2}(t)-p_r^{2}(t)}\delta _{mj}, \quad
G_1^{(1)}(k,t)_{jj}=0. \end{equation}  For technical reasons, we
introduce parameter $\alpha $ in front of the potential $W_1$,
$H_{\alpha }^{(1)}=(-\Delta )^l +\alpha W_1$, $0\leq \alpha\leq 1$.
We denote the operator $H_\alpha ^{(1)}$ with $\alpha =1$ simply by
$H^{(1)}$.

 \begin{thm}\label{Thm:main-1}
 Suppose $t$ belongs to the $(k^{2\beta-2-s_1-2\delta })$-neighborhood
 in $K_1$ of the non-resonant set $\chi _1(\lambda ,\beta ,s_1,\delta )$,
 $0<\delta <s_1$, $8s_1+6\delta <2\beta <1-15s_1-8\delta $. Then for sufficiently large $k $,
 $k >k _0(\|W_1\|,s_1,\beta , \delta )$ and for all $\alpha $, $-1\leq \alpha \leq 1$, there exists a single
 eigenvalue of the operator $H_{\alpha }^{(1)}(t)$ in the interval
 $\varepsilon_1 (k,\delta ):= (k^{2}-k^{2\beta-1-s_1-\delta }, k^{2}+k^{2\beta-1-s_1-\delta })$. It is given by the series
  \beq\label{E:lambda_j^1}
  \lambda_j^{(1)} (\alpha  ,t)=p_j^{2}(t)+\sum _{r=2}^{\infty}\alpha ^rg_r^{(1)}(k,t),
  \eeq
 converging absolutely, where the
 index $j$ is  defined as in Lemma \ref{Lem:Geometric-1}. The spectral projection corresponding to
 $\lambda_j^{(1)} (\alpha  ,t)$ is given by the series
  \beq\label{E:E_j^1}
  E_j^{(1)} (\alpha  ,t)=E_j+\sum _{r=1}^{\infty }\alpha^rG_r^{(1)}(k,t),
  \eeq
 which converges in the trace class $\mathbf{S_1}$.

 Moreover, $g_r^{(1)}(k,t)$ and  $G_r^{(1)}(k,t)$ satisfy the
 estimates:
  \beq\label{InE:g_r}
  | g_r^{(1)}(k,t) |<k^{2\beta-1-s_1-\delta }\bigl(c\|W_1\|k^{-\beta+4s_1 +2\delta
  }\bigr)^r, \eeq
  \beq \label{InE:G_r}
  \|G_r^{(1)}(k,t)\| _1< r\bigl(c k^{-\beta+4s_1+2\delta }\bigr)^r,
  \eeq
    $\|\cdot \|_1$ being the norm in the trace
    class $\mathbf{S_1}$.
   In addition,
  \beq\label{InE:g_2-g_3}
  | g_2^{(1)}(k,t) |<c\|W_1\|^2k^{-2+10s_1+4\delta},\quad
  |g_3^{(1)}(k,t)| < c\|W_1\|^3 k^{-2+10s_1+4\delta},
  \eeq
  \beq\label{InE:g_2>0}
   g_2^{(1)}(k,t) >0,\mbox {  when  } \|W_1\|\neq 0,
  \eeq

  Operators $G_r^{(1)}(k,t)$ are finite dimensional:
  \begin{equation} G_r^{(1)}(k,t)_{il}=0\ \ \mbox{when }|i-j|>rR_0\mbox{ or
  }|l-j|>rR_0. \label{finitedim} \end{equation}
 The series (\ref{E:lambda_j^1}) and (\ref{E:E_j^1}) converge uniformly with
 respect to $\alpha $ in the complex disk $| \alpha | \leq 1$. \end{thm}

 \begin{cor}\label{Cor:main-1}
 The perturbed eigenvalue and its spectral projection satisfy
  \begin{equation}
  | \lambda_j^{(1)} (\alpha ,t)-p_j^{2}(t)|
   \leq \alpha^2 C(\|W_1\|) k^{-1-2\beta +15s_1+11\delta } \label{InE:lambda_j^1},\eeq
  \beq \|E_j^{(1)} (\alpha ,t)-E_j\|_1\leq c | \alpha|  \|W_1\| k^{-\beta +4s_1 +3\delta }.\label{InE:E_j^1}
  \eeq

 \end{cor}

% \begin{remark} \label{R:May10}
% The theorem states that $\lambda_j^{(1)}(\alpha ,t)$ is the unique
% eigenvalue in the interval $\varepsilon_1 (k,\delta )$. This means
% that $|\lambda_j^{(1)} (\alpha,t)-k^{2}|<k^{-4s_1-\delta }$.
%  Formula \eqref{2.20} provides a stronger estimate on the location
% of $\lambda_j^{(1)} (\alpha,t)$, the right-hand side of
% \eqref{2.20} being smaller than  of $\varepsilon _1$.
% \end{remark}

 \bpf
 The proof of the theorem is based on expanding the resolvent\break
 $(H_{\alpha}^{(1)}(t)-z)^{-1}$ in a perturbation series for $z$
 belonging to the contour $C_1$ about the unperturbed eigenvalue
 $p_j^{2}(t)$. It is completely analogous to the proofs of Theorems 2.1 and 3.1 in \cite{19}. Indeed, it is obvious that
  \begin{align}
  &(H_{\alpha }^{(1)}(t)-z)^{-1}=(H_0^{(1)}(t)-z)^{-1/2}(I-\alpha A_1)^{-1}(H_0^{(1)}(t)-z)^{-1/2},
  \label{E:resolvent-1}\\
  &A_1 = A_1(z,t) := -(H_0^{(1)}(t)-z)^{-1/2}W_1(H_0^{(1)}(t)-z)^{-1/2}. \label{E:A_1}
  \end{align}
It follows from Corollary \ref{cor:geometric-1}, estimates
(\ref{w_q}) and $R_0<k^{s_1}$,   that
  \begin{gather}
  \|(H_0^{(1)}(t)-z)^{-1}\| < 2k^{-2\beta +1+s_1+\delta},\ \ z\in C_1,\quad \label{InE:resolvent-0}\\
  \|A_1\| <16\|W_1\| k^{-\beta+4s_1+2\delta } \label{InE:A_1}.
  \end{gather}
Thus, $\|A_1\| <1$ for sufficiently large $k$,
 $k>k_0(\|W_1\|,\beta , s_1, \delta )$. Expanding
 $(I-\alpha A_1)^{-1}$  in powers of $\alpha A_1$, we obtain
  \beq\label{E:perturbation}
  (H_{\alpha }^{(1)}(t)-z)^{-1}-(H_0^{(1)}(t)-z)^{-1}=\sum_{r=1}^{\infty }\alpha ^r (H_0^{(1)}(t)-z)^{-1/2}A_1^r(H_0^{(1)}(t)-z)^{-1/2}.
  \eeq
 Note that $(H_0^{(1)}(t)-z)^{-1}\in \mathbf{S_2}$. Taking into account
 estimates~\eqref{InE:A_1}, we see that the series~\eqref{E:perturbation}
 converges in the class $\mathbf{S_1}$ uniformly with respect to
 $\alpha $ in the whole complex disk $|\alpha | \leq 1$. For real $\alpha$-s we substitute the series
 into the following formula for a spectral projection
 \begin{equation}E_j^{(1)} (\alpha ,t)=-\frac{1}{2\pi i}\oint _{C_1}(H_{\alpha
 }^{(1)}(t)-z)^{-1}dz . \label{2.2.23} \end{equation}
 Integrating termwise, we arrive at \eqref{E:E_j^1}.
Next, we prove estimate~(\ref{InE:G_r}) for $G_r^{(1)}(k,t)$.
Indeed, it is easy to see that \begin{equation} G_r^{(1)}(k,t)=
-\frac{1}{2\pi i}\oint
_{C_1}(H_0^{(1)}(t)-z)^{-1/2}A_1^r(H_0^{(1)}(t)-z)^{-1/2}dz.
\label{2.2.25} \end{equation} We introduce the operator
$A_0=(I-E_j)A_1(I-E_j)$. It is obvious that $\| A_0\| \leq \|
A_1\|$. In addition  \begin{equation} \oint
_{C_1}(H_0^{(1)}(t)-z)^{-1/2}A_0^r(H_0^{(1)}(t)-z)^{-1/2}dz=0,
\label{2.2.26} \end{equation}  since the integrand is holomorphic
inside the circle. Thus,  \begin{equation} G_r^{(1)}(k,t)=
-\frac{1}{2\pi i}\oint
_{C_1}(H_0^{(1)}(t)-z)^{-1/2}(A_1^r-A_0^r)(H_0^{(1)}(t)-z)^{-1/2}dz.
\label{2.2.27} \end{equation} Since
$A_1-A_0=E_jA_1(I-E_j)+(I-E_j)A_1E_j$, $E_jA_1E_j=0$ and $E_j$ is a
one- dimensional projection, we get \begin{equation} \| A_1-A_0\|
_1\leq \| A_1\| . \label{2.2.29} \end{equation} Since
\begin{equation} \| A^r_1-A_0^r\| _1\leq r\| A_1-A_0\| _1\| A_1\|
^{r-1}, \label{2.2.28} \end{equation} we obtain from
relation~(\ref{InE:A_1}) that  \begin{equation} \| A_1^r-A_0^r\|
_1\leq r\left(c\|W_1\|k^{-\beta+4s_1+2\delta }\right)^r
.\label{2.2.30} \end{equation}  Noting that the length of $C_1$ is
equal to $2\pi k^{2\beta -1-s_1-\delta }$, we obtain from
formula~(\ref{2.2.27}):  \begin{equation} \| G^{(1)}_r(k,t)\| _1\leq
k^{2\beta -1-s_1-\delta }\| (H_0^{(1)}(t)-z)^{-1/2}\| ^2\|
A_1^r-A_0^r\|_1 \label{2.2.31} \end{equation}  Using
inequalities~(\ref{InE:resolvent-0}) and (\ref{2.2.30}), we
get~(\ref{InE:G_r}). Convergence of the series in the complex disk
$|\alpha |<1$ easily follows.  Note that $(A_{1})_{il}=0$ if
$|i-l|>R_0$. Now it is not difficult to check \eqref{finitedim}, for
details see \cite{19}.

We show now that operator $H_{\alpha }^{(1)}$, $-1<\alpha <1$,  has
a single eigenvalue $(n_{\alpha }=1)$  in the interval $\varepsilon
(k, \delta )$. It is easy to see that $n_{\alpha }$ is determined
from the formula $ n_{\alpha }-n_0=\mbox{Tr}(E_j^{(1)}-E_j). $
Considering formula~(\ref{E:E_j^1}), we obtain  $$ \mid n_{\alpha
}-n_0\mid <\sum _{r=1}^{\infty }\| G_r^{(1)}(k,t)\| _1=o(1).$$ Since
$n_{\alpha }$ and $n_0$ are integers, and $n_0=1$ by the hypothesis
of the theorem, the operator $H_{\alpha }^{(1)}(t)$ for all $\alpha
$, $-1\leq \alpha \leq 1$, has a single eigenvalue in $\varepsilon
(k,\delta )$.

Further, we use the well-known formula:  \begin{equation}
\frac{\partial \lambda ^{(1)}(\alpha ,t)}{\partial \alpha }
=-\frac{1}{2\pi i}\mbox{Tr}\oint _{C_1}W_1(H_{\alpha
}^{(1)}(t)-z)^{-1}dz \label{2.2.34} \end{equation} Using
formula~(\ref{E:perturbation}) and considering that
$\mbox{Tr}(W_1(H_0^{(1)}(t)-z)^{-1})=0$, we obtain: \begin{equation}
\frac{\partial \lambda ^{(1)}(\alpha ,t)}{\partial \alpha }= \sum
_{r=2}^{\infty }r\alpha ^{r-1}g_r^{(1)}(k,t). \label{2.2.34a}
\end{equation} Integrating the last relation with respect to $\alpha
$ and noting that $\lambda ^{(1)}(0,t)=p_j^{2}(t)$, we get
formula~(\ref{E:lambda_j^1}). To prove estimate~(\ref{InE:g_r}) we
note that $g_r^{(1)}=-(2\pi ir)^{-1}\mbox{Tr}\oint _{C_1}A_1^rdz$,
and, therefore, \begin{equation} \mid g_r ^{(1)}(k,t) \mid \leq
r^{-1}k^{2\beta -1-s_1-\delta }\| A^r_1-A_0^r\|_1, \label{2.2.31a}
\end{equation}
 Using estimate~(\ref{2.2.30}), we obtain
inequality~(\ref{InE:g_r}). Estimates (\ref{InE:g_2-g_3}) are better
than those provided for $g_2$, $g_3$ by the general estimate
\eqref{InE:g_r}.  We obtain estimate (\ref{InE:g_2-g_3}) for
$g_2(k,t)$ using the second part of
 formula (\ref{2.14}) and the estimate (\ref{InE:Geometric-1-2}).
 Estimate \eqref{InE:g_2>0} easily follows from the same formulas.
 The estimate for $g_3(k,t)$ can be obtained in the analogous way. For details, see \cite{19}, Theorem 3.2.
 \epf
The series \eqref{E:lambda_j^1} and \eqref{E:E_j^1} can be extended
as holomorphic functions of $t$ in a complex neighborhood of $\chi
_1(k,\beta ,s_1,\delta)$. They can be differentiated  with respect
to $t$ any number of times with retaining their asymptotic
character. Before  presenting exact  statements, we introduce the
following notations:
 \begin{gather*}
 T(m) :=\frac{\partial ^{|m|}}{\partial t_1^{m_1}
 \partial t_2^{m_2}}, \quad T(0)f:=f, \\
 m=(m_1,m_2),\quad |m|:=m_1+m_2,\quad m !:=m_1 ! m_2 !.
 \end{gather*}

 \begin{lem}\label{Lem:derivative-1}
 Coefficients $g_r^{(1)}(k,t),$ and  $G_r^{(1)}(k,t)$ can be
 continued as holomorphic functions of two variables from the real
 $(k^{2\beta-2-s_1-2\delta })$-neighborhood of the non-resonance set
 $\chi_1(k,\beta, s_1, \delta)$ to its complex $(k^{2\beta-2-s_1-2\delta })$-neighborhood.
 In this complex neighborhood the following estimates hold.
  \begin{gather}
  |T(m)g_r^{(1)}(k,t)| <m!k^{2\beta-1-s_1-\delta }\bigl(c\|W_1\| k^{-\beta+4s_1+2\delta}\bigr)^r
   k^{|m|(2-2\beta +s_1+2\delta)}, \label{InE:g_r^1-derivative}\\
  \| T(m)G_r^{(1)}(k,t)\|<m!r\bigl(c\|W_1\| k^{-\beta+4s_1+2\delta}\bigr)^r
   k^{| m|(2-2\beta +s_1+2\delta)}. \label{InE:G_r^1-derivative}
  \end{gather}
  The coefficients $g_r(k,t)$, $r<k^{s_1}R_0^{-1}$ can be  continued to the complex $(k^{-3s_1-2\delta
  })$-neighborhood of
  $\chi_1(k,\beta, s_1, \delta)$ . They obey the estimates:
  \beq\label{InE:g_2-der}
  | T(m) g_2^{(1)}(k,t) |<cm!\|W_1\|^2k^{-2+10s_1+4\delta}k^{| m|(3s_1+2\delta)},
  \eeq
  \beq\label{InE:g_3-der}
  |T(m) g_3^{(1)}(k,t)| < cm!\|W_1\|^3 k^{-2+10s_1+4\delta}k^{|
  m|(3s_1+2\delta)}.
  \eeq
  \beq\label{InE:g_r-der}
  |T(m) g_r^{(1)}(k,t)| < m!k^{2\beta-1-s_1-\delta }\bigl(c\|W_1\|
  k^{-\beta+4s_1+2\delta}\bigr)^r
   k^{|
  m|(3s_1+2\delta)},\ \ 4\leq r<k^{s_1}R_0^{-1}.
  \eeq
 \end{lem}
 \begin{proof}
  Coefficients $g_r^{(1)}(k,t),$ and  $G_r^{(1)}(k,t)$ can be extended analytically
  to the $(k^{2\beta-2-s_1-2\delta })$-neighborhood of $\chi_1(k,\beta,s_1,\delta)$,
  since  estimates \eqref{c-1}, \eqref{InE:circle-1} are stable in this neighborhood.
  The coefficients $g_r(k,t)$, $r<k^{s_1}R_0^{-1}$, can be extended to a bigger
  neighborhood, since only \eqref{InE:Geometric-1-2} is required to estimate them.
 The estimates for the derivatives are obtained by means of Cauchy integrals.
  For details, see Theorem 3.3 and Corollary 3.3 in
  \cite{19}.\end{proof}
  Lemma \ref{Lem:derivative-1} implies the following theorem.

 \begin{thm}\label{Thm:derivative-1}
 The series \eqref{E:lambda_j^1} and \eqref{E:E_j^1} can be continued as
 holomorphic functions of two variables from the real
 $(k^{2\beta-2-s_1-2\delta })$-neighborhood of the non-resonance set
 $\chi_1(k,\beta,s_1,\delta)$ to its complex $(k^{2\beta-2-s_1-2\delta })$-neighborhood.
 The following estimates hold in the complex neighborhood:
  \begin{gather}
  | T(m)(\lambda_j^{(1)} (\alpha ,t)-p_j^{2}(t))|
   <m!C(W_1)\alpha^2
   k^{-1-2\beta +15s_1+11\delta +|m|(2-2\beta +s_1+2\delta)},\label{InE:lambda_j^1-derivative}\\
  \| T(m)(E_j^{(1)} (\alpha ,t)-E_j)\|
   <cm!\|W_1\|\alpha k^{-\beta +4s_1 +3\delta +|m| (2-2\beta +s_1+2\delta))}.\label{InE:E_j^1-derivative}
  \end{gather}
 There are stronger estimates for $|m|=1,2$:
\begin{equation}
  \left|\nabla \lambda_j^{(1)} (\alpha ,t)-
   2\vec p_j(t)\right|<C(W_1)\alpha ^2k^{-1 -2\beta +18s_1+13\delta }, \label{E:grad-lambda_j^1}
   \end{equation}
 \begin{equation} \left|T(m)\left(\lambda_j^{(1)} (\alpha ,t)-p_j^2\right)\right|<C(W_1)\alpha ^2k^{-1 -2\beta +21s_1+15\delta }, \mbox{ if } |m|=2.
  \label{InE:second-derivative-lambda_j^1}
  \end{equation}
 \end{thm}
 %%%%%%%%%O\left(k^{-1-2\beta+21s_1+15\delta }\right)

The next lemma is used in the second step of approximation, where
the operator $H^{(1)}(t)$ plays a role of the initial
(unperturbed) operator.

 \begin{lem}\label{Lem:resolvent-1}
 For any $z$ on the circle $C_1$ given in \eqref{E:circle-1} and $t$ in
 the  $(k^{2\beta-2-s_1-2\delta })$-neighborhood of $\chi_1(k,\delta)$ in $K_1$,
  \beq\label{InE:resolvent-1}
  \|(H^{(1)}(t)-z)^{-1}\| \leq 2k^{-2\beta +1+s_1+\delta }.
  \eeq
 \end{lem}
 \bpf
 The estimate follows from (\ref{E:resolvent-1}),
 (\ref{InE:resolvent-0}) and (\ref{InE:A_1}).
   \epf

\subsection{Non-resonant part of the Isoenergetic Set of $H^{(1)}$.}
In this subsection we choose $\beta $ to have the biggest
   possible value $\beta =\beta _0$, $2\beta _0=1-15s_1-9\delta $. Let $S_1(\lambda )$  be the isoenergetic surface
of the perturbed operator $H_{\alpha}^{(1)}$, i.e.,
\begin{equation}\label{2.65.1} S_1(\lambda )=\{t \in K_1 : \exists n
\in \N\ \text{s.t.}\ \lambda_n^{(1)}(\alpha,t)=\lambda  \},
\end{equation} where $\{\lambda_n^{(1)}(\alpha,t)\}_{n=1}^{\infty }$
is the complete set of eigenvalues of $H_{\alpha}^{(1)}(t)$. We
construct a ``non-resonance'' subset  $\chi _1^*(\lambda )$ of
$S_1(\lambda)$, which corresponds to the non-resonance eigenvalues
$\lambda_{j}^{(1)}(\alpha , t)$ given by the perturbation series. By
Lemma~\ref{Lem:Geometric-1},  for every $t$ belonging to the
non-resonant set $\chi_1(\lambda,\beta ,s_1,\delta )$, there is a
single $j\in\Z^2$ such that $p_j(t)=k$, $k=\sqrt \lambda $. This
means that formula (\ref{chi _1}) establishes a one-to-one
correspondence between $\chi_1(\lambda,\beta ,s_1,\delta )$ and
$\mathcal{D}_0(\lambda )_{nonres}$. Let
$\vec{\varkappa}\in\mathcal{D}_0(\lambda )_{nonres}$ and $j\in
\Z^2$, $t\in \chi _1(\lambda,\beta ,s_1,\delta)$ are defined by the
relation $\vec{\varkappa }=\vec p_j(t)$.\footnote{Usually the vector
$\vec p_j(t)$ is denoted by $\vec k$, the corresponding plane wave
being $e^{\langle \vec k,x \rangle }$. We use the  less common
notation $\vec{\varkappa }$, since we already have other $k$'s in
the text.}
 According to Theorem
\ref{Thm:main-1}, for sufficiently large $k$, there exists an
eigenvalue of the operator $H^{(1)}_{\alpha}(t)$, $t={\mathcal
K_1}\vec\varkappa$, $0\leq\alpha\leq 1$, given by
\eqref{E:lambda_j^1}. It is convenient here to denote
$\lambda^{(1)}_j(\alpha,t)$ by
$\lambda^{(1)}(\alpha,\vec{\varkappa})$; we can do this since there
is a one-to-one correspondence between $\vec{\varkappa}$ and the
pair $(t,j)$. We rewrite \eqref{E:lambda_j^1} in the form
\begin{equation}\label{2.66}
\lambda^{(1)}(\alpha,\vec{\varkappa})=\varkappa^{2}+f_1(\alpha,\vec{\varkappa}),
    \quad  \varkappa =|\vec \varkappa |, \;\;\  f_1(\alpha,\vec{\varkappa})=\sum
_{r=2}^{\infty} \alpha^{r}g_r^{(1)}(\vec{\varkappa}), \end{equation}
    $g_r^{(1)}(\vec{\varkappa})$ being defined by \eqref{E:g_r} with
$j$ and $t$ such that $\vec{p}_j(t)=\vec{\varkappa}$. By Theorem
\ref{Thm:derivative-1}, $f_1(\alpha,\vec{\varkappa})$ satisfies the
following  estimates when $2\beta =1-15s_1-9\delta $:
\begin{gather}\label{2.67} |f_1(\alpha,\vec \varkappa)|\leq
\alpha^2C(W_1)k^{-2+30s_1+20\delta}, \\ \label{2.67a}
|T(m)f_1(\alpha,\vec \varkappa)|\leq
\alpha^2C(W_1)k^{-2+30s_1+20\delta +|m|(1+16s_1+11\delta)}.
\end{gather} Estimates \eqref{E:grad-lambda_j^1},
\eqref{InE:second-derivative-lambda_j^1} yield: \begin{gather}
\nabla f_1 (\alpha ,\vec{\varkappa})
   =O\left(k^{-2+ 33s_1+22\delta
   }\right),
    \label{E:grad-f_1} \\
  \nabla \lambda^{(1)} (\alpha ,\vec{\varkappa})
   =2\vec{\varkappa}+O\left(k^{-2+ 33s_1+22\delta
   }\right),
    \label{E:grad-lambda_j^1*} \\
  T(m)f_1 (\alpha ,\vec{\varkappa})=O\left(k^{-2+36s_1+24\delta
  }\right),\quad T(m)\lambda^{(1)} (\alpha ,\vec{\varkappa})=O(1),\
  \
  \mbox{ if } |m|=2. \label{InE:second-derivative-lambda_j^1*}
  \end{gather}
By Theorem \ref{Thm:derivative-1}, the series \eqref{2.66} converges
 in the
$(k^{-1-16s_1-11\delta})$-neighborhood of $\mathcal{D}_0(\lambda
)_{nonres}$, the estimate
\eqref{2.67}--\eqref{InE:second-derivative-lambda_j^1*} hold.

Let us recall that operator $H^{(1)}(t)$ is defined in $l^2$, i.e.,
just by the matrix depending on $t$. We define $H^{(1)}(\vec
\varkappa )$ by substituting $\vec \varkappa
 $ instead of $\vec p_j(t)$ into $H^{(1)}(t)$. For any $t\in \chi _1(k,\beta _0,s_1,\delta )$
  matrices $H^{(1)}(t)$ and $H^{(1)}(\vec
\varkappa )$ are the same, up to the shift of indices by $j$.
Further we consider  $H^{(1)}(\vec \varkappa )$ for
$\vec{\varkappa}= \varkappa \vec \nu $, when $\varphi $ is in the
complex $2k^{-2-16s_1-11\delta }$ neighborhood of $\varPhi _1$ and
$\varkappa $ is in the complex $2k^{-1-16s_1-11\delta }$
neighborhood of $k$.  Using  Lemma \ref{Lem:lowerbounds*},
 we easily obtain the estimates analogous to
 \eqref{InE:resolvent-0},
  \eqref{InE:A_1}, which
 provide the convergence of the series for the resolvent. Operator $H^{(1)}(\vec
\varkappa )$ is not self-adjoint for complex $\varphi $ and we do
not try to make  spectral analysis of it. However, we notice that
 the series  \eqref{2.66} converges when $\varphi $ is in the
complex $2k^{-2-16s_1-11\delta }$ neighborhood of $\varPhi _1$ and
$\varkappa $ is in the complex $2k^{-1-16s_1-11\delta }$
neighborhood of $k$. Lemma \ref{Lem:resolvent-1} admits the
following generalization.
\begin{lem}\label{Lem:resolvent-1'}
 For any $z$ on the circle $C_1$ given in \eqref{E:circle-1} and $\vec{\varkappa}= \varkappa \vec \nu $, $\varphi $
 being
in the complex $2k^{-2-16s_1-11\delta }$ neighborhood of $\varPhi
_1$, $\varkappa $ being in the complex $2k^{-1-16s_1-11\delta }$
neighborhood of $k$,
  \beq\label{InE:resolvent-1'}
  \|(H^{(1)}(\vec \varkappa )-z)^{-1}\| \leq 2k^{-2\beta +1+s_1+\delta }.
  \eeq
 \end{lem}

   Let $\mathcal{B}_1(\lambda)$ be a set of unit
vectors corresponding to $\mathcal{D}_0(\lambda)_{nonres}$: $$
\mathcal{B}_1(\lambda)=\{\vec{\nu} \in S_1 : k \vec{\nu} \in
 \mathcal{D}_0(\lambda)_{nonres}\}=\{\vec{\nu} \in S_1 : \vec{\nu} =(\cos \varphi
,\sin \varphi ),\ \varphi \in \Theta _1\}. $$ It is easy to see that
$\mathcal{B}_1(\lambda)$ is a unit circle with holes, centered at
the origin.
 Since formulas
\eqref{2.66}--\eqref{InE:second-derivative-lambda_j^1*} hold in the
$(k^{-1-16s_1-11\delta})$-neighborhood of $\mathcal{D}_0(\lambda
)_{nonres}$,  they hold for any $\varkappa\vec{\nu}$ such that
$\vec{\nu} \in \mathcal{B}_1(\lambda),\
|\varkappa-k|<k^{-1-16s_1-11\delta}$.
  We define $\mathcal{D}_1(\lambda )$ as the level set of
the function $\lambda^{(1)}(\alpha,\vec{\varkappa})$ in this
neighborhood: \begin{equation}\label{2.68} \mathcal{D}_1(\lambda
):=\{\vec{\varkappa }=\varkappa \vec{\nu}:
\vec{\nu}\in\mathcal{B}_1(\lambda), \
|\varkappa-k|<k^{-1-16s_1-11\delta},\
\lambda^{(1)}(\alpha,\vec{\varkappa})=\lambda \} . \end{equation} We
prove in Lemma \ref{L:2.13} that $\mathcal{D}_1(\lambda )$ is a
distorted circle with holes, which is close to  the circle of radius
$k$; see \ Fig.\ \ref{F:1}. First, we prove that the equation
$\lambda^{(1)}(\alpha,\vec{\varkappa})=\lambda $ is solvable with
respect to $\varkappa =|\vec\varkappa|$ for any $\vec\nu=\frac{\vec
\varkappa}{\varkappa} \in\mathcal{B}_1(\lambda)$.

\begin{lem}\label{L:2.12} For every
$\vec{\nu}\in\mathcal{B}_1(\lambda)$ and every $\alpha$,
$0\leq\alpha\leq 1$, and sufficiently large $\lambda $, there is a
unique $\varkappa _1 =\varkappa _1(\lambda ,\vec{\nu})$ in the
interval $$
I_1:=[k-k^{-1-16s_1-11\delta},k+k^{-1-16s_1-11\delta}],\quad
k^{2}=\lambda , $$ such that \begin{equation}\label{2.70}
\lambda^{(1)}(\alpha,\varkappa _1\vec{\nu})=\lambda . \end{equation}
Furthermore, $|\varkappa _1 - k| \leq C(W_1)k^{-3+30s_1+20\delta}$.
\end{lem} \bpf Formula \eqref{E:grad-lambda_j^1*} yields
\begin{equation}\frac{\partial \lambda^{(1)}(\alpha,\varkappa
)}{\partial \varkappa }=2\varkappa ,\label{pderiv}\end{equation}
when $|\varkappa-k|<k^{-1-16s_1-11\delta}$. Now the  lemma easily
follows from \eqref{2.67}. For details see Lemma 2.10 in
\cite{KL}.\epf

Let us introduce new notations. Let $\hat \varPhi _1$ be the
$(k^{-2-16s_1-11\delta})$-neighborhood of $\varPhi _1$. Note that
 \eqref{2.66}--\eqref{InE:second-derivative-lambda_j^1*} hold for $\varphi$ in $\hat \varPhi _1$
and even its $(k^{-2-16s_1-11\delta})$-neighborhood when $\vec
\varkappa =\varkappa \vec \nu $, $|\varkappa -k
|<k^{-1-16s_1-11\delta}$. Let
\begin{equation} \varkappa _1(\varphi):=\varkappa _1 (\lambda,
\vec{\nu}),\;\: \vec{\nu}= (\cos\varphi , \sin \varphi ),\;\:  \vec
\varkappa _1 (\varphi) =\varkappa _1(\varphi) \vec{\nu},\;\:
h_1(\varphi )=\varkappa _1(\varphi) -k, \label{honda1}
\end{equation} $\varkappa _1 (\lambda, \vec{\nu})$ being defined by
Lemma \ref{L:2.12}.

\begin{lem} \label{L:2.13} \begin{enumerate} \item For sufficiently
large $\lambda $, the set $\mathcal{D}_1(\lambda )$ is a distorted
circle with holes; it can be described by the formula
\begin{equation} \mathcal{D}_1(\lambda )=\bigl\{\vec \varkappa \in
\R^2: \vec \varkappa =\varkappa_1 (\varphi)\vec \nu,\ \ \vec{\nu}
\in \mathcal{B}_1(\lambda) \},\label{May20} \end{equation} where $
\varkappa_1 (\varphi)=k+h_1(\varphi)$ and $h_1(\varphi)$ obeys the
inequalities \begin{equation}\label{2.75}
 |h_1|<C(W_1)k^{-3+ 30s_1+20\delta
   },\quad
 \left|\frac{\partial h_1}{\partial \varphi} \right| <
C(W_1)k^{-1+33s_1+22\delta }. \end{equation}

\item The total length of $\mathcal{B}_1(\lambda)$ satisfies the
estimate \begin{equation}\label{theta1}
L\left(\mathcal{B}_1\right)=2\pi(1+O(k^{-\delta })). \end{equation}

\item The function $h_1(\varphi)$ can be extended as a holomorphic
function of $\varphi $ from $\Theta _2$ to $\hat \varPhi _1$.
Estimates \eqref{2.75} hold in $\hat \varPhi _1$ too.

\item The curve $\mathcal{D}_1(\lambda)$ has a length which is
asymptotically close to that of the whole circle in the  sense that
\begin{equation}\label{2.77}
L\bigl(\mathcal{D}_1(\lambda)\bigr)\underset{\lambda \rightarrow
\infty}{=}2\pi k\bigl(1+O(k^{-\delta})\bigr),\quad \lambda =k^{2}.
\end{equation} \end{enumerate} \end{lem}

\begin{proof} Here are the main points of the proof, for details see
Lemma 2.11 in \cite{KL}. \begin{enumerate} \item Inequalities
(\ref{2.75}) easily follow from \eqref{2.66} -- \eqref{2.67a} and
the definition of $\mathcal{D}_1(\lambda )$, Implicit function
theorem being applied.

\item By definition, $\mathcal{B}_1(\lambda)$ is the set of
directions corresponding to $\mathcal{D}_0(\lambda)_{nonres}$, the
latter set being a subset of the sphere of radius k. Formula
\eqref{chi _1} establishes a one-to-one correspondence between $\chi
_1(\lambda, \beta, s_1, \delta )$ and
$\mathcal{D}_0(\lambda)_{nonres}$, their lengths being equal.
Considering \eqref{E:Geometric-1-full}, we obtain
$L\left(\mathcal{D}_0(\lambda)_{nonres}\right)=2\pi
k\bigl(1+O(k^{-\delta})\bigr)$. Hence, \eqref{theta1} holds.

\item The series  \eqref{2.66} converges for $\vec \varkappa
=\varkappa \vec \nu $, when  $\varphi$ in $\hat \varPhi _1$ and even
its $(k^{-2-16s_1-11\delta})$-neighborhood and $\varkappa \in \omega
$, $\omega $ being the complex
$(k^{-1-16s_1-11\delta})$-neighborhood of $k$, $\omega =\{\varkappa
\in \C:|\varkappa -k|<k^{-1-16s_1-11\delta}\}$.  Function $\lambda
^{(1)}(\alpha, \vec \varkappa )$ is analytic in $\varkappa $ and
$\varphi $. The estimate (\ref{2.67}) holds. Thus, $ |\lambda
^{(1)}(\alpha ,\vec \varkappa \nu )-\varkappa
^2|<C(W_1)k^{-2+30s_1+20\delta },$ and $ |\varkappa ^2-k^2|\approx
k^{-16s_1-11\delta}$ when $\varkappa \in
\partial \omega.$ Applying Rouche's theorem in $\omega $, we obtain
that the equations $\lambda ^{(1)}(\vec \varkappa \nu )=k^2$ and
$\varkappa ^2=k^2$ have the same number of solutions $\varkappa $ in
$\omega $ for every $\varphi \in \hat \varPhi _1$. Obviously,
$\varkappa ^2=k^2$ has just one solution. Therefore, $\lambda
^{(1)}(\alpha, \vec \varkappa \nu )=k^2$ also has just one solution
$\varkappa _1(\varphi )$. Using \eqref{pderiv} and Implicit function
Theorem, we obtain that $\varkappa _1(\varphi )$ is an analytic
function of $\varphi $ in $\hat \varPhi _1$ and estimates
(\ref{2.75}) hold.

\item Estimate (\ref{2.77}) follows from (\ref{2.75}) and
(\ref{theta1}). \end{enumerate}
 \end{proof}
 Next, we define the non-resonance subset $\chi_1^*(\lambda )$ of
the isoenergetic set $S_1(\lambda )$ as the parallel shift of
$\mathcal{D}_1(\lambda )$ into $K_1$:\begin{equation}\label{2.81}
\chi_1^*(\lambda ):=\mathcal{K}_1\mathcal{D}_1(\lambda ).
\end{equation}

\begin{lem} \label{Apr4} The set $\chi_1^*(\lambda )$ belongs to the
$(k^{-3+30s_1+20\delta})$-neighborhood of $\chi_1(\lambda)$ in
$K_1$. If $t\in \chi_1^*(\lambda )$, then the operator
$H^{(1)}_{\alpha }(t)$ has a simple eigenvalue
$\lambda_n^{(1)}(\alpha ,t)$, $n\in \N$, equal to $\lambda $, no
other eigenvalues being in the interval $\varepsilon_1 (k,\delta )$,
$\varepsilon_1 (k,\delta ):=(k^{2}-k^{-16s_1-11\delta },
k^{2}+k^{-16s_1-11\delta })$. This eigenvalue is given by the
perturbation series \eqref{E:lambda_j^1}, where $j$ is uniquely
defined by $t$ from the relation $p_j^{2}(t)\in \varepsilon_1
(k,\delta )$. \end{lem}

\begin{proof} By Lemma \ref{L:2.13}, $\mathcal{D}_1(\lambda )$ is in
the $\left(k^{-3+30s_1+20\delta}\right)$-neighborhood of
$\mathcal{D}_0(\lambda )$. Since $\chi_1(\lambda )=\mathcal
K_1\mathcal{D}_0(\lambda )$ and $\chi_1^*(\lambda )=\mathcal
K_1\mathcal{D}_1(\lambda )$, we immediately obtain that
$\chi_1^*(\lambda )$ is in the $(ck^{-3+30s_1+20\delta
})$-neighborhood of $\chi_1(\lambda)$. The size of this neighborhood
is less than $k^{-1-16s_1-11\delta }$, here $-1-16s_1-11\delta
=2\beta _0 -2-s_1-2\delta $ for $2\beta _0=1-15s_1-9\delta $. Hence,
Theorem~\ref{Thm:main-1} holds  for any $t\in \chi_1^*(\lambda )$:
there is a single eigenvalue of $H^{(1)}_{\alpha }(t)$ in the
interval $\varepsilon_1 (k,\delta )$. Since $\chi_1^*(\lambda
)\subset S_1(\lambda )$, this eigenvalue is equal to $\lambda $. By
the theorem, the eigenvalue is given by the series
\eqref{E:lambda_j^1}, $j $ being uniquely defined by $t$ from the
relation $p_j^{2}(t)\in \varepsilon_1 (k,\delta )$. \end{proof}

\begin{lem}\label{L:May10a} Formula \eqref{2.81} establishes a
one-to-one correspondence between $\chi_1^*(\lambda )$ and
$\mathcal{D}_1(\lambda)$. \end{lem}

\begin{rem} From the  geometric point of view, this means that
$\chi_1^*(\lambda)$ does not have self-intersections. \end{rem}

\begin{proof} Suppose there is a pair $\vec \varkappa_{1,1},
\vec\varkappa_{1,2} \in \mathcal{D}_1(\lambda)$, such that $\mathcal
K_1\vec \varkappa_{1,1}=\break\mathcal K_1\vec \varkappa_{1,2}=t$,
$t\in \chi_1^*(\lambda )$. By the definition (\ref{2.68}) of
$\mathcal{D}_1(\lambda)$, we have $\lambda^{(1)}(\alpha,\vec
\varkappa_{1,1})=\lambda^{(1)}(\alpha,\vec \varkappa _{1,2})=\lambda
$, i.e., the eigenvalue $\lambda $ of $H^{(1)}_{\alpha }(t)$ is not
simple. This contradicts  the previous lemma. \end{proof}

\subsection{Preparation for the Next Approximation.} In the next
steps of approximations we will need  estimates for the resolvent
$\left(H^{(1)}\bigl(\vec \varkappa _1(\varphi )+\vec
b\bigr)^{-1}-k^2\right)^{-1}$, $\vec b\in K_1\setminus \{0\}$,
$\varphi  \in \varPhi _1 $. Let $b_0$ be the distance of the point
$\vec b$ to the nearest corner of $K_1$:
\begin{equation}\label{3.2.3.1}
b_0=\min_{m=(0,0),(0,1),(1,0),(1,1)}|\vec b-2\pi m/a|.
\end{equation} We will consider separately two cases: $b_0\geq
 k^{-1-16s_1-12\delta }$ and
$0<b_0<k^{-1-16s_1-12\delta }$. Our goal is to prove Lemmas
\ref{norm**}
 and \ref{L:july9c}, which give the estimates
for the resolvent on a subset of $\varPhi _1$ for the cases $b_0\geq
 k^{-1-16s_1-12\delta }$ and
$0<b_0<k^{-1-16s_1-12\delta }$, respectively. In this subsection we
choose $\beta $ to be relatively small, $100s_1<\beta
<1/12-28s_1-14\delta $. We denote such $\beta $ by $\beta
_1$.\subsubsection{The case $b_0\geq k^{-1-16s_1-12\delta }$.} The
goal is to construct a set $\mathcalO^{(1)}_s(\vec
 b)$ such that   Lemma \ref{norm**} holds. The
general scheme of considerations  is similar to that for the case
$H=(-\Delta )^l+V$, $l\geq 6$, see \cite{KL}, Section 3.5.1. First,
we describe shortly the scheme in \cite{KL}. Further, we explain
what changes have to be made to adjust it for the present case.
Indeed, in \cite{KL} we constructed a set $\mathcalO(\vec b)$, which
consists of small discs centered at the poles of the resolvent
\begin{equation} \Bigl(H^{(1)}_0(\vec k(\varphi )+\vec
b)-k^2\Bigr)^{-1}. \label{resolvent_0} \end{equation} However, the
size of discs was chosen sufficiently large to ensure that
\begin{equation} \Bigl\|\Bigl(H^{(1)}_0(\vec k(\varphi )+\vec
b)-k^2\Bigr)^{-1}\Bigr\| <k^{-\mu }, \ \mu >0, \mbox{ when }\varphi
\not \in \mathcalO(\vec b).\label{resolvent_0-norm} \end{equation}
 It was shown that the number of
 poles of \eqref{resolvent_0} in $\varPhi _0$ does not exceed $c_0k^{2+2s_1}$. Using
 perturbation arguments we showed that the estimate analogous to
 \eqref{resolvent_0} holds for the perturbed resolvent
 \begin{equation} \Bigl(H^{(1)}(\vec \varkappa _1(\varphi )+\vec
b)-k^2\Bigr)^{-1}\label{resolvent} \end{equation} in $\varPhi
_1\setminus \mathcalO(\vec b)$: \begin{equation}
\Bigl\|\Bigl(H^{(1)}(\vec \varkappa _1(\varphi )+\vec
b)-k^2\Bigr)^{-1}\Bigr\| <k^{-\mu }, \ \mu >0.
\label{resolvent-norm} \end{equation} Next, suppose
$\mathcalO_c(\vec
 b)$ is a connected component of $\mathcalO(\vec
 b)$, which is entirely in $\varPhi _1$: $\mathcalO_c(\vec
 b)\subset \varPhi _1$. Then \eqref{resolvent} is an analytic function in $\mathcalO_c(\vec
 b)$. Using determinants and Rouche's theorem, we
showed that \eqref{resolvent_0} and \eqref{resolvent} have the same
number of poles inside each connected component of $\mathcalO(\vec
 b)$. Therefore, the total number of poles of \eqref{resolvent} in such components does
 not exceed $c_0k^{2+2s_1}$. Next, we replaced the discs in $\mathcalO(\vec
 b)$ by much smaller discs around the poles of \eqref{resolvent}.
 By doing this, we obtained a much smaller set $\mathcalO_s(\vec
 b)$, $\mathcalO_s(\vec
 b)\subset \mathcalO(\vec
 b)$.  Using analyticity of \eqref{resolvent} inside $\mathcalO(\vec
 b)$ and the maximum principle, we arrived at the estimate:
 \begin{equation} \Bigl\|\Bigl(H^{(1)}(\vec \varkappa _1(\varphi )+\vec
b)-k^2\Bigr)^{-1}\Bigr\| <ck^{-\mu
}\left(\frac{R_b}{r^{(1)}}\right)^J,\ \
J=c_0k^{2+2s_1},\label{resolvent-norm*} \end{equation}
 where $R_b$ is the size of the biggest component of $\mathcalO(\vec
 b)$ and $r^{(1)}$ is the radius of the discs constituting $\mathcalO_s(\vec
 b)$. Thus, we shrank the set $\mathcalO(\vec
 b)$, but still obtained a bound for the norm of the resolvent, which is essential for the next
 step of approximation in \cite{KL}. It may happen, however, that $\mathcalO_c(\vec
 b)\cap \varPhi _1 \neq \emptyset$, but $\mathcalO_c(\vec
 b)\not \subset \varPhi _1 $. In this case we showed that the radii of the discs in $\mathcalO_c(\vec
 b)$  were sufficiently small to make sure that the function $\varkappa _1(\varphi )$
 can be analytically extended from $\varPhi _1$ to the interior each $\mathcalO_c(\vec
 b)$ when $\mathcalO_c(\vec
 b)\cap \varPhi _1\neq \emptyset$ even if $\mathcalO_c(\vec
 b)\not \subset \varPhi _1 $. For $l\geq 6$ this condition on radii does
 not contradict to \eqref{resolvent_0-norm}.

 In the present case $l=1$ the plan is basically the same. However,
 we have essential technical complications. The set $\mathcalO(\vec
 b)$, constructed for $l\geq 6$, is now too small to provide convergence of perturbation series for \eqref{resolvent}
   when
 $\varphi \not \in  \mathcalO(\vec
 b)$. Therefore, we need to construct a bigger set $\mathcalO^{(1)}(\vec
 b)$, such that  the series converges when $\varphi \not \in  \mathcalO^{(1)}(\vec
 b)$. It can be constructed by analogy with \eqref{O^1}. Let $\mathcalO^{(1)}_c(\vec
 b)$ be a connected component of $\mathcalO^{(1)}(\vec
 b)$. If $\mathcalO^{(1)}_c(\vec
 b)\subset  \varPhi _1$, then considering as in  \cite{KL}, we obtain estimates similar to \eqref{resolvent-norm*}.
 A difficult case  is $\mathcalO^{(1)}_c(\vec
 b)\cap \varPhi _1 \neq \emptyset$, but $\mathcalO^{(1)}_c(\vec
 b)\not \subset \varPhi _1$. On one hand,
the function $\vec \varkappa
 _1(\varphi )$, included in \eqref{O^1}, is defined  on the set $\varPhi _1$, which has holes as
 small as $\varepsilon =k^{-2-12s_1-9\delta }$. These holes are essential in the
 construction of $\varPhi _1$. Hence, the function $\vec \varkappa
 _1(\varphi )$ can be analytically extended  no further than $c\varepsilon $, $c<1$, neighborhood of $\varPhi
 _1$. Thus, to extend $\vec \varkappa
 _1(\varphi )$ analytically into a connected component $\mathcalO^{(1)}_c(\vec b)$ of  a set $\mathcalO^{(1)}(\vec b)$,
 such that  $\mathcalO^{(1)}_c(\vec
 b) \not \subset \varPhi _1$, we have to
 make sure that the size of $\mathcalO^{(1)}_c(\vec
 b)$  is  smaller than $\varepsilon $. On the other hand, deleting only such a
 small set from $\varPhi _1$ cannot provide convergence of
 perturbations series for \eqref{resolvent} in $\varPhi _1\setminus \mathcalO^{(1)}(\vec
 b)$. The properties of the smallest set $\mathcalO^{(1)}(\vec
 b)$, which we can construct to provide convergence of the perturbation series, is
 given in Lemma \ref{hope}. Its connected component still can have a size up
 to $k^{-11/6+\beta _1+12s_1+4\delta }$, at least this is the strongest estimate we can prove.
 Obviously, the size of such a component is still greater then
 the size  $\varepsilon $ of the neighborhood of $\varPhi _1$
 where $\varkappa _1(\varphi )$ is holomorphic. Thus, we can not
 construct the set $\mathcalO^{(1)}(\vec
 b)$ which fits into $\varepsilon $-neighborhood of $\varPhi _1$, even
 though it is much smaller then \eqref{O^1}.
To overcome this difficulty we need additional considerations. We
represent $\vec \varkappa
 _1$ in the form $\vec \varkappa
 _1(\varphi )=\vec \varkappa
 _*(\varphi )+O\left(k^{-ck^{3/4}}\right)$, where $\vec \varkappa
 _*(\varphi )$ is an analytic function in a set $\tilde \varPhi _0 $
 larger then $\varPhi _1 $,
$\tilde \varPhi _0 \supset \varPhi _1$, see Lemma \ref{L:varkappa
_*}. The smallest hole in
 $\tilde \varPhi _0$ has the size $k^{-7/4-12s_1-9\delta }$. Since the size of each connected component of
  $\mathcalO^{(1)}(\vec
 b)$ is smaller than $k^{-11/6+\beta _1+12s_1+4\delta }$, $\vec \varkappa
 _*(\varphi )$  is holomorphic  in every $\mathcalO^{(1)}_c(\vec
 b)$, such that $ \mathcalO^{(1)}_c(\vec
 b)\cap \varPhi _1 \neq \emptyset $. We prove an estimate for the norm of
 \begin{equation} \Bigl(H^{(1)}(\vec \varkappa _*(\varphi )+\vec
b)-k^2\Bigr)^{-1}\label{resolvent*} \end{equation} outside
$\mathcalO_c^{(1)}(\vec
 b)$. Since, the set $\mathcalO^{(1)}(\vec
 b)$ is rather small, the proof of an estimate for the norm is quite technical (Lemmas
 \ref{hope}, \ref{norm}, \ref{hope1}, \ref{norm*}). Further
 we
  construct  a set $\mathcalO^{(1)}_s(\vec
 b)
$ taking small discs around the poles of \eqref{resolvent*}. Using
the analyticity of the resolvent and the maximum principle, we
obtain (Lemma \ref{norm*}) an estimate of the type
\eqref{resolvent-norm*} for \eqref{resolvent*}: \begin{equation}
\Bigl\|\Bigl(H^{(1)}(\vec \varkappa _*(\varphi )+\vec
b)-k^2\Bigr)^{-1}\Bigr\| <ck^{\mu
_*}\left(\frac{R_b}{r^{(1)}}\right)^J,\ \
J=c_0k^{2+2s_1},\label{resolvent-norm**},\ \ \mu _*>0.\end{equation}
The size of each connected component of $\mathcalO^{(1)}_s(\vec
 b)$ is much less than $k^{-2-12s_1-9\delta }$. Therefore, the function
$\varkappa _1(\varphi )$  can be analytically extended into interior
of every connected component of $\mathcalO^{(1)}_s(\vec
 b)$
 intersecting with  $\varPhi _1$. Using the estimate $\vec \varkappa
 _1(\varphi )=\vec \varkappa
 _*(\varphi )+O\left(k^{-ck^{3/4}}\right)$ in the $\varepsilon $ neighborhood of $\varPhi _1$, we prove that
 \eqref{resolvent} obeys an estimate similar to
 \eqref{resolvent-norm**}, see Lemma \ref{norm**}. This lemma is used in the second and step of approximation.

 One more technical difficulty is related to the fact that the
 resolvent \eqref{resolvent_0} is
 not from the trace class when $l=1$, while it is from the trace
 class when $l>1$. This means that the determinant of the operator
$\left(H^{(1)}(\vec k(\varphi )+\vec
b)-k^2\right)\left(H_0^{(1)}(\vec k(\varphi )+\vec
b)-k^2\right)^{-1}$, which we considered in \cite{KL} as a complex
function of $\varphi $ for $l\geq 6$, is not defined for the present
case $l=1$. To overcome this difficulty, we take a family of finite
dimensional projections $P_N$, $P_N\to _{s}I$ as $N$ goes to
infinity. We consider  a finite dimensional operator $H^{(1)}_N
=P_NH^{(1)}P_N$ and  the determinant of $\left(H_N^{(1)}(\vec
k(\varphi )+\vec b)-k^2\right)\left(H_{0,N}^{(1)}(\vec k(\varphi
)+\vec b)-k^2\right)^{-1}$. We prove all necessary results for
$H_N^{(1)}$ and, then, the analogous results for $H^{(1)}$ by
sending $N$ to infinity.

\begin{definition} \label{def:0'} Let \begin{equation}\varPhi _0'=\varPhi
_0\setminus \left(\cup _{q:\ 0<p_q<4k^{s_1},\pm }\ \mathcalO_q^{\pm
}\right),\label{0'} \end{equation} Obviously, $\varPhi _1\subset
\varPhi _0'\subset \varPhi _0$.
 Note that the circles $\mathcalO_q^{\pm }$
deleted from $\varPhi _0$ to obtain $\varPhi _0'$ are relatively
large. Their radius is larger  that $\frac{1}{4}k^{-4s_1-\delta }$,
see \eqref{r_m}. These circles are essentially bigger that other
$\mathcalO_m^{\pm }$ constituting $\mathcal O^{(1)}$. Properties of
$\varPhi _0'$ are stable in its $k^{-4s_1-2\delta }$  neighborhood.
We denote such neighborhood of $\varPhi _0'$ by $\hat \varPhi _0'$.
\end{definition} \begin{definition}
\label{newdef} The set $\mathcal O ^{(1)}(\vec b)$ is defined by the
formula: \begin{equation}\mathcal O ^{(1)}(\vec b)=\cup
_{0<p_m<4k,\pm }\ \mathcal O_m^{\pm}(\vec
b),\label{O^1(b)}\end{equation} where $\mathcal O_m ^{\pm}(\vec b)$
are discs in complex plane centered at $\varphi ^{\pm }_m(\vec b)$,
which are zeros of $|\vec k (\varphi )+\vec p_m(\vec b)|_*^2-k^2$.
The radii of the discs are given by Definition \ref{Def:r_m(b)}.
\end{definition}
\begin{definition} The total size of $\mathcalO^{(1)}(\vec b)$ is
the sum of the sizes of its connected components. \end{definition}

\begin{lem} \label{hope} Let $100s_1<\beta
_1<1/12-28s_1-14\delta $. \begin{enumerate} \item If $\varphi \in
\hat \varPhi _0'\setminus \mathcal O^{(1)}(\vec b)$ and $m$ is such
that $\big ||\vec k (\varphi )+\vec p_m(\vec b)|_*^2-k^2\big|\
<k^{\beta _1}$, then
\begin{equation} \min _{0<p_q<k^{s_1}}\big||\vec k(\varphi )+\vec
p_{m+q}(\vec b)|_*^2-k^2\big| >k^{\beta _1},\label{dop1}
\end{equation}
\begin{equation} \label{defO_*}\Big ||\vec k (\varphi )+\vec
p_m(\vec b)|_*^2-k^2\Big|\Big||\vec k(\varphi )+\vec p_{m+q_1}(\vec
b)|_*^2-k^2\Big|\Big ||\vec k(\varphi )+\vec p_{m+q_2}(\vec
b)|_*^2-k^2\Big|>\frac{1}{64}k^{2\beta _1}
\end{equation} when
 $0<p_{q_1},p_{q_2}<k^{s_1}$. This property
is preserved in the $k^{-4+2\beta _1-2s_1-\delta }$ neighborhood of
$\hat \varPhi _0'\setminus \mathcal O^{(1)}(\vec b)$.

\item  The size of each connected component  $\mathcal O
_c^{(1)}(\vec b)$ of $\mathcal O ^{(1)}(\vec b)$ is less than
$k^{-\gamma }$, $\gamma =11/6-\beta _1-12s_1-4\delta $. Each
component contains no more than $c_1k^{2/3+s_1}$, $c_1=c_1(d_1,d_2)$
discs. The total size of $\mathcal O ^{(1)}(\vec b)$ does not exceed
$2\pi c_1k^{-5/6+\beta _1+12s_1+4\delta }$. The set $\mathcal O
^{(1)}(\vec b)$ contains less that $c_0k^{2+2s_1}$ discs.
\end{enumerate}
\end{lem}

\begin{cor} \label{Cor:1} For every $\varphi $  in the $k^{-4+2\beta _1-2s_1-\delta  }$ neighborhood of
$\hat \varPhi _0'\setminus \mathcal O^{(1)}(\vec b)$ and for every
$m\in Z^2 $ \begin{equation} \label{el}\bigl||\vec k(\varphi )+\vec
p_m(\vec b)|^2-k^2\bigr|>k^{-2+2\beta _1-2s_1}.\end{equation}
\end{cor}
 \bpf[Proof of Corollary
\ref{Cor:1}] Obviously, $$\left|\big |\vec k(\varphi )+\vec
p_{m+q_{1,2}}(\vec b)|_*-k^2\right|\leq \left|\big |\vec k(\varphi
)+\vec p_{m}(\vec b)|_*-k^2\right|+2\left|\langle \vec k(\varphi
)+\vec p_{m}(\vec b),\vec p_{q_{1,2}}\rangle _*\right| +p_q^2.$$ If
$\left|\big |\vec k(\varphi )+\vec p_{m}(\vec b)|-k^2\right|<1$,
then, clearly, $\left|\big |\vec k(\varphi )+\vec p_{m+q_{1,2}}(\vec
b)|-k^2\right|\leq ck^{1+s_1}.$ Now the corollary follows from
\eqref{defO_*} and the last estimate. \epf
\begin{cor}\label{Cor:2} If $\varphi $  in the $k^{-4+2\beta _1-2s_1-\delta }$ neighborhood of $\hat \varPhi _0'\setminus \mathcal O^{(1)}(\vec b)$,
then the following estimates holds:
\begin{equation} \left\|\left(H^{(1)}_0\bigl(\vec k(\varphi )+\vec
b\bigr)^{-1}-k^2)\right)^{-1}\right\|<ck^{2-2\beta _1+2s_1}.
\label{normresH_0} \end{equation}The resolvent has no more than
$c_1k^{2/3+s_1}$ poles in each connected component of $O^{(1)}(\vec
b)$. The total number of poles in $\varPhi _0$ is less than
$c_0k^{2+2s_1}$.\end{cor} \begin{cor} If $\mathcal O ^{(1)}_c(\vec
b)\cap \varPhi _0'\not =\emptyset $, then $\mathcal O ^{(1)}_c(\vec
b)\subset \hat \varPhi _0'$.\end{cor} This corollary follows from
the statement that the size of $\mathcal O ^{(1)}_c(\vec b)$ does
not exceed $k^{-\gamma }$ and the definition of $\hat \varPhi _0'$
as the $k^{-4s_1-2\delta }$ neighborhood of $\varPhi _0'$.

The lemma is proven is Section \ref{5.2}. Note that the total size
$k^{-5/6+\beta _1+12s_1+4\delta }$ of the set $\mathcal O
^{(1)}(\vec b)\cap \hat \varPhi _0'$ is small comparing even with
the smallest of the circle $\mathcalO_q$, $p_q<k^{s_1}$.

\begin{lem}\label{norm} If $\varphi  \in \hat \varPhi _0'\setminus \mathcal
O ^{(1)}(\vec b)$, then \begin{equation} \left\|\bigl(H (\vec
k(\varphi )+\vec b)-k^{2}\bigr)^{-1}\right\|<ck^{4}.
\label{normres}\end{equation} The only possible singularities of the
resolvent  $(H (\vec k(\varphi )+\vec b)-k^{2}\bigr)^{-1}$ in each
connected $\mathcalO^{(1)}_c(\vec b)$ component of
$\mathcalO^{(1)}(\vec b)$, such that $\mathcalO^{(1)}_c(\vec b)\cap
\varPhi _0'\neq \emptyset $, are poles. The number of poles
(counting multiplicity) inside each  component
 does not exceed $c_1k^{2/3+s_1}$. The total number of such poles  is less then $c_0k^{2+2s_1}$.\end{lem}
\begin{rem} In Lemma \ref{norm} we could prove convergence of perturbation series and the estimate for
the resolvent analogous
 to (\ref{normresH_0}) by somewhat longer considerations.
However, (\ref{normres}) is good enough for our purposes. The main
reason that  convergence holds, in spite of weaker conditions on
$\varphi $ are formulas of the type \eqref{2.14}. \end{rem}

\begin{proof} Let $\varphi  \in \hat \varPhi _0'\setminus \mathcal O
^{(1)}(\vec b)$. We define the set $\Omega (\varphi)\subset \Z^2$ as
follows: $$\Omega =\left\{m\in \Z^2: \bigl||\vec k(\varphi )+\vec
p_m(\vec b)|^2-k^2\bigr|<k^{\beta _1}\right\}.$$ Let $P_0$ be the
diagonal projection, corresponding to $\Omega $: $P_{0mm}=1$ if and
only if $m\in \Omega $. Let $P_1=I-P_0$.

It follows from Lemma \ref{hope} that $P_0W_1P_0=0$. Indeed, suppose
it is not so. Then, there is a pair $m,m+q\in \Omega $ such that
$w_q\neq 0$. Function $W_1$ is a trigonometric polynomial, see
\eqref{w_q}. Hence $0<p_q<k^{s_1}$. This contradicts to
\eqref{dop1}.

%%%%%%%Let

%%%%%%%$$Q=\bigl(H (\vec k(\varphi )+\vec b)-k^{2}\bigr)^{-1}W_1,\ \ \
%%%%%%%B_0=P0QP_1QP_0.$$

Next, we consider  Hilbert identity:
$$(H^{(1)}-k^2)^{-1}=(H^{(1)}_0-k^2)^{-1}-(H^{(1)}_0-k^2)^{-1}W_1(H^{(1)}-k^2)^{-1},$$
abbreviations $H^{(1)}=H^{(1)}\left(\vec k(\varphi )+\vec b\right)$,
$H^{(1)}_0=H^{(1)}_0\left(\vec k(\varphi )+\vec b\right)$ being
used. Applying $P_1$ from two sides and solving for
$P_1(H^{(1)}-k^2)^{-1}P_1$, we obtain: \begin{equation}
\label{P_1resP_1} P_1(H^{(1)}-k^2)^{-1}P_1=(I+\mathcal
E)^{-1}P_1(H^{(1)}_0-k^2)^{-1}\left(I-W_1P_0(H^{(1)}-k^2)^{-1}P_1\right),
\end{equation} where $\mathcal E: P_1l^2\to P_1l^2$, $ \mathcal
E=P_1(H^{(1)}_0-k^2)^{-1}W_1P_1.$ By the definition of $P_1$,
\begin{equation} \|P_1(H^{(1)}_0-k^2)^{-1}\|\leq k^{-\beta _1},
\label{<}\end{equation}
 \begin{equation}
 \|\mathcal E\|\leq \|W_1\|k^{-\beta _1}.\label{<<}
 \end{equation}
  Next, we
apply $P_0$ from the left and $P_1$ from the right to the Hilbert
identity. Considering that $P_0W_1P_0=0$ and
$P_0(H^{(1)}_0-k^2)^{-1}P_1=0$, we obtain: \begin{equation}
\label{P_0resP_1}
P_0(H^{(1)}-k^2)^{-1}P_1=-P_0(H^{(1)}_0-k^2)^{-1}W_1P_1(H^{(1)}-k^2)^{-1}P_1.
\end{equation} Substituting (\ref{P_1resP_1}) into (\ref{P_0resP_1})
gives:
\begin{equation}\label{subst}P_0(H^{(1)}-k^2)^{-1}P_1=BP_0(H^{(1)}-k^2)^{-1}P_1-C,
\end{equation} $$B=P_0QP_1(I+\mathcal E)^{-1}P_1QP_0,\ \ \
C=P_0QP_1(I+\mathcal E)^{-1}P_1(H_0^{(1)}-k^2)^{-1},$$
$$Q=\bigl(H_0^{(1)}-k^{2}\bigr)^{-1}W_1.$$ It follows from
\eqref{normresH_0} that \begin{equation} \label{|Q|}
\|Q\|<C(W_1)k^{2-2\beta _1 +2s_1}. \end{equation} Using also
(\ref{<}) and (\ref{<<}) we get: \begin{equation} \label{|C|} \|C\|<
C(W_1)k^{2-3\beta _1+2s_1}.\end{equation} Let us prove that
\begin{equation} \label{|B|} \|B\|<C(W_1)k^{-2\beta _1+8s_1}.
\end{equation} Using $(I+\mathcal E)^{-1}=I+\sum
_{r=1}^{N}(-\mathcal E)^r+(-\mathcal E)^{N+1}(I+\mathcal E)^{-1}$,
$N=[40\beta _1^{-1}]$, we expand $B$ into sum of $N+2$ operators:
$$B=\sum _{r=0}^{N+1}B_r,\ \ \ \ B_0=P_0QP_1QP_0, \ \
B_r=(-1)^rP_0QP_1(QP_1)^rQP_0, \ r=1,...,N,$$ $$
B_{N+1}=(-1)^{N+1}P_0QP_1(QP_1)^{N+1}(I+\mathcal E)^{-1}P_1QP_0.$$
Let us estimate $\|B_r\|$, $r=0,...,N$. First,  we prove that $B_0$
is diagonal and \begin{equation} \|B_0\|<k^{-2\beta _1+6s_1}.
\label{B_0} \end{equation} Indeed, suppose it is not diagonal. Then
there is a pair $m,m+q\in \Omega $, $q\neq 0$, such that
$B_{0m,m+q}=0$. Considering that $W_1$ is a trigonometric
polynomial, we obtain that $p_q<2R_0<k^{s_1}$. Now we easily arrive
to contradiction with \eqref{dop1}, Lemma \ref{hope}. Thus,
$\|B_0\|=\sup _{m\in \Omega }|B_{0mm}|,$ $$B_{0mm}=\sum _{q\in \Z^2,
0<p_q<R_0}\frac{|w_{q}|^2}{\bigl(|\vec k(\varphi )+\vec p_m(\vec
b)|^2-k^2\bigr)\ \bigl(|\vec k(\varphi )+\vec p_{m+q}(\vec
b)|^2-k^2\bigr)}.$$ Considering that $w_q=\bar w_{-q}$,  we easily
show that the right-hand side is equal to $$ \sum _{ \small
\begin{array}{c}q\in \Z^2,\\
0<p_q<R_0\end{array}}\frac{|w_{q}|^2\left(|\vec k(\varphi )+\vec
p_m(\vec b)|^2-k^2+p_q^2\right)}{\bigl(|\vec k(\varphi )+\vec
p_m(\vec b)|^2-k^2\bigr)\ \bigl(|\vec k(\varphi )+\vec p_{m+q}(\vec
b)|^2-k^2\bigr)\bigl(|\vec k(\varphi )+\vec p_{m-q}(\vec
b)|^2-k^2\bigr)}.$$ Using \eqref{dop1} and (\ref{defO_*}), we obtain
$B_{0mm}=O(k^{-2\beta _1+6s_1})$. Therefore, (\ref{B_0}) holds.
Similar argument yields that $B_1$ is diagonal and $\|B_1\|=\sup
_{m\in \Omega }|B_{1mm}|,$ $$B_{1mm}=$$ $$\sum _{{ \small
\begin{array}{c}q_1,q_2\in \Z^2,\\ 0<|p_{q_{1,2}}|<R_0,\\
q_1+q_2\neq
0\end{array}}}\frac{-w_{q_1}w_{q_2}w_{-q_1-q_2}}{\bigl(|\vec
k(\varphi )+\vec p_m(\vec b)|^2-k^2\bigr)\ \bigl(|\vec k(\varphi
)+\vec p_{m+q_1}(\vec b)|^2-k^2\bigr)\bigl(|\vec k(\varphi )+\vec
p_{m+q_1+q_2}(\vec b)|^2-k^2\bigr)}.$$ Applying \eqref{defO_*} in
Lemma \ref{hope}, we obtain
\begin{equation} \|B_1\|<\|W_1\|^3k^{-2\beta _1+8s_1}. \label{normB_1}
\end{equation} Similar estimates hold for all $B_r$, $r=2,...N$:
\begin{equation} \|B_r\|\leq (c\|W_1\|)^{r+2}k^{-(\beta
_1-4s_1)(r+1)}.\end{equation}
 To estimate $B_{N+1}$, we note
that $\|P_1Q\|<\|W_1\|k^{-\beta _1}$. Considering that
$\|P_0Q\|<ck^{2-2\beta _1+2s_1}$ and $N$ is sufficiently large, we
arrive to $\|B_{N+1}\|<ck^{-2\beta _1+8s_1}$. Hence, (\ref{|B|}) is
proven. Next, considering \eqref{subst} and  (\ref{|C|}),
(\ref{|B|}), we obtain: \begin{equation} \label{normP_0resP_1}
\left\|P_0(H^{(1)}-k^2)^{-1}P_1\right\|<ck^{2-3\beta _1
+2s_1}.\end{equation} Using the last estimate in (\ref{P_1resP_1}),
we get \begin{equation} \label{normP_1resP_1} \left\|P_1
(H^{(1)}-k^2)^{-1}P_1\right\|<ck^{2-4\beta _1+2s_1}.\end{equation}
Taking into account that $P_1(H^{(1)}-k^2)^{-1}P_0$ corresponding to
a $\varphi $ is the adjoint of $P_0(H^{(1)}-k^2)^{-1}P_1$
corresponding to $\bar \varphi $ and the set $\mathcalO^{(1)}(\vec
b)$ is symmetric with respect to real axis, we obtain:
\begin{equation} \label{normP_1resP_0}
\left\|P_1(H^{(1)}-k^2)^{-1}P_0\right\|<ck^{2-3\beta
_1+2s_1}.\end{equation} Applying $P_0$ to both parts of Hilbert
equations, using $P_0W_1P_0=0$ and (\ref{normresH_0}),
(\ref{normP_1resP_0}), we obtain:
\begin{equation}\label{normP_0resP_0}
\left\|P_0(H^{(1)}-k^2)^{-1}P_0\right\|<ck^{4-5\beta
_1+4s_1}<ck^4.\end{equation}
 Combining (\ref{normP_0resP_1}) --
(\ref{normP_0resP_0}), we obtain (\ref{normres}). Note that
(\ref{normres}) holds on the boundary $\partial
\mathcalO^{(1)}_c(\vec b)$ of each connected component
$\mathcalO^{(1)}_c(\vec b)$, such that $\mathcalO^{(1)}_c(\vec
b)\cap \varPhi _0'\neq \emptyset $. It is true, since the size of
$\mathcalO^{(1)}_c(\vec b)$ is less than $ck^{-\gamma }$, i.e. much
smaller than $k^{-4s_1-2\delta }$.

It remains to show that all singularities of $(H^{(1)}-k^2)^{-1}$
inside  each connected component  $\mathcalO^{(1)}_c(\vec b)$, such
that $\mathcalO^{(1)}_c(\vec b)\cap \varPhi _0'\neq \emptyset $, are
poles and the number of poles, counting multiplicity, does not
exceed $c_1k^{2/3+s_1}$. We follow here the approach developed in
\cite{KL} for $(-\Delta )^l+V$, $l>6$, The plan in \cite{KL} is the
following. We consider the operator $I+A:=
 (H^{(1)}-k^2)(H^{(1)}_0-k^2)^{-1}$ and ,
 use a well-known relation
 (see e.g. \cite{RS}):
    \begin{equation}\label{3.2.27}
    \Bigl|\det(I+A)-1\Bigr| \leq \|A\|_1
    e^{\|A\|_1+2},\ A \in \mathbf{S}_1.
    \end{equation} on the boundary of $\mathcalO_c^{(1)}(\vec b)$. Using Rouche's theorem, we arrive at the conclusion that
    $I+A$ has the same number of zeros and
 poles inside $\mathcalO_c^{(1)}(\vec b)$. Therefore, the resolvents
 $(H^{(1)}-k^2)^{-1}$ and $(H^{(1)}_0-k^2)^{-1}$ have the same
 number of poles. Using Corollary \ref{Cor:2}, we obtain that the
number of poles of $(H^{(1)}_0-k^2)^{-1}$ does not exceed
$c_1k^{2/3+s_1}$.
    However, there is a technical
 obstacle on the way of this proof in the present case $l=1$: the determinant of $(H^{(1)}-k^2)(H^{(1)}_0-k^2)^{-1}$ is
 not defined, since the resolvent of $H_0$ is not from $\mathbf{S}_1$.  To overcome this obstacle, we introduce a family of
 expanding diagonal finite dimensional projections $P_{N}: P_N\to I$. We consider a finite dimensional analog of $H^{(1)}$ with a
 multiplier $\alpha $ in front of $W_1$:
 \begin{equation}H^{(1)}_{\alpha ,N}= H^{(1)}_0P_N+\alpha P_NW_1P_N,\ \ \ 0\leq \alpha \leq
 1. \label{alpha} \end{equation}
 First, we show that all $(H^{(1)}_{\alpha ,N}-k^2)^{-1}$, $0\leq \alpha \leq 1$, have the same number of
 poles  inside each $\mathcalO ^{(1)}_c(\vec b)$. Indeed, let $\alpha ,\alpha _0
 \in [0,1]$. We introduce  operator $A$ by the formula:
 $I+(\alpha -\alpha _0)A=(H^{(1)}_{\alpha ,N}-k^2)(H^{(1)}_{\alpha _0,N}-k^2)^{-1}.$
 Obviously,
 $A=P_NW_1P_N(H^{(1)}_{\alpha _0,N}-k^2)^{-1}.$
 Applying formula \eqref{3.2.27} to $I+(\alpha -\alpha _0)A$, we obtain:
 \begin{equation}\left|\det (H^{(1)}_{\alpha ,N}-k^2)(H^{(1)}_{\alpha
 _0,N}-k^2)^{-1}-1\right|\leq |\alpha -\alpha
 _0|\|A\|_1e^{|\|A\|_1+2}. \label{determinant}
 \end{equation}
 Clearly,
 $\|A\|\leq \|W_1\|\|(H^{(1)}_{\alpha _0,N}-k^2)^{-1}\|.$
 Considering as in the proof of \eqref{normres}, we obtain
 $\|(H^{(1)}_{\alpha _0,N}(\varphi )-k^2)^{-1}\|<ck^{4},$
 when $\varphi $ is on the boundary of $\mathcalO c^{(1)}(\vec
 b)$.
 Since $A$ is an $N$ dimensional operator:
 $\|A\|_1\leq \|A\|N.$
 Combining the last three inequalities, we get:
 $\|A\|_1<ck^{4}N.$ Using this estimate in \eqref{determinant}, we obtain:
 \begin{equation}\left|\det (H^{(1)}_{\alpha ,N}(\varphi )-k^2)(H^{(1)}_{\alpha
 _0,N}(\varphi )-k^2)^{-1}-1\right|\leq c|\alpha -\alpha
 _0|Nk^{4}, \label{determinant1}
 \end{equation}
 when $\varphi $ is on the boundary of $\mathcalO _c^{(1)}(\vec
 b)$.
 If $|\alpha -\alpha _0|$ is sufficiently small, then the right-hand
 side of \eqref{determinant1} is less than 1. By Rouche's theorem $\det (H^{(1)}_{\alpha ,N}-k^2)(H^{(1)}_{\alpha
 _0,N}-k^2)^{-1}$ has the same number of poles and zeros inside $\mathcalO _c^{(1)}(\vec
 b)$. Therefore, $\det (H^{(1)}_{\alpha ,N}-k^2)$ and $\det (H^{(1)}_{\alpha
 _0,N}-k^2)$ have the same number of zeros. Covering $[0,1]$ by
 small intervals and using a finite step induction, we obtain that
$\det (H^{(1)}_{\alpha ,N}-k^2)$ has the same number of zeros as
$\det (H^{(1)}_{0,N}-k^2)$ for any $\alpha \in [0,1]$.

\label{21} Obviously the zeros of $\det (H^{(1)}_{0,N}-k^2)$ in
$\mathcalO^{(1)}_c(\vec b)$ stay the same after $N$ surpasses a
certain number. They are solutions of the equations $|\vec k(\varphi
)+\vec p_m|^2=k^2$ in $\mathcalO^{(1)}_c(\vec b)$. Hence, the number
of zeros of $\det (H^{(1)}_{1,N}-k^2)$ is the same for all
sufficiently large $N$.  It is equal to the number of discs in
$\mathcalO^{(1)}_c(\vec b)$.We denote this number by $M$. By Lemma
\ref{hope}, $M<c_1k^{2/3+s_1}$. We denote the zeros of $\det
(H^{(1)}_{1,N}-k^2)$ as $\varphi _1^{(N)},...\varphi _M^{(N)}$,
multiplicity being taking into account. Obviously, the function
$\prod _{n=1}^M\left(\varphi -\varphi
_n^{(N)}\right)(H^{(1)}_{1,N}-k^2)^{-1}$ is holomorphic in
$\mathcalO^{(1)}(\vec b)$. For a fixed $n$ the sequence
$\big\{\varphi _n^{(N)}\big\}_{N=1}^{\infty }$ has an accumulation
point. We choose a subsequence $N_{i}$ such that each $\big\{\varphi
_n^{(N_i)}\big\}_{i=1}^{\infty }$  converges. With slight abuse of
notations we drop the index $i$ and consider that each
$\big\{\varphi _n^{(N)}\big\}_{N=1}^{\infty }$ has a limit $\varphi
_n $. Considering as in the proof of \eqref{normres}, we obtain:
\begin{equation}\label{111a} \left\|(H^{(1)}_{1,N}(\varphi
)-k^2)^{-1}\right\|<ck^{4}, \ \varphi \in \partial
\mathcalO^{(1)}_c(\vec b) \end{equation} Therefore,
\begin{equation}\label{111} \left\|\prod _{n=1}^M\left(\varphi
-\varphi _n^{(N)}\right)(H^{(1)}_{1,N}(\varphi
)-k^2)^{-1}\right\|<cr^{M}k^{4},\end{equation} when $ \varphi \in
\partial \mathcalO^{(1)}(\vec b),$  $r $ being the size of
$\mathcalO^{(1)}(\vec b)$, $r<k^{-11/6+\beta _1+12s_1+4\delta }$. By
the maximum principle, the above estimate holds inside
$\mathcalO^{(1)}_c(\vec b)$ too. Suppose $|\varphi -\varphi
_n|>2\varepsilon $ for all $n$ and some $\varepsilon >0$. Then
$|\varphi -\varphi _{n,N}|>\varepsilon $ for all $n$ and
sufficiently large $N$. Using \eqref{111}, we obtain:
\begin{equation}\label{112} \left\|(H^{(1)}_{1,N}(\varphi
)-k^2)^{-1}\right\|<cr^{M}k^{4}\varepsilon ^{-M}.\end{equation}
Thus, all resolvents are bounded uniformly in $N$, when $\varphi
\not = \varphi _n$, $n=1,...M$. Now, it is easy to show now that
$(H^{(1)}_{1,N}(\varphi )-k^2)^{-1}$ tends to $(H^{(1)}_{1}(\varphi
)-k^2)^{-1}$ in the class of bounded operators when $\varphi \not =
\varphi _n$, $n=1,...M$. Taking the limit in (\ref{111}), we obtain:
\begin{equation}\label{111*} \left\|\prod _{n=1}^M\left(\varphi
-\varphi _n\right)(H^{(1)}_{1}(\varphi
)-k^2)^{-1}\right\|<cr^{M}k^{4}.\end{equation} This means that the
only possible singularities of of $(H^{(1)}_{1}(\varphi )-k^2)^{-1}$
are poles at the points $\varphi  = \varphi _n$, $n=1,...M$. The
number of poles, counting multiplicity does not exceed
$M=c_1k^{2/3+s_1}$.
\end{proof}

Let us recall that  function $\varkappa _1(\varphi )$, defined by
Lemma \ref{L:2.12}, is holomorphic in $\hat \varPhi _1$, see Lemma
\ref{L:2.13}. Now we prove existence of a function $\varkappa
_{*}(\varphi )$, which is a good approximation of $\varkappa
_1(\varphi )$ in $\hat \varPhi _1$ and defined on a larger set $\hat
\varPhi _*$.. Let
 $$\varPhi _*=\varPhi _0\setminus \tilde{\mathcal O}
^{(1)},\ \ \ \tilde {\mathcal O }^{(1)}=\cup _{i\in Z^2,
0<p_i<k^{\bf 3/4}}\mathcal O_{i}.$$ The size of  the discs
constituting $\tilde {\mathcal O }^{(1)}$ is at least
$k^{-7/4-16s_1-9\delta }$. Let $\hat \varPhi _* $ be the
$k^{-7/4-16s_1-10\delta }-$neighborhood of $ \varPhi _* $. Since the
size of the neighborhood is much smaller than the size of the discs,
$\varPhi _*$ and $\hat \varPhi _*$ are the sets of the same type.
The following relations hold: , $\varPhi _1\subset \varPhi _*\subset
\varPhi _0'\subset \varPhi _0$, $\hat \varPhi _1\subset \hat \varPhi
_*\subset \hat \varPhi _0'\subset \varPhi _0$.

\begin{lem} \label{L:varkappa _*} Let $100s_1<\beta _1<1/12-28s_1-14\delta $. There is a function $\varkappa
_{*}(\varphi )$ holomorphic in $\hat \varPhi _*$, which satisfies
the estimates analogous to (\ref{2.75}) on this set:
\begin{equation}\label{varkappa _*}
 |\varkappa _{*}-k|<C(W_1)k^{-3+30s_1+20\delta},\quad
 \left|\frac{\partial \varkappa _{*}}{\partial \varphi} \right| <
C(W_1)k^{-1+33s_1+22\delta }. \end{equation} It satisfies the
following estimate when $\varphi \in \hat \varPhi _1:$
\begin{equation} |\varkappa _1(\varphi )-\varkappa _{1*}(\varphi
)|<ck^{-\frac{1}{4}k^{3/4}} \label{35}\end{equation} \end{lem}
\begin{cor} \label{varkappa} Function $\varkappa _{*}$ is
holomorphic and obeys (\ref{varkappa _*}) inside each connected
component  $\mathcalO_c^{(1)}(\vec b)$ of  $\mathcalO^{(1)}(\vec
b)$, such that $\mathcalO_c^{(1)}(\vec b)\cap \varPhi _* \not
=\emptyset $.\end{cor} \underline{Proof of the corollary.}
 By Lemma \ref{hope}, $\mathcalO_c^{(1)}(\vec b)$ is in the
$ck^{-\gamma }$-neighborhood of $\varPhi _*$. Considering that
$-7/4-16s_1-10\delta
>-\gamma $, we obtain $\mathcalO_c^{(1)}(\vec b)\subset \hat \varPhi _*$.
%%%%%%%%\begin{definition} We denote by $\mathcalO_c^{(1)}(\vec b)$ a
%%%%%%connected component of $\mathcalO^{(1)}(\vec b)$, such that
%%%%%%%$\mathcalO_c^{(1)}(\vec b)\cap \tilde \varPhi _0 \not =\emptyset
%%%%%%%%%$.\label{O_c}\end{definition}

%%%%%%%The set $\tilde O^{(1)}$ is defined in
%%%%%%%such a way that any of its disc has a radius no smaller than
%%%%%%%$k^{2\beta -17/6 }$. By lemma (\ref{.}), each connected component
%%%%%%%has a much smaller length $O(k^{2\beta -17/6 })$. Therefore,
%%%%%%%$\varkappa _{*}$ analytically extended from $O_*^{(1)}(\vec
%%%%%%%b)\setminus \tilde O^{(1)}\not \empty $ to the whole component of
%%%%%%%$O_*^{(1)}(\vec b)$, estimates (\ref{.}), (\ref{.}) being preserved.

 \begin{proof} Let us consider the function:
\begin{equation}\label{2.66a}
\lambda^{(1)}_*(\alpha,\vec{\varkappa})=\varkappa^{2}+f_{1*}(\alpha,\vec{\varkappa}),
    \quad  \varkappa =|\vec \varkappa |, \;\; f_{1*}(\alpha,\vec{\varkappa})=\sum
_{r=2}^{\lfloor k^{3/4}\rfloor +1}
\alpha^{r}g_r^{(1)}(\vec{\varkappa}).
\end{equation} Obviously $f_{1*}(\alpha,\vec{\varkappa})$ is a
finite sum for the series $f_{1}(\alpha,\vec{\varkappa})$, see
(\ref{2.66}). It satisfies the  estimates analogous to (\ref{2.67}),
 when $\varphi \in \varPhi _1$
    \begin{equation} |f_{1*}(\alpha,\vec
\varkappa)|\leq 2\alpha^2k^{-2+30s_1+20\delta},\ \ \
|T(m)f_{1*}(\alpha,\vec \varkappa)|\leq 2
\alpha^2k^{-2+30s_1+20\delta +|m|(1+16s_1+11\delta)}.\label{2.67ab}
\end{equation}  By the
construction of $\varPhi _*$, the estimates
\eqref{InE:Geometric-1-1} and \eqref{InE:Geometric-1-2} hold for all
$\varphi \in \varPhi _*$ and $p_i\leq k^{3/4}$. Now it is easy to
see that all coefficients $g_r^{(1)}(\vec{\varkappa})$, $2\leq r\leq
k^{3/4}$, are holomorphic functions in  $\varPhi _*$ and even to its
$2k^{-7/4-16s_1-10\delta }$ neighborhood. The estimates
(\ref{InE:g_r}), (\ref{InE:g_2-g_3}), (\ref{InE:g_r^1-derivative}),
\eqref{InE:g_2-der}--\eqref{InE:g_r-der} hold when $r\leq k^{3/4}$.
Therefore, the finite sum can be analytically extended to
$2k^{-7/4-16s_1-10\delta }$ neighborhood of $\varPhi _*$, the
estimates (\ref{2.67ab}) being valid. Estimating the tail of the
series, we get
\begin{equation*} \label{f_1-f_{1*}} |f_1-f_{1*}|<k^{-\frac{1}{4} k^{3/4}}.\end{equation*} Solving the equation
$\lambda^{(1)}_*(\alpha,\vec{\varkappa})=k^2$ for $\varkappa $, we
obtain that there is a function $\varkappa _{*}(\varphi )$ defined
in the $\hat \varPhi _*$, such that (\ref{35}) holds. This function
obviously obeys estimates (\ref{varkappa _*}).
\end{proof}

\begin{lem} \label{hope1} \begin{enumerate} \item Let $100s_1<\beta
_1<1/12-28s_1-14\delta $.  If $\varphi \in \hat \varPhi _*\setminus
\mathcal O^{(1)}(\vec b)$, and $m$ is such that $\big ||\vec
\varkappa _* (\varphi )+\vec p_m(\vec b)|_*^2-k^2\big|\
<\frac{1}{2}k^{\beta _1}$, then
\begin{equation} \min _{0<p_q<k^{s_1}}\big||\vec \varkappa _*
(\varphi )+\vec p_{m+q}(\vec b)|_*^2-k^2\big|
>\frac{1}{2}k^{\beta _1},\label{dop1a}
\end{equation}\begin{equation} \label{defO_**}\big ||\vec
\varkappa _* (\varphi )+\vec p_m(\vec b)|_*^2-k^2\big|\ \big||\vec
\varkappa _* (\varphi )+\vec p_{m+q_1}(\vec b)|_*^2-k^2\big|\ \big
||\vec \varkappa _* (\varphi )+\vec p_{m+q_2}(\vec
b)|_*^2-k^2\big|>\frac{1}{128}k^{2\beta _1} \end{equation} when
 $0<p_{q_1},p_{q_2}<k^{s_1}$. This property is preserved in the
$\frac{1}{2}k^{-4+2\beta _1-2s_1-\delta }$ neighborhood of $\hat
\varPhi _*\setminus \mathcal O^{(1)}(\vec b)$.

\item Each equation $|\vec \varkappa _* (\varphi )+\vec p_m(\vec
b)|_*^2=k^2$ has the same number of solutions  as the ``unperturbed"
equation $|\vec k (\varphi )+\vec p_m(\vec b)|_*^2=k^2$ inside every
$\mathcal O_c^{(1)}(\vec b)$,  $\mathcalO_c^{(1)}(\vec b)\cap
\varPhi _* \not =\emptyset $.\end{enumerate}\end{lem} \begin{cor}The
number of points $\varphi $  satisfying one of the equations $|\vec
\varkappa _* (\varphi )+\vec p_m(\vec b)|_*^2=k^2$, $m\in Z^2$, in
$\mathcal O_c^{(1)}(\vec b)$, $\mathcalO_c^{(1)}(\vec b)\cap \varPhi
_* \not =\emptyset $, does not exceed $c_1k^{2/3+s_1}$. The total
number of such points is less then $c_0k^{2+2s_1}$. \end{cor}

The corollary follows from the above lemma and the last statement of
Lemma \ref{hope}.

\begin{proof}Let $\varphi \in \hat \varPhi _*\setminus \mathcal
O^{(1)}(\vec b)$. Noticing that $\hat \varPhi _*\setminus \mathcal
O^{(1)}(\vec b)\subset \hat \varPhi _0'\setminus \mathcal
O^{(1)}(\vec b)$, we obtain that $\varphi $ satisfies the conditions
of Lemma \ref{hope}. Considering the first inequality in
\eqref{varkappa _*}, we conclude: \begin{equation}\label{July14b}
 \big ||\vec \varkappa _* (\varphi )+\vec
p_m(\vec b)|_*^2-|\vec k (\varphi )+\vec p_m(\vec
b)|_*^2\big|=O(k^{-2+ 30s_1+20\delta})\ \ \mbox{for any }m\in
Z^2.\end{equation} Taking into account that  $30s_1+20\delta <2\beta
_1-2s_1$ and \eqref{el}, we get \begin{equation}\label{July14c}
 \big ||\vec \varkappa _* (\varphi )+\vec
p_m(\vec b)|_*^2-|\vec k (\varphi )+\vec p_m(\vec
b)|_*^2\big|<\frac{1}{2} \left||\vec k (\varphi )+\vec p_m(\vec
b)|_*^2-k^2\right|\ \ \mbox{for any }m\in Z^2.\end{equation}
Therefore,
 $2\big||\vec \varkappa _* (\varphi )+\vec p_m(\vec
b)|_*^2-k^2\big|>\big||\vec k (\varphi )+\vec p_m(\vec
b)|_*^2-k^2\big|$ for all $m\in Z^2$. Now, the first statement of
the lemma easily follows from the first statement of Lemma
\ref{hope}.

In particular, \eqref{July14c} holds on the boundary of each
$\mathcal O_c^{(1)}(\vec b)$,  $\mathcalO_c^{(1)}(\vec b)\cap
\varPhi _* \not =\emptyset $, see Corollary \ref{varkappa}. By
Rouche's theorem, each equation $|\vec \varkappa _* (\varphi )+\vec
p_m(\vec b)|_*^2=k^2$ has the same number of solutions inside every
 $\mathcal O_c^{(1)}(\vec b)$ as the
``unperturbed" equation $|\vec k (\varphi )+\vec p_m(\vec
b)|_*^2=k^2$.\end{proof}

\begin{lem}\label{norm*}   Let $100s_1<\beta
_1<1/12-28s_1-14\delta  $. If $\varphi \in \hat \varPhi _*\setminus
\mathcal O^{(1)}(\vec b)$, then \begin{equation} \left\|\bigl(H
(\vec \varkappa _*(\varphi )+\vec
b)-k^{2}\bigr)^{-1}\right\|<ck^{4}. \label{normres*}\end{equation}
This estimate is stable in the $\frac{1}{2}k^{-4+2\beta
_1-2s_1-\delta }$ neighborhood of $\hat \varPhi _*\setminus \mathcal
O^{(1)}(\vec b)$. The only possible singularities of the resolvent
$(H (\vec \varkappa _*(\varphi )+\vec b)-k^{2}\bigr)^{-1}$ inside
each $\mathcalO^{(1)}_c(\vec b)$, $\mathcalO_c^{(1)}(\vec b)\cap
\varPhi _* \not =\emptyset $, are poles. The number of poles,
counting multiplicity, inside each such $\mathcalO^{(1)}_c(\vec b)$
does not exceed $c_1k^{2/3+s_1}$. The total number of such poles is
less then $c_0k^{2+2s_1}$. \end{lem}
\begin{proof} The proof of The lemma is analogous to that of Lemma
\ref{norm} up to replacement of Lemma \ref{hope} by Lemma
\ref{hope1}. \end{proof} \begin{definition}\label{D:small} Let us
numerate all components $\mathcalO^{(1)}_c(\vec b)$,
$\mathcalO_c^{(1)}(\vec b)\cap  \varPhi _* \not =\emptyset $, by
index $i$, $i=1,...,I$. We denote the poles of $(H^{(1)} (\vec
\varkappa _*(\varphi )+\vec b)-k^{2}\bigr)^{-1}$ in a component
$\mathcalO^{(1)}_c(\vec b)_i$ by $\varphi _{i,n_i}$,
$n_i=1,...,M_i$, $M_i<c_1k^{2/3+s_1}$.  Let us consider the discs
$\mathcalO_{s}(\vec b)_{i,n_i}$ of the radius
$r^{(1)}=k^{-4-6s_1-3\delta}$ around these poles.  This radius is
much less then the size of discs constituting
$\mathcalO_c^{(1)}(\vec b)_i$. Let $\mathcalO_{s}^{(1)}(\vec b)_i$
be the union of all small discs corresponding to a component
$\mathcalO^{(1)}_c(\vec b)_i$:
\begin{equation} \mathcalO_{s}^{(1)}(\vec b)_i=\cup
_{n_i=1}^{M_i}\mathcalO_{s}^{(1)}(\vec
b)_{i,n_i}.\label{small}\end{equation}  Obviously,
$\mathcalO_{s}^{(1)}(\vec b)_i\subset \mathcalO_{c}^{(1)}(\vec b)_i$
and, therefore, $\mathcalO_{s}^{(1)}(\vec b)_i\cap
\mathcalO_{s}^{(1)}(\vec b)_{i'}=\emptyset $ if $i\neq i'$. Let
\begin{equation} \mathcalO_{s}^{(1)}(\vec b)=\cup
_{i=1}^I\mathcalO_{s}^{(1)}(\vec b)_i. \label{small+}\end{equation}
\end{definition}

\begin{lem} \label{size}The size of each connected component of
$\mathcalO_{s}^{(1)}(\vec b)$ does not exceed
$c_1k^{-3\frac{1}{3}-5s_1-3\delta }$. The total size of
$\mathcalO_{s}^{(1)}(\vec b)$ does not exceed $c_0k^{-2-4s_1-3\delta
}$.\end{lem}
\begin{proof} Since each disc has the radius $r^{(1)}=k^{-4-6s_1-3\delta}$
and $M_i<c_1k^{2/3+s_1}$, the size of a connected component of
$\mathcalO_{s}^{(1)}(\vec b)$ does not exceed
$c_1k^{-3\frac{1}{3}-5s_1-3\delta }$. The total number of discs in
$\mathcalO_{s}^{(1)}(\vec b)$ does not exceed  $c_0k^{2+2s_1}$. Now
the second statement
 of the lemma easily follows. \end{proof}

 Further we consider only
those connected components $\mathcalO_{sc}^{(1)}(\vec
 b)$ of $\mathcalO_{s}^{(1)}(\vec
 b)$ who have non-empty intersection with  $ \varPhi _1$: $\mathcal O_{sc}^{(1)}(\vec
 b)_i\cap  \varPhi _1\neq \emptyset $.

 \begin{lem} Function $\varkappa _1(\varphi )$ is analytic in every
 $\mathcal O_{sc}^{(1)}(\vec b)$ such that $\mathcal O_{sc}^{(1)}(\vec
 b)\cap \varPhi _1\neq \emptyset $. The estimates \eqref{varkappa _*} hold inside
 $\mathcal O_{sc}^{(1)}(\vec b)$.\end{lem}
\begin{proof}By Lemma \ref{size}, the length of each connected component of
$\mathcalO_{s}^{(1)}(\vec b)$ is less than
$c_1k^{-3\frac{1}{3}-5s_1-\delta }$. Thus, each connected component
of $\mathcalO_{s}^{(1)}(\vec b)$, which has a non-empty intersection
with $\varPhi _1$, is, in fact, in $\hat \varPhi _1$.
 By Lemma \ref{L:2.13}(3), $\varkappa _1(\varphi )$ is analytic in in $\hat \varPhi
 _1$ and, therefore, in $\mathcalO_{s}^{(1)}(\vec b)$.
\end{proof}

\begin{lem}\label{norm**}  If $100s_1<\beta
_1<1/12-28s_1-14\delta $ and $\varphi \in \hat \varPhi _1\setminus
\mathcal O_{s}^{(1)}(\vec b)$, then \begin{equation} \left\|\bigl(H
(\vec \varkappa _1(\varphi )+\vec
b)-k^{2}\bigr)^{-1}\right\|<k^{J^{(1)}},\ \ J^{(1)}=5c_1k^{2/3+s_1}.
\label{normres**}\end{equation} This estimate is stable in the
$k^{-4-6s_1-4\delta }$ neighborhood of $\hat \varPhi _1\setminus
\mathcal O_s^{(1)}(\vec b)$. The resolvent $(H (\vec \varkappa _1
(\varphi )+\vec b)-k^{2}\bigr)^{-1}$ is an analytic function of
$\varphi $ in every component of $\mathcal O_{s}^{(1)}(\vec b)$,
whose intersection with $\varPhi _1$ is not empty. The only
singularities of the resolvent are poles. The number of poles in
each connected component does not exceed $c_1k^{2/3+s_1}$. The total
number of poles in all components of $\mathcal O_{s}^{(1)}(\vec b)$,
whose intersection with $\varPhi _1$ is not empty, is less then
$c_0k^{2+2s_1}$.\end{lem} \begin{proof} Suppose  $\varphi \in
\mathcal O_{c}^{(1)}(\vec b)_i$. Using Lemma \ref{norm*}, we obtain
\begin{equation} \left\|\prod _{n_i=1}^{M_i}(\varphi -\varphi
_{i,n_i})(H (\vec \varkappa _*(\varphi )+\vec
b)-k^{2}\bigr)^{-1}\right\|<cr^{c_1k^{2/3}}k^{4}, \label{us}
\end{equation}
$r$ being the maximal size of the components $\mathcal
O_{c}^{(1)}(\vec b)_i$, $r=k^{-\gamma }$. Let $\varphi \in \mathcal
O_{c}^{(1)}(\vec b)_i\setminus \mathcal O_{s}^{(1)}(\vec b)$.
Considering that $|\varphi -\varphi _{i,n_i}|>k^{-4-6s_1-3\delta}$,
 we obtain
\begin{equation}2\left\|(H (\vec \varkappa _*(\varphi )+\vec
b)-k^{2}\bigr)^{-1}\right\|<k^{5c_1k^{2/3+s_1}}. \label{us1}
\end{equation} Combining the last estimate with \eqref{normres*}, we
obtain that
\begin{equation}2\left\|(H (\vec \varkappa _*(\varphi )+\vec
b)-k^{2}\bigr)^{-1}\right\|<k^{5c_1k^{2/3+s_1}}\ \ \mbox{when
}\varphi \in \hat \varPhi _1 \setminus \mathcal O_{s}^{(1)}(\vec b).
\label{us1a}
\end{equation}
           Hilbert identity yields:
\begin{equation}\bigl(H^{(1)} (\vec \varkappa _1 (\varphi )+\vec
b)-k^{2}\bigr)^{-1}- \bigl(H^{(1)} (\vec \varkappa _*(\varphi )+\vec
b)-k^{2}\bigr)^{-1}= \bigl(H^{(1)} (\vec \varkappa _1 (\varphi
)+\vec b)-k^{2}\bigr)^{-1}\mathcalE, \label{HilbertIdentity}
\end{equation} $$\mathcalE=\bigl(H_0 (\vec \varkappa _*(\varphi
)+\vec b)-H_0(\vec \varkappa _1 (\varphi )+\vec b)\bigr)\bigl(
H^{(1)} (\vec \varkappa _*(\varphi )+\vec b)-k^{2}\bigr)^{-1}.$$ Let
us show that $\|\mathcalE \|<\frac{1}{2}$. Indeed, we write
$\mathcalE $ in the form $\mathcalE =\mathcalE _1A,$ $$\mathcalE
_1=\Bigl(H_0 (\vec \varkappa _*(\varphi )+\vec b)-H_0(\vec \varkappa
_1 (\varphi )+\vec b)\Bigr)\bigl(H _0(\vec \varkappa _*(\varphi
)+\vec b)+k^{2}\bigr)^{-1},$$ $$ A=\bigl(H _0(\vec \varkappa
_*(\varphi )+\vec b)+k^{2}\bigr)\bigl( H^{(1)} (\vec \varkappa
_*(\varphi )+\vec b)-k^{2}\bigr)^{-1}.$$ It is easy to show that
$\mathcalE _1$ is a diagonal operator and $\|\mathcalE _1\|\leq
|\varkappa _*(\varphi )-\varkappa _1 (\varphi )|.$ Using (\ref{35}),
we obtain:
$$\|\mathcalE _1\|\leq ck^{-1-(\beta _0-4s_1-4\delta)k^{3/4}}.$$  It
easily follows from formula for $A$ that $\|A\|\leq
1+2k^2\|(\bigl(H^{(1)} (\vec \varkappa _*(\varphi )+\vec
b)-k^{2}\bigr)^{-1}\|.$ Using \eqref{us1}, we get $\|A\|\leq
k^{6c_1k^{2/3+s_1}}.$ Multiplying the estimates for the norms
$\mathcalE_1$ and $\|A\|$, we get $\|\mathcalE \|=o(1)$. Now
\eqref{HilbertIdentity} yields: \begin{equation}\left\|\bigl(H^{(1)}
(\vec \varkappa _1 (\varphi )+\vec b)-k^{2}\bigr)^{-1}\right\|\leq
2\left\|\bigl(H^{(1)} (\vec \varkappa _*(\varphi )+\vec
b)-k^{2}\bigr)^{-1}\right\|<k^{5c_1k^{2/3+s_1}}. \label{plane}
\end{equation}
Considering that the discs in $\mathcal O^{(1)}(\vec b)$ have the
radius $k^{-4-6s_1-3\delta }$, we easily get that the estimate
\eqref{normres**} is stable in the $k^{-4-6s_1-4\delta }$
neighborhood of $\hat \varPhi _*\setminus \mathcal O_s^{(1)}(\vec
b)$.

 Next, we use  determinants to estimate the number of
poles of $\bigl(H^{(1)} (\vec \varkappa _1 (\varphi )+\vec
b)-k^{2}\bigr)^{-1}$. We will follow the scheme established in Lemma
\ref{norm}.  First, we note that \eqref{HilbertIdentity} and
\eqref{plane} hold not only for $\vec \varkappa _1 (\varphi )$, but
also for all points in the segment between $\vec \varkappa _*
(\varphi )$ and $\vec \varkappa _1 (\varphi )$, i.e., for $\vec
\varkappa _* (\varphi )+\alpha \bigl(\vec \varkappa _1 (\varphi
)-\vec \varkappa _* (\varphi )\bigr)$, $\alpha \in [0,1]$. Second,
we consider the family of projectors $P_N$ and the operator
$H^{(1)}_N=P_NH^{(1)}P_N$. We use analogs of (\ref{us1}),
\eqref{HilbertIdentity} for operators $H_N$ and $\alpha \in [0,1]$.
Using a a multistep procedure similar to that in Lemma \ref{norm},
see \eqref{alpha}--\eqref{determinant1}, we obtain that
$\bigl(H^{(1)}_{N}(\vec \varkappa _{*}(\varphi )+\vec
b)-k^{2}\bigr)^{-1}$ and $\bigl(H^{(1)}_{N}(\vec \varkappa
_{1}(\varphi )+\vec b)-k^{2}\bigr)^{-1}$ have the same number of
poles inside each connected component of $\mathcalO_{s}^{(1)}(\vec
b)$, whose intersection with $\varPhi _1$ is not empty. Considering
exactly as in Lemma \ref{norm}, see page \pageref{21}, we show that
all operators $\bigl(H^{(1)}_{N}(\vec \varkappa _{*}(\varphi )+\vec
b)-k^{2}\bigr)^{-1}$ with sufficiently large $N$ have the same
number of poles $M$, $M<c_1k^{2/3+s_1}$, inside a component of of
$\mathcalO_{s}^{(1)}(\vec b)$, whose intersection with $\varPhi _1$
is not empty. Hence, the same is true for $\bigl(H^{(1)}_{N}(\vec
\varkappa _{1}(\varphi )+\vec b)-k^{2}\bigr)^{-1}$. Considering
further as in the proof of Lemma \ref{norm}, we obtain that
$\bigl(H^{(1)}(\vec \varkappa _{1}(\varphi )+\vec
b)-k^{2}\bigr)^{-1}$ has no more that $c_1k^{2/3}$ poles.
\end{proof}

\subsubsection{The set ${\mathcal O}_s^{(1)}(\vec b)$ for small $b
_0$}\label{SS:3.7.1}

Everything  we  considered so far is valid for $\vec b$ obeying the
inequality $b_0 \geq k^{-1-16s_1-12\delta}$, here $b_0$ is the
distance from $\vec b$ to the nearest vertex of $K_1$.
  However, in the next section and later, $b_0$ will be taken
smaller, since the reciprocal lattice is getting finer with each
step. To prepare for this, let us consider $\vec b$ being close to a
vertex of $K_1$: \begin{equation} 0<b_0 < k^{-1-16s_1-12\delta }.
\label{july9d} \end{equation} We show that for such $\vec b$ the
resolvent $\bigl(H^{(1)}(\vec y(\varphi ))-k^{2}\bigr)^{-1}$, $\vec
y(\varphi )=\vec \varkappa _1(\varphi )+\vec b$ has no more than two
poles $\varphi ^{\pm}$ in $\hat \varPhi _1$. We surround these poles
by two contours $\gamma ^{(1)\pm}$ and obtain estimate
\eqref{3.7.1.10.1/2} for $\bigl(H^{(1)}(\vec y(\varphi
))-k^{2}\bigr)^{-1}$ when
 $\varphi $ is outside $\gamma ^{(1)\pm}$.

%%%%%%%%%In fact, it easily follows from (\ref{july9d}) that:
%%%%%%%%%\begin{equation} b_0=_{k \to \infty}o(k^{-1-4s_1-2\delta}).
%%%%%%%%%\label{Jan11} \end{equation}

Suppose $|\vec b |=b_0$, i.e.,  the closest vertex of $K_1$ for
$\vec b$ is $(0,0)$. The perturbation series (\ref{2.66}) converge
for both $\lambda ^{(1)}\left(\vec \varkappa _1(\varphi )\right)$
and $\lambda ^{(1)}\left(\vec y(\varphi )\right)$ when $\varphi \in
\hat \varPhi _1$, and both functions are holomorphic in $\hat
\varPhi _1$
 (Lemma \ref{L:2.13}). Note that
$\lambda ^{(1)}\left(\vec \varkappa _1(\varphi )\right)=k^{2}$ for
all $\varphi \in \hat \varPhi _1$. We base our further
considerations on these perturbation series expansions. For $\vec b$
being close to a vertex $\vec e$ other than $(0,0)$, we take $\vec
y(\varphi )=\vec \varkappa _1(\varphi )+\vec b-\vec e$.

We define $\varphi _b\in [0,2\pi )$ by the formula $\vec b=b_0(\cos
\varphi _b, \sin \varphi _b)$ when  $|\vec b |=b_0$, and by the
analogous formula $\vec b-\vec e=b_0(\cos \varphi _b, \sin \varphi
_b)$ when $\vec b$ is close to a vertex $\vec e$ other than $(0,0)$.

\begin{lem}\label{L:3.7.1.1} If $\vec b$ satisfies (\ref{july9d})
and $|\epsilon_0|<b_0k^{-1-16s_1-12\delta}$, then the equation
    \begin{equation}\label{3.7.1.2}
    \lambda ^{(1)}\left(\vec y (\varphi)\right)=k^{2}+\epsilon_0
    \end{equation}
 has no more than two
solutions, $\varphi^{\pm}_{\epsilon _0}$, in $\hat \varPhi _1$. They
satisfy the inequality
    \begin{equation}
    \left|\varphi^{\pm}_{\epsilon _0}-(\varphi_b \pm \pi
    /2)\right|<\frac{1}{8}k^{-2-16s_1-11\delta}. \label{july5a}
    \end{equation}
\end{lem} \begin{proof}  Suppose $W_1=0$ and $|\vec b|=b_0$, i.e.,
the closest vertex of $K_1$ for $\vec b$ is $(0,0)$. Then the
equation (\ref{3.7.1.2}) has the form $|k\vec \nu +\vec
b|_*^{2}=k^{2}+\epsilon_0$, $\vec \nu =(\cos \varphi ,\sin \varphi
)$. It is easy to show that it has two solutions
$\varphi^{\pm}_{\epsilon _0}$ satisfying (\ref{july5a}). Applying
perturbative arguments and Rouch\'{e}'s
    theorem, we prove the lemma for nonzero $W_1$. A detailed proof can be found in  Appendix 1.
    In the case
    when $\vec b$ is close to a vertex other than $(0,0)$, the
    considerations are the same up to a parallel shift.
 \end{proof}
\begin{lem}\label{L:july5a} Suppose $\vec b$ satisfies
(\ref{july9d}), $\varphi \in \hat \varPhi _1$ and obeys the
inequality analogous to (\ref{july5a}): $\left|\varphi-(\varphi_b
\pm \pi
    /2)\right|<k^{-2-16s_1-11\delta}$. Then,
    \begin{equation}\label{3.7.1.6.1/2}
    \frac{\partial}{\partial \varphi}\lambda^{(1)}\left(\vec
    y(\varphi)\right)=_{k\to \infty}\pm 2b_0k\bigl(1+o(1)\bigr).
    \end{equation}
\end{lem} \begin{proof} Let $W_1=0$ and $|\vec b|=b_0$. Then
$\lambda^{(1)}(\vec
    y(\varphi))=|k\vec \nu +\vec b|^{2}_*$ and
    $$\frac{\partial}{\partial \varphi}|k\vec \nu +\vec b|^{2}_*=
    2 \langle k\vec \nu +\vec b,k\vec \mu \rangle_*, \ \ \vec \mu
    =\frac{\partial \vec \nu }{\partial \varphi }
    =(-\sin \varphi, \cos \varphi ).$$
    For $\varphi $ close to $\varphi_b \pm \pi
    /2$, we have $\langle \vec b,\vec \mu
\rangle =\pm b_0(1+o(1))$. Considering also that $\langle \vec \mu,
\vec \nu \rangle =0$, we obtain $\frac{\partial}{\partial
\varphi}|k\vec \nu +b|^{2}=\pm 2b_0k(1+o(1))$. Applying perturbative
arguments, we get a similar formula for nonzero $W_1$. For a
detailed proof see Appendix 2. In the case
    when $\vec b$ is close to a vertex other than $(0,0)$, the
    considerations are the same up to a parallel shift.
 \end{proof}

\begin{definition} \label{SN} Let $\Gamma^{(1)\pm}(\vec b)$ be two
open disks centered at $\varphi_0^{\pm}\in \hat \varPhi _1$
\footnote{$\varphi_0^{\pm}$ is $\varphi_{\epsilon _0}^{\pm}$ for
$\epsilon _0 =0$.} with the radius
 $r^{(1)}=k^{-4 -6s_1-3\delta }$; $\gamma^{(1)\pm}(\vec
b)$ be their boundary circles and ${\mathcal O}_s^{(1)}(\vec
b)=\Gamma^{(1)+}\cup \Gamma^{(1)-}$. \end{definition}

\begin{lem}\label{L:3.7.1} For any $\varphi$ in $ \hat \varPhi _1\setminus
{\mathcal O}_s^{(1)}(\vec b)$,
    \begin{equation}\label{3.7.1.10}
    \bigl|\lambda^{(1)}\bigl(\vec
    y(\varphi)\bigr)-k^{2}\bigr|\geq b_0kr^{(1)}.
    \end{equation}
    This estimate is stable in the
$k^{-4-6s_1-4\delta }$ neighborhood of $\hat \varPhi _1\setminus
{\mathcal O}_s^{(1)}(\vec b)$.
\end{lem} \begin{proof} Suppose (\ref{3.7.1.10}) does not hold for
some $\varphi$ in $\hat \varPhi _1\setminus {\mathcal
O}_s^{(1)}(\vec b)$. This means that $\varphi$ satisfies  equation
(\ref{3.7.1.2}) with some $\varepsilon _0$: $|\varepsilon _0|
<b_0kr^{(1)}$. By Lemma \ref{L:3.7.1.1}, $\varphi $ obeys
(\ref{july5a}). Thus $\varphi$ could be either $\varphi_{\epsilon
_0}^{+}$ or $\varphi_{\epsilon _0}^{-}$. Without loss of generality,
assume $\varphi =\varphi_{\epsilon _0}^{+}$. Obviously, the
$r^{(1)}$-neighborhood of $\varphi_{\epsilon _0}^{+}$ also satisfies
conditions of Lemma \ref{L:july5a}, since $r^{(1)}$ is small
considering with the size $k^{-2-16s_1-11\delta }$ of the
neighborhood $\hat \varPhi _1$.
     Using (\ref{3.7.1.6.1/2}), we obtain that $\bigl|\lambda^{(1)}\bigl(\vec
    y(\varphi)\bigr)-k^{2}-\varepsilon _0\bigr|=2kb_0r^{(1)}(1+o(1))$ on the boundary of this neighborhood. Applying Rouche's
    Theorem, we obtain that there is a point
$\varphi_{0}^{+}$ in this neighborhood.
 It immediately follows that $\left|\varphi
    -\varphi_0^{+}\right|<r^{(1)}$, i.e., $\varphi \in \Gamma ^{(1)+} \subset O_s(\vec b)$, which contradicts the assumption
     $\varphi \in \varPhi _1\setminus {\mathcal O}_s^{(1)}(\vec
    b)$.
    \end{proof}

\begin{lem}\label{L:july9c} Let $0<b_0<k^{-1-16s_1-12\delta }$. For
any $\varphi $ in  $\hat \varPhi _1\setminus {\mathcal
O}_s^{(1)}(\vec b)$ or its $k^{-4-6s_1-4\delta }$ neighborhood
    \begin{equation}\label{3.7.1.10.1/2}
    \Bigl\|\Bigl(H^{(1)}\bigl(\vec
    y(\varphi)\bigr)-k^{2}\Bigr)^{-1}\Bigr\|<\frac{4k^{3+6s_1+4\delta }}{b_0}.
    \end{equation}
    The resolvent  is an analytic
function of $\varphi $ in every component of $\mathcal
O_{s}^{(1)}(\vec b)$, whose intersection with $\varPhi _1$ is not
empty. The only singularities of the resolvent are poles. The number
of poles in each connected component does not exceed two.
\end{lem}
%%%%%%%%%%%\begin{cor}\label{C:new} If $\epsilon _1k^{-2l+1-2\delta
%%%%%%%%%%%}<b_0<k^{-2l+9+12s_1+7\delta}$ and $\varphi \in \varPhi _1\setminus
%%%%%%%%%%%{\mathcal O}_s(\vec b)$, then
%%%%%%%%%%%    \begin{equation}\label{3.7.1.10.1/2++}
 %%%%%%%%%%%   \Bigl\|\Bigl(H^{(1)}(\vec
 %%%%%%%%%%%   y(\varphi))-k^{2}\Bigr)^{-1}\Bigr\|<\frac{1}{\epsilon _1^2},
 %%%%%%%%%%%   \end{equation}
 %%%%%%%%%%%   \begin{equation}\label{3.7.1.13++}
 %%%%%%%%%%%   \Bigl\|\Bigl(H^{(1)}(\vec y(\varphi))-k^{2}\Bigr)^{-1}\Bigr\|_1<
%%%%%%%%%%%    \frac{c_0k^{2+2s_1}}{\epsilon _1^2}
 %%%%%%%%%%%   \end{equation}
%%%%%%%%%%%\end{cor} Corollary follows from the condition on $b_0$ and the
%%%%%%%%%%%estimate $r'k^{-3\delta}>16\epsilon _1$, which is obviously valid
%%%%%%%%%%%for sufficiently large $k$.
\begin{proof} Suppose we have proven that
\begin{equation}\label{16}\left\|\Bigl(\lambda^{(1)}\bigl(\vec
y(\varphi)\bigr)-k^{2}\Bigr)
    \Bigl(H^{(1)}\bigl(\vec y(\varphi)\bigr)-k^{2}\Bigr)^{-1}\right\| <
    2\end{equation}
for all $\varphi$ in $ \hat \varPhi _1$. Then, using
(\ref{3.7.1.10}) with $r^{(1)}=k^{-4-6s_1-3\delta }$, we easily get
(\ref{3.7.1.10.1/2}). Let us prove \eqref{16}. By Lemma
\ref{L:2.13}, $\vec \varkappa _1(\varphi )$ is a holomorphic
function in $\hat \varPhi _1$ and estimates \eqref{2.75} hold.
Therefore,  $\vec y(\varphi )=\vec k(\varphi )
+O(k^{-1-16s_1-12\delta })$. Using \eqref{InE:resolvent-1'} with
$2\beta = 1-15s_1-9\delta $, we obtain
\begin{equation}\Bigl\|\Bigl(H^{(1)}_{\alpha }(\vec
    y(\varphi))-z\Bigr)^{-1}\Bigr\|<k^{16s_1+10\delta }, \ \ z\in
    C_1,\ \ \alpha \in [0,1],
    \label{est} \end{equation} $C_1$ being given by \eqref{E:circle-1}. Using
considerations similar to those in Lemma \ref{norm} (see
\eqref{alpha} and further), we prove that $\Bigl(H_{\alpha }^{(1)}
\bigl(\vec y(\varphi )\bigr)-z\Bigr)^{-1}$ has at most the same
number of poles $z$ inside $C_1$ as $\Bigl(H_{0 }^{(1)} \bigl(\vec
y(\varphi )\bigr)-z\Bigr)^{-1}$, i.e., at most one pole. The pole
obviously exists and is located at the point $z=\lambda
^{(1)}\bigl(\vec y(\varphi )\bigr)$. Hence,
$\Bigl(\lambda^{(1)}(\vec y(\varphi))-z\Bigr)
    \Bigl(H^{(1)}(\vec y(\varphi))-z\Bigr)^{-1}$  is a holomorphic
    function of $z$ inside $C_1$ for a fixed $\varphi $.
    From the definition of $C_1$ and estimate \eqref{InE:lambda_j^1} it
    follows that $\left|\lambda^{(1)}(\vec
y(\varphi))-z\right|<2k^{-16s_1-10\delta }$. Multiplying \eqref{est}
and the last  estimate, we get $$\left\|\Bigl(\lambda^{(1)}(\vec
y(\varphi))-z\Bigr)
    \Bigl(H^{(1)}(\vec
    y(\varphi))-z\Bigr)^{-1}\right\|<2,\ \ z\in C_1.$$
    Using the maximum principle, we obtain the same estimate inside
    the circle and, therefore, for $z=k^2$. Thus, we have proved  \eqref{3.7.1.10.1/2} for $\varphi \in
    \hat \varPhi _1\setminus {\mathcal
O}_s^{(1)}(\vec b)$. It is easy to see that estimate
\eqref{3.7.1.10.1/2} is stable with respect
    to a perturbation of $\varphi $ of order $k^{-4-6s_1-4\delta }$.

    Suppose $\mathcal
O_{s}^{(1)}(\vec b)\cap \varPhi _1\not = \emptyset .$ This means
that $O_{s}^{(1)}(\vec b) \subset \hat \varPhi _1$. Considering
\eqref{16}, we see that the only poles of the resolvent are the
zeros of the function $\lambda^{(1)}\bigl(\vec
y(\varphi)\bigr)-k^{2}$. By Lemma \ref{L:3.7.1.1}, the number of
zeros does not exceed two.
\end{proof}

\section{The Second Approximation} \label{chapt4}

  Let us start with establishing a
 lower bound for $k$. Let $\eta >3\cdot 10^4$.
 Since $\eta s_1>2+4s_1$, there is a number $k_*>e$ such that
\begin{equation}
 C_*(1+s_1)k^{2+4s_1}\ln k <k^{\eta s_1},\  C_*=400(c_0+c_1+1)^2,\ c_0=32d_1d_2.\label{k}
 \end{equation}
  for any $k>k_*$. Assume also, that $k_*$ is sufficiently
 large to ensure validity of all estimates in the first  step
 for any $k>k_*$. In particular we assume that all $o(1)$ in the
 first step satisfy the estimate $|o(1)|<10^{-2}$ when $k>k_*$.

\setcounter{equation}{0}

\subsection{Operator $H_{\alpha}^{(2)}$}

Choosing $s_2=2s_1$, we define the second operator
$H_{\alpha}^{(2)}$ by the formula:
    \begin{equation}\label{3.1}
     H_{\alpha }^{(2)}=H^{(1)}+\alpha W_2,\quad     ( 0\leq \alpha \leq
     1), \qquad
     W_2=\sum_{r=M_1+1}^{M_2}V_r,
     \end{equation}
where $H^{(1)} $ is defined by (\ref{E:H_1}), $M_2$ is chosen in
such a way that $2^{M_2} \approx k^{s_2}$. Obviously, the periods of
$W_2$ are $2^{M_2-1} (d_1,0)$ and $2^{M_2-1} (0,d_2)$. We will write
them in the form: $N_1(a_1,0)$ and $N_1(0,a_2)$, where $a_1,a_2$ are
the periods of $W_1$ and $N_1=2^{M_2-M_1},\ \frac{1}{4}k^{s_2-s_1} <
N_1 < 4 k^{s_2-s_1}$. Note that
    \begin{equation}\|W_2\|_{\infty} \leq \sum_{r=M_1+1}^{M_2}
    \|V_r\|_{\infty} \leq \sum_{r=M_1+1}^{M_2}\exp(-2^{\eta r})
    <\exp(-k^{\eta s_1}). \label{W2}
    \end{equation}
\subsubsection{Multiple Periods of $W_1(x)$}\label{S:2.2}

 The operator $H^{(1)}=H_0 +
 W_1(x)$
has the periods $a_1, a_2$. The corresponding family of operators,
$\{H^{(1)}(t)\}_{t \in K_1}$, acts in $L_2(Q_1)$, where $Q_1=[0,a_1]
\times [0,a_2]$ and $K_1=[0, 2\pi/a_1)\times [0, 2\pi/a_2)$.
Eigenvalues of $H^{(1)}(t)$ are denoted by $\lambda_n^{(1)}(t)$, $n
\in \N$, and its spectrum by $\Lambda ^{(1)}(t)$. Now let us
consider the same $W_1(x)$ as a periodic function with the periods
$N_1a_1, N_1a_2$. Obviously, the definition of the operator
$H^{(1)}$ does not depend on the way how we define the periods of
$W_1$. However, the family of operators $\{H^{(1)}(t)\}_{t \in K_1}
$ does change, when we replace the periods $a_1, a_2$ by $N_1a_1,
N_1a_2$. The family of operators $\{H^{(1)}(t)\}_{t \in K_1}$ has to
be  replaced by a family of operators $\{ \tilde{H}^{(1)}(\tau)
\}_{\tau \in K_2}$ acting in $L_2(Q_2)$, where $Q_2=[0,N_1a_1]
\times [0,N_1a_2]$ and $K_2=[0, 2\pi/N_1a_1)\times [0,
2\pi/N_1a_2)$. We denote eigenvalues of $\tilde{H}^{(1)}(\tau)$ by
$\tilde{\lambda}_n^{(1)}(\tau)$, $n \in \N$ and its spectrum by
$\tilde{\Lambda}^{(1)}(\tau)$. The next lemma establishes a
connection between spectra of operators $H^{(1)}(t)$ and
$\tilde{H}^{(1)}(\tau)$. It easily follows from Bloch theory (see
e.g. \cite{RS}). \begin{lem}\label{L:3.1} For any $\tau \in K_2$,
    \begin{equation}\label{3.3}
    \tilde{\Lambda}^{(1)}(\tau)=\bigcup_{p \in P} \Lambda^{(1)}(t_p),
    \end{equation}
 where \begin{equation}
  P=\{p=(p_1,p_2) \in \Z^2 : 0 \leq p_1 \leq N_1-1,\ 0 \leq p_2
\leq N_1-1\} \label{May10a} \end{equation}
 and $t_p=(t_{p,1}, t_{p,2})=(\tau _1 +2\pi
p_1/N_1a_1, \tau_2+2\pi p_2/N_1a_2) \in K_1$, see Fig. 6. \end{lem}
\begin{figure}\label{F:6} \centering
    \psfrag{2p/a1}{\hspace{2mm}\small{$\frac{2\pi}{a_1}$}} \psfrag{2p/a2}{\small{$\frac{2\pi}{a_2}$}}
   \psfrag{2p/Na1}{\small{$\frac{2\pi}{N_1a_1}$}}
    \psfrag{2p/Na2}{\hspace{2mm}\small{$\frac{2\pi}{N_1a_2}$}}
\includegraphics[totalheight=.2\textheight]{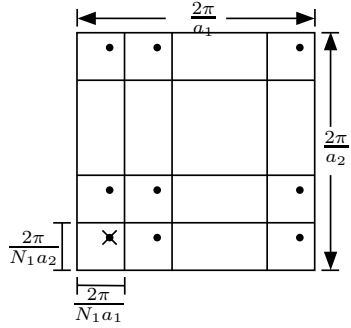}
\caption{Relation between $\tau \ (\times )$ and $t_p \ (\cdot )$}
\end{figure}

We defined  isoenergetic set $S_1(\lambda ) \subset K_1$ of
$H^{(1)}$ by formula (\ref{2.65.1}). Obviously, this definition is
directly associated with the family of operators $H^{(1)}(t)$ and,
therefore, with periods $a_1, a_2$, which we assigned to $W_1(x)$.
Now, assuming that the periods are equal to $N_1a_1, N_1a_2$, we
give an analogous definition of the isoenergetic set
$\tilde{S}_1(\lambda )$ in $K_2$:
   \begin{equation}
   \tilde{S}_1(\lambda ):=\{\tau \in K_2: \exists n \in \N:\ \
    \tilde{\lambda}_n^{(1)}(\tau)=\lambda \}.\label{8888}
    \end{equation}
By Lemma \ref{L:3.1}, $\tilde{S}_1(\lambda )$ can be expressed as
follows:
   \begin{equation}
    \tilde{S}_1(\lambda )=\Bigl\{\tau \in K_2: \exists n \in \N,\ p \in P:\
    \lambda_n^{(1)}\bigl(\tau+2\pi p/N_1a\bigr)=\lambda
    \Bigr\},
\label{8888a} \end{equation} $$ 2\pi
    p/N_1a=\left(\frac{2\pi p_1}{N_1a_1},\frac{2\pi
    p_2}{N_1a_2}\right).$$ The relation between $S_1 (\lambda)$
and $\tilde{S}_1 (\lambda)$ can be easily understood from the
geometric point of view as
    \begin{equation}
    \tilde{S}_1 (\lambda)=\mathcal{K}_2 S_1 (\lambda), \label{8888b}
    \end{equation}
where $\mathcal{K}_2$ is the parallel shift into $K_2$, i.e.,
    \begin{equation}
    \mathcal{K}_2:\R^2 \rightarrow K_2,\   \mathcal{K}_2(\tau+2\pi
    m/N_1a)=\tau,\ m \in \Z^2,\ \tau \in K_2.\label{8888c}
\end{equation}
    Thus, $\tilde{S}_1(\lambda )$ is obtained from $S_1 (\lambda)$ by
    cutting $S_1 (\lambda)$ into pieces of the size $K_2$ and shifting them
    together in $K_2$.

     \begin{definition} \label{D:May6}
    We say that $\tau $ is a point of
    self-intersection of $\tilde{S}_1(\lambda )$,  if there is a
    pair $m,\hat m\in N$, $m\neq \hat m$ such that $\tilde{\lambda}_m^{(1)}(\tau)=
    \tilde{\lambda}_{\hat m}^{(1)}(\tau)=\lambda $.\end{definition}
    \begin{remark} \label{R:May6} By Lemma \ref{L:3.1}, $\tau $ is a point of
    self-intersection of $\tilde{S}_1(\lambda )$,  if there is a
    pair $p,\hat p\in P$ and a pair $n,\hat n \in N$ such that $|p-\hat
    p|+|n-\hat n|\neq 0$ and $\lambda_n^{(1)}(\tau+2\pi
    p/N_1a)=
    \lambda_{\hat n}^{(1)}(\tau+2\pi
    \hat p/N_1a)=\lambda $.
    \end{remark}
Now let us recall that the isoenergetic set $S_1(\lambda )$ consists
of two parts:
    $S_1(\lambda )=\chi_1^*(\lambda )\cup \bigl(S_1(\lambda )
    \setminus \chi_1^*(\lambda ) \bigr),$
where $\chi_1^*(\lambda )$ is the first non-resonance set given by
(\ref{2.81}). Obviously $\mathcal{K}_2\chi_1^*(\lambda )\subset
\mathcal{K}_2S_1(\lambda)=\tilde{S}_1(\lambda )$ and can be
described by the formula: $$\mathcal{K}_2\chi_1^*(\lambda )=
\left\{\tau \in K_2:\ \exists p\in P: \tau +2\pi p/N_1a \in
\chi_1^*(\lambda )\right\}.$$ Let us consider only those
self-intersections of $\tilde{S}_1$ which belong to
$\mathcal{K}_2\chi_1^*(\lambda )$, i.e., we consider the points of
intersection of $\mathcal{K}_2\chi_1^*(\lambda )$ both with itself
and with $\tilde{S}_1(\lambda )\setminus
\mathcal{K}_2\chi_1^*(\lambda )$. \begin{lem}\label{L:May6} A
self-intersection $\tau $ of $\tilde{S}_1(\lambda )$ belongs to
$\mathcal{K}_2\chi_1^*(\lambda )$ if and only if there are a pair
$p,\hat p\in P$, $p\neq \hat p$ and a pair $n,\hat n\in N$ such that
$\tau +2\pi p/N_1a \in \chi_1^*(\lambda )$ and $\lambda
_n^{(1)}(\tau +2\pi p/N_1a)=\lambda _{\hat n}^{(1)}(\tau +2\pi \hat
p/N_1a)=\lambda$, the eigenvalue $\lambda _n^{(1)}(\tau +2\pi
p/N_1a)$ being given by the series (\ref{E:lambda_j^1}) with $t=\tau
+2\pi p/N_1a $ and
    $j $  uniquely defined by $t$ from the relation
    $p_j^{2}(t)\in \varepsilon _1$,
    \begin{equation}\varepsilon _1=
    (k^{2}-3k^{-16s_1-11\delta }, k^{2}+3k^{-16s_1-11\delta }).\label{varepsilon
    _1}
    \end{equation}
    \end{lem}
    \begin{proof} Suppose $\tau $ is a point of
     self-intersection  of $\tilde{S}_1(\lambda )$
belonging to $\mathcal{K}_2\chi_1^*(\lambda )$. Since, $\tau \in
\mathcal{K}_2\chi_1^*(\lambda )$, there is a $p\in P$ such that
$\tau +2\pi p/N_1a \in \chi_1^*(\lambda )$. By Lemma \ref{Apr4},
there is a single eigenvalue $\lambda _n^{(1)}(\tau +2\pi
p/N_1a)=\lambda$ of $H^{(1)}\left(\tau +2\pi p/N_1a \right)$
    in
    $\varepsilon _1$. It is given by the series (\ref{E:lambda_j^1}) with $t=\tau
+2\pi p/N_1a $ and
    $j $  uniquely defined by $t$ from the relation
    $p_j^{2}(t)\in \varepsilon _1$. Uniqueness means:
    $
    \lambda _n^{(1)}(\tau +2\pi
    p/N_1a)\neq \lambda _{\hat n}^{(1)}(\tau +2\pi
 p/N_1a)$ when $\hat
 n\neq n$.
 Since $\tau $ is a point of self-intersection of $\tilde S_1(\lambda )$,
 $\lambda
_n^{(1)}(\tau +2\pi p/N_1a)=\lambda _{\hat n}^{(1)}(\tau +2\pi \hat
p/N_1a)=\lambda$ for some $\hat p\neq p$. The converse part of the
lemma is trivial. \end{proof}

To obtain a new non-resonance set $\chi_2(\lambda )$ we remove from
$\mathcal{K}_2\chi_1^*(\lambda )$  a neighborhood of its
intersections (quasi-intersections) with the whole isoenergetic
surface $\tilde{S}_1(\lambda )$ given by
(\ref{8888})--(\ref{8888b}). More precisely, we remove from
$\mathcal{K}_2\chi_1^*(\lambda )$ the following set:
    \begin{multline} \label{3.2.1}
     \Omega _1(\lambda )=\{\tau \in \mathcal{K}_2\chi_1^*(\lambda ): \exists n,\hat{n} \in
    \N,\ p,\hat{p} \in P,\  p\neq \hat{p}:\ \lambda_{n}^{(1)}(\tau+2\pi
    p/N_1a)=\lambda ,\\ \tau+2\pi p/N_1a \in \chi_1^*(\lambda ),\
     |\lambda_{n}^{(1)}(\tau+2\pi p/N_1a)-\lambda_{\hat{n}}^{(1)}(\tau+2\pi
    \hat{p}/N_1a)| \leq \epsilon_1  \},
    \end{multline}
    where \footnote{Note that $\varepsilon _1$ is an interval, while $\epsilon _1$ is a number.}\begin{equation}
    \epsilon_1 =e^{-\frac{1}{4}k^{\eta
    s_1}}. \label{99s}
    \end{equation}
We define $ \chi_2(\lambda )$  by the formula:
    \begin{equation}\label{3.5.1}
     \chi_2(\lambda )=\mathcal{K}_2 \chi _1^*(\lambda )
    \ \setminus \Omega _1(\lambda ).
    \end{equation}

\subsection{Perturbation Formulae}\label{S:3.3}

Before proving the main result, we formulate  Geometric Lemma:
\begin{lem}[Geometric Lemma]\label{L:3.2}If $\lambda >k^{2}_*$,
there exists a non-resonance set $\chi _2(\lambda ,\delta ) \subset
\mathcal{K}_2\chi_1^*$ such that: \begin{enumerate} \item For any
$\tau \in \chi _2$, the following conditions hold: \begin{enumerate}
    \item There exists a unique $p \in P$
    such that $\tau + 2\pi p/N_1a \in \chi_1^*$.
    \footnote{From geometric point of view this means that $\chi _2 (\lambda)$
    does not have self-intersections.}
    \item The following relation holds:
    $$ \lambda _j^{(1)}(\tau + 2\pi p/N_1a)=k^{2},$$
    where $\lambda _j^{(1)}(\tau + 2\pi p/N_1a)$ is  given  by the perturbation series
    (\ref{E:lambda_j^1}) with $\alpha
    =1$, $j$ being uniquely defined by $t=\tau + 2\pi p/N_1a $ and
    the relation
    $p_j^{2}(t)\in \varepsilon _1$.
    \item The eigenvalue $\lambda _j^{(1)}(\tau + 2\pi p/N_1a)$ is
    a simple eigenvalue of $\tilde H^{(1)}(\tau )$ and
its distance from all other eigenvalues $\lambda _{\hat
n}^{(1)}(\tau + 2\pi \hat{p}/N_1a),\
    {\hat n} \in
    \N$ of $\tilde{H}_1(\tau )$ is greater than $\epsilon_1 = e^{-\frac{1}{4}k^{\eta s_1}}$:
    \begin{equation}\label{3.6}
    |\lambda _j^{(1)}(\tau + 2\pi p/N_1a)-\lambda _{\hat n}^{(1)}(\tau + 2\pi
    \hat{p}/N_1a)|>\epsilon_1.
    \end{equation}
\end{enumerate}
\item For any $\tau$ in the real $(\epsilon_1 k^{-1-\delta
})$-neighborhood   of $\chi_2$, there exists a unique $p \in P$ such
that $\tau +2\pi p/N_1a$ is in the $(\epsilon_1 k^{-1-\delta
})$-neighborhood  of $\chi_1^*$ and
    \begin{equation}\label{3.7}
    | \lambda_j^{(1)}(\tau +2\pi p/N_1a)-k^{2}| < 2\epsilon_1 k^{-\delta
    },
    \end{equation}
    $j$ being uniquely defined by $t=\tau + 2\pi p/N_1a $ $j$ and
    the relation
    $p_j^{2}(t)\in \varepsilon _1$. An estimate analogous to
    \eqref{3.6} holds:
    \begin{equation}\label{3.6a}
    2|\lambda _j^{(1)}(\tau + 2\pi p/N_1a)-\lambda _{\hat n}^{(1)}(\tau + 2\pi
    \hat{p}/N_1a)|>\epsilon_1.
    \end{equation}
    \item The second non-resonance set $\chi_2$ has an asymptotically
full measure in $\chi_1^*$ in the following sense:
\begin{equation}\label{3.9} \frac{L(\mathcal{K}_2\chi_1^* \setminus
\chi _2))}{L(\chi_1^*)}<c_0k^{-2-2s_1}. \end{equation}
\end{enumerate} \end{lem} \begin{cor}\label{C:3.3} If $\tau$ belongs
to the complex $(\epsilon_1 k^{-1-\delta})-$neighborhood of the
second non-resonance set $\chi_2(\lambda,\delta)$, then for any
$z\in \C$ lying on the circle \begin{equation} C_2=\{ z: |z-k^{2}|
=\epsilon_1/2 \}, \label{C_2} \end{equation}the following
inequalities hold: \begin{equation}
\|(\tilde{H}^{(1)}(\tau)-z)^{-1}\| \leq
\frac{4}{\epsilon_1},\label{3.10} \end{equation}
%%%%%%%\\
%%%%%%%&\|(\tilde{H}^{(1)}(\tau)-z)^{-1}\|_1 \leq
%%%%%%%\frac{c_0k^{2+2s_2}}{\epsilon_1}, \ \ \ c_0=32d_1d_2.
%%%%%%%\label{3.11} \end{align}
\end{cor} Corollary is proven in Appendix 3.

\begin{proof} Let us consider
\begin{equation}\label{O^2}\mathcalO^{(2)}=\cup _{p'\in P\setminus
\{0\}}\mathcalO_{s}\left(\frac{2\pi p'}{N_1a}\right),\end{equation}
where $\mathcalO_{s}(\cdot )$ is defined by (\ref{small+}) in
Definition \ref{D:small}, $\vec b=\frac{2\pi p'}{N_1a}$. Note, that
the definition make sense, since $|\frac{2\pi
p'}{N_1a}|>d_{max}^{-1}k^{-s_2}>k^{-1-16s_1-12\delta }$,
$d_{max}=\max\{d_1,d_2\}$. Let
\begin{equation}\varPhi _2=\varPhi _1\setminus \mathcalO^{(2)},\ \ \
\Theta _2=\varPhi _2\cap [0,2\pi ). \label{3.19a} \end{equation} By
Lemma \ref{norm**},
\begin{equation} \left\|\left(H ^{(1)}\left(\vec \varkappa
_1(\varphi )+\frac{2\pi
p'}{N_1a}\right)-k^{2}\right)^{-1}\right\|<k^{J^{(1)}},
J^{(1)}=c_1k^{2/3+s_1}.\label{normres***}\end{equation} for all
$p'\in P\setminus\{0\}$ and $\varphi \in \varPhi _2$. Estimate
\eqref{k} yields $k^{J^{(1)}}<\epsilon _1^{-1}$.
 \begin{figure}
 \centering
\psfrag{Phi_3}{$\varPhi_3$}
\includegraphics[totalheight=.2\textheight]{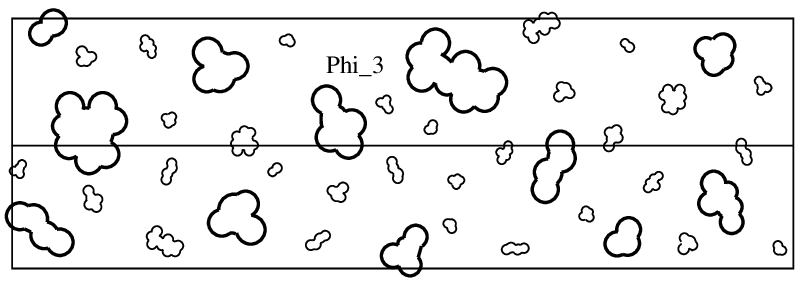}
\caption{The set $\varPhi
  _3$.}\label{F:8}
\end{figure}
\label{pageF:8} We consider $\mathcalD_{1,nonres}\subset
\mathcalD_{1}$:
\begin{equation} \label{D_{1,nonres}}
\mathcalD_{1,nonres}=\left\{\vec \varkappa _1(\varphi ), \ \varphi
\in \Theta _2\right\}.\end{equation}We define $\chi _2$ by the
formula: \begin{equation} \label{chi_2} \chi
_2=\mathcalK_2\mathcalD_{1,nonres}.\end{equation} By definition of
$\mathcalK_2$, for every $\tau $ in $\chi _2$, there are $p\in P$
and $j\in Z^2$ such that \begin{equation} \label{tau}\tau +
\frac{2\pi p}{N_1a}+\frac{2\pi j}{a}=\vec \varkappa _1(\varphi ),\ \
\vec \varkappa _1(\varphi )\in \mathcalD_{1,nonres}.\end{equation}
Considering \eqref{normres***},  the estimate $k^{J^{(1)}}<\epsilon
_1^{-1}$ and the definition of $\mathcalD_{1,nonres}$, we obtain:
\begin{equation} \left\|\left(H ^{(1)}\left(\tau+\frac{2\pi \hat
p}{N_1a}\right)-k^{2}\right)^{-1}\right\|<\epsilon _1^{-1}, \ \ \
\hat p=p+p'. \label{normres****}\end{equation} Note that the index
$j$ does not play a role, since it just produces the shift $j$ of
indices of the matrix elements of the resolvent. Considering that
$p'$ can be any but zero, we obtain that (\ref{normres****}) holds
for all $\hat p\in P\setminus\{p\}$. Taking into account that
$\lambda _j\left(\tau + \frac{2\pi p}{N_1a}\right)=k^{2}$ and
inequality \eqref{normres****}, we arrive at \eqref{3.6} for all
$\hat p\neq p$. It remains to check \eqref{3.6} for $p=\hat p$. Let
$t=\tau +\frac{2\pi p}{N_1a}$. By (\ref{tau}), $t\in \mathcalK_1
\mathcalD _1=\chi _1$. By Theorem \ref{Thm:main-1} $\lambda
_j\left(\tau + \frac{2\pi p}{N_1a}\right)$ is a holomorphic function
of $\tau $ in $(\epsilon_1 k^{-1-\delta})-$neighborhood of
 $\chi_2(\lambda,\delta)$ and
 $$|\lambda _j^{(1)}(\tau + 2\pi p/N_1a)-\lambda _{\hat n}^{(1)}(\tau
+ 2\pi
    p/N_1a)|>k^{2\beta -2-2s_1}>\epsilon_1.$$

Part 2 follows from stability of all estimates with respect to
perturbation of $\tau $ smaller then $(\epsilon_1 k^{-1-\delta})$.
Indeed, suppose $\tau $ is in the real $(\epsilon_1
k^{-1-\delta})-$neighborhood of
 $\chi_2(\lambda,\delta)$. Then, there is a $\tau _0 \in
 \chi_2(\lambda,\delta)$, such that $|\tau -\tau _0|<\epsilon_1
 k^{-1-\delta}$. Let $p,j$ be defined by $\tau _0$ as in Part 1.
 Obviously,
 perturbation series for $\lambda _j^{(1)}(\tau + 2\pi p/N_1a)$
 converges and \eqref{3.7} holds. Using \eqref{3.6}, we easily obtain
 \eqref{3.6a}. Therefore, $p$, $j$ are defined uniquely by \eqref{3.7}.

    Let us estimate the size of $\mathcalO_2$. According to Lemma
    \ref{size}, the total size of each $\mathcalO_{s}\left(\frac{2\pi
p'}{N_1a}\right)$ is less then $c_0k^{-2-4s_1}$. Considering that
the number of $p$ does not exceed $4k^{2s_1}$, we obtain that the
total size of $\mathcalO_2$ is less then $c_0k^{-2-2s_1}$. Using
\eqref{2.75}, we arrive at (\ref{3.9}). \end{proof}
%%%%%%%%%%%%55\ref{A:1}.
\begin{remark} \label{R:11} Note that every point $\frac{2\pi
m}{N_1a}$ ($m\in \Z^2$) of a dual lattice corresponding to the
larger periods $N_1a_1, N_1a_2$ can be uniquely represented in the
form $\frac{2\pi m}{N_1a}=\frac{2\pi j}{a}+\frac{2\pi p}{N_1a}$,
where $m=N_1j+p$ and $\frac{2\pi j}{a}$ is a point of a dual lattice
for periods $a_1$, $a_2$, while $p\in P$ is responsible for refining
the lattice.

Let us consider a normalized eigenfunction $\psi_n(t,x)$ of
$H^{(1)}(t)$ in $L_2(Q_1)$. We extend it quasiperiodically to $Q_2$,
renormalize in $L_2(Q_2)$ and denote the new function by
$\tilde{\psi}_n(\tau,x)$, $\tau ={\mathcal K}_2t$. The Fourier
representations of $\psi_n(t,x)$ in $L_2(Q_1)$ and
$\tilde{\psi}_n(\tau,x)$ in $L_2(Q_2)$ are simply related. If we
denote Fourier coefficients of $\psi_n(t,x)$ with respect to the
basis of exponential functions
    $\frac{1}{|Q_1|^{1/2}}e^{i \langle t+\frac{2 \pi j}{a},x \rangle}$, $j \in \Z^2,$
in $L_2(Q_1)$  by $C_{nj}$, then,  the Fourier coefficients
$\tilde{C}_{nm}$ of $\tilde{\psi}_n(\tau,x)$ with respect to the
basis of exponential functions
    $\frac{1}{|Q_2|^{1/2}}e^{i \langle \tau + \frac{2\pi m}{N_1a },x \rangle}$, $m \in \Z^2,$
in $L_2(Q_2)$ are given by the formula:
    \begin{equation*}
    \tilde{C}_{nm}=
        \begin{cases}
        C_{nj},  &\text{if $m=jN_1+p$;}\\
        0,       &\text{otherwise},
        \end{cases}
    \end{equation*}
    $p$ being defined from the relation
$t=\tau+\frac{2\pi p}{N_1a },\ p \in P$. Hence, matrices of the
projections on $\psi_n(t,x)$ and $\tilde{\psi}_n(\tau,x)$ with
respect to the above bases are simply related:
    \begin{equation*}
    (\tilde{E}_n)_{j\hat{j}}=
        \begin{cases}
        (E_n)_{m\hat{m}},  &\text{if $m=jN_1+p,\ \hat{m}=\hat{j}N_1+p$;}\\
        0,       &\text{otherwise},
        \end{cases}
    \end{equation*}
$\tilde{E}_n$ and $E_n$ being projections in $L_2(Q_2)$ and
$L_2(Q_1)$, respectively.

Let us denote by $\tilde{E}_j^{(1)}\bigl(\tau+\frac{2\pi p}{N_1a }
\bigr)$ the spectral projection ${E}_j^{(1)}(\alpha, t)$ (see
(\ref{E:E_j^1})) with $\alpha =1$ and $t=\tau+\frac{2\pi p}{N_1a }$,
``extended" from $L_2(Q_1)$ to $L_2(Q_2)$. \end{remark}
 By analogy with (\ref{E:g_r}), (\ref{E:G_r}), we define functions $g_r^{(2)}(k,\tau)$ and
operator-valued functions $G_r^{(2)}(k,\tau)$, $r=1, 2, \cdots$, as
follows: \begin{equation}\label{3.13}
 g_r^{(2)}(k,\tau)=\frac{(-1)^r}{2\pi ir}\mbox{Tr}\oint _{C_2}
 \Bigl(\bigl(\tilde{H}^{(1)}(\tau)-z\bigr)^{-1}W_2\Bigr)^rdz,
 \end{equation}
\begin{equation}\label{3.14}
G_r^{(2)}(k,\tau)=\frac{(-1)^{r+1}}{2\pi i}\oint
_{C_2}\Bigl(\bigl(\tilde{H}^{(1)}(\tau)-z\bigr)^{-1}W_2\Bigr)^r
\bigl(\tilde{H}^{(1)}(\tau)-z\bigr)^{-1}dz,
 \end{equation}
 $\tilde{H}^{(1)}(\tau)$ being defined at the beginning of Section \ref{S:2.2}, $C_2$ being given by \eqref{C_2}.
We consider the operators $H_{\alpha}^{(2)}=H^{(1)}+\alpha W_2$ and
the family $H_{\alpha}^{(2)}(\tau)$, $\tau \in K_2$, acting in
$L_2(Q_2)$. By (\ref{W2}) and (\ref{99s}) \begin{equation}
\|W_2\|<\epsilon _1^4, \label{99t} \end{equation} $\|W_2\|$ being
the norm of the operator here. \begin{thm}\label{T:3.4}
 Suppose $\tau$ belongs to the
$(\epsilon_1 k^{-1-\delta })$-neighborhood in $K_2$ of the
 second non-resonance set $\chi _2(\lambda ,\delta)$, $0<\delta
<s_1$, $\epsilon_1=e^{-\frac{1}{4}k^{\eta s_1}}$. Then, for
sufficiently large $\lambda $, $\lambda >k_*^2$ and for all $\alpha
$, $0 \leq \alpha \leq 1$, there exists a unique eigenvalue of the
operator $H_{\alpha}^{(2)}(\tau)$
 in
the interval $\varepsilon_2 (k):= (k^{2}-\epsilon_1 /2,
k^{2}+\epsilon_1 /2)$. It is given by the series:
\begin{equation}\label{3.15} \lambda_{\tilde{j}}^{(2)} (\alpha
,\tau)=\lambda_j^{(1)}\bigl( \tau+2\pi p/N_1a \bigr)+\sum
_{r=1}^{\infty }\alpha ^r g_r^{(2)}(k,\tau),\ \ \ \ \tilde
j=j+p/N_1, \end{equation} converging absolutely in the disk
$|\alpha|  \leq 1$, where $p\in P$ and $j\in \Z^2$ are described as
in Geometric Lemma \ref{L:3.2}. The spectral projection
corresponding to $\lambda_{\tilde{j}}^{(2)} (\alpha ,\tau)$ is given
by the series: \begin{equation}\label{3.16} E_{\tilde{j}}^{(2)}
(\alpha ,\tau)=\tilde{E}_j^{(1)}\bigl(\tau+2\pi p/N_1a \bigr)+\sum
_{r=1}^{\infty }\alpha ^rG_r^{(2)}(k,\tau), \end{equation} which
converges in the trace class $\mathbf{S_1}$ uniformly with respect
to $\alpha $ in the disk  $| \alpha | \leq 1$.

 The
following estimates hold for coefficients $g_r^{(2)}(k,\tau)$,
$G_r^{(2)}(k,\tau)$, $r\geq 1$: \begin{equation}\label{3.17} \bigl|
g_r^{(2)}(k,\tau) \bigr|<\frac{3\epsilon_1}{2}(4\epsilon_1^3)^r,\ \
\ \ \bigl\| G_r^{(2)}(k,\tau)\bigr\| _1< 6r (4 \epsilon_1^3)^r.
\end{equation} \end{thm} \begin{cor}\label{C:3.5} The following
estimates hold for the perturbed eigenvalue and its spectral
projection: \begin{equation}\label{3.19} \Bigl|
\lambda_{\tilde{j}}^{(2)} (\alpha ,\tau)-\lambda_j^{(1)}
\bigl(\tau+2\pi p/N_1a\bigr)\Bigr| \leq  12 \alpha \epsilon_1 ^4,
\end{equation} \begin{equation}\label{3.20}
\Bigl\|E_{\tilde{j}}^{(2)} (\alpha
,\tau)-\tilde{E}_j^{(1)}\bigl(\tau+2\pi p/N_1a\bigr) \Bigr\|_1\leq
48 \alpha \epsilon_1^3 . \end{equation} \end{cor} \begin{remark}
\label{R:May10a} The theorem states that $\lambda_{\tilde{j}}^{(2)}
(\alpha ,\tau )$ is a single eigenvalue in the interval $\varepsilon
_2 (k,\delta )$. This means that $\bigl|\lambda_{\tilde{j}}^{(2)}
(\alpha ,\tau )-k^{2}\bigr|<\epsilon _1/2$. Formula (\ref{3.19})
provides a stronger estimate on the location of
$\lambda_{\tilde{j}}^{(2)} (\alpha ,\tau )$.\end{remark}

\begin{proof} The proof of the theorem  is based on expanding the
resolvent $(H_{\alpha}^{(2)}(\tau)-z)^{-1}$ in a perturbation series
for $z \in C_2 $. Integrating the resolvent yields the formulae for
an eigenvalue of $H_{\alpha}^{(2)}$ and its spectral projection. In
fact, it is obvious that \begin{equation}\label{3.21} (H_{\alpha
}^{(2)}(\tau)-z)^{-1}=(\tilde{H}^{(1)}(\tau)-z)^{-1}(I-\alpha
A_2)^{-1}, \ \  A_2 := -W_2(\tilde{H}^{(1)}(\tau)-z)^{-1}.
\end{equation} Suppose $z\in C_2$. Using Corollary \ref{C:3.3} and
estimate (\ref{99t}),
 we obtain:
\begin{equation}\label{3.22} \|(\tilde{H}^{(1)}(\tau)-z)^{-1}\| \leq
\dfrac{4}{\epsilon_1},\ \ \ \|A_2\| \leq \dfrac{4
\|W_2\|}{\epsilon_1}<4\epsilon_1^3<1. \end{equation} The last
inequality makes it possible to expand $(I-\alpha A_2)^{-1}$ in the
series in powers of $\alpha A_2$.  Integrating the series for the
resolvent and considering as in the proof of Theorem
\ref{Thm:main-1} we obtain formulae (\ref{3.15}), (\ref{3.16}).
Estimates (\ref{3.17}) follow from the estimates (\ref{3.22}).
\end{proof}

Next, we show that  the series~(\ref{3.15}),~(\ref{3.16}) can be
extended as holomorphic functions of $\tau $ in a complex
neighborhood of $\chi _2$; they can be differentiated   any number
of times with respect to $\tau$ and retain their asymptotic
character.

\begin{lem}\label{T:3.6} The following estimates hold for the
coefficients $g_r^{(2)}(k,\tau)$ and $G_r^{(2)}(k,\tau)$ in the
complex $(\frac{1}{2}\epsilon_1 k^{-1-\delta })$-neighborhood  of
the non-resonance set $\chi_2$:
    \begin{align}
    |T(m)g_r^{(2)}(k,\tau)| &< m!\cdot 3 \cdot 2^{2r-1+|m|}
    \epsilon_1^{3r+1-|m|}k^{|m|(1+\delta)},\label{3.39}\\
    \| T(m)G_r^{(2)}(k,\tau)\| &<
    m!\cdot 3r \cdot 2^{2r+1+|m|} \epsilon_1^{3r-|m|} k^{|m|(1+\delta)}.\label{3.40}
    \end{align}
\end{lem} \begin{proof} Since~(\ref{3.10}) is valid in the complex
$(\epsilon_1 k^{-1-\delta })$- neighborhood of the second
non-resonance set, it is not hard to see that the coefficients
$g_r^{(2)}(k,\tau)$ and $G_r^{(2)}(k,\tau)$ can be continued from
the real $(\epsilon_1 k^{-1-\delta })$-neighborhood of $\tau$ to the
complex $(\epsilon_1 k^{-1-\delta })$-neighborhood as holomorphic
functions of two variables and inequalities~(\ref{3.17}) are hereby
preserved. Estimating, by means of the Cauchy integral formula, the
value of the derivative with respect to $\tau$ in terms of the value
of the function itself on the boundary of the
$(\frac{1}{2}\epsilon_1 k^{-1-\delta })$-neighborhood of $\tau$
(formulas~(\ref{3.17})), we obtain~(\ref{3.39}) and~(\ref{3.40}).
 \end{proof}

From this lemma the following theorem easily follows.

\begin{thm}\label{T:3.5a} Under the conditions of Theorem
\ref{T:3.4} the series~(\ref{3.15}), ~(\ref{3.16})  can be continued
as holomorphic functions of two variables from the real
$(\epsilon_1k^{-1-\delta})$-neighborhood of the non-resonance set
$\chi_2$ to its complex $(\epsilon_1k^{-1-\delta})$-neighborhood and
the following estimates hold in the complex neighborhood:
\begin{align}
    \left| T(m)\left(\lambda_{\tilde{j}}^{(2)} (\alpha ,\tau)
    -\lambda_j^{(1)}(\tau+2\pi p/N_1a)\right)\right| &<\alpha
    C_m\epsilon_1^{4-|m|}k^{|m|(1+\delta)},
    \label{3.41}\\
   \left \| T(m)\left(E_{\tilde{j}}^{(2)} (\alpha ,\tau)-\tilde{E}_j^{(1)}(\tau+2\pi p/N_1a)
   \right)\right\|
    &< \alpha C_m\epsilon_1^{3-|m|}k^{|m|(1+\delta)},
    \label{3.42}
    \end{align}
    here and below $C_m=48m!2^{|m|}$.
\end{thm} \begin{cor}\label{C:3.6} \begin{equation}
  \left|\nabla \lambda_j^{(2)} (\alpha ,\tau)-
   2\vec k\right|<2C(W_1)k^{-1 -2\beta +15s_1+11\delta }, \ \ \vec k=\vec p_j(\tau +2\pi p/N_1a)\label{2.20**}
   \end{equation}
 \begin{equation}
  \left|T(m)\lambda_j^{(2)} (\alpha ,\tau)\right|<2+2C(W_1)k^{-1 -2\beta +21s_1+15\delta },
  \mbox{ if } |m|=2.
  \label{2.20**a}
  \end{equation}
  \end{cor}

%%%%%%%%%O\bigl(k^{-1 -2\beta _0+21s_1+15\delta }\bigr)

The next lemma will be used in the third step of approximation. The
operator $H^{(2)}(\tau)$ is $H^{(2)}_{\alpha }(\tau)$ with $\alpha
=1$. It will play a role of the initial (unperturbed) operator in
the third step.

\begin{lem}\label{L:3.5.1/2} For any $z$ on the circle $C_2$ and
$\tau$
 in the complex $(\epsilon_1k^{-1-\delta})-$ neighborhood of $\chi_2$,
    \begin{equation}\label{3.40.2}
    \|(H^{(2)}(\tau )-z)^{-1}\| \leq \frac{8}{\epsilon_1}.
    \end{equation}
\end{lem} \begin{proof} Considering the  Hilbert relation
    $$(H^{(2)}(\tau)-z)^{-1}=(\tilde{H}^{(1)}(\tau)-z)^{-1}
    +(\tilde{H}^{(1)}(\tau)-z)^{-1}(-W_2)(H^{(2)}(\tau)-z)^{-1},$$
and  the estimate (\ref{3.10}), together with the estimate
(\ref{99t}),  we obtain:
    \begin{equation}
    \|(H^{(2)}(\tau)-z)^{-1}\|  \leq
    \frac{\|(\tilde{H}^{(1)}(\tau)-z)^{-1}\|}{1-\|(\tilde{H}^{(1)}(\tau)-z)^{-1}W_2\|}
     \leq 2\|(\tilde{H}^{(1)}(\tau)-z)^{-1}\|
     \leq \frac{8}{\epsilon_1}.\label{3.40.4}
    \end{equation}
 \end{proof}

 \subsection{Non-resonant part of the isoenergetic set of
$H_{\alpha }^{(2)}$}\label{S:3.4}

Let $S_2(\lambda )$ be an isoenergetic set of the operator
$H_{\alpha}^{(2)}$:
    $S_2(\lambda)=\{\tau \in K_2 : \exists n \in \N:\
    \lambda_n^{(2)}(\alpha,\tau)=\lambda  \}$,
    here $\{\lambda_n^{(2)}(\alpha,\tau)\}_{n=1}^{\infty }$ is the
    spectrum of $H_{\alpha}^{(2)}(\tau)$.
Now we construct a non-resonance subset $\chi _2^*(\lambda )$ of
$S_2(\lambda
    )$. It
    corresponds to  non-resonance eigenvalues $\lambda
    _{\tilde j}^{(2)}(\tau )$ given by the perturbation series (\ref{3.15}).
    Recall that $\mathcal \mathcalD_1(\lambda )_{nonres}$ and $\chi _2(\lambda )$  are defined by the
    formulae \eqref{D_{1,nonres}} and \eqref{chi_2}
    respectively.
    Recall also that $\chi _2\subset \mathcal{K}_2\chi_1^*(\lambda )$ (see
    the Geometric Lemma) and $\chi_1^*(\lambda
    )=\mathcal{K}_1\mathcal D_1(\lambda )$, see (\ref{2.81}). Hence, $\chi _2\subset \mathcal{K}_2\mathcal D_1(\lambda
    )$.
\begin{lem}\label{L:3.7.1/2} The formula
$\mathcal{K}_2\mathcal{D}_1(\lambda )_{nonres}=\chi_2$ establishes
 one-to-one correspondence between
$\mathcal{D}_1(\lambda )_{nonres}$ and $\chi_2$. \end{lem}
\begin{proof} Suppose there is a pair $\vec \varkappa _1, \vec
\varkappa _{1*}\in \mathcal D_1(\lambda )_{nonres}$ such that
$\mathcal{K}_2\vec \varkappa _1= \mathcal{K}_2\vec \varkappa
_{1*}=\tau $, $\tau \in \chi _2$.  We introduce also
$t_1=\mathcal{K}_1\vec \varkappa _1$ and $t_{1*}=\mathcal{K}_1\vec
\varkappa _{1*}$.   The definition (\ref{2.81}) of $\chi_1^*(\lambda
    )$ implies that $t_1,t_{1*} \in \chi_1^*(\lambda
    )$, since $\vec \varkappa _1, \vec \varkappa _{1*}\in
\mathcal D_1(\lambda )_{nonres}\subset \mathcal D_1(\lambda )$.
Clearly, $\mathcal{K}_2t_1=\mathcal{K}_2t_{1*}=\tau $ and, hence,
$t_1=\tau +2\pi p_1/N_1a$, $t_{1*}=\tau +2\pi p_2/N_1a$ for some
$p_1,p_2\in P$. Now, by Part 1a of Geometric Lemma \ref{L:3.2},
$p_1=p_2$, and,
    therefore, $t_1=t_{1*}$. Next, by Lemma \ref{L:May10a}, $\vec \varkappa _1=\vec \varkappa
    _{1*}$.
     \end{proof}
We define $\mathcal B_2 (\lambda)$ as the set of directions
corresponding to the set $\Theta _2$ given by \eqref{3.19a}:
$$\mathcal B_2(\lambda)=\{\vec{\nu} \in \mathcal B_1(\lambda) :
   \varphi \in \Theta _2\}.$$
Note that $\mathcal B_2 (\lambda)$ is a unit circle with holes,
centered at the origin, and $\mathcal B_2(\lambda ) \subset
    \mathcal B_1(\lambda )$.

Let $\vec{\varkappa } \in \mathcal{D}_1(\lambda )_{nonres}$. By
(\ref{chi_2}), $\tau \equiv {\mathcal K}_2\vec \varkappa \in \chi
_2(\lambda )$. According to Theorem \ref{T:3.4}, for sufficiently
large $\lambda $, there exists an eigenvalue of the operator
$H^{(2)}_{\alpha}(\tau )$, given by (\ref{3.15}). It is convenient
here to denote $\lambda_{\tilde{j}}^{(2)} (\alpha ,\tau)$ by
$\lambda^{(2)}(\alpha,\vec{\varkappa })$.
 We can do this, since, by Lemma \ref{L:3.7.1/2},
there is one-to-one correspondence between $\vec{\varkappa }\in
\mathcal{D}_1(\lambda )_{nonres}$ and the pair $(\tau, \tilde j)$, $
\vec \varkappa =2\pi \tilde j/a+\tau $. We rewrite (\ref{3.15}) in
the form:
    \begin{equation}\label{3.66}
    \lambda^{(2)}(\alpha,\vec{\varkappa })
    =\lambda^{(1)}(\vec{\varkappa })+f_2(\alpha,\vec{\varkappa }),\
    \ \
f_2(\alpha,\vec{\varkappa  })=\sum _{r=1}^{\infty}
\alpha^{r}g_r^{(2)}(\vec{\varkappa }), \end{equation} here
$g_r^{(2)}(\vec{\varkappa })$ is given by (\ref{3.13}). The function
$f_2(\alpha,\vec \varkappa )$ satisfies the estimates: \begin{align}
    |f_2(\alpha,\vec \varkappa )|&\leq 12 \alpha \epsilon_1
    ^4,\label{3.67}\\
    |\nabla f_2(\alpha,\vec \varkappa )|&
    \leq 96 \alpha \epsilon_1 ^3 k^{1+\delta}.\label{3.68}
    \end{align}
By Theorems \ref{T:3.4} and \ref{T:3.5a}, the formulas (\ref{3.66})
-- (\ref{3.68}) hold even in the real $(\epsilon_1
k^{-1-\delta})$-neighborhood of $\mathcal{D}_1(\lambda )_{nonres}$,
i.e., for any $\vec \varkappa =\varkappa \vec{\nu}$ such that
$\vec{\nu} \in \mathcal B_2(\lambda)$ and $|\varkappa -\varkappa
_1(\lambda,\vec \nu)|<\epsilon_1 k^{-1-\delta}$. We define
$\mathcal{D}_2(\lambda)$ as a level set for
$\lambda^{(2)}(\alpha,\vec{\varkappa })$ in this neighborhood:
    \begin{equation}
    \mathcal{D}_2(\lambda ):=\left\{\vec{\varkappa }=\varkappa \vec \nu: \vec \nu \in
    \mathcal B_2,\ \
    \bigl|\varkappa  -\varkappa _1(\lambda,\vec
\nu) \bigr|<\epsilon_1 k^{-1-\delta},\
\lambda^{(2)}(\alpha,\vec{\varkappa })=\lambda
\right\}.\label{May14m} \end{equation} Next two lemmas are to prove
that $\mathcal{D}_2(\lambda)$ is a distorted circle with holes.
\begin{lem}\label{L:3.8} For every $\vec{\nu} \in \mathcal B_2$ and
every $\alpha$, $0 \leq \alpha \leq 1$, there is a unique $\varkappa
=\varkappa _2 (\lambda,\vec{\nu})$ in the interval
$I_2:=\bigl[\varkappa _1 (\lambda,\vec \nu)-\epsilon_1
k^{-1-\delta},\varkappa _1 (\lambda,\vec \nu)+\epsilon_1
k^{-1-\delta}\bigr]$ such that
    \begin{equation}\label{3.70}
    \lambda^{(2)}(\alpha,\varkappa _2 \vec{\nu})=\lambda .
    \end{equation}
Furthermore,
    \begin{equation}\label{3.71}
    |\varkappa _2(\lambda,\vec
\nu) -\varkappa _1(\lambda,\vec \nu)| \leq 12\alpha \epsilon_1
^4k^{-1}.
    \end{equation}
\end{lem} The proof is based on \eqref{3.66}, \eqref{3.67},
\eqref{2.20**} and completely analogous to that of Lemma 3.41 in
\cite{KL}, set $l=1$.

 Further, we use the
notations: \begin{equation} \varkappa _2(\varphi )\equiv \varkappa
_2(\lambda ,\vec \nu ),\ \ \ h_2(\varphi )=\varkappa _2(\varphi
)-\varkappa _1(\varphi ),\ \ \ \vec \varkappa _2(\varphi )=\varkappa
_2(\varphi )\vec \nu .\label{green1} \end{equation}
\begin{lem}\label{L:3.9} Let $10^{-4}<s_1<10^{-3}$, $\eta >10^4$.
Then the following statements hold for $\lambda >k_*^2$:
\begin{enumerate} \item The set $\mathcal{D}_2(\lambda )$ is a
distorted circle with holes: it can be described by the formula:
\begin{equation} \mathcal{D}_2(\lambda )=\bigl\{\vec \varkappa \in
\R^2: \vec \varkappa =\vec \varkappa _2(\varphi ),\ \ {\varphi } \in
\Theta _2(\lambda )\bigr\},\label{May20a} \end{equation} where
    $ \varkappa _2(\varphi )=\varkappa _1(\varphi)+h_2(\varphi)$,
$\varkappa _1(\varphi )$ is the ``radius" of $\mathcal{D}_1(\lambda
)$ and $h_2(\varphi )$ satisfies the estimates
    \begin{equation}\label{3.75}
    |h_2|\leq 12\alpha \epsilon_1 ^4k^{-1},\ \ \
    \left|\frac{\partial h_2}{\partial \varphi} \right| \leq
    96\alpha \epsilon_1^3 k^{1+\delta}.
    \end{equation}
     \item The total length of $\mathcal{B}_2(\lambda)$ satisfies the estimate:
\begin{equation}\label{theta2}
    L\left(\mathcal{B}_1\setminus \mathcal{B}_2\right)<c_0 k^{-2-2s_1}.
    \end{equation}
\item The function $h_2(\varphi )$ can be extended as a holomorphic
function of $\varphi $ into the complex non-resonance set $\varPhi
_2$ and its $k^{-4-6s_1-4\delta }$ neighborhood $\hat \varPhi _2$,
estimates (\ref{3.75}) being preserved. \item The curve
$\mathcal{D}_2(\lambda )$ has a length which is asymptotically close
to that of $\mathcal{D}_1(\lambda )$ in the following sense:
    \begin{equation}\label{3.77}
     L\Bigl(\mathcal{D}_2(\lambda )\Bigr)\underset{\lambda \rightarrow
     \infty}{=}L\Bigl(\mathcal{D}_1(\lambda
     )\Bigr)\Bigl(1+O\bigl(k^{-2-2s_1}\bigr)\Bigr),
     \end{equation}
     where $O\bigl(k^{-2-2s_1})=(1+o(1))c_0k^{-2-2s_1}$,
     $|o(1)|<10^{-2}$ when $k>k_*$.
     \end{enumerate}
\end{lem} \begin{proof} The proof is completely analogous to that of
Lemma 3.42 in \cite{KL}, set $l=1$. Here we give just a short
version. Indeed, the first inequality in \eqref{3.75} is equivalent
to \eqref{3.71}. Differentiating the identity $\lambda
^{(2)}\left(\vec \varkappa _2 (\varphi )\right)=\lambda
^{(1)}\left(\vec \varkappa _1 (\varphi )\right)=k^2$ with respect to
$\varphi $ and using \eqref{2.20**}, \eqref{3.66}, \eqref{3.68},
\eqref{3.71}, we easily obtain the second estimate in \eqref{3.75}.
Estimate \eqref{theta2} valid, since the total size of $O^{(2)}$ is
less than  $c_0k^{-2-2s_1}$. To prove the analyticity of
$h_2(\varphi )$ in $\varPhi _2$ we check the convergence of the
perturbation series for $\lambda ^{(2)}\left(\vec \varkappa _1
(\varphi )\right)$, $\varphi \in \varPhi _2$. It is enough to show
that
\begin{equation} \label{Sep10.09}
 \Bigl\|\Bigl(\tilde H^{(1)}\bigl(\vec \varkappa _1(\varphi)\bigr)-z\Bigr)^{-1}\Bigr\| \leq \frac{4}{\varepsilon
 _1},\ \ \ z\in C_2.
 \end{equation}
 This inequality immediately follows from
 \begin{equation} \label{Sep10.09a}
 \left\|\left( H^{(1)}\Bigl(\vec \varkappa _1(\varphi)+\frac{2\pi p}{N_1a}\Bigr)-z\right)^{-1}
 \right\| \leq \frac{4}{\varepsilon _1}
 \end{equation}
 proven for all $p\in P$. Let $p=0$. By Lemma \ref{Lem:resolvent-1'},
$$
  \|(H^{(1)}(\vec \varkappa _1(\varphi))-z)^{-1}\| \leq 2k^{-2\beta +1+s_1+\delta
  },\ \ \ \mbox{when } z\in C_1,\ \ 2\beta =1-15s_1-9\delta . $$
  It is not difficult to show that the resolvent has a single pole
  inside $C_1$ at the point $z=k^2$.
  The circle $C_2$ has the same centrum and a smaller radius
  $\epsilon _1/2$. Hence
  \beq\label{InE:resolvent-1++} \|(H^{(1)}(\vec \varkappa _1(\varphi))-z)^{-1}\| \leq \frac{4}{\epsilon _1},
  \ \ \ \mbox{when } z\in C_2. \eeq
  Next we use \eqref{normres**} with $\vec b= \frac{2\pi p}{N_1a}$.
  The right-hand side of
  \eqref{normres**} is less that $\epsilon _1^{-1}$, since $s_1\eta
  >1$. Hence, \eqref{Sep10.09a} holds for $p\neq 0$ too. Convergence
  of perturbation series follows from \eqref{Sep10.09} and
  \eqref{99t}. Note that the smallest circle in $\mathcal O^{(2)}$ has the size $k^{-4-6s_1-3\delta }$. Since
  $k^{-4-6s_1-4\delta }$ is much smaller than the radius of circles, all estimates are stable in the
  $k^{-4-6s_1-4\delta }$ of $\varPhi _2$. Now, using Rouche's and Implicit function theorems we
  easily show the equation $\lambda ^{(2)}(\varkappa \vec \nu )=k^2$
  has a solution $\varkappa =\varkappa _2(\varphi )$ which is
  holomorphic in $\varPhi _2$ and coincides with $\varkappa _1(\varphi
  )+h_2(\varphi )$ for real $\varphi $. Estimate \eqref{3.77}
  follows from \eqref{3.75} and \eqref{theta2}.

   Let us record a remark for the sequel.
Convergence of the series
 for the resolvent $\left( H^{(2)}\left(\vec \varkappa _1(\varphi
 )\right)-z\right)^{-1}$, $z\in C_2$,  means that the
 resolvent  has a single pole $z= \lambda ^{(2)}(\vec \varkappa _1(\varphi
 )\vec \nu
 )$ inside $C_2$. Similar result holds when we replace $\vec \varkappa _1(\varphi
 )$ by $\vec \varkappa _2(\varphi
 )$, since $\vec \varkappa _1(\varphi
 )$ and $\vec \varkappa _2(\varphi
 )$ are close: $\bigl|\vec \varkappa _2(\varphi
 )-\vec \varkappa _1(\varphi
 )|=o(\epsilon _1)$. Considering that $\lambda ^{(2)}(\vec \varkappa
 _2(\varphi )
 )=\lambda $, we obtain that $(z-\lambda )\left( H^{(2)}\left(\vec \varkappa _2(\varphi
 )\right)-z\right)^{-1}$ is holomorphic inside $C_2$ and the estimate
following holds:
 \begin{equation}
 \Bigl\|(z-\lambda )\left(H^{(2)}\bigl(\vec \varkappa _2(\varphi
 )\bigr)-z\right)^{-1}\Bigr\|<32. \label{16**}
 \end{equation}
\end{proof}

 Now define the non-resonance set $\chi_2^*(\lambda )$ in
$S_2(\lambda )$ by
    \begin{equation}\label{3.77.7}
    \chi_2^*(\lambda ):=\mathcal{K}_2\mathcal{D}_2(\lambda ).
    \end{equation}
    \begin{lem} \label{Apr4a}
    The set $\chi_2^*(\lambda )$ belongs to the
    $\left(12\alpha \epsilon _1^4k^{-1}\right)$-neighborhood of $\chi_2(\lambda
    )$ in $K_2$. If $\tau \in \chi_2^*(\lambda )$, then the operator
    $H^{(2)}_{\alpha }(\tau )$ has a simple eigenvalue
   equal to $\lambda $. This
    eigenvalue is given by the perturbation series (\ref{3.15}),
     where $p\in
P,j\in \Z^2$ are uniquely defined by $\tau $ as it is described in
Geometric Lemma \ref{L:3.2}, part 2. \end{lem}

\begin{proof} By Lemma \ref{L:3.9}, $\mathcal D_2(\lambda )$ is in
the $\left( 12\alpha \epsilon _1^4k^{-1}\right)$ neighborhood of
$\mathcal D_1(\lambda )_{nonres}$. Considering that
$\chi_2^*(\lambda )=\mathcal K_2\mathcal D_2(\lambda )$ and
$\chi_2(\lambda )=\mathcal K_2\mathcal D_1(\lambda )_{nonres}$ (see
(\ref{chi_2})), we immediately obtain that $\chi_2^*(\lambda )$ is
in the
    $(12\alpha \epsilon _1^4k^{-1})$-neighborhood of $\chi_2(\lambda
    )$. The size of this neighborhood is
less than $\epsilon _1k^{-1-\delta }$, hence Theorem \ref{T:3.4}
holds in it, i.e., for any $\tau \in \chi_2^*(\lambda )$ there is a
single eigenvalue of $H^{(2)}_{\alpha }(\tau )$ in the interval
$\varepsilon_2 (k,\delta )$. Since $\chi_2^*(\lambda )\subset
S_2(\lambda )$, this eigenvalue is equal to $\lambda $. By the
theorem, the eigenvalue is given by the series (\ref{3.15}),
     where $p\in
P,j\in \Z^2$ are are uniquely defined by $\tau $ as it is described
in Geometric Lemma \ref{L:3.2}, part 2. \end{proof}

    \begin{lem}\label{L:May10a*} Formula (\ref{3.77.7}) establishes
    one-to-one correspondence between $\chi_2^*(\lambda )$ and $\mathcal{D}_2(\lambda
    )$.\end{lem}
    \begin{remark} \label{R:May14}
    From geometric point of view this means that $\chi_2^*(\lambda
    )$ does not have self-intersections.\end{remark}
    \begin{proof} Suppose there is a pair $\vec \varkappa , \vec \varkappa
    ^*
    \in \mathcal{D}_2(\lambda)$ such that $\mathcal K_2\vec \varkappa =\mathcal
    K_2\vec \varkappa
    ^*
    =\tau $, $\tau \in \chi_2^*(\lambda )$. By the definition (\ref{May14m}) of
    $\mathcal{D}_2(\lambda
    )$, we have $\lambda^{(2)}(\alpha,\vec{\varkappa })=\lambda^{(2)}(\alpha,\vec{\varkappa
    }^*)=
    \lambda $, i.e., the eigenvalue $\lambda $ of
    $H^{(2)}_{\alpha }(\tau )$ is not simple. This contradicts to the
    previous lemma.
     \end{proof}
\subsection{Preparation for the Next Approximation} \label{S:3.6}
Let $\vec b^{(2)} \in K_2$ and $b_0^{(2)}$ be the distance of the
point $\vec b^{(2)}$ to the nearest corner of $K_2$:
\begin{equation}\label{3.2.3.1-2}
b_0^{(2)}=\min_{m=(0,0),(0,1),(1,0),(1,1)}|\vec b-2\pi m/N_1a|.
\end{equation}
 We assume $b_0^{(2)}=|\vec b^{(2)}|$. In the case
    when $\vec b$ is closer to a vertex other than $(0,0)$, the
    considerations are the same up to a parallel shift. We  consider  two cases:
$b_0^{(2)}\geq
 \epsilon _1k^{-1-2\delta }$ and
$0<b_0^{(2)}<\epsilon _1k^{-1-2\delta }$. Let \begin{equation}\vec
y^{(1)}(\varphi)=\vec \varkappa _1(\varphi )+\vec
    b^{(2)}. \label{y_1} \end{equation}\subsubsection{The case $b_0^{(2)}\geq \epsilon _1k^{-1-2\delta
}$.}\begin{definition} \label{D:*} We  define the set ${\mathcal
O}^{(2)}(\vec b^{(2)})$ by the formula: \begin{equation} {\mathcal
O}^{(2)}(\vec b^{(2)})=\cup _{p\in P} \mathcalO_{s}^{(1)}(2\pi
    p/N_1a+\vec b^{(2)} ). \label{O_2(b_2)} \end{equation}
In the above formula we assume $\mathcalO_{s}^{(1)}\cap \hat \Phi _1
\not =\emptyset $. \end{definition}
    \begin{lem}\label{L:3.7.2.2*mm} If
$\varphi \in \hat \varPhi _1\setminus {\mathcal O}^{(2)}(\vec
b^{(2)})$ or its $k^{-4-6s_1-4\delta }$ neighborhood, then
\begin{equation}\label{3.7.2.2*mm}
    \Bigl\|\Bigl(\tilde H^{(1)}\bigl(\vec y^{(1)}(\varphi)\bigr)-k^{2}\Bigr)^{-1}\Bigr\| \leq
    \frac{k^{4+6s_1+5\delta}}{\epsilon _1}
    \end{equation}
    The resolvent  is an analytic function of
$\varphi $ in every connected component  $ {\mathcal O}_c^{(2)}(\vec
b^{(2)})$ of $ {\mathcal O}^{(2)}(\vec b^{(2)})$, whose intersection
with $\varPhi _1$ is not empty. The only singularities of the
resolvent in such a component are poles. The number of poles
(counting multiplicity) of the resolvent
    inside ${\mathcal O}_c^{(2)}(\vec b^{(2)})$ is less that
    $c_0k^{1+2s_2}$. The size of each connected component is less
    than $k^{-3-4s_1-2\delta }$.
    The total number of poles (counting multiplicity) of the resolvent
    inside ${\mathcal O}^{(2)}(\vec b^{(2)})$ is less than
    $c_0k^{2+2s_2}$. The total size of $\mathcalO_2(\vec b^{(2)})$ is less then
$c_0k^{-2-2s_1}$.
    \end{lem}\begin{proof}
    Combining Lemmas \ref{norm**} and \ref{L:july9c} for $\vec b^{(1)}=2\pi
    p/N_1a+\vec b^{(2)}$, we obtain:
    \begin{equation}\label{comb1}
    \Bigl\|\Bigl(H^{(1)}\bigl(\vec \varkappa _1(\varphi )+2\pi
    p/N_1a+\vec b^{(2)}\bigr)-k^{2}\Bigr)^{-1}\Bigr\| \leq
    \max \left\{k^{J^{(1)}},\frac{4k^{3+6s_1+4\delta}}{b_0^{(2)}}\right\},
\end{equation} $$J^{(1)}=5c_1k^{2/3+s_1}<c_0k^{2+2s_1},\ p\in P,$$
for any $\varphi \in \hat \varPhi _1 \setminus \mathcalO_2(\vec
b^{(2)})$.
    Using \eqref{k}, we easily obtain $k^{J^{(1)}}< \epsilon _1^{-1}$.
    Considering the condition  $b_0^{(2)}\geq
 \epsilon _1k^{-1-\delta }$ and the definition of $\tilde H^{(1)}$, we
 obtain \eqref{3.7.2.2*mm} in $\hat \varPhi _1\setminus {\mathcal O}^{(2)}(\vec
b^{(2)})$ The estimate is stable with respect of the perturbation of
order $k^{-4-6s_1-4\delta }$, since the size of the discs in
$\mathcalO_2(\vec b^{(2)})$ is $k^{-4-6s_1-3\delta }$.

 Now we prove the second part of the lemma. Let $\Delta _*$ be a rectangle
in $\C$: $\Delta _*=\{\varphi : | \Re \varphi -\varphi _*|\leq
k^{-1},
  \ |\Im \varphi |<k^{-\delta }\}$ for some $\varphi _*\in [0,2\pi
  )$.  It is shown
  in the proof of Lemma \ref{Lem:calO+} (see the estimate for the number of points in $I_*$) that the total
  number of points $\varphi _m^{\pm }(\vec b)$ in $\Delta _*$ does not exceed
  $5c_0k^{1+2s_1}$ for any $\vec b\in K_1$. Therefore the number of discs of ${\mathcal O}^{(1)}(\vec b
  )$ which intersect $\Delta _*$ does not exceed
  $5c_0k^{1+2s_1}$. Therefore, the number of discs of  ${\mathcal O}^{(1)}_s(\vec b
  )$ which intersect $\Delta _*$ does not exceed
  $5c_0k^{1+2s_1}$ for any $\vec b\in K_1$. Hence, the number of discs of ${\mathcal O}^{(2)}(\vec b ^{(2)})$
   which intersect $\Delta _*$ does not exceed
  $5c_0k^{1+2s_2}$. Considering that the size of each disc is less then
  $k^{-4-6s_1-3\delta }$, we obtain that the total size of ${\mathcal O}^{(2)}(\vec b
  ^{(2)})\cap \Delta _*$ is less than $k^{-3-4s_1-2\delta }$. It is
  obvious now that the size of each connected component of ${\mathcal O}^{(2)}(\vec b
  ^{(2)})$ does not
  exceed $k^{-3-4s_1-2\delta }$. If ${\mathcal O}_c^{(2)}(\vec b
  ^{(2)})\cap \varPhi _1\not =\emptyset ,$ than ${\mathcal O}_c^{(2)}(\vec b
  ^{(2)})\subset \hat \varPhi _1$. Therefore, the resolvent is an
  analytic function in this ${\mathcal O}_c^{(2)}(\vec b
  ^{(2)})$. By construction, the poles of the resolvent are in the
  centra of the discs in ${\mathcal O}^{(2)}(\vec b
  ^{(2)})$. Therefore, the number of poles in each connected
  component does not exceed $6c_0k^{1+2s_2}$.

By Lemmas \ref{norm**} and \ref{L:july9c} the number of poles
 of the resolvent
 $\Bigl(H^{(1)}\bigl(\vec \varkappa _1(\varphi )+2\pi
    p/N_1a+\vec b^{(2)}\bigr)-k^{2}\Bigr)^{-1}$ inside  $\mathcalO_{s}(2\pi
    p/N_1a+\vec b^{(2)} )$ is less than $c_0k^{2+2s_1}$ for any $p\in
    P$. Considering that there are no poles in $\varPhi _1$ outside $\mathcalO_{s}(2\pi
    p/N_1a+\vec b^{(2)} )$, we obtain that the number of poles of
    the resolvent $\Bigl(H^{(1)}\bigl(\vec \varkappa _1(\varphi )+2\pi
    p/N_1a+\vec b^{(2)}\bigr)-k^{2}\Bigr)^{-1}$ inside  $\mathcalO^{(2)}(\vec b^{(2)}
    )$ is less than $c_0k^{2+2s_1}$.
    Taking into account that the number of $p\in P$ does not exceed $ck^{2(s_2-s_1)}$, $s_2=2s_1$, we obtain the estimate for the number of poles
    for the resolvent $\Bigl(\tilde H^{(1)}\bigl(\vec y^{(1)}((\varphi)\bigr)-k^{2}\Bigr)^{-1}$.

     We estimate the size of $\mathcalO ^{(2)}(\vec b^{(2)})$ the same way as we estimated the size of
  $\mathcalO ^{(2)}$. Indeed,  according to Lemma
    \ref{size}, the total size of each $\mathcalO_{s}\left(\frac{2\pi
p}{N_1a}+\vec b^{(2)}\right)$ is less then $c_0k^{-2-4s_1}$.
Considering that the number of $p$ does not exceed $4k^{2s_1}$, we
obtain that the total size of $\mathcalO_2(\vec b^{(2)})$ is less
then $c_0k^{-2-2s_1}$.

 \end{proof}

\begin{definition}\label{D:**} We denote the poles of the resolvent $\Bigl(\tilde
H^{(1)}\bigl(\vec y^{(1)}(\varphi)\bigr)-k^{2}\Bigr)^{-1}$ in
${\mathcal O}^{(2)}(\vec b^{(2)})$ as $\varphi ^{(2)}_n$,
$n=1,...,M^{(2)}$, $M^{(2)}<c_0k^{2+2s_2}$. Let us consider the
discs ${\mathcal O}_n^{(2)}(\vec b^{(2)})$ of the radius
$r^{(2)}=k^{-2-4s_2-\delta }r^{(1)}=k^{-6-7s_2-4\delta }$ centered
at these poles. Let \begin{equation}{\mathcal O}^{(2)}_{s}(\vec
b^{(2)})=\cup _{n=1}^{M^{(2)}}{\mathcal O}_n^{(2)}(\vec b^{(2)}).
\label{O_2}\end{equation}
\end{definition} \begin{lem} \label{L:totalsize} The total size of
${\mathcal O}^{(2)}_{s}(\vec
    b^{(2)})$ is less then $c_0k^{-2s_2-\delta }r^{(1)}=c_0k^{-4-5s_2-4\delta }$.\end{lem}
    \begin{proof} The lemma easily follows from the formula $r^{(2)}=k^{-2-4s_2-\delta
}r^{(1)}=k^{-6-7s_2-4\delta }$
    and the estimate $M^{(2)}<c_0k^{2+2s_2}$. \end{proof}

\begin{lem}\label{L:3.7.2.2*mmm} If $\varphi \in \hat \varPhi _1\setminus
{\mathcal O}^{(2)}_{s}(\vec b^{(2)})$, then
\begin{equation}\label{3.7.2.2*mmm}
    \Bigl\|\Bigl(\tilde H^{(1)}\bigl(\vec y^{(1)}(\varphi)\bigr)-k^{2}\Bigr)^{-1}\Bigr\| \leq
    \frac{1}{\epsilon _1^2}
    \end{equation} The estimate is stable in the $\bigl(r^{(2)}k^{-\delta }\bigr)$-neighborhood of
    $\varPhi _1\setminus
{\mathcal O}^{(2)}_{s}(\vec b^{(2)})$. The resolvent  is an analytic
function of $\varphi $ in every component of $ {\mathcal
O}^{(2)}_{s}(\vec b^{(2)})$, whose intersection with $\varPhi _1$ is
not empty. The only singularities of the resolvent are poles.
    The number of poles (counting multiplicity) of the resolvent
    inside ${\mathcal O}^{(2)}_{s}(\vec b^{(2)})$ is less than
    $ck^{2+2s_2}$.
    \end{lem}
    \begin{proof} By the definition of ${\mathcal
O}^{(2)}_{s}(\vec b^{(2)})$, the number of poles (counting
multiplicity) of the resolvent
    inside this set is less than
    $ck^{2+2s_2}$. Considering as in Lemma \ref{norm**} we obtain
    \begin{equation}\label{3.7.2.2*mmmm}
    \Bigl\|\Bigl(\tilde H^{(1)}\bigl(\vec y^{(1)}(\varphi)\bigr)-k^{2}\Bigr)^{-1}\Bigr\| \leq
    \nu ^{-M^{(2)}}\frac{k^{4+\delta}}{\epsilon _1},
    \end{equation}
    where $\nu $ is the coefficient of contraction, when we reduce ${\mathcal O}^{(2)}(\vec
    b^{(2)})$ to
    ${\mathcal
O}^{(2)}_{s}(\vec b^{(2)})$. Namely $\nu $ is the ratio of $r^{(2)}$
to the maximal size of ${\mathcal O}^{(2)}(\vec b^{(2)})$. By Lemma
\ref{L:3.7.2.2*mm} and the definition of $r^{(2)}$, $\nu
=k^{-3-5s_2-2\delta } $.  It is not difficult to show that that $\nu
^{-M^{(2)}}k^{4+\delta}<\epsilon _1 ^{-1}$, when $k>k_*$, see
\eqref{k}. The estimate  \eqref{3.7.2.2*mmm} easily follows.

\end{proof}

Obviously the total size of ${\mathcal O}^{(2)}_{s}(\vec b^{(2)})$
 is much smaller the
smallest disc in $O^{(2)}$. Therefore, the function $\vec \varkappa
_2(\varphi )$ is holomorphic inside each connected component of
${\mathcal O}^{(2)}_{s}(\vec b^{(2)})$ which has a non-empty
intersection with $\varPhi _2$. Let
    \begin{equation}\vec y ^{(2)}(\varphi )=\vec \varkappa _2(\varphi )+\vec
    b^{(2)}. \label{y_2}\end{equation}

    \begin{lem} \label{L:3.23} If
$\varphi \in \hat \varPhi _2\setminus {\mathcal O}^{(2)}_{s}(\vec
b^{(2)})$, then \begin{equation}\label{3.7.2.2*m}
    \Bigl\|\Bigl(H^{(2)}\bigl(\vec y^{(2)}(\varphi)\bigr)-k^{2}\Bigr)^{-1}\Bigr\| \leq
    \frac{2}{\epsilon _1^2}
    \end{equation}
    The estimate is stable in the $\bigl(r^{(2)}k^{-\delta }\bigr)$-neighborhood of
    $\varPhi _2\setminus
{\mathcal O}^{(2)}_{s}(\vec b^{(2)})$. The resolvent  is an analytic
function of $\varphi $ in every component of $ {\mathcal
O}^{(2)}_{s}(\vec b^{(2)})$, whose intersection with $\varPhi _2$ is
not empty. The only singularities of the resolvent are poles.
    The number of poles (counting multiplicity) of the resolvent
    inside ${\mathcal O}^{(2)}_{s}(\vec b^{(2)})$ is less than
    $ck^{2+2s_2}$.
    \end{lem} \begin{proof}
We use the Hilbert identity
    $$
    \Bigl( H^{(2)}\bigl(\vec y^{(2)}\bigr)-k^{2}\Bigr)^{-1}-
    \Bigl(\tilde H^{(1)}\bigl(\vec y^{(1)}\bigr)-k^{2}\Bigr)^{-1}=
\mathcalE\Bigl( H^{(2)}\bigl(\vec y^{(2)}\bigr)-k^{2}\Bigr)^{-1},
    $$
    where $y^{(1)},y^{(2)}$ are given by \eqref{y_1}, \eqref{y_2},
    $ \mathcalE=\mathcalE _1+\mathcalE_2 \mathcalE_3, $
     $$ \mathcalE_1=-\Bigl(\tilde H^{(1)}\bigl(\vec y^{(1)}\bigr)-k^{2}\Bigr)^{-1}
    W_2, \ \
  %%%%%%%%  $$\mathcalE_2=\Bigl(\tilde H^{(1)}\bigl(\vec y^{(1)}\bigr)-k^{2}\Bigr)^{-1}T_1,\ \
%%%%%%%%    T_1=\tilde H^{(1)}_0\bigl(\vec
 %%%%%%   y^{(2)}\bigr)-\tilde H^{(1)}_0\bigl(\vec y^{(2)}\bigr).$$
%%%%%%%%
%%%%%%%% Using the estimates (\ref{99t}) and (\ref{3.7.2.2*mmm}), we see
%%%%%%%%    that $\|\mathcalE_1\|=o(1)$. Next we represent $\mathcalE_2$ in
%%%%%%%%    the form:
    \mathcalE_2 =\Bigl(\tilde H^{(1)}\bigl(\vec y^{(1)}\bigr)-k^{2}\Bigr)^{-1}\Bigl(\tilde H^{(1)}_0\bigl(\vec
    y
    ^{(1)}\bigr)+k^2\Bigr),$$ $$ \mathcalE_3=\Bigl(\tilde H^{(1)}_0\bigl(\vec
    y^{(1)}\bigr)+k^2\Bigr)^{-1}\left(\tilde H^{(1)}_0\bigl(\vec
    y^{(2)}\bigr)-\tilde H^{(1)}_0\bigl(\vec y^{(2)}\bigr)\right).$$
    Using the estimates (\ref{99t}) and (\ref{3.7.2.2*mmm}), we see
  that $\|\mathcalE_1\|<\epsilon _1^2$.
    Using \eqref{3.7.2.2*mmm}, we get
    $\|\mathcalE_2\|<1+\|W_1-2k^2\|\varepsilon _1^{-2}<3k^2\varepsilon _1^{-2}.$
    It is easy to show that $\|\mathcalE_3\|<k^{-1}|\varkappa _1-\varkappa
    _2|$. Considering (\ref{3.71}), we get $\|\mathcalE_3\|<12\epsilon _1^4k^{-2}$. Therefore,
    $\|\mathcalE\|<4\epsilon _1^2<1/2$ when $k>k_*$. It is easy to
    see now that
    $$
    \left\|\Bigl( H^{(2)}\bigl(\vec y^{(2)}\bigr)-k^{2}\Bigr)^{-1}\right\|<
    2\left\|\Bigl(\tilde H^{(1)}\bigl(\vec y^{(1)}\bigr)-k^{2}\Bigr)^{-1}\right\|<\frac{2}{\epsilon _1^2}.$$
    Thus,  we have
    obtained \eqref{3.7.2.2*m}.
    Introducing operators $P_N$ and using the same technique as in the proof of Lemma
    \ref{norm} (see \eqref{alpha} and further)
    we prove that the number of poles of $\Bigl( H^{(2)}\bigl(\vec
    y^{(2)(\varphi )}\bigr)-k^{2}\Bigr)^{-1}$ obeys the same estimate as the number of poles of
    $\Bigl( \tilde H^{(1)}\bigl(\vec
    y^{(1)}(\varphi )\bigr)-k^{2}\Bigr)^{-1}$, i.e., it does not exceed $c_0k^{2+2s_2}$.\end{proof}

\subsubsection{The set ${\mathcal O_s^{(2)}}\left(\vec
b^{(2)}\right)$ for small $\vec b^{(2)}$} \label{SS:3.7.1m}

Everything  we  considered so far is valid if
    $b_0 ^{(2)}>\epsilon _1k^{-1-2\delta }$.
 However, in the next section
and later, $b_0^{(2)}$ is taken smaller, since the reciprocal
lattice is getting finer with each step. To prepare for this, let us
consider $\vec b^{(2)}$ being close to a vertex of $K_2$: $0<b_0
^{(2)}\leq \epsilon _1k^{-1-2\delta }$. We  show  that for such
$\vec b^{(2)}$, the resolvent $\Bigl(H^{(2)}\bigl(\vec
y^{(2)}(\varphi )\bigr)-k^{2}\Bigr)^{-1}$ has no more than two poles
in $\hat \varPhi _2$. We  surround these poles by two contours
$\gamma ^{\pm \ (2)}$ and prove an estimate for the norm of
$\Bigl(H^{(2)}\left(\vec y^{(2)}(\varphi )\right)-k^{2}\Bigr)^{-1}$
when $\varphi $ is outside these contours.

Suppose $|\vec b^{(2)} |=b_0^{(2)}$, i.e., the closest vertex of
$K_2$ for $\vec b^{(2)}$ is $(0,0)$. We consider the functions
$\lambda ^{(1)}(\vec y ^{(1)}(\varphi ))$ and $\lambda ^{(2)}(\vec y
^{(2)}(\varphi ))$ defined  by perturbation series (\ref{2.66}) and
(\ref{3.66}) for $\varphi \in \hat \varPhi _2$. The convergence of
these series can be easily justified. In fact, by Lemmas
\ref{L:2.13} and \ref{L:3.9}, $\vec \varkappa _1(\varphi )$ and
$\vec \varkappa _2(\varphi )$ are holomorphic functions of $\varphi
$ in $\hat \varPhi _2$.
%%%%%%%% It is easy to show that the lemmas hold
%%%%%%%even in $\tilde \varPhi _2$, since all estimates involved are
%%%%%%%preserved in $\tilde \varPhi _2$.
 The perturbation series
(\ref{2.66}) and (\ref{3.66}) converge for $\lambda ^{(1)}(\vec
\varkappa _1(\varphi ))$ and $\lambda ^{(2)}(\vec \varkappa
_2(\varphi ))$, respectively. Since the estimates involved are
stable with respect to a change of $\vec \varkappa _{1,2}$ not
exceeding $\epsilon _1k^{-1-\delta }$, the perturbation series for
$\lambda ^{(1)}\bigl(\vec y^{(1)}((\varphi )\bigr)$ and $\lambda
^{(2)}\bigl(\vec y ^{(2)}(\varphi )\bigr)$ also converge, both
functions being holomorphic in $\hat \varPhi _2$.
 We base our further
considerations on these perturbation series expansions. For $\vec
b^{(2)}$ being close to a vertex $\vec e$ other than $(0,0)$, we
take $\vec y^{(2)}(\varphi )=\vec \varkappa _2(\varphi )+\vec
b^{(2)}-\vec e$.

From now on, we denote the solutions $\varphi ^{\pm }_{\epsilon _0}$
of the equation $\lambda ^{(1)}\bigl(\vec y^{(1)}((\varphi
)\bigr)=k^{2}+\varepsilon _0$, introduced in Lemma \ref{L:3.7.1.1},
by $\varphi ^{(1)\pm }_{\epsilon _0}$. We set
$r^{(2)}=r^{(1)}k^{-2-4s_2-\delta }$. \begin{lem}\label{L:3.7.1.1m}
If $0<b_0 ^{(2)}\leq \epsilon _1k^{-1-2\delta }$ and $
|\epsilon_0|<b_0^{(2)}k^{1-\delta}r^{(2)}$, then the equation
    \begin{equation}\label{3.7.1.2*}
   \lambda ^{(2)}\left(\vec y ^{(2)}(\varphi)\right)=k^{2}+\epsilon_0
   \end{equation}
 has no more than two
solutions $\varphi^{(2)\pm}_{\epsilon _0}$ in $\hat \varPhi _2$. For
any $\varphi^{(2)\pm }_{\epsilon _0}\in \hat \varPhi _2$ there is
$\varphi^{(1) \pm }_{0}\in \hat \varPhi _1$ such that
    \begin{equation}
    \left|\varphi^{(2)\pm }_{\epsilon _0}-\varphi^{ (1)\pm}_{0}\right|
    <r^{(2)}/4, \label{july5a*}
    \end{equation}
    here and below $\varphi^{(1)\pm }_{0}$ is $\varphi ^{(1)\pm }_{\epsilon
    _0}$ for $\epsilon _0=0$.
\end{lem} \begin{proof}  First, we
expand $\lambda ^{(2)}(\vec y ^{(2)}(\varphi))-\lambda ^{(1)}(\vec y
^{(1)}(\varphi))$  near the point $\vec b^{(2)}=0$ and consider that
$\lambda ^{(2)}(\vec \varkappa _2(\varphi))=\lambda ^{(1)}(\vec
\varkappa _1(\varphi))=\lambda$. Then, using (\ref{3.68}),
(\ref{InE:second-derivative-lambda_j^1*}), (\ref{3.71}) and
(\ref{2.20**a}), we check that \begin{equation}\left|\lambda
^{(2)}(\vec y ^{(2)}(\varphi))-\lambda ^{(1)}(\vec y
^{(1)}(\varphi))\right|<b_0^{(2)}\epsilon _1 \label{Nov16a}
\end{equation} in $\hat \varPhi _2$ and even in its
$r^{(2)}$-neighborhood, the neighborhood being a subset of $\hat
\varPhi _1$. Suppose (\ref{3.7.1.2*}) holds for some $\varphi \in
\hat \varPhi _2$. By (\ref{Nov16a}), $\lambda ^{(1)}(\vec y
^{(1)}(\varphi))=k^{2}+\varepsilon _0'$, $\varepsilon
_0'<\varepsilon _0+b_0^{(2)}\epsilon
_1<2b_0^{(2)}k^{1-\delta}r^{(2)}$ when $k>k_*$. Hence, $\varphi $
satisfies conditions of Lemmas \ref{L:3.7.1.1} and \ref{L:july5a}.
Surrounding $\varphi $ by a circle $C$ of the radius $r^{(2)}/4$ and
using \eqref{3.7.1.6.1/2}, we see that $|\lambda ^{(1)}(\vec y
^{(1)}(\varphi))-k^{2}-\varepsilon _0'|\approx
\frac{1}{4}kb_0^{(2)}r^{(2)}>>|\varepsilon _0'|$ on this circle.
Applying Rouch\'{e}'s theorem, we obtain that there is a solution of
$\lambda ^{(1)}(\vec y ^{(1)}(\varphi))=k^{2}$ inside this circle.
Thus, any solution of (\ref{3.7.1.2*}) is in the circle of the
radius $r^{(2)}/4$ surrounding $\varphi^{\pm (1)}_{0}$, the point
$\varphi^{\pm (1)}_{0}$ being in the $(r^{(2)}/4)$-neighborhood of
$\hat \varPhi _2$. It remains to check that (\ref{3.7.1.2*}) has no
more than two solutions and no more than one in a vicinity of each
$\varphi^{\pm (1)}_{0}$. We construct the disk of the radius
$r^{(2)}/4$ centered at $\varphi^{\pm (1)}_{0}$ and note that
$|\lambda ^{(1)}(\vec y
^{(1)}(\varphi))-k^{2}|>\frac{1}{4}kb_0^{(2)}r^{(2)}$ outside the
circle. Using (\ref{Nov16a}) and
 Rouch\'{e}'s Theorem, we obtain that there is only one solution of
 (\ref{3.7.1.2*}) in the disk.
     \end{proof}
\begin{lem}\label{L:july5a*} Suppose $0<b_0 ^{(2)}\leq \epsilon
_1k^{-1-2\delta }$  and $\varphi \in \hat \varPhi _2$ obeys the
inequality analogous to (\ref{july5a*}): $\left|\varphi
-\varphi^{\pm (1)}_{0}\right|
    <r^{(2)}$.
 Then,
    \begin{equation}\label{3.7.1.6.1/2+}
    \frac{\partial}{\partial \varphi}\lambda^{(2)}\left(\vec
    y^{(2)}(\varphi)\right)=_{k\to
    \infty}\pm 2b_0^{(2)}k\bigl(1+o(1)\bigr),
    \end{equation}
    where $|o(1)|<10^{-2}+\epsilon _1$ when $k>k_*$.
\end{lem} The proof is completely analogous to that of Lemma
\ref{L:july5a}.

\begin{definition}\label{SN1} Let $\Gamma^{ (2)\pm}\left(\vec
b^{(2)}\right)$ be the open disks centered at $\varphi_0^{(2)\pm }$
with radius $r^{(2)}$; $\gamma^{ (2)\pm}\bigl(\vec b^{(2)}\bigr)$ be
their boundary circles and ${\mathcal O}_s^{(2)}\bigl(\vec
b^{(2)}\bigr)=\Gamma^{(2)+}\cup \Gamma^{(2)-}$.\end{definition}

\begin{lem}\label{L:3.7.1*} For any $\varphi$ in $\hat \varPhi
_2\setminus {\mathcal O}_s^{(2)}(\vec b^{(2)})$,
    \begin{equation}\label{3.7.1.10*}
    |\lambda^{(2)}(\vec
    y^{(2)}(\varphi))-k^{2}|\geq b_0^{(2)}k^{1-\delta}r^{(2)}.
    \end{equation}
\end{lem}
 The proofs of this and the next lemma
are analogous to those of Lemma \ref{L:3.7.1}, \ref{L:july9c}.

\begin{lem}\label{L:july9c*} For any $\varphi \in \hat \varPhi
_2\setminus {\mathcal O}_s^{(2)}\left(\vec b^{(2)}\right)$,
    \begin{equation}\label{3.7.1.10.1/2*}
    \Bigl\|\Bigl(H^{(2)}\left(\vec
    y^{(2)}(\varphi)\right)-k^{2}\Bigr)^{-1}\Bigr\|<\frac{16}{b_0^{(2)}r^{(2)}k^{1-\delta}},
    \end{equation}
  The estimate is stable in the $\bigl(r^{(2)}k^{-\delta }\bigr)$-neighborhood of
    $\varPhi _2\setminus
{\mathcal O}^{(2)}_{s}(\vec b^{(2)})$.  The resolvent  is an
analytic function of $\varphi $ in every component of $ {\mathcal
O}^{(2)}_{s}(\vec b^{(2)})$, whose intersection with $\varPhi _2$ is
not empty. The only singularities of the resolvent are poles.
     The resolvent has at most two poles
    inside ${\mathcal O}^{(2)}_{s}(\vec b^{(2)})$.
\end{lem}

%%%%%%%%\begin{cor} \label{C:new1} If $\epsilon _2k^{-1-2\delta
%%%%%%%%}<b_0^{(2)}\leq  \epsilon _1k^{-1-2\delta }$, and $\varphi \in
%%%%%%%%\varPhi _2\setminus {\mathcal O}_s^{(2)}\left(\vec b^{(2)}\right)$,
%%%%%%%%then
%%%%%%%%    \begin{equation}\label{3.7.1.10.1/2**}
%%%%%%%%    \Bigl\|\Bigl(H^{(2)}\left(\vec
%%%%%%%%    y^{(2)}(\varphi)\right)-k^{2}\Bigr)^{-1}\Bigr\|<\frac{1}{\epsilon _1^2},
%%%%%%%%    \end{equation}

%%%%%%%%    \end{cor}
%%%%%%%%    The corollary follows from the condition on $b_0^{(2)}$, for
%%%%%%%%    the formula $r^{(3)}=r^{(2)}k^{-2-4s_3-\delta }=k^{-8-6s_1-4s_2-4s_3-5\delta
%%%%%%%%    }$ and estimate (\ref{k}) ???????.
%%%%%%%%    The proof of the lemma is analogous to that of Lemma
%%%%%%%%\ref{L:july9c}.

\section{The Third Approximation} \label{chapt5}

\setcounter{equation}{0} Squaring both sides of \eqref{k}, we easily
obtain the relation which coincide with \eqref{k} up to the
substitution of $s_2$ instead of $s_1$: \begin{equation}
 C_*(1+s_2)k^{2+4s_2}\ln k <k^{\eta s_2},\  \ C_*=400(c_0+1)^2,\ \ c_0=32d_1d_2,\ \ s_2=2s_1.\label{k*}
 \end{equation}
  for any $k>k_*$. We will use \eqref{k*} in the next step.

\subsection{Operator $H_{\alpha}^{(3)}$}

Choosing $s_3=2s_2$, we define the third operator $H_{\alpha}^{(3)}$
by the formula:
 \begin{equation}\label{4.1}
     H_{\alpha }^{(3)}=H^{(2)}+\alpha W_3,\quad      (0\leq \alpha \leq
     1),\qquad
     W_3=\sum_{r=M_2+1}^{M_3}V_r, \notag
     \end{equation}
 where $M_3$ is chosen
in such a way that $2^{M_3}\approx k^{s_3}$. Obviously, the periods
of $W_3$ are $2^{M_3-1} (d_1,0)$ and $2^{M_3-1} (0,d_2)$. We write
them in the form: $N_2N_1(a_1,0)$ and $N_2N_1(0,a_2)$, here
$N_2=2^{M_3-M_2}$, $\frac{1}{4}k^{s_3-s_2}<N_2< 4k^{s_3-s_2}$. Note
that
    \begin{align*}\|W_3\|_{\infty} \leq \sum_{r=M_2+1}^{M_3}
    \|V_r\|_{\infty} \leq \sum_{r=M_2+1}^{M_3}\exp(-2^{\eta r})
    <\exp(-k^{\eta s_2}).\end{align*}
\subsection{Multiple Periods of $W_2(x)$} The operator,
    $H^{(2)}=H_1 + W_2(x)$,
has the periods $N_1a_1, N_1a_2$. The corresponding family of
operators, $\{H^{(2)}(\tau)\}_{\tau \in K_2}$, acts in $L_2(Q_2)$,
where $Q_2=[0,N_1a_1] \times [0,N_1a_2]$ and
 $K_2=[0, 2\pi/N_1a_1)\times [0, 2\pi/N_1a_2)$.  Since now on we
denote  quasimomentum $t$ from the first step by $t^{(1)}$,
quasimomentum $\tau $ from the second  step by $t^{(2)}$.
Correspondingly, quasimomentum for $H_{\alpha }^{(3)}$ we denote by
$t^{(3)}$. Eigenvalues of $H^{(2)}\left(t^{(2)}\right)$ are denoted
by $\lambda_n^{(2)}\left(t^{(2)}\right)$, $n \in \N$ and its
spectrum by $\Lambda ^{(2)}\left(t^{(2)}\right)$.

Next, let us consider  $W_2(x)$ as a periodic function with the
periods $N_2N_1a_1$, $N_2N_1a_2$. When changing the periods, the
family of operators
$\bigl\{H^{(2)}\left(t^{(2)}\right)\bigr\}_{t^{(2)} \in K_2}$ is
replaced by the family of operators $\bigl\{\tilde
H^{(2)}\left(t^{(3)}\right)\bigr\}_{ t^{(3)}\in K_3}$, acting in
$L_2(Q_3)$, where $Q_3=[0,N_2N_1a_1] \times [0,N_2N_1a_2]$ and
$K_3=[0, 2\pi/N_2N_1a_1)\times [0,2\pi/N_2N_1a_2)$. We denote
eigenvalues of $\tilde H^{(2)}\left(t^{(3)}\right)$ by
$\tilde{\lambda}_n^{(2)}\left(t^{(3)}\right)$, $n \in \N$, and its
spectrum by $\tilde{\Lambda}^{(2)}\left(t^{(3)}\right)$.  We denote
now by $P^{(1)}$ the set $P$, introduced by (\ref{May10a}), its
elements being  $p^{(1)}$. By Bloch theory (see e.g.\cite{RS}), for
any $t^{(3)} \in K_3$,
    \begin{equation}\label{4.3}
    \tilde{\Lambda}^{(2)}\left(t^{(3)}\right)=\bigcup_{p^{(2)}\in P^{(2)}}
    \Lambda^{(2)}\left(t^{(3)}+2\pi p^{(2)}/N_2N_1a\right),
    \end{equation}
where $P^{(2)}=\left\{p^{(2)}=\bigl(p^{(2)}_1,p^{(2)}_2\bigr) \in
\Z^2 : 0 \leq p^{(2)}_1 \leq N_2-1,\ 0 \leq p^{(2)}_2 \leq
N_2-1\right\}$, $$2\pi p^{(2)}/N_2N_1a=\left(\frac{2\pi p_1^{(2)}}
{N_2N_1a_1},\frac{ 2\pi p_2^{(2)}}{N_2N_1a_2}\right).$$ An
isoenergetic set $\tilde{S}_2(\lambda _0)\subset K_3$ of the
operator $\tilde H^{(2)}$ is defined by the formula:
    \begin{align*}
    \tilde{S}_2(\lambda )&=\left\{t^{(3)} \in K_3: \exists n \in \N: \
    \tilde{\lambda}_n^{(2)}\bigl(t^{(3)}\bigr)=\lambda \right\}\\
    &=\left\{t^{(3)} \in K_3: \exists n \in \N,\ p^{(2)} \in P^{(2)}:
    \lambda_n^{(2)}\left(t^{(3)}+2\pi p^{(2)}/N_2N_1a\right)=\lambda \right\}.
    \end{align*}
Obviously,
    $\tilde{S}_2=\mathcal{K}_3 S_2, $ where $\mathcal{K}_3$ is the
    parallel shift into $K_3$, that is,
    $$\mathcal{K}_3:\R^2 \rightarrow K_3,\ \ \mathcal{K}_3\left(t^{(3)}+2\pi
    m/N_2N_1a\right)=t^{(3)},\ \  m \in \Z^2,\ \ t^{(3)} \in K_3.$$
    We denote  index $j$, introduced in Part 1 of
Lemma \ref{Lem:Geometric-1} (Geometric Lemma for the First
approximation), by $j^{(1)}$ and $\tilde j $,
    introduced  in Part 1 of
Lemma \ref{L:3.2} (Geometric Lemma for the Second approximation), by
$j^{(2)}$, $j^{(2)}=j^{(1)}+p^{(1)}/N_1.$

\subsection{Perturbation Formulae}

\begin{lem}[Geometric Lemma]\label{L:4.2} For a sufficiently large
$\lambda $, $\lambda >k_*^{2}$, there exists a non-resonance set
$\chi _3(\lambda, \delta) \subset \mathcal{K}_3\chi_2^*$
 such that:
\begin{enumerate} \item For any point $t^{(3)}\in \chi _3$, the
following conditions hold: \begin{enumerate}
    \item There exists a unique $p^{(2)} \in P^{(2)}$
    such that $t^{(3)} + 2\pi p^{(2)}/N_2N_1a \in \chi_2^*$.
    \item The following relation holds:
    $$ \lambda _{{j}^{(2)}}^{(2)}\left(t^{(3)} + 2\pi
    p^{(2)}/N_2N_1a\right)=\lambda ,$$
      where $\lambda _{{j}^{(2)}}^{(2)}\left(t^{(3)} + 2\pi p^{(2)}/N_2N_1a\right)$
      is
      given  by the perturbation series (\ref{3.15}) with $\alpha =1$ and
      ${j}^{(2)}={j}+p/N_1$, here ${j}$ and
      $p$
       are
       defined by the point $\tau = t^{(3)} + 2\pi p^{(2)}/N_2N_1a \ $ as it is described in
        Part 1(b)
      of the Geometric Lemma for the previous step.
    \item The eigenvalue $\lambda _{{j}^{(2)}}^{(2)}\bigl(t^{(3)} + 2\pi
    p^{(2)}/N_2N_1a\bigr)$ is
    a simple eigenvalue of $\tilde H^{(2)}\left(t^{(3)}\right)$ and its distance to all other eigenvalues $\lambda _n^{(2)}\left(t^{(3)} + 2\pi
    \hat{p}^{(2)}/N_2N_1a\right)$ of $\tilde H^{(2)}\bigl(t^{(3)}\bigr)$ is greater than
    $\epsilon_2 =e^{-\frac{1}{4}k^{\eta s_2}}$:
    \begin{equation}\label{4.6}
    \Bigl|\lambda _{j^{(2)}}^{(2)}\left(t^{(3)} + 2\pi p^{(2)}/N_2N_1a\right)-
    \lambda _n^{(2)}\left(t^{(3)} + 2\pi
    \hat{p}^{(2)}/N_2N_1a\right)\Bigl|>\epsilon_2.
    \end{equation}
\end{enumerate} \item For any $t^{(3)}$ in the real $(\epsilon_2
k^{-1-\delta })$-neighborhood of  $\chi_3$, there exists a unique
$p^{(2)} \in P^{(2)}$ such that $t^{(3)} +2\pi p^{(2)}/N_2N_1a$ is
in the  $(\epsilon_2 k^{-1-\delta })$-neighborhood  of $\chi_2^*$
and
    \begin{equation}\label{4.7}
    \Bigl| \lambda_{j^{(2)}}^{(2)}\left(t^{(3)} +2\pi p^{(2)}/N_2N_1a\right)-k^{2}\Bigr| < \epsilon_2 k^{-\delta
    },
    \end{equation}${j}^{(2)}={j}+p/N_1$, here ${j}$ and
      $p$
       are
       defined by the point $\tau = t^{(3)} + 2\pi p^{(2)}/N_2N_1a \ $
        as it is described in
        Part 2
      of the Geometric Lemma for the previous step.
       An estimate analogous to
    \eqref{4.6} holds:
    \begin{equation}\label{4.6a}
    |\lambda _j^{(2)}(t^{(3)} +2\pi p^{(2)}/N_2N_1a)-\lambda _{\hat n}^{(1)}(t^{(3)} +2\pi \hat p^{(2)}/N_2N_1a)|>
    \bigl(1+o(1)\bigr)\epsilon_2.
    \end{equation}
\item The third nonresonance set $\chi_3$ has an asymptotically full
measure on $\chi_2^*$ in the following sense:
\begin{equation}\label{4.9} \frac{L\left(\mathcal{K}_3\chi_2^*
\setminus \chi _3)\right)}{L\left(\chi_2^*\right)}<
c_0k^{-4-2s_1-2s_2}. \end{equation} \end{enumerate} \end{lem}
\begin{proof} The proof of the lemma is analogous to that for
Geometric Lemma in the second step. Indeed, let us consider
\begin{equation}\label{O^3}\mathcalO^{(3)}=\cup _{\tilde p^{(2)}\in
P^{(2)}\setminus \{0\}}\mathcalO^{(2)}_{s}\left(\frac{2\pi \tilde
p^{(2)}}{N_2N_1a}\right),\end{equation} where
$\mathcalO_{s}^{(2)}\left(\frac{2\pi \tilde
p^{(2)}}{N_2N_1a}\right)$ is defined by (\ref{O_2}) with $\vec
b^{(2)}=\frac{2\pi \tilde p^{(2)}}{N_2N_1a}$. Note, that the
definition make sense, since $|\frac{2\pi \tilde
p^{(2)}}{N_2N_1a}|>d_{max}^{-1}k^{-s_3}>\epsilon _2k^{-1-2\delta }$.
Let \begin{equation} \varPhi _3=\varPhi _2\setminus
\mathcalO^{(3)},\ \ \ \Theta _3=\varPhi _3\cap [0,2\pi ).
\label{D:theta _3} \end{equation}By Lemma \ref{L:3.23},
\begin{equation} \left\|\left(H ^{(2)}\Bigl(\vec \varkappa
_2(\varphi )+\frac{2\pi \tilde
p^{(2)}}{N_2N_1a}\Bigr)-k^{2}\right)^{-1}\right\|<\frac{2}{\epsilon
_1^2} \label{normres****a}\end{equation}  for all $\tilde p^{(2)}\in
P^{(2)}\setminus\{0\}$ and $\varphi \in \hat \varPhi _3$, here and
below $\hat \varPhi _3$ is the $\left(r^{(2)}k^{-\delta
}\right)$-neighborhood of $\varPhi _3$. We consider
$\mathcalD_{2,nonres}\subset \mathcalD_{2}$:
\begin{equation} \label{D_{2,nonres}}
\mathcalD_{2,nonres}=\left\{\vec \varkappa _2(\varphi ), \ \varphi
\in \Theta _3\right\}.\end{equation}We define $\chi _3$ by the
formula: \begin{equation} \label{chi_3} \chi
_3=\mathcalK_3\mathcalD_{2,nonres}.\end{equation} By the definition
of $\mathcalK_3$, for every $t^{(3)}$ in $\chi _3$, there are
$p^{(1)}\in P^{(1)}$, $p^{(2)}\in P^{(2)}$ and $j\in Z^2$ such that
\begin{equation} \label{tau+}t^{(3)} + \frac{2\pi
p^{(2)}}{N_2N_1a}+\frac{2\pi p^{(1)}}{N_1a}+\frac{2\pi j}{a}=\vec
\varkappa _2(\varphi ),\ \ \vec \varkappa _2(\varphi )\in
\mathcalD_{2,nonres}.\end{equation} Considering the definition of
$\mathcalD_{2,nonres}$, we obtain: \begin{equation} \left\|\left(H
^{(2)}\Bigl(t^{(3)} + \frac{2\pi \hat
p^{(2)}}{N_2N_1a}\Bigr)-k^{2}\right)^{-1}\right\|<\frac{2}{\epsilon
_1^2}, \label{normres****b}\end{equation} $\hat p^{(2)}=
p^{(2)}+\tilde p^{(2)}.$ Note that the indices $j$, $p^{(1)}$ do not
play a role, since they just produce a shift of indices of the
matrix elements of the resolvent. Considering that $\tilde p^{(2)}$
can be any but zero, we obtain that (\ref{normres****b}) holds for
all $\hat p ^{(2)}\in P^{(2)}\setminus\{p^{(2)}\}$. Taking into
account that $\lambda _j^{(2)}(t^{(3)} +2\pi p^{(2)}/N_2N_1a)=k^{2}$
and $\epsilon _2<\epsilon _1^2$, we arrive at \eqref{4.6a} for all
$\hat p\neq p$. It remains to check \eqref{4.6a} for $p=\hat p$. Let
$t^{(2)}=t^{(3)} +2\pi p^{(2)}/N_2N_1a$. By (\ref{tau+}),
$t^{(2)}\in \mathcalK_2 \mathcalD _2$. Using \eqref{3.77.7}, we get
$t^{(2)}\in \chi _2^*$. By Theorem \ref{T:3.4}, $\lambda
_j^{(2)}(t^{(3)} +2\pi p^{(2)}/N_2N_1a)$ is the only eigenvalue of
$H ^{(2)}\left(t^{(3)} +2\pi p^{(2)}/N_2N_1a\right)$ in the interval
$\varepsilon _2$. Hence,
 $$|\lambda _j^{(2)}(t^{(3)} +2\pi p^{(2)}/N_2N_1a)-\lambda _{\hat n}^{(2)}(t^{(3)} +2\pi p^{(2)}/N_2N_1a)|
 >\epsilon_1.$$

    Part 2 holds, since all estimates are stable with respect to the
    perturbation of $t^{(3)}$ less then $\epsilon _2k^{-1-\delta }$.

    Let us estimate the size of $\mathcalO ^{(3)}$. According to Lemma
    \ref{L:totalsize}
the total size of each ${\mathcal O}^{(2)}_{s}(2\pi
p^{(2)}/N_2N_1a)$ is less then $c_0k^{-4-5s_2-4\delta }$.
     Considering that the number of
$p^{(2)}$ does not exceed $k^{2(s_3-s_2)}=k^{2s_2}$, $s_2=2s_1$, we
arrive at (\ref{4.9}).

\end{proof}

\begin{cor}\label{C:4.3} If $t^{(3)}$ belongs to the complex
$(\epsilon_2 k^{-1-\delta})-$neighborhood of
$\chi_3(\lambda,\delta)$, then for any $z$ lying on the circle
$C_3=\{ z: |z-k^{2}| =\epsilon_2/2 \} $, the following inequality
holds: \begin{align} \Bigl\|\bigl(\tilde
H^{(2)}(t^{(3)}\bigr)-z)^{-1}\Bigr\| & < \frac{4}{\epsilon_2}.
\label{4.11} \end{align} \end{cor} \begin{remark} Every point $2\pi
q/N_2N_1a $ ($q\in \Z ^2$) of the dual lattice  for periods
$N_2N_1a_1$, $N_2N_1a_2$ can be uniquely represented in the form:
$2\pi q/N_2N_1a =2\pi m/N_1a+2\pi p^{(2)}/N_2N_1a$, where $m\in
\Z^2$, $p^{(2)}\in P^{(2)}$. Note that $2\pi m/N_1a$ is a point of a
dual lattice for periods $N_1a_1$, $N_1a_2$ and $p^{(2)}\in P^{(2)}$
is responsible for refining the lattice. By Remark \ref{R:11}, $2\pi
q/N_2N_1a $ also can be uniquely represented as $2\pi q/N_2N_1a
=2\pi j/a+2\pi p^{(1)}/N_1a+2\pi p^{(2)}/N_2N_1a$, here $j\in \Z^2$,
$p^{(1)}\in P^{(1)}$, $p^{(2)}\in P^{(2)}$.

 Let us
consider a normalized eigenfunction $\psi_n\left(t^{(2)},x\right)$
of $H^{(2)}\left(t^{(2)}\right)$ in $L_2(Q_2)$. We extend it
quasiperiodically to $L_2(Q_3)$, renormalize and denote the new
function by $\tilde{\psi}_n\left(t^{(3)},x\right)$,
$t^{(3)}={\mathcal K}_3t^{(2)}$. The Fourier representations of
$\psi_n\left(t^{(2)},x\right)$ in $L_2(Q_2)$ and
$\tilde{\psi}_n(t^{(3)},x)$ in $L_2(Q_3)$ are simply related. If we
denote Fourier coefficients of $\psi_n\left(t^{(2)},x\right)$ with
respect to the basis
    $|Q_2|^{-1/2}e^{i\left(2\pi m/N_1a+t^{(2)},x\right)}$, $m \in \Z^2,$
in $L_2(Q_2)$  by $C_{nm}$, then, obviously, the Fourier
coefficients $\tilde{C}_{nq}$ of
$\tilde{\psi}_n\left(t^{(3)},x\right)$ with respect to the basis
    $|Q_3|^{-1/2}e^{i\left(2\pi q/N_2N_1a+t^{(3)},x\right)}$, $q \in
    \Z^2$,
in $L_2(Q_3)$ are given by the formula:
    \begin{equation*}
    \tilde{C}_{nq}=
        \begin{cases}
        C_{nm},  &\text{if $q=mN_2+p^{(2)}$;}\\
        0,       &\text{otherwise},
        \end{cases}
    \end{equation*}
    $p^{(2)}$ being defined from the relation
    $t^{(2)}=t^{(3)}+2\pi p^{(2)}/N_1N_2a$.
Correspondingly, matrices of the projections on $\psi_n(\tau,x)$ and
$\tilde{\psi}_n(t^{(3)},x)$ with respect to the above bases are
simply related:
    \begin{equation*}
    (\tilde{E}_n)_{q\hat{q}}=
        \begin{cases}
        (E_n)_{m\hat{m}},  &\text{if $q=mN_2+p^{(2)},\ \hat{q}=\hat{m}N_2+p^{(2)}$;}\\
        0,       &\text{otherwise},
        \end{cases}
    \end{equation*}
$\tilde{E}_n$ and $E_n$ being projections in $L_2(Q_3)$ and
$L_2(Q_2)$, respectively. \end{remark}

 We define functions $g_r^{(3)}(k,t^{(3)})$ and
operator-valued functions $G_r^{(3)}(k,t^{(3)})$, $r=1, 2, \cdots ,$
as follows: \begin{equation*}
 g_r^{(3)}\bigl(k,t^{(3)}\bigr)=\frac{(-1)^r}{2\pi ir}\mbox{Tr}\oint _{C_3}
 \left(\Bigl(\tilde H^{(2)}\bigl(t^{(3)}\bigr)-z\Bigr)^{-1}W_3\right)^rdz,
 \end{equation*}
\begin{equation*}
G_r^{(3)}\bigl(k,t^{(3)}\bigr)=\frac{(-1)^{r+1}}{2\pi i}\oint
_{C_3}\left(\Bigl(\tilde
H^{(2)}\bigl(t^{(3)}\bigr)-z\Bigr)^{-1}W_3\right)^r \Bigl(\tilde
H^{(2)}\bigl(t^{(3)}\bigr)-z\Bigr)^{-1}dz.
 \end{equation*}
\begin{thm}\label{T:4.4}
 Suppose $t^{(3)}$ belongs to the
$(\epsilon_2 k^{-1-\delta })$-neighborhood in $K_3$ of the third
nonresonance set $\chi _3(\lambda ,\delta )$, $0<\delta <s_1$,
$\epsilon_2=e^{-\frac{1}{4}k^{\eta s_2}}$. Then,  $\lambda >k_*^{2}$
and for all $\alpha $, $0 \leq \alpha \leq 1$, there exists a unique
eigenvalue of the operator $H_{\alpha}^{(3)}\left(t^{(3)}\right)$
 in
the interval $\varepsilon_3 (k):= (k^{2}-\epsilon_2 /2,
k^{2}+\epsilon_2 /2)$. It is given by the series:
\begin{equation}\label{4.15} \lambda_{{j^{(3)}}}^{(3)} \bigl(\alpha
,t^{(3)}\bigr)=\lambda_{j^{(2)}}^{(2)}\left( t^{(3)}+2\pi
p^{(2)}/N_2N_1a \right)+\sum _{r=1}^{\infty }\alpha ^r
g_r^{(3)}\bigl(k,t^{(3)}\bigr), \end{equation} converging absolutely
in the disk $|\alpha|  \leq 1$, where
${j^{(3)}}:=j^{(2)}+p^{(2)}/N_2N_1$, $p^{(2)}$, $j^{(2)}$ being
described in Geometric Lemma  \ref{L:4.2}. The spectral projection,
corresponding to $\lambda_{{j^{(3)}}}^{(3)} (\alpha ,t^{(3)})$, is
given by the series: \begin{equation}\label{4.16}
E_{{j^{(3)}}}^{(3)} \bigl(\alpha
,t^{(3)}\bigr)=\tilde{E}_{j^{(2)}}^{(2)}\left(t^{(3)}+2\pi
p^{(2)}/N_2N_1a \right)+\sum _{r=1}^{\infty }\alpha
^rG_r^{(3)}\bigl(k,t^{(3)}\bigr), \end{equation} which converges in
the trace class $\mathbf{S_1}$ uniformly with respect to $\alpha $
in the disk  $| \alpha | \leq 1$.

The following estimates hold for coefficients
$g_r^{(3)}(k,t^{(3)})$,  $G_r^{(3)}(k,t^{(3)})$:
\begin{equation}\label{4.17} \Bigl\| g_r^{(3)}\bigl(k,t^{(3)}\bigr)
\Bigr|<\frac{3\epsilon_2}{2}\left(4\epsilon_2^3\right)^r, \ \ \ \ \
 \Bigl\| G_r^{(3)}\bigl(k,t^{(3)}\bigr)\Bigr\| _1< 6r \left(4 \epsilon_2^3\right)^r.
\end{equation} \end{thm} \begin{cor}\label{C:4.5}
 The
following estimates hold for the perturbed eigenvalue and its
spectral projection: \begin{equation}\label{4.19} \Bigl|
\lambda_{{j^{(3)}}}^{(3)} \bigl(\alpha
,t^{(3)}\bigr)-\lambda_{j^{(2)}}^{(2)} \bigl(t^{(3)}+2\pi
p^{(2)}/N_2N_1a\bigr)\Bigr| \leq 12 \alpha \epsilon_2 ^4,
\end{equation} \begin{equation}\label{4.20}
\Bigl\|E_{{j^{(3)}}}^{(3)} \bigl(\alpha
,t^{(3)}\bigr)-\tilde{E}_{j^{(2)}}^{(2)}\bigl(t^{(3)}+2\pi
p^{(2)}/N_2N_1a\bigr) \Bigr\|_1\leq 48 \alpha \epsilon_2^3 .
\end{equation} \end{cor} Proof of the theorem is analogous to that
of the Theorem \ref{T:3.4}. The series~(\ref{4.15}),~(\ref{4.16})
can be extended as holomorphic functions of $t^{(3)}$ in the complex
$\left(\epsilon _2k^{-1-\delta }\right)$-neighborhood of $\chi _3$;
they can be differentiated any number of times with respect to
$t^{(3)}$ and retain their asymptotic character. The results
analogous to Lemma \ref{T:3.6}, Theorem \ref{T:3.5a}, Corollary
\ref{C:3.6} and Lemma \ref{L:3.5.1/2} hold.

    \subsection{Nonresonant part of the isoenergetic set of
$H_{\alpha }^{(3)}$}\label{S:3.4*}

This section is analogous to Section \ref{S:3.4} for the second
step. Indeed, let $S_3(\lambda )$ be an isoenergetic set of the
operator $H_{\alpha}^{(3)}$:
    $S_3(\lambda)=\{t^{(3)} \in K_3 : \exists n \in \N:
    \lambda_n^{(3)}(\alpha,t^{(3)})=\lambda  \}$,
    here $\{\lambda_n^{(3)}(\alpha,t^{(3)})\}_{n=1}^{\infty }$ is the
    spectrum of $H_{\alpha}^{(3)}(t^{(3)})$.
Now we construct a non-resonance subset $\chi _3^*(\lambda )$ of
$S_3(\lambda
    )$. It
    corresponds to  non-resonance eigenvalues $\lambda
    _{j^{(3)}}^{(3)}(t^{(3)} )$ given by the perturbation series
    (\ref{4.15}). Recall that
    $\mathcal{D}_2(\lambda )_{nonres}$ and $\chi _3$ are defined by
    formulas \eqref{D_{2,nonres}} and \eqref{chi_3}.
Recall also that $\chi _3\subset \mathcal{K}_3\chi_2^*(\lambda )$
(see
     Geometric Lemma \ref{L:4.2}) and $\chi_2^*(\lambda
    )=\mathcal{K}_2\mathcal{D}_2(\lambda )$, see (\ref{3.77.7}). Hence, $\chi _3\subset \mathcal{K}_3\mathcal{D}_2(\lambda
    )$.
The following lemma is analogous to Lemma \ref{L:3.7.1/2} in the
second step. \begin{lem}\label{L:3.7.1/2*} The formula
$\mathcal{K}_3\mathcal{D}_2(\lambda )_{nonres}=\chi_3$ establishes
 one-to-one correspondence between
$\mathcal{D}_2(\lambda )_{nonres}$ and $\chi_3$. \end{lem}

\begin{proof} The proof is analogous to that of Lemma
\ref{L:3.7.1/2} up to the shift of indices by 1, i.e., $\chi _2\to
\chi _3$, $\chi_1^*(\lambda )\to \chi_2^*(\lambda )$, $\tau
=t^{(2)}\to  t^{(3)}$; we use  formula (\ref{3.77.7}) instead of
(\ref{2.81}), Part 1a of the Geometric Lemma for the third step
instead of Part 1a of the Geometric Lemma for the second step, and
Lemma  \ref{L:May10a*} instead of  \ref{L:May10a}.
     \end{proof}

We define $\mathcal{B}_3(\lambda )$ as the set of directions
corresponding to $\Theta _3$, $\Theta _3$ being given by
\eqref{D:theta _3}: $$\mathcal{B}_3(\lambda)=\{\vec{\nu} \in
\mathcal B_2(\lambda):
   \varphi  \in \Theta _3\}.$$
Note that $\mathcal B_3 (\lambda)$ is a unit circle with holes,
centered at the origin, and $\mathcal B_3(\lambda ) \subset
    \mathcal B_2(\lambda )\subset \mathcal{B}_1(\lambda )$.
We define $\mathcal{D}_3(\lambda)$ as a level set for
$\lambda^{(3)}(\alpha,\vec{\varkappa })$ in the $\left(\epsilon
_2k^{-1-\delta }\right)$-neighborhood of $\mathcal{D}_2(\lambda
)_{nonres}$:
    \begin{equation*}
    \mathcal{D}_3(\lambda ):=\left\{\vec{\varkappa  }=\varkappa  \vec \nu: \vec \nu \in
    \mathcal{B}_3(\lambda ),\ \
    \bigl| \varkappa  -\varkappa _2(\lambda,\vec
\nu) \bigr|<\epsilon_2k^{-1-\delta},\
\lambda^{(3)}(\alpha,\vec{\varkappa  })=\lambda
\right\}.\label{May14m*} \end{equation*} Next two lemmas are to
prove   that $\mathcal{D}_3(\lambda )$ is a distorted circle with
holes. Their formulations and proofs are analogous to those of
Lemmas \ref{L:3.8} and \ref{L:3.9}. \begin{lem}\label{L:3.8*} For
every $\vec{\nu} \in \mathcal{B}_3(\lambda )$ and every $\alpha$, $0
\leq \alpha \leq 1$, there is a unique $\varkappa   =\varkappa _3
(\lambda,\vec{\nu})$ in the interval $I_3:=\bigl[\varkappa _2
(\lambda,\vec \nu)-\epsilon_2 k^{-1-\delta},\varkappa _2
(\lambda,\vec \nu)+\epsilon_2 k^{-1-\delta}\bigr]$ such that
    \begin{equation}\label{3.701}
    \lambda^{(3)}(\alpha,\varkappa _3 \vec{\nu})=\lambda .
    \end{equation}
Furthermore,
    \begin{equation}\label{3.71m}
    |\varkappa _3 (\lambda,\vec
\nu)-\varkappa _2(\lambda,\vec \nu)| \leq 2\alpha \epsilon_2
^4k^{-1}.
    \end{equation}
\end{lem}  Further we use the notations $\varkappa _3(\varphi
)\equiv \varkappa _3 (\lambda,\vec{\nu})$, $h_3(\varphi )\equiv
\varkappa _3(\varphi )-\varkappa _2(\varphi )$, $\vec \varkappa
_3(\varphi )=\varkappa _3(\varphi )\vec \nu $.
\begin{lem}\label{L:3.9*} \begin{enumerate}
The following statements hold  for $\lambda >k^2_*$:
\item The set $\mathcal{D}_3(\lambda )$ is a distorted circle with
holes: it can be described by the formula: \begin{equation}
\mathcal{D}_3(\lambda )=\bigl\{\vec \varkappa \in \R^2: \vec
\varkappa =\vec \varkappa _3(\varphi ),\ \ \varphi \in \Theta
_3(\lambda )\bigr\},\label{May20a1} \end{equation} where
    $ \varkappa _3(\varphi )=
    \varkappa _2(\varphi)+h_3(\varphi ), $
$\varkappa _2(\varphi )$ is the ``radius" of $\mathcal{D}_2(\lambda
)$ and $h_3(\varphi )$ satisfies the estimates
    \begin{equation}\label{3.75*}
    |h_3|<12\alpha \epsilon_2 ^4k^{-1},\ \ \
    \left|\frac{\partial h_3}{\partial \varphi} \right| \leq
    96\alpha \epsilon_2^3 k^{1+\delta}.
    \end{equation}
    \item The total length of $\mathcal{B}_3(\lambda)$ satisfies the estimate:
\begin{equation}\label{theta2*s}
    L\left(\mathcal{B}_2\setminus \mathcal{B}_3\right)<4\pi k^{-4-2s_1-2s_2}.
    \end{equation}
\item The function $h _3(\varphi )$ can be extended as a holomorphic
function of $\varphi $ to the complex non-resonce set $\varPhi _3$
and its $\left(k^{-\delta }r^{(2)}\right)$-neighborhood $\hat
\varPhi _3$, estimates (\ref{3.75*}) being preserved.
\item The curve $\mathcal{D}_3(\lambda )$ has a length which is
asymptotically close to that of $\mathcal{D}_2(\lambda )$ in the
following sense:
    \begin{equation}\label{3.77m}
     L\Bigl(\mathcal{D}_3(\lambda )\Bigr)\underset{\lambda \rightarrow
     \infty}{=}L\Bigl(\mathcal{D}_2(\lambda
     )\Bigr)\Bigl(1+O(k^{-4-2s_1-2s_2})\Bigr),
     \end{equation}
     where $O\bigl(k^{-2-2s_1-2s_2})=(1+o(1))c_0k^{-2-2s_1-2s_2}$,
     $|o(1)|<10^{-2}$ when $k>k_*$.
     \end{enumerate}
\end{lem} \begin{proof} The proof is analogous to that of Lemma
\ref{L:3.9}. Note only that, in Part 2, when proving convergence  of
the series for the resolvent $\left(H^{(3)}(\vec \varkappa
_2(\varphi
 ))-z\right)^{-1}$, we use the estimate
 \begin{equation}
 \Bigl\|\left(\tilde H^{(2)}(\vec \varkappa _2(\varphi
 ))-z\right)^{-1}\Bigr\|<\frac{4}{\epsilon _2},\ \ z\in C_3,\label{July3b2}
 \end{equation}
 analogous to (\ref{Sep10.09}), the operator  $\tilde H^{(2)}$ acting in $L_2(Q_3)$. The estimate
 (\ref{July3b2}) follows from \eqref{normres****a} and
\eqref{3.40.2}.
 As
 a side result of these considerations, we obtain an estimate
 analogous to (\ref{16**}) for the new resolvent and $z$ being
 inside $C_3$.
      \end{proof}
We define the non-resonance set, $\chi_3^*(\lambda )$ in
$S_3(\lambda )$ by the formula analogous to (\ref{3.77.7}):
    \begin{equation}\label{3.77.7*}
    \chi_3^*(\lambda ):=\mathcal{K}_3\mathcal{D}_3(\lambda ).
    \end{equation}
    The
following lemmas are analogous to Lemmas \ref{Apr4a} and
\ref{L:May10a*}.
    \begin{lem} \label{Apr4a*}
    The set $\chi_3^*(\lambda )$ belongs to the
    $\left(12\alpha \epsilon _2^4k^{-1}\right)$-neighborhood of $\chi_3(\lambda
    )$ in $K_3$. If $t^{(3)}\in \chi_3^*(\lambda )$, then the operator
    $H^{(3)}_{\alpha }(t^{(3)})$ has a simple eigenvalue
   equal to $\lambda $. This
    eigenvalue is given by the perturbation series (\ref{4.15}).
\end{lem}

    \begin{lem}\label{L:May10a**} Formula (\ref{3.77.7*}) establishes
    one-to-one correspondence between $\chi_3^*(\lambda )$ and $\mathcal{D}_3(\lambda
    )$.\end{lem}
    \begin{remark} \label{R:May14*}
    From geometric point of view this means that $\chi_3^*(\lambda
    )$ does not have self-intersections.\end{remark}

\subsection{Preparation for the Next Approximation} \label{S:3.6*}
Let $\vec b^{(3)} \in K_3$ and $b_0^{(3)}$ be the distance of the
point $\vec b^{(3)}$ to the nearest corner of $K_3$:
\begin{equation}\label{3.2.3.1*}
b_0^{(3)}=\min_{m=(0,0),(0,1),(1,0),(1,1)}|\vec b^{(3)}-2\pi
m/N_2N_1a|. \end{equation}
 We assume $b_0^{(3)}=|\vec b^{(3)}|$. In the case
    when $\vec b ^{(3)}$ is closer to a vertex other than $(0,0)$, the
    considerations are the same up to a parallel shift. We  consider  two cases:
$b_0^{(3)}\geq
 \epsilon _2k^{-1-2\delta }$ and
$0<b_0^{(3)}<\epsilon _2k^{-1-2\delta }$. Let \begin{equation}\vec
y^{(2)}(\varphi)=\vec \varkappa _2(\varphi )+\vec
    b^{(3)}. \label{y_1*} \end{equation}\subsubsection{The case $b_0^{(3)}\geq \epsilon _2k^{-1-2\delta
}$.}\begin{definition} \label{D:**hh}We  define the set ${\mathcal
O}^{(3)}(\vec b^{(3)})$ by the formula: \begin{equation} {\mathcal
O}^{(3)}(\vec b^{(3)})=\cup _{p^{(2)}\in P^{(2)}}
\mathcalO_{s}^{(2)}(2\pi
    p^{(2)}/N_2N_1a+\vec b^{(3)} ), \label{O_3(b_3)}\end{equation}
    set $\mathcalO_{s}^{(2)}$  being defined by formula \eqref{O_2}.
    We assume $\mathcalO_{s}^{(2)}\cap \hat \varPhi _2 \not
    =\emptyset $.
\end{definition}
The set ${\mathcal O}^{(3)}(\vec b^{(3)})$ consists  the disks with
the radius $r^{(2)}=r^{(1)}k^{-2-4s_2-\delta }$ centered at  poles
of the resolvent $\Bigl(\tilde H_2\bigl(\vec y^{(2)}(\varphi
)\bigr)-k^2\Bigr)^{-1}$.
    \begin{lem} \label{L:3.7.2.2*mm3}If
$\varphi \in \hat \varPhi _2\setminus {\mathcal O}^{(3)}(\vec
b^{(3)})$, then
\begin{equation}\label{3.7.2.2*mm3}
    \Bigl\|\Bigl(\tilde H^{(2)}\bigl(\vec y^{(2)}(\varphi)\bigr)-k^{2}\Bigr)^{-1}\Bigr\| \leq
    \frac{17r^{(2)}k^{3\delta}}{\epsilon _2}
    \end{equation}
    The estimate is stable in the $\bigl(r^{(2)}k^{-\delta }\bigr)$-neighborhood of
    $\hat \varPhi _2\setminus
{\mathcal O}^{(3)}(\vec b^{(3)})$.
    The resolvent  is an analytic function of
$\varphi $ in every connected component  $ {\mathcal O}_c^{(3)}(\vec
b^{(3)})$ of $ {\mathcal O}^{(3)}(\vec b^{(3)})$, whose intersection
with $\varPhi _2$ is not empty. The only singularities of the
resolvent in such a component are poles.
    The number of poles (counting multiplicity) of the resolvent
    inside ${\mathcal O}^{(3)}(\vec b^{(3)})$ is less than
    $ck^{2+2s_3}$. The total size of ${\mathcal O}^{(3)}(\vec b^{(3)})$
    does not exceed $c_0k^{-4-2s_1-2s_2}$.
    \end{lem}\begin{proof} The proof is analogous to that of Lemma
    \ref{L:3.7.2.2*mm}. To obtain \eqref{3.7.2.2*mm3} we combine
   Lemmas \ref{L:3.23} and \ref{L:july9c*} for $\vec b^{(2)}=2\pi
    p^{(2)}/N_2N_1a+\vec b^{(3)}$, $p^{(2)}\in P^{(2)}$, and take into account that
   $\epsilon _1^2 >\epsilon _2$.
    By  the same lemmas  the number of poles of $\Bigl(H^{(2)}\bigl(\vec \varkappa _2(\varphi )+2\pi
    p^{(2)}/N_2N_1a+\vec b^{(3)}\bigr)-k^{2}\Bigr)^{-1}$ inside ${\mathcal O}_s^{(2)}(2\pi
    p^{(2)}/N_2N_1a+\vec b^{(3)})$ is less than $ck^{2+2s_2}$.
    Considering that this resolvent does not have poles outside ${\mathcal O}_s^{(2)}(2\pi
    p^{(2)}/N_2N_1a+\vec b^{(3)})$, we obtain that the total number
    of poles of $\Bigl(H^{(2)}\bigl(\vec \varkappa _2(\varphi )+2\pi
    p^{(2)}/N_2N_1a+\vec b^{(3)}\bigr)-k^{2}\Bigr)^{-1}$ inside ${\mathcal O}^{(3)}(\vec
    b^{(3)})$ is less than $c_0k^{2+2s_2}$.
    Taking into account that the number of $p^{(2)}\in P^{(2)}$ does not exceed $ck^{2(s_3-s_2)}$, $s_3=2s_2$,
    we obtain the estimate for the number of poles
    for the resolvent $\Bigl(\tilde H^{(2)}\bigl(\vec y^{(2)}((\varphi)\bigr)-k^{2}\Bigr)^{-1}$.

  We estimate the size of $\mathcalO ^{(3)}(\vec b^{(3)})$ the same way as we estimated the size of
  $\mathcalO ^{(3)}$. Indeed, according to Lemma
    \ref{L:totalsize}
the total size of each ${\mathcal O}^{(2)}_{s}(2\pi
p^{(2)}/N_2N_1a)$ is less then $c_0k^{-4-5s_2-4\delta }$.
     Considering that the number of
$p^{(2)}$ does not exceed $k^{2(s_3-s_2)}=k^{2s_2}$, $s_2=2s_1$, we
arrive at the estimate for  $\mathcalO ^{(3)}(\vec
b^{(3)})$.\end{proof}

\begin{definition} \label{D:**h}We denote the poles of the resolvent $\Bigl(\tilde
H^{(2)}\bigl(\vec y^{(2)}((\varphi)\bigr)-k^{2}\Bigr)^{-1}$ in
${\mathcal O}^{(3)}(\vec b^{(3)})$ as $\varphi ^{(3)}_n$,
$n=1,...,M^{(3)}$ $M^{(3)} <c_0k^{2+2s_3}$. Let us consider the
circles ${\mathcal O}_n^{(3)}(\vec b^{(3)})$ of the radius
$r^{(3)}=k^{-2-4s_3-\delta }r^{(2)}$ around these poles. Let
\begin{equation}{\mathcal O}^{(3)}_{s}(\vec b^{(3)})=\cup
_{n=1}^{M^{(3)}}{\mathcal O}_n^{(3)}(\vec
b^{(3)}).\label{O_3}\end{equation} \end{definition}\begin{lem}
\label{L:totalsize3} The total size of ${\mathcal O}^{(3)}_{s}(\vec
    b^{(3)})$ is less then $c_0k^{-2s_3-\delta }r^{(2)}$.\end{lem}
    \begin{proof} The lemma easily follows from the formula $r^{(3)}=k^{-2-4s_3-\delta
}r^{(2)}$
    and the estimate $M^{(3)}<c_0k^{2+2s_3}$. \end{proof}

\begin{lem}\label{L:3.7.2.2*mmm3} If $\varphi \in \hat \varPhi _2\setminus
{\mathcal O}^{(3)}_{s}(\vec b^{(3)})$, then
\begin{equation}\label{3.7.2.2*mmm3}
    \Bigl\|\Bigl(\tilde H^{(2)}\bigl(\vec y^{(2)}(\varphi)\bigr)-k^{2}\Bigr)^{-1}\Bigr\| \leq
    \frac{1}{\epsilon _2^2}
    \end{equation}The estimate is stable in the $\bigl(r^{(3)}k^{-\delta }\bigr)$-neighborhood of
    $\hat \varPhi _2\setminus
{\mathcal O}^{(3)}_{s}(\vec b^{(3)})$.
    The resolvent  is an analytic function of
$\varphi $ in every connected component  $ {\mathcal
O}_{sc}^{(3)}(\vec b^{(3)})$ of $ {\mathcal O}^{(3)}_s(\vec
b^{(3)})$, whose intersection with $\varPhi _2$ is not empty. The
only singularities of the resolvent in such a component are poles.
    The number of poles (counting multiplicity) of the resolvent
    inside ${\mathcal O}^{(3)}_{s}(\vec b^{(3)})$ is less than
    $c_0k^{2+2s_3}$.
    \end{lem}
    \begin{proof} By the definition of ${\mathcal
O}^{(3)}_{s}(\vec b^{(3)})$, the number of poles (counting
multiplicity) of the resolvent
    inside this set is less than
    $ck^{2+2s_3}$. Considering as in Lemma \ref{norm**} we obtain
    \begin{equation}\label{3.7.2.2*mmmm-2}
    \Bigl\|\Bigl(\tilde H^{(2)}\bigl(\vec y^{(2)}(\varphi)\bigr)-k^{2}\Bigr)^{-1}\Bigr\| \leq
    \nu ^{-M^{(3)}}\frac{17r^{(2)}k^{3\delta}}{\epsilon _2},
    \end{equation}
    where $\nu $ is the coefficient of contraction, when we reduce ${\mathcal O}^{(3)}(\vec
    b^{(3)})$ to
    ${\mathcal
O}^{(3)}_{s}(\vec b^{(3)})$. Namely $\nu $ is the ratio of $r^{(3)}$
to the maximal size of ${\mathcal O}^{(3)}(\vec b^{(3)})$. By Lemma
\ref{L:3.7.2.2*mm3} and the definition of $r^{(3)}$, $\nu
=k^{-4-4s_3-4s_2-2\delta } $.  Considering that $\nu
^{-M^{(3)}}r^{(2)}k^{\delta}<\epsilon _2^{-1}$, we obtain
\eqref{3.7.2.2*mmm3}.
\end{proof}

By Lemma \ref{L:totalsize3}, the total size of ${\mathcal
O}^{(3)}_{s}(\vec b^{(3)})$ is less then $r^{(2)}$. Since the
smallest circle in  ${\mathcal O}^{(3)}$ has the size $r^{(2)}$, the
function $\vec \varkappa _3(\varphi )$ is holomorphic inside each
connected component of ${\mathcal O}^{(3)}_{s}(\vec b^{(3)})$ which
has non-empty intersection with $\varPhi _3$. Let
    \begin{equation}\vec y ^{(3)}(\varphi )=\vec \varkappa _3(\varphi )+\vec
    b^{(3)}. \label{y_3}\end{equation}

    \begin{lem} \label{L:3.23-3} If
$\varphi \in \hat \varPhi _3\setminus {\mathcal O}^{(3)}_{s}(\vec
b^{(3)})$, then \begin{equation}\label{3.7.2.2*m3}
    \Bigl\|\Bigl(H^{(3)}\bigl(\vec y^{(3)}(\varphi)\bigr)-k^{2}\Bigr)^{-1}\Bigr\| \leq
    \frac{2}{\epsilon _2^2}
    \end{equation}  The estimate is stable in the $\bigl(r^{(3)}k^{-\delta }\bigr)$-neighborhood of
    $\hat \varPhi _3\setminus
{\mathcal O}^{(3)}_{s}(\vec b^{(3)})$.  The resolvent  is an
analytic function of $\varphi $ in every component of $ {\mathcal
O}^{(3)}_{s}(\vec b^{(3)})$, whose intersection with $\varPhi _3$ is
not empty. The only singularities of the resolvent are poles.
    The number of poles (counting multiplicity) of the resolvent
    inside ${\mathcal O}^{(3)}(\vec b^{(3)})$ is less than
    $c_0k^{2+2s_3}$.
    \end{lem}
    \begin{proof} The proof is  analogous to that of Lemma
    \ref{L:3.23} up to the shift of indices by one. We use Lemma \ref{L:3.7.2.2*mmm3}, $\|W_3\|_{\infty} <\epsilon
    _2^4$,
    and the first estimate in \eqref{3.75*}.\end{proof}

\subsubsection{The set ${\mathcal O_s^{(3)}}\left(\vec
b^{(3)}\right)$ for small $\vec b^{(3)}$} \label{SS:3.7.1m-3}
 The considerations of this section is  analogous to those of
 Section \ref{SS:3.7.1m} up to the shift of indices by one.
 The following  lemmas and the definitions are
completely analogous to \ref{L:3.7.1.1m} -- \ref{L:july9c*}.
\begin{lem}\label{L:3.7.1.1m+} If $0<b_0 ^{(3)}\leq \epsilon
_2k^{-1-2\delta }$ and $ |\epsilon_0|<b_0^{(3)}k^{1-\delta}r^{(3)}$,
then the equation
    \begin{equation}\label{3.7.1.2*3}
   \lambda ^{(3)}\left(\vec y ^{(3)}(\varphi)\right)=k^{2}+\epsilon_0
   \end{equation}
 has no more than two
solutions $\varphi^{\pm\ (3)}_{\epsilon _0}$ in $\hat \varPhi _3$.
For any $\varphi^{\pm\ (3)}_{\epsilon _0}$ there is $\varphi^{\pm
(2)}_{0}\in \hat \varPhi _2$ such that
    \begin{equation}
    \left|\varphi^{\pm (3)}_{\epsilon _0}-\varphi^{\pm (2)}_{0}\right|
    <r^{(3)}/4, \label{july5a*3}
    \end{equation}
    here and below $\varphi^{\pm (2)}_{0}$ is $\varphi ^{\pm (2)}_{\epsilon
    _0}$ for $\epsilon _0=0$.
\end{lem} \begin{lem}\label{L:july5a*-3} Suppose $0<b_0 ^{(3)}\leq
\epsilon _2k^{-1-2\delta }$  and $\varphi \in \hat \varPhi _3$ obeys
the inequality analogous to (\ref{july5a*3}): $\left|\varphi
-\varphi^{\pm (2)}_{0}\right|
    <r^{(3)}$.
 Then,
    %%%%%\begin{equation}\label{3.7.1.6.1/2+}
    $\frac{\partial}{\partial \varphi}\lambda^{(3)}\left(\vec
    y^{(3)}(\varphi)\right)=_{k\to
    \infty}\pm 2b_0^{(3)}k\bigl(1+o(1)\bigr),$   where $|o(1)|<10^{-2}+\epsilon _1+\epsilon _2$ when $k>k_*$.
    %%%%%%\end{equation}
\end{lem}

\begin{definition}\label{SN1-3} Let $\Gamma^{(3)\pm }\left(\vec
b^{(3)}\right)$ be the open disks centered at $\varphi_0^{(3)\pm }$
with radius $r^{(3)}$; $\gamma^{(3)\pm }_s\bigl(\vec b^{(3)}\bigr)$
be their boundary circles and ${\mathcal O}_s^{(3)}\bigl(\vec
b^{(2)}\bigr)=\Gamma^{(3)+}\cup \Gamma^{(3)-}$.\end{definition}

\begin{lem}\label{L:3.7.1*-3} For any $\varphi$ in $\hat \varPhi
_3\setminus {\mathcal O}_s^{(3)}(\vec b^{(3)})$,
    %%%%%%%%\begin{equation}\label{3.7.1.10*}
    $|\lambda^{(3)}(\vec
    y^{(3)}(\varphi))-k^{2}|\geq b_0^{(3)}k^{1-\delta}r^{(3)}.$
%%%%%%%%    \end{equation}
\end{lem}

\begin{lem}\label{L:july9c**} For any $\varphi \in \hat \varPhi
_3\setminus {\mathcal O}_s^{(3)}\left(\vec b^{(3)}\right)$,
    \begin{equation}\label{3.7.1.10.1/2*-3}
    \Bigl\|\Bigl(H^{(3)}\left(\vec
    y^{(3)}(\varphi)\right)-k^{2}\Bigr)^{-1}\Bigr\|<\frac{16}{b_0^{(3)}r^{(3)}k^{1-\delta}}.
    \end{equation}
     The estimate is stable in the $\bigl(r^{(3)}k^{-\delta }\bigr)$-neighborhood of
    $\varPhi _3\setminus
{\mathcal O}^{(3)}_{s}(\vec b^{(3)})$.  The resolvent  is an
analytic function of $\varphi $ in every component of $ {\mathcal
O}^{(3)}_{s}(\vec b^{(3)})$, whose intersection with $\varPhi _3$ is
not empty. The only singularities of the resolvent are poles.
     The resolvent has at most two poles
    inside ${\mathcal O}^{(3)}_{s}(\vec b^{(3)})$.
\end{lem}

\section{The $n$-th Step of Approximation. Swiss Cheese Method.}
 \setcounter{equation}{0} \subsection{Introduction.}
\label{Section 5} On the $n$-th step, $n\geq 4$, we choose
$s_{n}=2s_{n-1}$  and define the operator $H_{\alpha}^{(n)}$ by the
formula:
 \begin{equation}\label{4.1m}
     H_{\alpha }^{(n)}=H^{(n-1)}+\alpha W_{n},\quad      (0\leq \alpha \leq
     1),\qquad
     W_{n}=\sum_{r=M_{n-1}+1}^{M_n}V_r, \notag
     \end{equation}
where $M_{n}$ is chosen in such a way that $2^{M_{n}}\approx
k^{s_{n}}$. Obviously, the periods of $W_{n}$ are $2^{M_{n}-1}
(d_1,0)$ and $2^{M_{n}-1} (0,d_2)$. We  write them in the form:
$N_{n-1}\cdot...\cdot N_1(a_1,0)$ and $N_{n-1}\cdot...\cdot
N_1(0,a_2)$, here  $N_{n-1}=2^{M_{n}-M_{n-1}}$,
$\frac{1}{4}k^{s_{n}-s_{n-1}}<N_{n-1}< 4k^{s_n-s_{n-1}}$. Note that
    $\|W_n\|_{\infty} \leq \sum_{r=M_{n-1}+1}^{M_n}
    \|V_r\|_{\infty} <
    \exp(-k^{\eta s_{n-1}}).$ By analogy with the definition of $\epsilon _1$, $\epsilon _2$
    we introduse the notation $\epsilon _n=\exp(-k^{\eta
    s_{n}})$.

    The $n$-th step is analogous to the second step up to replacement
    the indices 3 by $n$, 3 by $n-1$, the producr $N_2N_1$ by
    $N_{n-1}\cdot ...\cdot N_1$, etc.

We note that  $k$, satisfying (\ref{k}), obeys the analogous
    condition for with  any $s_n$ instead of $s_1$
    \begin{equation}
  C_*(1+s_{n})k^{2+4s_{n}}\ln k <k^{\eta s_{n}}\label{k**}
 \end{equation}
with the same constant $ C_*$. The inequality (\ref{k**}) can be
obtained from (\ref{k}) by  induction. This is an important fact: it
means that the lower bound for $k$ does not grow with $n$, i.e., all
steps hold uniformly in $k$ for $k>k_*$, $k_*$ being introduced by
(\ref{k}). Further we assume $k>k_*$.

The formulation of the geometric lemma for $n$-th step is the same
as that for Step 2 up to a shift of indices, we skip it here for
shortness. Note only that in the lemma we use the set
$\chi_{n-1}^*(\lambda)$ to define $\chi_{n}(\lambda)$. In fact,  we
started with the
    definition of $\chi _1(\lambda )$ and then use it to define $\chi _1^*(\lambda
    )$ (Step 1). Considering $\chi _1^*(\lambda
    )$, we constructed $\chi _2(\lambda
    )$  (Lemma \ref{L:3.2}) and later used it to define $\chi _2^*(\lambda
    )$ (Section \ref{S:3.4}). Using $\chi _2^*(\lambda
    )$, we introduced $\chi _3(\lambda )$ (Lemma \ref{L:4.2}) and, then $\chi _3^*(\lambda
    )$ (Section \ref{S:3.4*}).
    Thus, the process goes like $\chi _1\to \chi _1^*\to \chi
    _2\to \chi _2^*\to \chi _3\to \chi _3^*$.
    Here we start with the  set $\chi_{n-1}^*(\lambda)$  defined by
(\ref{3.77.7*}) for $n=4$ and by (\ref{3.77.7*+}) for $n>4$.
\footnote{Strictly speaking we assume that there is a subset
$\chi_{n-1}^*(\lambda)$ of the isoenergetic surface
$S_{n-1}(\lambda)$ of $H^{(n-1)}$ such that perturbation series of
the type (\ref{4.15}), (\ref{4.16}) converges for $t^{(n-1)}\in
\chi_{n-1}^*(\lambda)$ and $\chi_{n-1}^*(\lambda)$ has properties
described in Section \ref{S:3.4*} up to replacement of $3$ by $n-1$.
In particular, we assume that
$\chi_{n-1}^*(\lambda)=\mathcal{K}_{n-1}\mathcal{D}_{n-1}(\lambda
)$, where $\mathcal{D}_{n-1}(\lambda)$ satisfies the analog of Lemma
\ref{L:3.9*} and that the analogs of
 Lemmas \ref{Apr4a*} and \ref{L:May10a**} hold too. Here, $\mathcal{K}_{n-1}$ is
 the parallel shift into $K_{n-1}$. Further in this section we
 describe the next set
 $\chi_{n}^*(\lambda)$ which has analogous properties.}
 The estimate (\ref{4.9}) for $n$-th step takes the form:
\begin{equation}\label{4.9*}
\frac{L\left(\mathcal{K}_n\chi_{n-1}^*\setminus \chi
_n)\right)}{L\left(\chi_{n-1}^*\right)}<k^{-S_n},\ \ \ \ S_n=2\sum
_{i=1}^{n-1}(1+s_i). \end{equation} The formulation of the main
results (perturbation formulae) for $n$-th step is the same as for
the second and third step: Theorems analogous up to the shift of
indices to Theorem \ref{T:3.4}/\ref{T:4.4}, Lemma \ref{T:3.6},
Theorem \ref{T:3.5a}, Corollary \ref{C:3.6} and Lemma
\ref{L:3.5.1/2} hold.

\subsection{Proof of Geometric Lemma.}
 The proof of the lemma is analogous to that for
Geometric Lemma in the second step. Let us consider
\begin{equation}\label{O^n}\mathcalO^{(n)}=\cup _{\tilde
p^{(n-1)}\in P^{(n-1)}\setminus
\{0\}}\mathcalO^{(n-1)}_{s}\left(\frac{2\pi \tilde
p^{({n-1})}}{N_{n-1}...N_1a}\right),\end{equation} where
$\mathcalO_{s}^{(n-1)}\left(\frac{2\pi \tilde
p^{(2)}}{N_{n-1}...N_1a}\right)$ is defined by Definition
\ref{D:**h} with $\vec b^{(3)}=\frac{2\pi \tilde
p^{(3)}}{N_3N_2N_1a}$ when $n=4$. When $n>4$ we use Definition
\ref{D:****} with $n-1$ instead of $n$ and take
$b^{(n-1)}=\frac{2\pi \tilde p^{({n-1})}}{N_{n-1}...N_1a}$. The
radius $r^{(n-1)}$ of $\mathcalO_{s}^{(n-1)}$ is
 defined by the recurrent formula:
$r^{(n-1)}=k^{-2-4s_{n-1}-\delta }r^{(n-2)}$. This means
\begin{equation}r^{(n-1)}=k^{-2n-4\sum _{k=1}^{n-1}s_n -2s_1-2\delta
}=k^{-2n-2\sum _{k=1}^{n}s_n -2\delta }=k^{-S_{n+1}-2\delta
},\label{r_n} \end{equation} $S_n$ being defined by \eqref{4.9*}.
Note, that the definition make sense, since $|\frac{2\pi \tilde
p^{(n-1)}}{N_{n-1}...N_1a}|>d _{max}^{-1}k^{-s_n}>\epsilon
_{n-1}k^{-1-2\delta }$,  when $k>k_*$, the estimate \eqref{k**} has
been used. The last inequality can be easily proved by induction.
Let
\begin{equation} \varPhi _n=\varPhi _{n-1}\setminus \mathcalO^{(n)},\ \ \
\Theta _n=\varPhi _n\cap [0,2\pi ), \label{D:theta _n}
\end{equation} $\varPhi _3$, $\Theta _3$ are given by \eqref{D:theta
_3}. By Lemmas \ref{L:3.23-3}, \ref{L:3.23-n} \begin{equation}
\left\|\left(H ^{(n-1)}\Bigl(\vec \varkappa _{n-1}(\varphi
)+\frac{2\pi \tilde
p^{(n-1)}}{N_{n-1}...N_1a}\Bigr)-k^{2}\right)^{-1}\right\|<\frac{2}{\epsilon
_{n-2}^2} \label{normres****an}\end{equation}  for all $\tilde
p^{(n-1)}\in P^{(n-1)}\setminus\{0\}$ and $\varphi \in \hat \varPhi
_n$, here and below $\hat \varPhi _n$ is $r^{(n-1)}k^{-\delta
}$-neighborhood of $\varPhi _n$. +We consider
$\mathcalD_{n-1,nonres}\subset \mathcalD_{n-1}$:
\begin{equation} \label{D_{n-1,nonres}}
\mathcalD_{n-1,nonres}=\left\{\vec \varkappa _{n-1}(\varphi ), \
\varphi \in \Theta _n\right\}.\end{equation}We define $\chi _n$ by
the formula: \begin{equation} \label{chi_n} \chi
_n=\mathcalK_n\mathcalD_{n-1,nonres}.\end{equation} By the
definition of $\mathcalK_n$, for every $t^{(n)}$ in $\chi _n$, there
are $p^{(1)}\in P^{(1)}$,... ,$p^{(n-1)}\in P^{(n-1)}$ and $j\in
Z^2$ such that \begin{equation} \label{tau+n}t^{(n)} + \frac{2\pi
p^{(n-1)}}{N_{n-1}...N_1a}+...+\frac{2\pi p^{(1)}}{N_1a}+\frac{2\pi
j}{a}=\vec \varkappa _{n-1}(\varphi ),\ \ \vec \varkappa
_{n-1}(\varphi )\in \mathcalD_{n-1,nonres}.\end{equation}
Considering the definition of $\mathcalD_{n-1,nonres}$ and
\eqref{normres****an}, we obtain:
\begin{equation} \left\|\left(H ^{(n-1)}\Bigl(t^{(n)} + \frac{2\pi
\hat
p^{(n-1)}}{N_{n-1}...N_1a}\Bigr)-k^{2}\right)^{-1}\right\|<\frac{2}{\epsilon
_{n-2}^2}, \label{normres****bn}\end{equation} $\hat p^{(n-1)}=
p^{(n-1)}+\tilde p^{(n-1)}.$ Note that the indices $j$,
$p^{(1)}$,...,$p^{(n-1)}$  do not play a role, since they just
produce a shift of indices of the matrix elements of the resolvent.
Considering that $\tilde p^{(n-1)}$ can be any but zero, we obtain
that (\ref{normres****bn}) holds for all $\hat p ^{(n-1)}\in
P^{(n-1)}\setminus\{p^{(n-1)}\}$. Taking into account that $\lambda
_j^{(n-1)}(t^{(n)} +2\pi p^{(n-1)}/N_{n-1}...N_1a)=k^{2}$ and
$\epsilon _{n-1}<\epsilon _{n-2}^2$. We arrive at the analog of
\eqref{4.6a} for all $\hat p^{(n-1)}\neq p^{(n-1)}$.This also proves
that $p^{(n-1)}$ is uniquely defined by \eqref{tau+n}. It remains to
check the analog of \eqref{4.6a} for $p^{(n-1)}=\hat p^{(n-1)}$. Let
$t^{(n-1)}:=t^{(n)} +2\pi p^{(n-1)}/N_{n-1}...N_1a$. By
(\ref{tau+n}), $t^{(n-1)}\in \mathcalK_{n-1} \mathcalD _{n-1}$.
Using \eqref{3.77.7*} for $n=4$ and \eqref{3.77.7*+} with $n-1$
instead of $n$ for $n>4$, we get $t^{({n-1})}\in \chi _{n-1}^*$. By
the analog of Theorem \ref{T:4.4} for step $n-1$, $\lambda
_j^{(n-1)}(t^{(n)} +2\pi p^{(n-1)}/N_{n-1}...N_1a)$ is the only
eigenvalue of $H ^{(n-1)}\left(t^{(n)} +2\pi
p^{(n-1)}/N_{n-1}...N_1a\right)$ in the interval $\varepsilon
_{n-1}$. Hence,
 $$|\lambda _j^{(n-1)}(t^{(3)} +2\pi p^{(n-1)}/N_{n-1}...N_1a)-\lambda _{\hat m}^{({n-1})}(t^{(n)} +2\pi p^{({n-1})}/
 N_{n-1}...N_1a)|
 >\epsilon_{n-2}.$$
Thus, the analog of \eqref{4.6} holds for all $p^{(n-1)}\in
P^{(n-1)}$.
    Part 2 holds, since all estimates are stable with respect to the
    perturbation of $t^{(3)}$ less then $\epsilon _{n-1}k^{-1-\delta }$.

    Let us estimate the size of $\mathcalO_{n}$. According to Lemma
    \ref{L:totalsize3}, \ref{L:totalsizen},
the total size of each ${\mathcal O}^{({n-1})}_{s}(2\pi
p^{({n-1})}/N_{n-1}...N_1a)$ is less then $c_0k^{-2s_{n-1}-\delta
}r^{(n-2)}$.
     Considering that the number of
$p^{({n-1})}$ does not exceed $k^{2(s_n-s_{n-1})}=k^{2s_{n-1}}$,
$s_{n}=2s_{n-1}$, we obtain that the size of $\mathcalO_{n}$ is less
than $k^{-\delta }r^{(n-2)}$. Using the formula for $r^{(n-2)}$, we
obtain that the total size of $\mathcalO_{n}$ does not exceed
$k^{-2(n-1)-2\sum _{k=1}^{n-1}s_k}$. Using this estimate
  we easily arrive at (\ref{4.9*}). It is easy
to see that $S_n=2(n-1)+\left(2^{n}-2\right)s_1$ and $S_n\approx
2^{n}s_1\approx s_n$ for large $n$. The lemma is proved.

\subsection{Nonresonant part of the isoenergetic set of $H_{\alpha
}^{(n)}$}

Now we construct a nonresonance subset  $\chi _n^*(\lambda )$ of the
isoenergetic surface $S_n(\lambda
    )$ of  $H_{\alpha }^{(n)}$ in $K_n$, $S_n(\lambda
    )\subset K_n.$ It
    corresponds to  nonresonance eigenvalues  given by  perturbation series.
     The sets $\chi _1^*(\lambda )$, $\chi _2^*(\lambda
    )$,
    $\chi _3^*(\lambda )$ are defined in the previous steps as well
    as the non-resonance sets $\chi _1(\lambda )$, $\chi _2(\lambda
    )$,
    $\chi _3(\lambda )$. Let us recall that we started with the
    definition of $\chi _1(\lambda )$ and then use it to define $\chi _1^*(\lambda
    )$. Considering $\chi _1^*(\lambda
    )$, we constructed $\chi _2(\lambda
    )$ (Step 2).  Next,  we defined $\chi _2^*(\lambda
    )$. Using $\chi _2^*(\lambda
    )$, we introduced $\chi _3(\lambda )$ and, then $\chi _3^*(\lambda
    )$.
    Thus, the process looks like $\chi _1\to \chi _1^*\to \chi
    _2\to \chi _2^*\to \chi _3\to \chi _3^*$. The geometric lemma in
    this section gives us $\chi _4$ and every next $\chi _n$ if $\chi _{n-1}^*$ is defined.
    To ensure the reccurent procedure  we show now how to define
     $\chi _n^*(\lambda )$ using  $\chi _n(\lambda
    )$.

We define $\mathcal{B}_n$ as the set of directions corresponding to
$\Theta _n$:
$$\mathcal{B}_n(\lambda)=\{\vec{\nu} \in S_1 :
    \varphi \in \Theta _n\}.$$
Note that $\mathcal{B}_n$ is a unit circle with holes centered at
the origin and $\mathcal{B}_n(\lambda ) \subset
    \mathcal{B}_{n-1}(\lambda )$.
We define $\mathcal{D}_n(\lambda)$ as a level set for
$\lambda^{(n)}(\alpha,\vec{\varkappa})$ in the $(\epsilon
_{n-1}k^{-1-\delta })$-neighborhood of $\mathcal{D}_{n-1,
nonres}(\lambda)$:
    $$\mathcal{D}_n(\lambda )=\left\{\vec{\varkappa }=\varkappa \vec \nu: \vec \nu \in
    \mathcal{B}_n,
    \bigl| \varkappa -\varkappa _{n-1}(\lambda,\vec
\nu)\bigr|<\epsilon_{n-1}k^{-1-\delta},
\lambda^{(n)}(\alpha,\vec{\varkappa })=\lambda \right\}.$$
Considering as in the previous step, we prove the analogs of Lemmas
\ref{L:3.8*} and \ref{L:3.9*}. For shortness, we provide  here only
the second lemma. By analogy with previous sections, we shorten
notations here:  $\varkappa _{n-1}(\varphi )\equiv \varkappa
_{n-1}(\lambda ,\vec{\nu})$,  $\vec \varkappa _{n-1}(\varphi )\equiv
\varkappa _{n-1}(\lambda ,\vec{\nu})\vec{\nu}$.
\begin{lem}\label{L:3.9*+} For $\lambda >k^2_*$: \begin{enumerate} \item The set
$\mathcal{D}_n(\lambda )$ is a distorted circle with holes: it can
be described by the formula: \begin{equation} \mathcal{D}_n(\lambda
)=\bigl\{\vec \varkappa  \in \R^2: \vec \varkappa
 =\vec \varkappa _{n}(\varphi ),\ \ \varphi
\in \Theta _n(\lambda)\bigr\},\label{May20a+} \end{equation} where
    $ \varkappa _{n}(\varphi )=
    \varkappa _{n-1}(\varphi )+h_n(\varphi), $
and $h_n(\varphi )$ satisfies the estimates
    \begin{equation}\label{3.75*+}
    |h_n|<12\alpha \epsilon_{n-1} ^4k^{-1},\ \ \
    \left|\frac{\partial h_n}{\partial \varphi} \right| \leq
    4\alpha \epsilon_{n-1}^3 k^{1+\delta}.
    \end{equation}
     \item The total length of $\mathcal{B}_n(\lambda)$ satisfies the estimate:
\begin{equation}\label{theta2*}
    L\left(\mathcal{B}_{n-1}\setminus \mathcal{B}_{n}\right)<4\pi k^{-S_n}.
    \end{equation}
\item Function $\varkappa _{n}(\varphi )$ can be extended as a
holomorphic function of $\varphi $ to  $\hat \varPhi _{n}$,
estimates (\ref{3.75*+}) being preserved. \item The curve
$\mathcal{D}_n(\lambda )$ has a length which is asymptotically close
to that of $\mathcal{D}_{n-1}(\lambda )$ in the following sense:
    \begin{equation}\label{3.77mm}
     L\Bigl(\mathcal{D}_n(\lambda )\Bigr)\underset{\lambda \rightarrow
     \infty}{=}L\Bigl(\mathcal{D}_{n-1}(\lambda )\Bigr)\Bigl(1+O(k^{-S_n})\Bigr).
     \end{equation}
     \end{enumerate}
\end{lem}

Now define the nonresonance set, $\chi_n^*(\lambda )$ in
$S_n(\lambda )$ by the formula analogous to (\ref{3.77.7*}). Indeed,
    \begin{equation}\label{3.77.7*+}
    \chi_n^*(\lambda ):=\mathcal{K}_n\mathcal{D}_n(\lambda ).
    \end{equation}
The following lemmas are analogous to Lemmas \ref{Apr4a*} and
\ref{L:May10a**}.

    \begin{lem}\label{Apr4a*+}
    The set $\chi_n^*(\lambda )$ belongs to the
    $\left(2\alpha \epsilon _{n-1}^4k^{-1}\right)$-neighborhood of $\chi_n(\lambda
    )$ in $K_n$. If $t^{(n)}\in \chi_n^*(\lambda )$, then the operator
    $H^{(n)}_{\alpha }(t^{(n)})$ has a simple eigenvalue
   equal to $\lambda $. This
    eigenvalue is given by the perturbation series analogous to (\ref{4.15}).
\end{lem}

    \begin{lem}\label{L:May10a**+} Formula (\ref{3.77.7*+}) establishes
    one-to-one correspondence between $\chi_n^*(\lambda )$ and $\mathcal{D}_n(\lambda
    )$.\end{lem}

\subsection{Preparation for the Next Approximation} \label{S:3.6*-n}
Let $\vec b^{(n)} \in K_n$ and $b_0^{(n)}$ be the distance of the
point $\vec b^{(n)}$ to the nearest corner of $K_n$. We assume
$b_0^{(n)}=|\vec b^{(n)}|$.  We  consider  two cases: $b_0^{(n)}\geq
 \epsilon _{n-1}k^{-1-2\delta }$ and
$0<b_0^{(n)}<\epsilon _{n-1}k^{-1-2\delta }$. Let
\begin{equation}\vec y^{({n-1})}(\varphi)=\vec \varkappa _{n-1}(\varphi
)+\vec
    b^{(n)}. \label{y_1*-n} \end{equation}\subsubsection{The case $b_0^{(n)}\geq \epsilon _{n-1}k^{-1-2\delta
}$.}\begin{definition} \label{D:***}We  define the set ${\mathcal
O}^{(n)}(\vec b^{(n)})$ by the formula: \begin{equation} {\mathcal
O}^{(n)}(\vec b^{(n)})=\cup _{p^{({n-1})}\in P^{({n-1})}}
\mathcalO_{s}^{({n-1})}\left(2\pi
    p^{({n-1})}/N_{n-1}...N_1a+\vec b^{(n)} \right), \label{O_n(b_n)}\end{equation}
    set $\mathcalO_{s}^{({n-1})}$  being defined by Definition \ref{D:***} for $n=4$ and by \ref{D:****} for $n>4$. \end{definition} The set ${\mathcal O}^{(n)}(\vec b^{(n)})$ consists
the disks with the radius
$r^{({n-1})}=r^{(n-2)}k^{-2-4s_{n-1}-\delta }$ centered at poles of
the resolvent $\Bigl(\tilde H_{n-1}\bigl(\vec y^{({n-1})}(\varphi
)\bigr)-k^2\Bigr)^{-1}$.
    \begin{lem}\label{L:3.7.2.2*mmn} If
$\varphi \in \hat  \varPhi _{n-1}\setminus {\mathcal O}^{(n)}(\vec
b^{(n)})$, then
\begin{equation}\label{3.7.2.2*mmn}
    \Bigl\|\Bigl(\tilde H^{({n-1})}\bigl(\vec y^{({n-1})}(\varphi)\bigr)-k^{2}\Bigr)^{-1}\Bigr\| \leq
    \frac{17r^{({n-1})}k^{3\delta}}{\epsilon _{n-1}}
    \end{equation}
     The estimate is stable in the $\bigl(r^{(n-1)}k^{-\delta }\bigr)$-neighborhood of
    $\hat \varPhi _{n-1}\setminus
{\mathcal O}^{(n)}(\vec b^{(n)})$.
    The resolvent  is an analytic function of
$\varphi $ in every connected component  $ {\mathcal O}_c^{(n)}(\vec
b^{(n)})$ of $ {\mathcal O}^{(n)}(\vec b^{(n)})$, whose intersection
with $\varPhi _{n-1}$ is not empty. The only singularities of the
resolvent in such a component are poles.
    The number of poles (counting multiplicity) of the resolvent
    inside ${\mathcal O}^{(n)}(\vec b^{(n)})$ is less than
    $ck^{2+2s_n}$. The total size of ${\mathcal O}^{(n)}(\vec b^{(n)})$
    does not exceed $c_0k^{-S_n}$.
    \end{lem}\begin{proof} The proof is analogous to that of Lemma
    \ref{L:3.7.2.2*mm}. Let $n=4$. To obtain \eqref{3.7.2.2*mmn} we combine
   Lemmas \ref{L:3.23-3} and \ref{L:july9c**} for $\vec b^{({n-1})}=2\pi
    p^{({n-1})}/N_{n-1}...N_1a+\vec b^{(n)}$ taking into account
   $\epsilon _{n-2}^2 >\epsilon _{n-1}$ and the estimate for $b_0^{(n)}$.
   If $n>4$, then using the recurrent procedure, we apply Lemmas
   \ref{L:3.23-n} and \ref{L:july9c**n}  with $n-1$ instead of $n$.
Moreover, the number of poles of $\Bigl(H^{({n-1})}\bigl(\vec
\varkappa _{n-1}(\varphi )+\vec b^{(n-1)}\bigr)-k^{2}\Bigr)^{-1}$
inside ${\mathcal O}^{(n)}(\vec b_{n})$ is less than
    $ck^{2+2s_{n-1}}$.
    Considering that the number of $p^{({n-1})}\in P^{({n-1})}$ does not exceed $ck^{2(s_n-s_{n-1})}$, $s_n=2s_{n-1}$,
    we obtain the estimate for the number of poles
    for the resolvent $\Bigl(\tilde H^{({n-1})}\bigl(\vec y^{({n-1})}((\varphi)\bigr)-k^{2}\Bigr)^{-1}$.
     We estimate the size of $\mathcalO ^{(n)}(\vec b^{(n)})$ the same way as we estimated the size of
  $\mathcalO ^{(n)}$.  \end{proof}

\begin{definition}\label{D:****} We denote the poles of the resolvent $\Bigl(\tilde
H^{({n-1})}\bigl(\vec y^{({n-1})}((\varphi)\bigr)-k^{2}\Bigr)^{-1}$
in ${\mathcal O}^{(n)}(\vec b^{(n)})$ as $\varphi ^{(n)}_m$,
$m=1,...,M^{(n)}$ $M^{(n)} <c_0k^{2+2s_n}$. Let us consider the
circles ${\mathcal O}_m^{(n)}(\vec b^{(n)})$ of the radius
$r^{(n)}=k^{-2-4s_n-\delta }r^{({n-1})}$ around these poles. Let
\begin{equation}{\mathcal O}^{(n)}_{s}(\vec b^{(n)})=\cup
_{m=1}^{M^{(n)}}{\mathcal O}_m^{(n)}(\vec
b^{(n)}).\label{O_n}\end{equation} \end{definition}
\begin{lem}
\label{L:totalsizen} The total size of ${\mathcal O}^{(n)}_{s}(\vec
    b^{(n)})$ is less then $c_0k^{-2s_n-\delta }r^{({n-1})}$.\end{lem}
    \begin{proof} The lemma easily follows from the formula $r^{(n)}=k^{-2-4s_n-\delta
}r^{({n-1})}$
    and the estimate $M^{(n)}<c_0k^{2+2s_n}$. \end{proof}

\begin{lem}\label{L:3.7.2.2*mmmn} If $\varphi \in \hat \varPhi _{n-1}\setminus
{\mathcal O}^{(n)}_{s}(\vec b^{(n)})$, then
\begin{equation}\label{3.7.2.2*mmmn}
    \Bigl\|\Bigl(\tilde H^{({n-1})}\bigl(\vec y^{({n-1})}(\varphi)\bigr)-k^{2}\Bigr)^{-1}\Bigr\| \leq
    \frac{1}{\epsilon _{n-1}^2}
    \end{equation}
     The estimate is stable in the $\bigl(r^{(n)}k^{-\delta }\bigr)$-neighborhood of
    $\hat \varPhi _{n-1}\setminus
{\mathcal O}^{(n)}_s(\vec b^{(n)})$.
    The resolvent  is an analytic function of
$\varphi $ in every connected component  $ {\mathcal
O}_{sc}^{(n)}(\vec b^{(n)})$ of $ {\mathcal O}^{(n)}_s(\vec
b^{(n)})$, whose intersection with $\varPhi _{n-1}$ is not empty.
The only singularities of the resolvent in such a component are
poles.
    The number of poles (counting multiplicity) of the resolvent
    inside ${\mathcal O}^{(n)}_s(\vec b^{(n)})$ is less than
    $ck^{2+2s_n}$.
    \end{lem}
    \begin{proof} By the definition of ${\mathcal
O}^{(n)}_{s}(\vec b^{(n)})$, the number of poles (counting
multiplicity) of the resolvent
    inside this set is less than
    $M^{(n)}=c_0k^{2+2s_n}$. Considering as in Lemma \ref{norm**} we obtain
    \begin{equation}\label{3.7.2.2*mmmm-n}
    \Bigl\|\Bigl(\tilde H^{({n-1})}\bigl(\vec y^{({n-1})}(\varphi)\bigr)-k^{2}\Bigr)^{-1}\Bigr\| \leq
    \nu ^{-M^{(n)}}\frac{17r^{({n-1})}k^{3\delta}}{\epsilon _{n-1}},
    \end{equation}
    where $\nu $ is the coefficient of contraction, when we reduce ${\mathcal O}^{(n)}(\vec
    b^{(n)})$ to
    ${\mathcal
O}^{(n)}_{s}(\vec b^{(n)})$. Namely $\nu $ is the ratio of $r^{(n)}$
to the maximal size of ${\mathcal O}^{(n)}(\vec b^{(n)})$. By Lemma
\ref{L:3.7.2.2*mmn} and formula \eqref{r_n}, $\nu
=k^{-4-4s_n-4s_{n-1}} $. Considering that $r^{({n-1})}k^{\delta}<1$
and {\bf using \eqref{k**} }, we obtain $\nu
^{-M^{(n)}}r^{({n-1})}k^{\delta}<\epsilon _{n-1}^{-1}$ when $k>k_*$.
Thus, we obtain \eqref{3.7.2.2*mmmn}.

\end{proof}

By Lemma \ref{L:totalsizen}, the total size of ${\mathcal
O}^{(n)}_{s}(\vec b^{(n)})$ is less then $r^{({n-1})}$. Therefore,
the function $\vec \varkappa _n(\varphi )$ is holomorphic inside
each connected component of ${\mathcal O}^{(n)}_{s}(\vec b^{(n)})$
which has non-empty intersection with $\varPhi _n$. Let
    \begin{equation}\vec y ^{(n)}(\varphi )=\vec \varkappa _n(\varphi )+\vec
    b^{(n)}. \label{y_n}\end{equation}

    \begin{lem} \label{L:3.23-n} If
$\varphi \in \hat \varPhi _n\setminus {\mathcal O}^{(n)}_{s}(\vec
b^{(n)})$, then \begin{equation}\label{3.7.2.2*mn}
    \Bigl\|\Bigl(H^{(n)}\bigl(\vec y^{(n)}(\varphi)\bigr)-k^{2}\Bigr)^{-1}\Bigr\| \leq
    \frac{2}{\epsilon _{n-1}^2}
    \end{equation}
     The estimate is stable in the $\bigl(r^{(n)}k^{-\delta }\bigr)$-neighborhood of
    $\hat \varPhi _{n}\setminus
{\mathcal O}^{(n)}_s(\vec b^{(n)})$.
    The resolvent  is an analytic function of
$\varphi $ in every connected component  $ {\mathcal
O}_{sc}^{(n)}(\vec b^{(n)})$ of $ {\mathcal O}^{(n)}_s(\vec
b^{(n)})$, whose intersection with $\varPhi _{n}$ is not empty. The
only singularities of the resolvent in such a component are poles.
    The number of poles (counting multiplicity) of the resolvent
    inside ${\mathcal O}^{(n)}(\vec b^{(n)})$ is less than
    $c_0k^{2+2s_n}$.
    \end{lem}
    \begin{proof} The proof is  analogous to that of Lemma
    \ref{L:3.23} up to the shift of indices by two. We use Lemma \ref{L:3.7.2.2*mmmn}, $\|W_n\|_{\infty} <\epsilon
    _{n-1}^4$,
    and the first estimate in \eqref{3.75*+}.\end{proof}
    \subsubsection{The set ${\mathcal O_s^{(n)}}\left(\vec
b^{(n)}\right)$ for small $\vec b^{(n)}$} \label{SS:3.7.1mn}
 The considerations of this section is  analogous to those of
  By analogy with
the previous subsection, we choose $r^{(n)}=k^{-2-2s_n-\delta
}r^{(n-1)}$ here. The following   Lemmas \ref{L:3.7.1.1m+n} and
\ref{L:july5a*n} are identical to \ref{L:3.7.1.1m+}  and
\ref{L:july5a*-3} up to replacement of indices $(3)$ by $(n)$, $(2)$
by $(n-1)$, etc. Next, Definition \ref{SN1n} is analogous to
\ref{SN1-3}.  Lemmas \ref{L:3.7.1*n}, \ref{L:july9c**n} are
analogous to \ref{L:3.7.1*-3}, \ref{L:july9c**}.
\begin{lem}\label{L:3.7.1.1m+n} If $0<b_0 ^{(n)}\leq \epsilon
_{n-1}k^{-1-2\delta }$ and $
|\epsilon_0|<b_0^{(n)}k^{1-\delta}r^{(n)}$, then the equation
    \begin{equation}\label{3.7.1.2*n}
   \lambda ^{(n)}\left(\vec y ^{(n)}(\varphi)\right)=k^{2}+\epsilon_0
   \end{equation}
 has no more than two
solutions $\varphi^{\pm\ (n)}_{\epsilon _0}$ in $\hat \varPhi _n$.
For any $\varphi^{\pm\ (n)}_{\epsilon _0}$ there is $\varphi^{\pm
(n-1)}_{0}\in \hat \varPhi _{n-1}$ such that
    \begin{equation}
    \left|\varphi^{\pm (n)}_{\epsilon _0}-\varphi^{\pm (n-1)}_{0}\right|
    <r^{(n)}/4, \label{july5a*n}
    \end{equation}
    here and below $\varphi^{\pm (n-1)}_{0}$ is $\varphi ^{\pm (n-1)}_{\epsilon
    _0}$ for $\epsilon _0=0$.
\end{lem} \begin{lem}\label{L:july5a*n} Suppose $0<b_0 ^{(n)}\leq
\epsilon _{n-1}k^{-1-2\delta }$  and $\varphi \in \hat \varPhi _n$
obeys the inequality analogous to (\ref{july5a*n}): $\left|\varphi
-\varphi^{\pm (n-1)}_{0}\right|
    <r^{(n)}$.
 Then,
    %%%%%\begin{equation}\label{3.7.1.6.1/2+}
    $\frac{\partial}{\partial \varphi}\lambda^{(n)}\left(\vec
    y^{(n)}(\varphi)\right)=_{k\to
    \infty}\pm 2b_0^{(n)}k\bigl(1+o(1)\bigr),$   where $|o(1)|<10^{-2}+\epsilon _1+...+\epsilon _{n-1}$ when $k>k_*$.
    %%%%%%\end{equation}
\end{lem}

\begin{definition}\label{SN1n} Let $\Gamma^{(n)\pm }\left(\vec
b^{(n)}\right)$ be the open disks centered at $\varphi_0^{(n)\pm }$
with radius $r^{(n)}$; $\gamma^{(n)\pm }_s\bigl(\vec b^{(n)}\bigr)$
be their boundary circles and ${\mathcal O}_s^{(n)}\bigl(\vec
b^{(n-1)}\bigr)=\Gamma^{(n)+}\cup \Gamma^{(n)-}$.\end{definition}

\begin{lem}\label{L:3.7.1*n} For any $\varphi$ in $\hat \varPhi
_n\setminus {\mathcal O}_s^{(n)}(\vec b^{(n)})$,
    %%%%%%%%\begin{equation}\label{3.7.1.10*}
    $|\lambda^{(n)}(\vec
    y^{(n)}(\varphi))-k^{2}|\geq b_0^{(n)}k^{1-\delta}r^{(n)}.$
%%%%%%%%    \end{equation}
\end{lem}

\begin{lem}\label{L:july9c**n} For any $\varphi \in \hat \varPhi
_{n-1}\setminus {\mathcal O}_s^{(n)}\left(\vec b^{(n)}\right)$,
    \begin{equation}\label{3.7.1.10.1/2*n}
    \Bigl\|\Bigl(H^{(n)}\left(\vec
    y^{(n)}(\varphi)\right)-k^{2}\Bigr)^{-1}\Bigr\|<\frac{16}{b_0^{(n)}r^{(n)}k^{1-\delta}},
    \end{equation}
     The estimate is stable in the $\bigl(r^{(n)}k^{-\delta }\bigr)$-neighborhood of
    $\varPhi _n\setminus
{\mathcal O}^{(n)}_{s}(\vec b^{(n)})$.  The resolvent  is an
analytic function of $\varphi $ in every component of $ {\mathcal
O}^{(n)}_{s}(\vec b^{(n)})$, whose intersection with $\varPhi _n$ is
not empty. The only singularities of the resolvent are poles.
     The resolvent has at most two poles
    inside ${\mathcal O}^{(n)}_{s}(\vec b^{(n)})$.
\end{lem}

\section{Limit-Isoenergetic Set and Eigenfunctions} \label{chapt7}

\setcounter{equation}{0}

\subsection{Limit-Isoenergetic Set and Proof of the Bethe-Sommerfeld
Conjecture} At every step $n$ we constructed  a subset
$\mathcal{B}_n(\lambda)$ of the unit circle,
  and a function $\varkappa _n(\lambda,\vec{\nu})$,
$\vec{\nu} \in \mathcal{B}_n(\lambda)$, with the following
properties. The sequence $\mathcal{B}_{n}(\lambda)$ is decreasing:
$\mathcal{B}_{n}(\lambda)\subset \mathcal{B}_{n-1}(\lambda)$. The
set $\mathcal{D}_{n}(\lambda )$ of vectors
$\vec{\varkappa}=\varkappa _n(\lambda ,\vec{\nu})\vec{\nu}$,
   $\vec{\nu} \in \mathcal{B}_{n}(\lambda )$,
    is a slightly distorted circle with holes,  see Fig.\ref{F:1}, Fig.\ref{F:2}, formula (\ref{Dn})
    and Lemmas \ref{L:2.13}, \ref{L:3.9},
\ref{L:3.9*}, \ref{L:3.9*+}. For any $\vec \varkappa
_n(\lambda,\vec{\nu})\in \mathcal{D}_{n}(\lambda )$ there is a
simple eigenvalue of
 $H^{(n)}(\vec \varkappa _n)$
equal to $\lambda $ and  given by a perturbation series.~
\footnote{The operator $H^ {(n)}(\vec \varkappa )$ is defined for
every $\vec \varkappa \in \R^2$. The perturbation series is given by
a formula analogous to (\ref{3.66}), which coincides with
(\ref{3.15}) up to a shift of indices corresponding to the parallel
shift of $\vec \varkappa $ into $K_n$.} Let
    $\mathcal{B}_{\infty}(\lambda)=\bigcap_{n=1}^{\infty}\mathcal{B}_n(\lambda).$
Since $\mathcal{B}_{n+1} \subset \mathcal{B}_n$ for every $n$,
$\mathcal{B}_{\infty}(\lambda)$ is a unit circle with the infinite
number of holes, more and more holes of smaller and smaller size
appearing at each step. \begin{lem} \label{L:Dec9} The length of
$\mathcal{B}_{\infty}(\lambda)$ satisfies estimate (\ref{B}) with
$\gamma _3=\delta $. \end{lem} \begin{proof}
 Using (\ref{theta1}),  (\ref{theta2}), (\ref{theta2*s})
and (\ref{theta2*}) and consideirng that $S_n\approx 2^ns_1$, we
easily conclude that
$L\left(\mathcal{B}_n\right)=\left(1+O(k^{-\delta})\right)$,
$k=\lambda ^{1/2}$ uniformly in $n$. Since $\mathcal{B}_n$ is a
decreasing sequence of sets, (\ref{B}) holds. \end{proof} Let us
consider
    $\varkappa _{\infty}(\lambda, \vec{\nu})=\lim_{n \to \infty}\varkappa _n(\lambda,\vec{\nu}),\quad
    \vec{\nu} \in \mathcal{B}_{\infty}(\lambda ).$
    \begin{lem} The limit $\varkappa _{\infty}(\lambda,
    \vec{\nu})$ exists for any $\vec{\nu} \in \mathcal{B}_{\infty}(\lambda
    )$ and the following estimates hold when $n\geq 1$:
    \begin{equation}\label{6.1}
    \left|\varkappa _{\infty}(\lambda, \vec{\nu})-\varkappa
    _n(\lambda,\vec{\nu})\right|<14\epsilon_{n} ^4k^{-1},\ \
    \epsilon _{n}=\exp(-\frac{1}{4}k^{\eta
    s_{n}}),\ \
    s_n=2^{n-1}s_1.
    \end{equation}
    \end{lem}
    \begin{cor}\label{Dec18}
    For every $\vec{\nu} \in \mathcal{B}_{\infty}(\lambda)$ estimate (\ref{h}) holds, where $\gamma
_4=3-30s_1-20\delta
    >0$.
    \end{cor}
  \begin{proof}  The lemma easily follows from
  the estimates (\ref{3.75}), (\ref{3.75*}) and (\ref{3.75*+}). To obtain corollary we use
 (\ref{2.75}).\end{proof}

    Estimates (\ref{3.75}), (\ref{3.75*}) and (\ref{3.75*+}) justify convergence of the series $\sum_{m=1}^{\infty}
    \frac{\partial h_n}{\partial \varphi },$ and hence,
    of the sequence $\frac{\partial \varkappa _n}{\partial \varphi }.$
    We denote the limit of this sequence by $\frac{\partial \varkappa _{\infty}}{\partial \varphi }.$
    \begin{lem} The  estimate (\ref{Dec9a}) with $\gamma
_5=1-33s_1-22\delta >0$ holds for any $\vec \nu \in
    \mathcal{B}_{\infty}(\lambda)$.
   \end{lem}
   \begin{proof}The lemma easily follows from (\ref{2.75}), (\ref{3.75}), (\ref{3.75*}) and
   (\ref{3.75*+}).\end{proof}

We define $\mathcal{D}_{\infty}(\lambda )$ by (\ref{D}). Clearly,
$\mathcal{D}_{\infty}(\lambda )$ is a slightly distorted circle of
radius $k$ with the infinite number of holes. We can assign a
tangent vector $\frac{\partial \varkappa _{\infty}}{\partial \varphi
}\vec \nu +\varkappa _{\infty}\vec \mu $, $\vec \mu =(-\sin \varphi
,\cos \varphi )$ to the curve $\mathcal{D}_{\infty}(\lambda )$, this
tangent vector being the limit of corresponding tangent vectors for
curves $\mathcal{D}_{n}(\lambda )$ at points $\vec \varkappa
_n(\lambda ,\vec \nu )$ as $n\to \infty $.

 \begin{remark} \label{R:Dec9} We easily see from (\ref{6.1}), that any $\vec{\varkappa}
\in \mathcal{D}_{\infty}(\lambda )$ belongs to the
$\left(14\epsilon_{n} ^4k^{-1}\right)$-neighborhood of
$\mathcal{D}_n(\lambda )$. Applying perturbation formulae for $n$-th
step, we easily obtain that  there is an
 eigenvalue  $\lambda^{(n)}(\vec \varkappa )$ of $H^{(n)}(\vec \varkappa )$
satisfying the estimate $\lambda^{(n)}(\vec \varkappa )=\lambda
+\delta _n$, $\delta _n=O\left(\epsilon _{n}^4\right)$, the
eigenvalue $\lambda^{(n)}(\vec \varkappa )$
     being given by a perturbation
series of the type (\ref{3.66}). Hence, for every $\vec{\varkappa}
\in \mathcal{D}_{\infty}(\lambda)$ there is a limit:
\begin{equation} \lim _{n\to \infty }\lambda^{(n)}(\vec \varkappa
)=\lambda.\label{6.2} \end{equation} \end{remark}
\begin{thm}[Bethe-Sommerfeld Conjecture] The spectrum of operator
$H$ contains a semi-axis. \end{thm} \begin{proof}
    By Remark \ref{R:Dec9}, there is a point of the spectrum of
$H_n$ in the $\delta _n$-neighborhood of $\lambda $ for every
$\lambda >k_*^{2}$, $k_*$ being introduced by (\ref{k}). Since
$\|H_n-H\|<\epsilon_{n}^4$, there is a point of the spectrum of $H$
in the $\delta _n^*$-neighborhood of $\lambda $,  $\delta
_n^*=\delta _n+\epsilon_{n}^4$. Since it is true for every $n$ and
the spectrum of $H$ is closed, $\lambda $ is in the spectrum of $H$.
\end{proof}

\subsection{Generalized Eigenfunctions of $H$} A plane wave is
usually denoted by $e^{i\langle \vec k, x
    \rangle}$, $\vec k \in \R^2$. Here we  use $\vec \varkappa $ instead of
    $\vec k$ to comply with our previous notations. We show that for
    every  $\vec \varkappa $ in
a set \begin{equation}\mathcal{G} _{\infty }=\cup _{\lambda \geq
\lambda _*}\mathcal{D}_{\infty}(\lambda ),\ \ \lambda
_*=k_*^{2},\label{Ginfty}\end{equation} there is a solution $\Psi
_{\infty }(\vec \varkappa , x)$ of the equation for eigenfunctions:
    \begin{equation} -\Delta \Psi _{\infty}(\vec \varkappa , x)+V(x)\Psi _{\infty }(\vec \varkappa ,
    x)=\lambda _{\infty}(\vec \varkappa )\Psi _{\infty }(\vec \varkappa , x),
    \label{6.2.1}
    \end{equation}
which can be represented in the form
    \begin{equation}
    \Psi _{\infty }(\vec \varkappa , x)=e^{i\langle \vec \varkappa , x
    \rangle}\Bigl(1+u_{\infty}(\vec \varkappa , x)\Bigr),\ \ \ \
    \bigl\|u_{\infty}(\vec \varkappa , x))\bigr\| _{L_{\infty }(\R^2)}<c(V)|\vec \varkappa |^{-\gamma_1},
   \label{6.2.1a}
    \end{equation}
where $u_{\infty}(\vec \varkappa , x)$ is a limit-periodic function,
 $\gamma _1=1/2-15s_1-8\delta $; the eigenvalue $\lambda _{\infty}(\vec \varkappa )$ satisfies the asymptotic
formula:
 \begin{equation}\lambda _{\infty}(\vec \varkappa )=|\vec \varkappa |^{2}+O(|\vec
\varkappa |^{-\gamma _2}), \ \ \ \gamma _2=2-30s_1-20\delta
.\label{6.2.4} \end{equation} We also show that the set $\mathcal{G}
_{\infty }$ satisfies (\ref{full}).

 In fact,   by (\ref{6.1}), any $\vec{\varkappa} \in \mathcal{D}_{\infty}(\lambda
)$ belongs to the $(\epsilon_n k^{-1-\delta})$-neighborhood of
$\mathcal{D}_n(\lambda )$. Applying \eqref{InE:lambda_j^1},
\eqref{InE:E_j^1} with $2\beta =2\beta _0=1-15s_1-9\delta $ and the
perturbation formulae proved for next steps, we obtain the following
inequalities:
    \begin{equation}\bigl\|E^{(1)}(\vec{\varkappa})-{E}^{(0)}(\vec{\varkappa})\bigr\|_1<c\|W_1\|k^{-1/2+12s_1+8\delta},
    \   \ \bigl\|E^{(n+1)}(\vec{\varkappa})-\tilde{E}^{(n)}(\vec{\varkappa})\bigr\|_1<
    48\epsilon_n^3, \quad n \geq 1,\label{6.2.2}
    \end{equation}
    \begin{equation}\bigl|\lambda ^{(1)}(\vec{\varkappa})-|\vec \varkappa |^{2}
     \bigr|
     <C(W_1)k^{-2+30s_1+20\delta }
    , \ \ \ \ \bigl|\lambda ^{(n+1)}(\vec \varkappa )-\lambda ^{(n)}(\vec \varkappa )\bigr|<12\epsilon _n^4,
     \quad n \geq 1,
    \label{6.2.3}
    \end{equation}
where $E^{(n+1)},\ \tilde{E}^{(n)}$ are one-dimensional spectral
projectors in $L_2(Q_{n+1})$ corresponding to potentials $W_{n+1}$
and $W_n$, respectively; $\lambda ^{(n+1)}(\vec \varkappa )$ is the
eigenvalue corresponding to $E^{(n+1)}(\vec{\varkappa})$,
${E}^{(0)}(\vec{\varkappa})$ corresponds to $V=0$ and the periods
$a_1$, $a_2$. This means that for properly chosen eigenfunctions
$\Psi _{n+1}(\vec \varkappa ,x)$: \begin{equation}\label{psi1}
    \|\Psi _{1}-\Psi _0\|_{L_2(Q_{1})}<c\|W_1\|k^{-1/2+12s_1+8\delta }|Q_1|^{1/2},\ \ \ \Psi _0(x)=e^{i \langle \vec \varkappa ,x
    \rangle},
    \end{equation}
    \begin{equation}
    \|\Psi _{n+1}-\tilde \Psi _n\|_{L_2(Q_{n+1})} < 100\epsilon_n^{3}|Q_{n+1}|
    ^{1/2},\label{Dec9c}
    \end{equation}
    where $\tilde \Psi _n $
    is $\Psi _n $ extended quasi-periodically from $Q_n$ to
    $Q_{n+1}$. Eigenfunctions $\Psi _{n}$, $n\geq 1$, are chosen to
    obey
    two
    conditions. First, $\|\Psi _{n}\|_{L_2(Q_{n})}=|Q_{n}|^{1/2}$; \footnote{The condition
    $\|\Psi _{n}\|_{L_2(Q_{n})}=|Q_{n}|^{1/2}$ implies
    $\|\tilde \Psi _{n}\|_{L_2(Q_{n+1})}=|Q_{n+1}|^{1/2}$.} second
    $
    (\Psi _{n},\tilde \Psi _{n-1})_n>0$, here    $(\cdot
    ,\cdot)_n$ is an inner product in $L_2(Q_{n})$. These two
    conditions, obviously, provide a unique choice of each $\Psi
    _n$.
    Considering that $\Psi _{n+1}$ and $\tilde \Psi _n$ satisfy  equations
for eigenfunctions and taking into account (\ref{6.2.3}),
(\ref{Dec9c}) we obtain: $\|\Psi _{n+1}-\tilde \Psi
_n\|_{W_2^{2}(Q_{n+1})} <
    2k^{2}\epsilon_n^{3}|Q_{n+1}|^{1/2}$, $n \geq
    1$,
and, hence,
    $\|\Psi _{n+1}-\tilde \Psi _n\|_{L_{\infty}(Q_{n+1})}
    < 6k^{2}\epsilon_n^{3}|Q_{n+1}|^{1/2}$.  Since $\Psi _{n+1}$ and $\tilde \Psi _n$ obey the same
quasiperiodic conditions, the same inequality holds in the whole
space $\R^2$:
    \begin{equation}
    \|\Psi _{n+1}-\Psi _n\|_{L_{\infty}(\R^2)}
    < 6k^{2}\epsilon_n^{3}|Q_{n+1}|^{1/2}, \ \ n \geq 1, \label{Dec10}
    \end{equation}
where $\Psi _{n+1},\Psi _n$ are quasiperiodically extended to
$\R^2$. Obviously, we have a Cauchy sequence in $L_{\infty}(\R^2)$.
Let $\Psi _{\infty }(\vec \varkappa ,x)=\lim_{n \to \infty}\Psi
_n(\vec \varkappa, x).$ This limit is defined pointwise uniformly in
$x$ and in $W_{2,loc}^{2}(\R^2)$.

\begin{thm} \label{T:Dec10} For every sufficiently large $\lambda $,
$\lambda >\lambda _0(\sum_{r=1}^{\infty }\|V_r\|,\delta )$, and
$\vec{\varkappa} \in \mathcal{D}_{\infty}(\lambda)$ the sequence of
functions $\Psi _n(\vec{\varkappa},x)$ converges  in
$L_{\infty}(\R^2)$ and $W_{2,loc}^{2}(\R^2)$. The limit function
$\Psi _{\infty }(\vec{\varkappa},x)$,
    $\Psi _{\infty }(\vec{\varkappa},x)=\lim_{n\to \infty }
    \Psi _n(\vec{\varkappa},x)$,
     satisfies the equation
    \begin{equation}\label{6.7}
     -\Delta \Psi _{\infty }(\vec{\varkappa}, x)+V(x)\Psi _{\infty }(\vec{\varkappa},
    x)= \lambda \Psi _{\infty }(\vec{\varkappa}, x).
    \end{equation}
 It can be represented in the form
   \begin{equation}\label{6.4}
    \Psi _{\infty }(\vec{\varkappa},x)=e^{i\langle \vec{\varkappa}, x
    \rangle}\bigl(1+u_{\infty}(\vec{\varkappa}, x)\bigr),
    \end{equation}
where $u_{\infty}(\vec{\varkappa}, x)$ is a limit-periodic function:
    \begin{equation}\label{6.5}
    u_{\infty}(\vec{\varkappa}, x)=\sum_{n=1}^{\infty}\tilde u_n(\vec{\varkappa},
    x),
    \end{equation}
 $\tilde u_n(\vec{\varkappa}, x)$  being a periodic function with the periods
$2^{M_n-1}d_1, 2^{M_n-1}d_2$ with $2^{M_n}\approx k^{2^{n-1}s_1}$,
\begin{equation}
    \|\tilde{u}_1\|_{L_{\infty}(\R^2)} <c(V)k^{-\gamma_1}, \ \ \
    \gamma _1=1/2-15s_1-8\delta , \label{6.6a}
    \end{equation}
    \begin{equation}\label{6.6}
    \|\tilde{u}_n\|_{L_{\infty}(\R^2)} <6k^{2}\epsilon_{n-1}^{3}|Q_{n}|^{1/2}, \ \ \ n\geq
    2.\end{equation}
\end{thm} \begin{cor} Function $u_{\infty}(\vec{\varkappa}, x)$
obeys the
 estimate (\ref{6.2.1a}).
\end{cor}

\begin{proof}
 Let us show that $\Psi _{\infty }$ is a
limit-periodic function. Obviously,
    $ \Psi _{\infty } =\Psi _0+\sum_{n=0}^{\infty}(\Psi_{n+1}-\Psi _n)
$, the series converging in $L_{\infty}(\R^2)$ by (\ref{Dec10}).
Introducing the notation $\tilde u_{n+1}=e^{-i \langle \vec
\varkappa ,x\rangle}(\Psi_{n+1}-\Psi _n)$, we arrive at (\ref{6.4}),
(\ref{6.5}). Note that
 $\tilde{u}_n$ is periodic with  the periods $2^{M_n-1}d_1,
 2^{M_n-1}d_2$. Estimate (\ref{6.6})  follows from (\ref{Dec10}). We check
 (\ref{6.6a}). Indeed,  by (\ref{psi1}), Fourier coefficients
 $(\tilde u_1)_m$, $m\in \Z^2$, satisfy the estimate
 $\left|(\tilde{u}_1)_m\right|<c(V)k^{-1/2+12s_1+8\delta }|Q_1|^{1/2}<c(V)k^{-1/2+13s_1+8\delta }$. We will use this estimate for
 $m:|m-j|<k^{s_1}$. Next, we obtain a stronger estimate for other $m$-s. Indeed, the
 inequality
 $\left|E^{(1)}(\varkappa )_{mj}\right|<\left( c(V)k^{-(1/2-12s_1-7\delta )}\right)^{|m-j|R_0^{-1}}$
  follows from \eqref{InE:G_r} and \eqref{finitedim}. Hence, similar estimates holds for
 Fourier coefficients of $\Psi _1$:
 $\left|(\tilde{u}_1)_m\right|< \left(c(V)k^{-(1/2-12s_1-7\delta
 )}\right)^{|m-j|R_0^{-1}}$.
 Summarizing the
 inequalities, we obtain that
 (\ref{6.6a}) holds. It remains to prove
(\ref{6.7}). Indeed, $\Psi _n(\vec \varkappa, x)$, $n\geq 1$,
satisfy equations for eigenfunctions: $H^{n)}\Psi _n=\lambda
^{(n)}(\vec \varkappa )\Psi _n $. Considering that $\Psi _n(\vec
\varkappa,x)$ converges to $\Psi (\vec \varkappa, x)$ in
$W_{2,loc}^2$ and relation (\ref{6.2}), we arrive at
(\ref{6.7}).\end{proof}

\begin{thm} \label{Thm:6.10} Formulae (\ref{6.2.1}), (\ref{6.2.1a}) and (\ref{6.2.4})
hold for every $\vec \varkappa \in \mathcal{G} _{\infty }$. The set
$\mathcal{G} _{\infty }$ is Lebesgue measurable and satisfies
  (\ref{full})  with $\gamma _3=\delta $.
\end{thm} \begin{proof} By Theorem~\ref{T:Dec10}, (\ref{6.2.1}),
(\ref{6.2.1a}) hold, where $\lambda _{\infty}(\vec \varkappa
)=\lambda $ for $\vec \varkappa \in \mathcal{D}_{\infty}(\lambda )$.
Using (\ref{h}), which is proven in Corollary~\ref{Dec18}, with
$\varkappa _{\infty }=|\vec \varkappa |$, we easily obtain
(\ref{6.2.4}). It remains to prove (\ref{full}). Let us consider a
small region $U_n(\lambda _0)$ around an isoenergetic surface
$\mathcal{D}_n(\lambda _0)$, $\lambda _0>k_*^{2}$. Namely,
$U_n(\lambda _0)=\cup _{|\lambda -\lambda _0|<\rho _n}
\mathcal{D}_{n}(\lambda )$, $\rho _n=\epsilon _{n-1}k^{-2\delta}$,
$k=\lambda _0^{1/2}$. Considering an estimate of the type
(\ref{E:grad-lambda_j^1*}) for $\lambda ^{(n)}(\vec \varkappa )$,
which holds in the
$\left(\epsilon_{n-1}k^{-1-2\delta}\right)$-neighborhood of
$\mathcal{D}_n(\lambda _0)$, we see that $U_n(\lambda _0)$ is an
open set (a distorted ring with holes) and the width of the ring is
of order $\epsilon_{n-1}k^{-1-2\delta}$.  By Part 4 of Lemmas
\ref{L:2.13}, \ref{L:3.9}, \ref{L:3.9*}, \ref{L:3.9*+}, the length
of $\mathcal{D}_n(\lambda _0)$ is $ 2\pi k \left(1+O(k^{-\delta
})\right)$. Hence, $|U_n(\lambda _0)|=2\pi k \rho
_n\left(1+O(k^{-\delta })\right)$. It easily follows from Lemma
\ref{L:3.9*+} that $U_{n+1}\subset U_n$. Definition of
$\mathcal{D}_{\infty }(\lambda _0)$ yield: $\mathcal{D}_{\infty
}(\lambda _0)=\cap _{n=1}^{\infty }U_n(\lambda _0)$. Hence,
$\mathcal{G}_{\infty }=\cap _{n=1}^{\infty }\mathcal{G}_n$, where
 \begin{equation}\mathcal{G}_n=\cup _{\lambda _0
>\lambda _*}U_n(\lambda _0)=\cup _{\lambda
>\lambda _*-\rho _n(\lambda _*)}\mathcal{D}_n(\lambda ). \label{Gn}\end{equation} Considering that
$U_{n+1}\subset U_n$ for every $\lambda _0\geq \lambda _*$, we
obtain $\mathcal{G}_{n+1}\subset \mathcal{G}_n$. Hence,
$\left|\mathcal{G} _{\infty }\cap
        \bf B_R\right|=\lim _{n\to \infty }\left|\mathcal{G} _{n}\cap
        \bf B_R\right|$. Summarizing volumes of the regions
        $U_{n}$, we easily conclude $\left|\mathcal{G} _{n}\cap
        \bf B_R\right|=|{\bf B_R}|\left(1+O(R^{-\delta })\right)$ uniformly in $n$. Thus, we have
        obtained (\ref{full}) with $\gamma _3=\delta $.
\end{proof}

\section{Proof of Absolute Continuity of the Spectrum}\label{chapt8}
The proof is completely analogous to that for the case $l\geq 6$.
Here we give only the list of lemmas and the main result. For
details, see \cite{cpde}.

\subsection{Projections $E_n(\mathcal{G}_n')$,
$\mathcal{G}_n'\subset \mathcal{G}_n$.}

Let us consider the open sets $\mathcal{G}_n$ given by (\ref{Gn}).
 There is a family of Bloch eigenfunctions $\Psi _n(\vec
    \varkappa ,x)$, $\vec \varkappa \in \mathcal{G}_n$, of the operator
    $H^{(n)}$, which are described by the perturbation formulas
    (\ref{na}).
    Let   $\mathcal{G}_n'$ be a Lebesgue measurable subset of  $\mathcal{G}_n$.
    We consider the spectral projection
    $E_n\left(\mathcal{G}_n'\right)$ of $H^{(n)}$, corresponding to
    functions $\Psi _n(\vec
    \varkappa ,x)$, $\vec \varkappa \in \mathcal{G}_n'.$
By \cite{6r}, $E_n\left( \mathcal{G}'_n\right): L_2(\R^2)\to
L_2(\R^2)$ can be presented by the formula:
    \begin{equation} E_n\left( \mathcal{G}'_n\right)F=\frac{1}{4\pi ^2}\int
    _{ \mathcal{G}'_n}\bigl( F,\Psi _n(\vec
    \varkappa )\bigr) \Psi _n(\vec
    \varkappa ) d\vec \varkappa \label{s}
    \end{equation}
    for any $F\in C_0^{\infty}(\R^2)$, here and below $\bigl( \cdot ,\cdot \bigr)$
    is the canonical scalar product in $L_2(\R^2)$, i.e.,
    $$\bigl( F,\Psi _n(\vec
    \varkappa )\bigr)=\int _{\R^2}F(x)\overline{\Psi _n(\vec
    \varkappa ,x)}dx.$$
The above formula can be rewritten in the form:
    \begin{equation} E_n\left(\mathcal{G}'_n\right)=S_n\left(\mathcal{G}'_n\right)T_n \left(
    \mathcal{G}'_n\right), \label{ST}
    \end{equation}
    %%%5555L_2(\R^2)
    $$T_n:C_0^{\infty}(\R^2) \to L_2\left(  \mathcal{G}'_n\right), \ \
    \ \ S_n:L_{\infty}\left( \mathcal{G}'_n\right)\to L_2(\R^2),$$
    \begin{equation}
    T_nF=\frac{1}{2\pi }\bigl( F,\Psi _n(\vec
    \varkappa )\bigr) \mbox{\ \ for any $F\in C_0^{\infty}(\R^2)$},
    \label{eq2}
    \end{equation}
    $T_nF$ being in $L_{\infty }\left(  \mathcal{G}'_n\right)$, and,
    \begin{equation}S_nf = \frac{1}{2\pi }\int _{  \mathcal{G}'_n}f (\vec \varkappa)\Psi _n(\vec
    \varkappa ,x)d\vec \varkappa  \mbox{\ \ for any $f \in L_{\infty }\left(
    \mathcal{G}'_n\right)$.} \label{ev}
    \end{equation}
%%%%%%%%%It is
%%%%%%%%%    easy to show that $T_nF\in L_{\infty }(\mathcal G_n)$, when $F\in C_0^{\infty }(\R^2)$.
%%%%%%%%%    Hence $E_n\left( \mathcal{G}'_n\right)$ can be described by formula
%%%%%%%%%    (\ref{s}) for $F\in C_0^{\infty }(\R^2)$.
 By \cite{6r}, $\|T_nF\|_{L_2\left( \mathcal{G}'_n\right)}\leq \|F\|_{L_2(\R^2)}$ and
$\|S_nf \|_{L_2(\R^2)}\leq \|f \|_{L_2\left(
\mathcal{G}'_n\right)}$.
    Hence, $T_n$, $S_n$ can be extended by
    continuity from $C_0^{\infty }(\R^2)$, $L_{\infty }\left(  \mathcal{G}'_n\right)$
    to $L_2(\R^2)$ and $L_2\left( \mathcal{G}'_n\right)$,
    respectively. Thus, the operator $E_n\left( \mathcal{G}'_n\right)$ is
    described by (\ref{ST}) in the whole space $L_2(\R^2)$.

Obviously, for every $\vec \varkappa $ in $\mathcal{G}_n$, there
exists a pair $(\lambda _n ,\varphi )$ such that $\lambda _n
=\lambda ^{(n)}(\vec \varkappa )$ and that $(\cos \varphi ,\sin
\varphi )=\frac{\vec \varkappa }{|\vec \varkappa |}$. Let us
introduce new coordinates $(\lambda _n ,\varphi )$ in
$\mathcal{G}_n$: $\lambda _n=\lambda ^{(n)}(\vec \varkappa )$,
$(\cos \varphi ,\sin \varphi )=\frac{\vec \varkappa }{|\vec
\varkappa |}$. \begin{lem} Every point  $\vec \varkappa $ in
$\mathcal{G}_n$ is represented by a unique pair $(\lambda _n,\varphi
)$, $\lambda _n>\lambda _*$, $\varphi \in [0,2\pi )$, such that
\begin{equation} \vec \varkappa (\lambda _n,\varphi )=\varkappa _n
(\lambda_n,\vec \nu )\vec \nu,\ \ \vec \nu =(\cos \varphi ,\sin
\varphi ), \label{yu6} \end{equation} $\varkappa _n (\lambda_n,\vec
\nu )$ being the ``radius" of the isoenergetic curve
$\mathcal{D}_n(\lambda _n)$ in the direction $\vec \nu $. %%%%%%%%,
where $\lambda _*=k_*^{2l}$. \end{lem}

For any function $f(\vec \varkappa)$ integrable on $\mathcal{G}_n$,
we use the new coordinates and write
    \begin{align*}
    \int_{\mathcal{G}_n} f(\vec \varkappa)d\vec \varkappa &= \int_{\R^2}\chi
    \left(\mathcal{G}_n,\vec \varkappa\right)f(\vec \varkappa)d\vec \varkappa\\
    &=\int_{\lambda _*}^{\infty }\int_0^{2\pi}
    \chi \bigl( \mathcal{G}_n, \vec \varkappa (\lambda_n,\varphi )\bigr)
    f\bigl(\vec \varkappa (\lambda_n,\varphi )\bigr)
    \frac{\varkappa _n(\lambda_n,\vec \nu)}{\frac{\partial \lambda_n}{\partial \varkappa }
    }d\varphi \ d\lambda_n,
    \end{align*}
where    $\chi \left(\mathcal{G}_n,\vec \varkappa\right)$ is the
characteristic function on $\mathcal{G}_n$, $\vec \varkappa (\lambda
_n,\varphi )$ is given by \eqref{yu6} and $\frac{\partial
\lambda_n}{\partial \varkappa }=\left(\nabla \lambda ^{(n)}(\vec
\varkappa ),\vec \nu \right)\left|_{\vec \varkappa =\vec \varkappa
_n(\lambda_n,\vec \nu)}.\right.$ Let \begin{equation}
\mathcal{G}_{n, \lambda}=\{ \vec \varkappa \in {\mathcal{G}}_n:
\lambda ^{(n)}(\vec \varkappa) < \lambda\}. \label{d} \end{equation}
 This set is Lebesgue measurable, since ${\mathcal{G}}_n $ is
open and $\lambda ^{(n)}(\vec \varkappa)$ is continuous on $
{\mathcal{G}}_n$.

\begin{lem}\label{L:abs.6}
$\left|{\mathcal{G}}_{n,\lambda+\varepsilon} \setminus
{\mathcal{G}}_{n,\lambda}\right| \leq 2\pi \varepsilon $ when $0\leq
\varepsilon \leq 1$. \end{lem}

 By (\ref{s}),
$E_n\left({\mathcal{G}}_{n,\lambda+\varepsilon}\right)-E_n\left({\mathcal{G}}_{n,\lambda}\right)
=E_n\left({\mathcal{G}}_{n,\lambda+\varepsilon}\setminus
    {\mathcal{G}}_{n,\lambda}\right)$. Let us obtain an estimate for
    this projection.
\begin{lem}\label{L:abs.7} For any $F \in C_0^{\infty}(\R^2)$ and
$0\leq \varepsilon \leq 1$, \begin{equation}
\left\|\bigl(E_n({\mathcal{G}}_{n,\lambda+\varepsilon})-E_n({\mathcal{G}}_{n,\lambda})\bigr)F\right\|^2_{L_2(\R^2)}
 \leq C( F) \epsilon , \label{tootoo1}
 \end{equation}
 where $C(F)$ is uniform with respect to $n$ and $\lambda$.
\end{lem}

\subsection{ Sets ${\mathcal{G}}_{\infty}$ and
${\mathcal{G}}_{\infty ,\lambda }$.} \label{S:8.1} The sets
$\mathcal{G}_{\infty}$, $\mathcal{G}_{n}$ are given by
\eqref{Ginfty}, \eqref{Gn}.
%%%%%%%By definition,
%%%%%%%$$\mathcal{G}_{\infty}=\bigcup_{\lambda > \lambda_*}\mathcal D _{\infty}(\lambda) \text{ and }
%%%%%%%\mathcal{G}_{n}=\bigcup _{\lambda >\lambda_*} \mathcal D_n(\lambda).$$
 As it was shown in the proof of Theorem \ref{Thm:6.10},
$
    \mathcal{G}_{n+1}\subset \mathcal{G}_n,$
    %%%%%%% \label{ac2}
   %%%%%%%%% \end{equation}
%%%%%%%%\begin{equation}\label{tt}
    $\mathcal{G}_{\infty}=\bigcap_{n=1}^{\infty}{\mathcal{G}}_n. $
%%%%%%%%%%%    \end{equation}
Therefore, the perturbation formulas for $\lambda ^{(n)}(\vec
\varkappa )$ and $\Psi _n(\vec \varkappa )$ hold in
$\mathcal{G}_{\infty}$ for all $n$. Moreover, coordinates
$(\lambda_n, \varphi )$ can be used in $\mathcal{G}_{\infty}$  for
every $n$. Let
     \begin{equation}
     \mathcal{G}_{\infty, \lambda }=\left\{\vec \varkappa \in
    \mathcal{G}_{\infty }: \lambda _{\infty }(\vec \varkappa )<\lambda
    \right\}. \label{dd}
    \end{equation}
The function $\lambda _{\infty }(\vec \varkappa )$ is a Lebesgue
measurable function, since it is a limit of the sequence of
measurable functions. Hence, the set  $\mathcal{G}_{\infty, \lambda
}$ is measurable.

\begin{lem}\label{add6*} The measure of the symmetric difference of
two sets $\mathcal{G}_{\infty, \lambda }$ and $\mathcal{G}_{n,
\lambda}$ converges
 to zero as $n \to
\infty$ uniformly in $\lambda$ in every bounded interval:
    $$\lim _{n\to \infty }\left|\mathcal{G}_{\infty, \lambda }\Delta \mathcal{G}_{n, \lambda
    }\right|=0.$$
 \end{lem}

    \subsection{Spectral Projections $E(\mathcal{G}_{\infty , \lambda })$.}

In this section, we show that spectral projections
$E_n(\mathcal{G}_{\infty , \lambda })$ have a strong limit
$E_{\infty }(\mathcal{G}_{\infty , \lambda })$ in $L_2(\R^2)$ as $n$
tends to infinity. The operator $E_{\infty }(\mathcal{G}_{\infty ,
\lambda })$ is a spectral projection of $H$. It can be represented
in the form $E_{\infty }(\mathcal{G}_{\infty , \lambda })=S_\infty
T_{\infty }$, where $S_{\infty }$ and $T_{\infty }$ are strong
limits of $S_n(\mathcal{G}_{\infty , \lambda })$ and
$T_n(\mathcal{G}_{\infty , \lambda })$, respectively.  For any $F\in
C_0^{\infty }(\R^2)$, we show: \begin{equation} E_{\infty }\left(
\mathcal{G}_{\infty , \lambda }\right)F=\frac{1}{4\pi ^2}\int
    _{ \mathcal{G}_{\infty , \lambda }}\bigl( F,\Psi _{\infty }(\vec
    \varkappa )\bigr) \Psi _{\infty }(\vec
    \varkappa ) d\vec \varkappa ,\label{s1}
    \end{equation}
    \begin{equation} HE_{\infty }\left(
\mathcal{G}_{\infty , \lambda }\right)F=\frac{1}{4\pi ^2}\int
    _{ \mathcal{G}_{\infty , \lambda }}\lambda _{\infty }(\vec
    \varkappa )\bigl( F,\Psi _{\infty }(\vec
    \varkappa )\bigr) \Psi _{\infty }(\vec
    \varkappa ) d\vec \varkappa .\label{s1uu}
    \end{equation}
   %%%%%%% $\bigl( F,\Psi _{\infty }(\vec
 %%%%%%%   \varkappa )\bigr) \Psi _{\infty }$ being an integral analogous
 %%%%%%%%   to the dot product in $L_2(R^2)$:
 %%%%%%%%    \begin{equation}\bigl( F,\Psi _{\infty}(\vec
%%%%%%%%     \varkappa )\bigr)=\int _{R^2}F(x)\overline{\Psi _{\infty }(\vec
%%%%%%%%     \varkappa ,x)}dx. \label{eq1}
%%%%%%%%     \end{equation}
Using properties of $E_{\infty }\left( \mathcal{G}_{\infty , \lambda
}\right)$, we  prove absolute continuity of the branch of the
spectrum corresponding to functions $\Psi _{\infty }(\vec
    \varkappa )$.

Now we consider the sequence of operators $T_n(\mathcal{G}_{\infty ,
\lambda
    })$ which are given by (\ref{eq2}) and act from $L_2(\R^2)$ to $L_2(\mathcal{G}_{\infty , \lambda
    })$. We prove that the sequence has a strong limit and
    describe its properties.

\begin{lem} \label{Lem1} The sequence $T_n(\mathcal{G}_{\infty ,
\lambda
    })$  has a strong limit $T_{\infty }(\mathcal{G}_{\infty , \lambda
    })$. The operator $T_{\infty }(\mathcal{G}_{\infty , \lambda
    })$ satisfies $\|T_{\infty }\|\leq 1$ and can be described by the
    formula
    $T_{\infty }F=\frac{1}{2\pi }\bigl( F,\Psi _{\infty }(\vec
    \varkappa )\bigr) $ for any $F\in C_0^{\infty }(\R^2)$.
    The convergence of $T_n(\mathcal{G}_{\infty , \lambda
    })F$ to $T_{\infty }(\mathcal{G}_{\infty , \lambda
    })F$ is uniform in $\lambda $ for every $F\in L_2(\R^2)$.
\end{lem}

Now we consider the sequence of operators $S_n(\mathcal{G}_{\infty ,
\lambda })$ which are given by (\ref{ev}) with $\mathcal
G_n'=\mathcal{G}_{\infty , \lambda
    }$:
    $$S_n(\mathcal{G}_{\infty , \lambda
    }):\ L_2(\mathcal{G}_{\infty , \lambda
    })\to L_2(\R^2).$$ We prove that the sequence has a strong limit and
    describe its properties.

\begin{lem} \label{Lem2} The sequence of operators
$S_n(\mathcal{G}_{\infty , \lambda
    })$ has a strong limit $S_{\infty }(\mathcal{G}_{\infty , \lambda
    })$. The operator $S_{\infty }(\mathcal{G}_{\infty , \lambda
    })$ satisfies $\|S_{\infty }\|\leq 1$ and can be described by the
    formula
    \begin{equation}
    (S_{\infty }f) (x)= \frac{1}{2\pi }\int _{\mathcal{G}_{\infty , \lambda
    }}f (\vec \varkappa)\Psi _{\infty }(\vec
    \varkappa ,x) d\vec \varkappa  \label{ev1}
    \end{equation}
for any $f \in L_{\infty }\left( \mathcal{G}_{\infty , \lambda
}\right)$. The convergence of $S_n(\mathcal{G}_{\infty , \lambda
    })f $ to $S_{\infty }(\mathcal{G}_{\infty , \lambda
    })f $  is uniform in $\lambda $ for every $f \in L_{2}\left(
    \mathcal{G}_{\infty }\right)$.
\end{lem}

\begin{lem}\label{May8} Spectral projections
$E_n(\mathcal{G}_{\infty , \lambda })$ have a strong limit
$E_{\infty }(\mathcal{G}_{\infty , \lambda })$ in $L_2(\R^2)$, the
convergence being uniform in $\lambda $ for every element. The
operator $E_{\infty }(\mathcal{G}_{\infty , \lambda })$ is a
projection. For any $F\in C_0^{\infty }(\R^2)$ it is given by
(\ref{s1}) and formula (\ref{s1uu}) holds.
\end{lem}

\begin{lem}\label{onemore} There is a strong limit
$E_\infty(\mathcal{G}_{\infty})$ of the projections $E_\infty
(\mathcal{G}_{\infty,\lambda })$ as $\lambda $ goes to infinity.
\end{lem} \begin{cor}\label{onemore1} The operator $E_{\infty
}(\mathcal{G}_{\infty})$ is a projection. \end{cor}

\begin{lem}\label{L:abs.9} Projections
$E_\infty(\mathcal{G}_{\infty,\lambda })$, $\lambda \in \R$, and
$E_\infty(\mathcal{G}_{\infty})$ reduce the operator $H$. \end{lem}

\begin{lem} The family of projections
$E_\infty(\mathcal{G}_{\infty},\lambda )$ is the resolution of the
identity of the operator $HE_\infty(\mathcal{G}_{\infty})$ acting in
$E_\infty(\mathcal{G}_{\infty})L_2(\R^2)$. \end{lem}

\subsection{Proof of Absolute Continuity.}

Now we show that the  branch of spectrum (semi-axis) corresponding
to $\mathcal G_{\infty }$  is absolutely continuous.

\begin{thm}\label{T:abs} For any $F\in C_0^{\infty }(\R^2)$ and
$0\leq \varepsilon \leq 1$,
    \begin{equation}
    \left| \left(E_\infty(\mathcal{G}_{\infty,\lambda+\varepsilon})F,F\right)-
    \left(E_\infty(\mathcal{G}_{\infty,\lambda})F,F
    \right) \right| \leq C_F\varepsilon .\label{May10*}
    \end{equation}

\end{thm} \begin{cor} The spectrum of the operator
$HE_\infty(\mathcal G_{\infty})$ is absolutely continuous. \end{cor}
\begin{proof} By formula (\ref{s1}),
    $$ | \left(E_\infty(\mathcal{G}_{\infty,\lambda+\varepsilon})F,F\right)-\left(E(\mathcal{G}_{\infty,\lambda})F,F
    \right) | \leq C_F\left| \mathcal{G}_{\infty , \lambda +\varepsilon
    }\setminus \mathcal{G}_{\infty , \lambda } \right| .$$
Applying Lemmas \ref{L:abs.6} and \ref{add6*}, we immediately get
(\ref{May10*}).

\end{proof}

\section{Geometrical Lemmas.}

%%%%%%%%%%%%%%%%%%%%%%%%%%%%%%%%%%%%%%%%%%%%%%%%%%%%%%%%%%%%%%%
\subsection{Proof of Lemma \ref{Lem:Geometric-1} (Geometric Lemma,
step I)} \label{sec:geometric-1}   Corollaries \ref{cor:t} and
\ref{C:2.1.2}, obtained in Sections \ref{S:7.1.1} and \ref{S:7.1.2},
together give Lemma \ref{Lem:Geometric-1} and its corollary.

\subsubsection{Set $\chi _1(\lambda, \beta ,s_1,\delta
)$.}\label{5.1.1} The set $\chi _1(\lambda, \beta ,s_1,\delta )$ is
given by formulas \eqref{chi _1}--\eqref{eqm}. It remains to define
radii $r_m$ of the discs $\mathcalO^{\pm}_m$. First, we need more
notations. Angle $\varphi _m$ is defined as a polar coordinate of
$\vec p_m$:
 $$
  \vec p_m := \vec p_m(0) =  p_m (\cos \vphi_m, \sin \vphi_m),\ m=(m_1,m_2) \in
  \Z^2.$$
 In addition,  let $\Pi_{m,q,\pm }$ be the distance in $\C$ between $|\vec
 k(\varphi )+\vec p_m|_*^2$ and $|\vec
 k(\varphi )+\vec p_{m+q}|_*^2$ when $\varphi =\varphi _m^{\pm }$,
 i.e., when $\varphi $ solves $|\vec
 k(\varphi )+\vec p_m|_*^2=k^2$. Obviously,
 \beq\label{E:pi_mq}
 \Pi_{m,q,\pm } :=  |\la \vec k(\vphi_m^\pm)+\vec p_m, \vec p_q \ra_*+p_q^2|.
 \eeq
 We will be interested in values of $ \Pi_{m,q,\pm }$ for
 $q:0<p_q<k^{s_1}$. Let
 \beq\label{E:pi_mq'}
 \Pi_{m,q,\pm }' :=  |\la \vec k(\vphi_m^\pm)+\vec p_m, \vec p_q \ra_*|+k^{2s_1}.
 \eeq
 Obviously, $\Pi_{m,q,\pm }'=\Pi_{m,q,\pm } +O\left(k^{2s_1}\right)$, $\Pi_{m,q,\pm }'\geq k^{2s_1}$. Next,
\beq\label{E:pi_m}
 \Pi_m := \min_{0<p_q<k^{s_1},\ \pm}\Pi_{m,q,\pm }'=\min_{0<p_q<k^{s_1},\ \pm} |\la \vec k(\vphi_m^\pm)+\vec p_m, \vec p_q \ra_*|+k^{2s_1}.
 \eeq
The value $\Pi_m $ characterizes the smallest distance between
$|k(\varphi )+\vec p_m|_*^2$ and $|\vec
 k(\varphi )+\vec p_{m+q}|_*^2$ when $|k(\varphi )+\vec p_m|_*^2=k^2$,
 $0<p_q<k^{s_1}$. It coincides with this  distance up to the
 value $O(k^{2s_1})$. Further we use that $\Pi_m \geq k^{2s_1}$.
 \begin{definition}\label{Def:r_m}
  The radius $r_m$ of the two open disks $\mathcalO_m^\pm$ in $\C$ centered at $\vphi_m^\pm$ is defined by
   \beq \label{r_m}
   r_m := \left\{
    \begin{array}{ll}
    \dfrac{k^{-3s_1-\delta}}{p_m} &
      \text{if } 0< p_m \le 4k^{s_1},\\ [1.5em]
     \dfrac{k^{2\beta -1}}{p_m \Pi_m \sqrt{\left|1-\frac{p_m^2}{4k^2}\right|}} &
      \text{if }  p_m>4k^{s_1} \ \mbox{and}\ \left|1-\frac{p_m^2}{4k^2}\right|\ge \frac{k^{2\beta-2}}{\Pi_m},\\ [2.0em]
     \dfrac{8 k^{\beta }}{p_m \sqrt{\Pi_m}} &
      \text{if } \left|1-\frac{p_m^2}{4k^2}\right| < \frac{k^{2\beta-2}}{\Pi_m}.
    \end{array}
   \right.
   \eeq
\end{definition}
 \begin{remark}
 If $p_m \ge 4k$, then the imaginary part of $\vphi_m^\pm$ is too large to be in $\varPhi_0$
 and so we do not need to remove corresponding disks from $\varPhi_0$ to construct $\varPhi_1$.
 \end{remark}
 \begin{lem}\label{Lem:r_m^2}
 For every $m$ with $0<p_m < 4k$, $r_m=o(1)$ as $k \to \infty$.
 \end{lem}
 \bpf
 Let us consider the following four cases:

 If $0< p_m \le 4k^{s_1}$, then $r_m <
 \frac{k^{-3s_1-\delta}}{k^{-s_1}}=k^{-2s_1-\delta}=o(1)$, since the
periods $a_1$, $a_2$ of $H^{(1)}$ are of order $k^{s_1}$.

 If $4k^{s_1} < p_m \le k$, then $r_m^2 < \frac{k^{4\beta-2}}{k^{2s_1}  k^{4s_1} 3/4}<2k^{4\beta-2-6s_1}=o(1)$,
 since $\Pi_m \ge k^{2s_1}$.

 If $k < p_m < 4k$ and $\left|1-\frac{p_m^2}{4k^2}\right| \ge \frac{k^{2\beta-2}}{\Pi_m}$,
 then $r_m^2 < \frac{k^{4\beta -2}}{k^2  \Pi_m^2 \frac{k^{2\beta-2}}{\Pi_m}} < k^{2\beta-2-2s_1}=o(1)$.

 If $\left|1-\frac{p_m^2}{4k^2}\right| < \frac{k^{2\beta-2}}{\Pi_m}$, then of course $p_m>k$
  and $r_m < \frac{8k^{\beta}}{k^{1+s_1}}=8k^{\beta-1-s_1}=o(1)$.
 \epf

 \subsubsection{Proof of the statements (i), (ii) in Lemma
 \ref{Lem:Geometric-1}.} \label{S:7.1.1}
\begin{lem}\label{Lem:lowerbounds}
 Let $4s_1 <2\beta <{1-15s_1-8\delta }$, $\vphi$ is in  $\varPhi_1$ or its $(k^{2\beta-3-s_1-2\delta })$-neighborhood,
  $m,q\in Z^2$, $m,q,m+q\neq 0$, $0<p_q<k^{s_1}$. Then
  \beq\label{InE:1st-lowerbound}
  2\bigl| |\vec k(\vphi)+\vec p_m|_*^2-k^2\bigr|
  \bigl| |\vec k(\vphi)+\vec p_{m+q}|_*^2-k^2\bigr| >k^{2\beta},
  \eeq
  \beq\label{InE:2nd-lowerbound}
  \bigl| |\vec k(\vphi)+\vec p_q|_*^2-k^2\bigr| > k^{1-3s_1-\delta}.
  \eeq
%%%%%%%%%%  and
%%%%%%%%%%  \beq\label{InE:3rd-lowerbound}
%%%%%%%%%%  2\bigl| |\vec k(\vphi)+\vec p_m|_*^2 - |\vec k(\vphi)+\vec p_{m+q}|_*^2 \bigr|
%%%%%%%%%%  > k^{2\beta} \bigl| |\vec k(\vphi)+\vec p_m|_*^2-k^2\bigr|^{-1},
%%%%%%%%%%  \eeq
%%%%%%%%%% where $m \neq 0,\ m+q \neq 0,\ 0<p_q<k^{s_1}$.
 \end{lem}
 \begin{cor}[Statements (i), (ii) in Lemma
 \ref{Lem:Geometric-1}, Corollary \ref{cor:geometric-1}] \label{cor:t} If $t\in \chi _1(\lambda ,\beta ,s_1,\delta )$, then there
 is a unique $j\in \Z^2$ such that $p_j(t)=k$ and \eqref{InE:Geometric-1-1}, \eqref{InE:Geometric-1-2} hold.
 For any $t$ in the $(k^{2\beta-2-s_1-2\delta })$-neighborhood of the non-resonance set in $\C^2$,
 there exists a unique $j\in \Z ^2$ such that
\eqref{InE:Geometric-1-nbh} and \eqref{InE:Geometric-1-1},
\eqref{InE:Geometric-1-2}, \eqref{c-1}
 \eqref{InE:circle-1} hold. \end{cor} \bpf[Proof of
Corollary \ref{cor:t}] If $t\in \chi _1(\lambda ,\beta ,s_1,\delta
)$, then, by the definition of $\chi _1$, there
 is $\varphi \in \Theta _1$ such that $t=\mathcalK_1\vec k(\varphi )$.
Hence, there exists a $j$ such that $p_j(t)=k$. Let us show
 that  $j$ is unique. Suppose there is $j'=j+m$, $m\neq 0$ such
 that $p_{j+m}(t)=k$. Then $\vec p_{j+m}(t)=\vec k(\varphi )+\vec p_{m}$
 and $|\vec k(\varphi )+\vec p_{m}|=k$. The last relation contradicts to
 \eqref{InE:1st-lowerbound}. Therefore, $j'=j$.

 Substituting $\vec k(\varphi )=\vec
 p_j(t)$ into \eqref{InE:1st-lowerbound},
 \eqref{InE:2nd-lowerbound} and using the notation $j+m=i$, we obtain \eqref{InE:Geometric-1-1} and
\eqref{InE:Geometric-1-2}. Similar arguments work when $t$ is in the
$(k^{2\beta-2-s_1-2\delta })$-neighborhood of $\chi _1(\lambda
,\beta ,s_1,\delta )$ in $\C^2$.

  The inequalities  (\ref{c-1}), (\ref{InE:circle-1}) are
 obvious, when $p_i>4k$. We assume now $p_i<4k $. Let us prove
 (\ref{c-1}). When $i=j$, (\ref{c-1}) follows directly from
 (\ref{E:circle-1}) and (\ref{InE:Geometric-1-nbh}). Suppose $i\neq j$.
 Clearly, $|p_i^{2}(t)-p_{i+q}^{2}(t)|=O(k^{1+s_1})$.
 This together with (\ref{InE:Geometric-1-1}) yields:
 $|p_i^{2}(t)-p_j^{2}(t)|>  ck^{2\beta-1-s_1}$. Considering (\ref{E:circle-1}), we
 obtain
 \begin{equation} 2|p_i^{2}(t)-z|\geq |p_i^{2}(t)-p_j^{2}(t)|.
 \label{t}
 \end{equation}
 The estimate (\ref{c-1}) easily follows.
Let us prove (\ref{InE:circle-1}). In the case of $i = j$,  using
\eqref{InE:Geometric-1-2} and the definition of $C_1$, we get
  \beq
  | p_{j+q}^{2}(t)-z| \geq | p_j^{2}(t)-p_{j+q}^{2}(t)|-k^{2\beta-1-s_1-\delta} > \frac{1}{2}k^{1-3s_1-\delta}.
  \eeq
 Using this together with the definition of $C_1$, we get \eqref{InE:circle-1} in the case $i=j$.
 In the case of $i \neq j$  \eqref{InE:Geometric-1-1} and
 (\ref{t}) yield \eqref{InE:circle-1}.
It is easy to see that all the estimates are stable under a
perturbation of $t$ of order $k^{2\beta-2-s_1-2\delta }$. Therefore,
the estimate \eqref{InE:circle-1} can be extended to the complex
$(k^{2\beta-2-s_1-2\delta })$-neighborhood of $\chi_1(k, \beta, s_1,
\delta )$. \epf
 \bpf[Proof of Lemma \ref{Lem:lowerbounds}]
 If $p_m>4k$ then \eqref{InE:1st-lowerbound} is obvious,
 since both factors are greater than $k^2$. Assume  $0<p_m<4k $. Noting that
  \beq
  |\vec k(\vphi)+\vec p_m|_*^2-k^2 = 2kp_m\cos (\vphi-\vphi_m)+p_m^2
  \label{08.20}
  \eeq
 and recalling that $\vphi_m^\pm$ are the solutions of
 $|\vec k(\vphi)+\vec p_m|_*^2 = k^2 $, we see:
  \beq\label{E:sin-cos}
  \cos (\vphi_m^\pm-\vphi_m) = -\frac{p_m}{2k},\quad
   |\sin (\vphi_m^\pm-\vphi_m)| = \sqrt{\left|1-\frac{p_m^2}{4k^2}\right|}.
  \eeq
 Now  let $\vphi$ be on the boundary of $\mathcalO_m^+$ or $\mathcalO_m^-$.
 Expanding (\ref{08.20}) around $\vphi_m^\pm$, we get:
  \beq\label{E:first-factor}
  |\vec k(\vphi)+\vec p_m|_*^2-k^2
  = 2k p_m \sin(\vphi_m^\pm-\vphi_m)r_m\left(1+O(r_m^2)\right)
   - k p_m \cos(\vphi_m^\pm-\vphi_m)r_m^2\left(1+O(r_m^2)\right).
  \eeq
Next, we  prove
  \beq\label{InE:first-factor}
  \bigl| |\vec k(\vphi)+\vec p_m|_*^2-k^2\bigr| > k^{2\beta}/\Pi_m.
  \eeq
 in three cases as in Definition \ref{Def:r_m}(i). In special, we
 get (\ref{InE:2nd-lowerbound}) in the first case.

Case\itemthm: $r_m = \frac{k^{-3s_1-\delta}}{p_m }$ when $0< p_m
\leq 4k^{s_1}$.
 The modulus of the first term in \eqref{E:first-factor} is
 $2k^{1-3s_1-\delta}\sqrt{1-\frac{p_m^2}{4k^2}} (1+o(1))> \frac{3}{2}k^{1-3s_1-\delta}$ and that of the second
  term is $O(k^{-6s_1-2\delta})$, which is much smaller than the first term. Thus, we obtain \eqref{InE:2nd-lowerbound}.
  Considering that $k^{1-3s_1-\delta}>\frac{2k^{2\beta}}{\Pi_m}$, we obtain \eqref{InE:first-factor}.

 Case\itemthm: $r_m = \frac{k^{2\beta-1}}{p_m  \Pi_m \sqrt{\left|1-\frac{p_m^2}{4k^2}\right|}}$
 when $4k^{s_1} < p_m < 4k  \text{ and }
 \left|1-\frac{p_m^2}{4k^2}\right|>\frac{k^{2\beta-2}}{\Pi_m}$. Substituting $r_m$ into (\ref{E:first-factor}), we get that
 the modulus of the first term is $\frac{2k^{2\beta}}{\Pi_m}\left(1+o(1)\right)$ and that of the second term is
 $\frac{k^{4\beta-2}}{\Pi_m^2 \left|1-\frac{p_m^2}{4k^2}\right|}(1+o(1))$.
 Using the condition $\left|1-\frac{p_m^2}{4k^2}\right|>\frac{k^{2\beta-2}}{\Pi_m}$, one can
  easily see that the former is at least twice greater than the latter. Thus,  we get \eqref{InE:first-factor}.

 Case\itemthm: $r_m = \frac{8 k^{\beta}}{p_m \sqrt{\Pi_m}}$ when $\left|1-\frac{p_m^2}{4k^2}\right| <
 \frac{k^{2\beta-2}}{\Pi_m}$. This time the modulus of the second term is $\frac{32k^{2\beta}}{\Pi_m}\left(1+o(1)\right)$ and
 that of the first is smaller than $\frac{16k^{2\beta}}{\Pi_m}(1+o(1))$. Therefore we again have \eqref{InE:first-factor}.

 Now we prove \eqref{InE:1st-lowerbound}. If $0<p_m,p_{m+q}\leq 4k^{s_1}$,  then
 \eqref{InE:1st-lowerbound}
easily follows from \eqref{InE:2nd-lowerbound} proven in the Case
(i) above for such $m,m+q$. Next, if both factors in the left hand
side of \eqref{InE:1st-lowerbound} is greater than $ k^{\beta}$,
  then  \eqref{InE:1st-lowerbound} is obvious. Therefore, without loss of generality, we can assume
  \begin{equation}\label{beta}
  \bigl| |\vec k(\vphi)+\vec p_m|_*^2-k^2\bigr| \le k^{\beta},\
\ \  p_m > 4k^{s_1}. \end{equation} Suppose we have proved  that
 \begin{equation}
  \Bigl| |\vec k(\vphi)+\vec p_{m+q}|_*^2-k^2\Bigr|>
  \frac{1}{2}\Pi_m \label{InE:p_m-p_m+q+}
  \end{equation}
 for every $m: \ \Pi_m > k^{\beta}$ and $\varphi $ satisfying
  \eqref{beta}. \footnote{The only conditions required for \eqref{InE:p_m-p_m+q+}
 are $\Pi_m > k^{\beta}$ and  \eqref{beta}. We will use \eqref{InE:p_m-p_m+q+} under
 this condition in Lemma \ref{Lem:lowerbounds*}.}
Then, notice that $\Pi_m >
  k^{\beta}$ follows from
\eqref{InE:first-factor}.
 Considering  \eqref{InE:first-factor} and \eqref{InE:p_m-p_m+q+} together, we obtain \eqref{InE:1st-lowerbound}.

 It remains to prove \eqref{InE:p_m-p_m+q+}. First, we check that
 \beq\label{InE:p_m-p_m+q}
  \bigl| |\vec k(\vphi)+\vec p_m|_*^2 - |\vec k(\vphi)+\vec p_{m+q}|_*^2\bigr|
  \ge \frac{3}{2}\Pi_m +O(k^{2s_1})
  \eeq
   for every $m: \ \Pi_m > k^{\beta}$ and $\varphi $ satisfying
  \eqref{beta}.
Indeed, taking into account \eqref{E:first-factor} and considering
as in the proof of \eqref{InE:first-factor},
 we conclude that the set of $\varphi $ satisfying \eqref{beta} is
 inside the circle of the radius $r_m'$ around $\varphi _m^{\pm }$,
 where \begin{equation}
 r_m'=\left\{\begin{array}{ll}\frac{k^{\beta -1
 }}{p_m\sqrt{1-\frac{p_m^2}{4k^2}}},&\mbox{when}\ \
 1-\frac{p_m^2}{4k^2}>k^{\beta -2}\\
 10k^{\beta /2-1}, &\mbox{otherwise.}\end{array}\right.\label{r_m'}
 \end{equation}
 It is easy to show that for any $\varphi $ inside those circles:
 \beq\label{InE:p_m-p_m+q*}
  \left|\Bigl| |\vec k(\vphi)+\vec p_m|_*^2 - |\vec k(\vphi)+\vec
  p_{m+q}|_*^2\Bigr|-\Bigl| |\vec k(\vphi_m^{\pm })+\vec p_m|_*^2 - |\vec k(\vphi_m^{\pm })+\vec
  p_{m+q}|_*^2\Bigr|\right|\leq
\eeq $$2kp_qr_m'<\frac{1}{2}k^{\beta }+o(1)<\frac{1}{2}\Pi
_m+o(1).$$ Considering the definition (\ref{E:pi_m}) of $\Pi _m$, we
easily get: \beq\label{InE:p_m-p_m+q**}
 \Bigl| |\vec k(\vphi_m^{\pm })+\vec p_m|_*^2 - |\vec k(\vphi_m^{\pm })+\vec
  p_{m+q}|_*^2\Bigr|\geq 2\Pi _m +O(k^{2s_1}).
\eeq Combining \eqref{InE:p_m-p_m+q*} and \eqref{InE:p_m-p_m+q**},
we arrive at \eqref{InE:p_m-p_m+q}. Now inequalities \eqref{beta}
and \eqref{InE:p_m-p_m+q}
 yield \eqref{InE:p_m-p_m+q+}.
Thus,
 \eqref{InE:1st-lowerbound} is proven.

 Note that all estimates
 are stable with respect the perturbation of $\varphi $ of order of
 $k^{2\beta -3-s_1-2\delta }$. Therefore,
 \eqref{InE:1st-lowerbound}, \eqref{InE:2nd-lowerbound} hold not
 only in $\varPhi _1$, but also in its $(k^{2\beta -3-s_1-2\delta
 })-$neighborhood.

%%%%%%%%%% Furthermore, \eqref{InE:first-factor} and \eqref{InE:p_m-p_m+q} yield
%%%%%%%%%%  \bdm
%%%%%%%%%%  k^{2\beta} \bigl| |\vec k(\vphi)+\vec p_m|_*^2-k^2\bigr|^{-1} < \Pi_m
%%%%%%%%%%  \le \bigl| |\vec k(\vphi)+\vec p_m|_*^2 - |\vec k(\vphi)+\vec p_{m+q}|_*^2
%%%%%%%%%%  \bigr|.
%%%%%%%%%%  \edm
 %%%%%%%%+Thus, \eqref{InE:3rd-lowerbound} holds.
 \epf
 \subsubsection{Proof of the statement (iii) in Lemma
 \ref{Lem:Geometric-1}.} \label{S:7.1.2}
 \begin{lem}\label{Lem:calO}
 \beq\label{InE:sum-r_m'}
  \sum_{\substack{m \in \Z^2 \\ 0< p_m \leq 4k^{s_1}}}r_m = O(k^{-\delta }),
  \eeq
  \beq\label{InE:sum-r_m}
  \widehat {\sum }_{\substack{m \in \Z^2 \\ 4k^{s_1}< p_m < 4k}}r_m = O(k^{-\gamma
  _0}),
  \eeq
  where $$\gamma _0=\left\{\begin{array}{ll} 1-2\beta -15 s_1-7\delta ,\ \ &
  \mbox{if }\beta \geq 1/6;\\5/6-\beta -15s_1-7\delta \ \ &
  \mbox{if }\beta <1/6,\end{array} \right.$$
  the sum $\widehat \sum $ in \eqref{InE:sum-r_m} includes only such $m$ that $\mathcalO _m\not \subset
  \cup _{\substack{ 0< p_q \leq
  4k^{s_1},\pm}}\mathcalO_q^{\pm }.$
\end{lem}
 \begin{cor} \label{C:2.1.2} If $2\beta <1-15s_1-8\delta $, then the total length of $\mathcal
 O^{(1)}\cap [0,2\pi )$ does not exceed $O(k^{-\delta
  })$. Estimate \eqref{E:Geometric-1-full} holds.
%%%%% \beq\label{InE:sum-r_m}
%%%%%   \sum_{\substack{m \in \Z^2 \\ 0< p_m < 4k}}r_m = O(k^{-\delta
%%%%%   }).
%%%%%   \eeq
  \end{cor}
 \bpf
 We split $m \in \Z^2$ with $0< p_m < 4k$ into five sets:
 \begin{align*}
  I_1 &:= \left\{m \in \Z^2 : 0< p_m \le 4k^{s_1} \right\},\\
  I_2 &:= \left\{m \in \Z^2 : 4k^{s_1} < p_m \le k^{1-4s_1-4\delta} \right\},\\
  I_3 &:= \left\{m \in \Z^2 : p_m > k^{1-4s_1-4\delta},\ 2k-p_m \ge k^{1-8s_1-4\delta}
  \right\},\\
I_4 &:= \left\{m \in \Z^2 : 2k-p_m < k^{1-8s_1-4\delta},
                         \left|1-\frac{p_m^2}{4k^2}\right| \ge \frac{k^{2\beta-2}}{4\Pi_m}
                         \right\},\\
  I_5 &:= \left\{m \in \Z^2 : \left|1-\frac{p_m^2}{4k^2}\right| < \frac{k^{2\beta-2}}{4\Pi_m}
  \right\}.
 \end{align*}
 and define $\Sigma_j := \sum_{m \in I_j} r_m,\ j=1,..,5$ and get an upper bound for each.

 \noindent \itemthm By \eqref{r_m}, \begin{equation}
 \Sigma_1 = \sum_{m \in I_1} \frac{k^{
 -3s_1-\delta}}{p_m }. \label{add3}
 \end{equation}
 Considering that the size of the lattice formed by $\vec p_m$ is
 $ck^{-s_1}$, we obtain:
 $$\sum_{m \in I_1} \frac{1}{p_m } \le ck^{2s_1}\int _0^{2\pi }\int_0^{k^{s_1}}
 \frac{1}{r} \cdot r\, dr d\varphi = ck^{3s_1}.$$  Substituting the last
 estimate into (\ref{add3}), we obtain
 (\ref{InE:sum-r_m'}).

 \noindent \itemthm Now we estimate $\Sigma _2$.
 Let us check that \begin{equation}
 \label{I_2 pi_m}\Pi_m \ge \frac{1}{2}k^{1-3s_1-3\delta}, \text{
 when  } m\in  I_2. \end{equation}
  Indeed, suppose
  $\Pi_m < \frac{1}{2}k^{1-3s_1-3\delta}$. Then for some $q \in \Z^2$ with
  $0<p_q<k^{s_1}$:
  $
  \bigl|\la \vec k(\vphi_m^\pm)+\vec p_m, \vec p_q\ra_*\bigr| + k^{2s_1}
  < \frac{1}{2}k^{1-3s_1-3\delta},
  $
 where $\pm $ means $+$ or $-$. Hence,
  $
  \bigl|\la \vec k(\vphi_m^\pm), \vec p_q\ra_*\bigr| <
  k^{1-3s_1-3\delta},
  $
 since $\bigl|\la \vec p_m, \vec p_q\ra_*\bigr| \le k^{1-4s_1-4\delta}\cdot k^{s_1}=k^{1-3s_1-4\delta}$ when
  $m\in I_2$.
 Therefore,
  $
   \bigl| |\vec k(\vphi_m^\pm)+\vec p_q|_*^2-k^2\bigr|
   = \bigl| 2\la \vec k(\vphi_m^\pm), \vec p_q\ra_* + p_q^2 \bigr|
   \le 2k^{1-3s_1-3\delta} + k^{2s_1} < 3k^{1-3s_1-3\delta}
  $,
 which means $\vphi_m^\pm \in \mathcalO_q^\pm$, see \eqref{InE:2nd-lowerbound} in Lemma
 \ref{Lem:lowerbounds}.
 Moreover, $\varphi _m^{\pm}$ is relatively close to the centrum of
 $\mathcalO_q^\pm$, its distance to the centrum of $\mathcalO_q^\pm$
 is less than $r_qk^{-2\delta }$.
 Using \eqref{r_m}, one can easily check that $r_m <<r_qk^{-2\delta}$ when $m\in I_2$.
  Thus $\mathcalO_m^\pm$ is completely contained in $\mathcalO_q^\pm$. We do not  consider
  such an $m$ in the sum \eqref{InE:sum-r_m}. Thus, we can use $\Pi_m \ge \frac{1}{2}k^{1-3s_1-3\delta}$.
By \eqref{r_m}
  \bdm
  \begin{split}
  \Sigma_2 = \sum_{m \in I_2}
  \frac{k^{2\beta -1}}{p_m  \Pi_m \sqrt{\left|1-\frac{p_m^2}{4k^2}\right|}}
  \le \frac{ck^{2\beta -1}}{ k^{1-3s_1-3\delta}} \sum_{m \in I_2} \frac{1}{p_m}
  \end{split} \edm
 We estimate the sum on the r.h.s. by the integral
  $k^{2s_1}\int _0^{2\pi }\int_{4k^{s_1}}^{k^{1-4s_1-4\delta}} \frac{1}{r}
  \cdot r\, dr d\varphi $. Computing the integral, we obtain :
   \begin{equation}
  \Sigma_2 \le ck^{2\beta-1+ s_1-\delta}. \label{add6}
  \end{equation}
\noindent \itemthm Let $m\in I_3$. First we note that
  \beq\label{InE:sqrt}
  \sqrt{1-\frac{p_m^2}{4k^2}} \ge \frac{1}{2}k^{-4s_1-2\delta}
  \eeq
 since $2k-p_m \ge k^{1-8s_1-4\delta}$. For the moment we assume to have
  \beq\label{InE:sum-1/pi_m}
  \sum_{m \in I_3} \frac{1}{\Pi_m} \le ck^{1+7s_1+\delta}.
  \eeq
 Then,
  \bdm
  \begin{split}
  \Sigma_3 = \sum_{m \in I_3}
  \frac{k^{2\beta -1}}{p_m  \Pi_m \sqrt{\left|1-\frac{p_m^2}{4k^2}\right|}}
  \le \frac{ck^{2\beta -1}}{k^{1-4s_1-4\delta}\cdot k^{-4s_1-2\delta}} \sum_{m \in I_3} \frac{1}{\Pi_m}
 %%%%%%%%%% \le ck^{2\beta-2+8s_1+6\delta} \cdot k^{1+4s_1+\delta}.
  \end{split}
  \edm
 where we used the definition of $I_3$ and \eqref{InE:sqrt}. Considering also
 \eqref{InE:sum-1/pi_m}, we obtain:
\begin{equation} \Sigma _3\leq   ck^{2\beta-1+15s_1+7\delta}.
\label{add7}\end{equation}
 Now we need to show \eqref{InE:sum-1/pi_m}. Obviously, $\varphi _m^{\pm }$ are real, when $m\in I_3$. Therefore,
 all $\langle \cdot ,\cdot \rangle_*$ and $|\cdot |_*^2$
 in the following formulas are real too. We note that $|\vec k(\vphi_m^\pm)|_*^2 = k^2$,
 $|\vec k(\vphi_m^\pm)+\vec p_m|_*^2 = k^2$ and
 $\bigl|\la \vec k(\vphi_m^\pm)+\vec p_m, \vec p_{q_m} \ra_*\bigr| + k^{2s_1} = \Pi_m$ for some
 $q_m,\ 0<p_{q_m}<k^{s_1}$ and $\vphi_m^{+}$ or/and $\vphi_m^{-}$. Assume for definiteness
 that the last equality holds for $\vphi_m^{+}$. Denoting $\vec k(\vphi_m^+)+\vec p_m$ by $\vec k_m$ gives
 us:
 \begin{equation}
 |\vec k_m-\vec p_m|_*^2 = k^2, \ \ \ \ |\vec k_m|_*^2 = k^2\ \ \ \ \label{add1}
 \end{equation}
\begin{equation} \bigl|\la \vec k_m, \vec p_{q_m} \ra_*\bigr| +
 k^{2s_1} = \Pi_m \label{add1add1}\end{equation}
 Obviously, $k^{2s_1}\leq \Pi _m\leq k^{1+s_1}$. Define $\Omega_j $ as the set of $\vec p_m$ in $I_3$ satisfying  the inequality :
 $ k^{2s_1+j\delta}<\Pi_m\leq k^{2s_1+(j+1)\delta}$,
 \ $j=0,\cdots,J,\ J=\lfloor \frac{1-s_1}{\delta} \rfloor$. Here $\lfloor r \rfloor = \max\{z \in \Z: z \le r\}$.
 Obviously, $\Omega_j \subset \cup _{0<p_q<k^{s_1}}\Omega_{j,q}$, where $\Omega_{j,q}$
 consists of $\vec p_m $ in $I_3$ satisfying (\ref{add1}) and
 \begin{equation}  k^{2s_1+j\delta}<\bigl|\la \vec k_m, \vec p_{q} \ra_*\bigr| +
 k^{2s_1}\leq k^{2s_1+(j+1)\delta}.\label{add5}
 \end{equation}
Let $\vec p_m\in \Omega_{j,q}$. By (\ref{add1}), (\ref{add5}), $\vec
p_m $ belongs to a circle of radius $k$ centered at a point $\vec
k_m$, where $\vec k_m$ belongs to a circle of the radius $k$
centered at the origin and satisfies the inequality
$k^{2s_1+j\delta}<\bigl|\la \vec k_m, \vec p_{q} \ra_*\bigr| +
 k^{2s_1}\leq k^{2s_1+(j+1)\delta}$, i.e. $\vec k_m$ belongs to one of two arcs.  Considering that $p_{q}>k^{-s_1}$,
 we easily see that all such points $\vec p_m $
 belong to a couple of rings of the same radii $k-k^{3s_1+(j+1)\delta},
 k+k^{3s_1+(j+1)\delta}$ centered at two points $\vec k^{q,\pm }$
 satisfying the conditions $|\vec k^{q,\pm }|=k, \langle \vec k^{q,\pm },\vec
 p_{q}\rangle =0$. Estimating the number of points $\vec p_m$ at this
 region by its area, we obtain, that the number of points in $\Omega
 _{j,q}$ does not exceed $ck^{1+5s_1+(j+1)\delta}$. Therefore, the number of
 points in $\Omega
 _{j}$ does not exceed $ck^{1+9s_1+(j+1)\delta}$.
 Thus, we obtain the estimate
  \bdm
  \sum_{m \in I_3} \frac{1}{\Pi_m} \le \sum_{j=0}^J \sum_{m \in \Omega_j}\frac{1}{\Pi_m}
  \le \sum_{j=0}^J c  k^{1+9s_1+(j+1)\delta} \cdot \frac{1}{k^{2s_1+j\delta}} \le
  ck^{1+7s_1+\delta},
  \edm
 which finishes this case.

 \noindent \itemthm Let $m\in I_4$. For the moment, we assume the following two estimates,
 \footnote{To prove \eqref{InE:pi_m} we use only the following
 conditions: $2k-p_m \leq
  -k^{1-8s_1-4\delta}$ or $|2k-p_m |\le k^{1-8s_1-4\delta}$ and at least one of $\varphi _m^{\pm}$ is  in
 $\varPhi _0'=\varPhi _0\setminus \cup _{0<p_q<4k^{s_1}\pm }\mathcalO _q^{\pm}$ or  its $2k^{-4s_1-2\delta }$-neighborhood. Under these conditions, we
  use estimate  \eqref{InE:pi_m} also in Lemmas
 \ref{Lem:lowerbounds*}, \ref{Lem:calO+}.}
  \beq\label{InE:pi_m}
  \Pi_m > \frac{1}{4}k^{1-9s_1-4\delta},
  \eeq
  \beq\label{InE:sum-2k-pm}
  \omega :=\sum_{m \in I_4}\frac{1}{\sqrt{|2k-p_m|}}
  <\left\{\begin{array}{ll}
  ck^{3/2+3s_1+\delta  },&\mbox{if  } \beta \geq 1/6;\\ck^{-\beta +5/3+3s_1+\delta  },&\mbox{if  } \beta
  <1/6.\end{array}\right.
  \eeq
 Then,
  \bdm
  \begin{split}
  \Sigma_4 = \sum_{m \in I_4}
  \frac{k^{2\beta -1}}{p_m  \Pi_m \sqrt{\left|1-\frac{p_m^2}{4k^2}\right|}}
  \le \frac{ck^{2\beta -1}}{k \cdot k^{1-9s_1-4\delta} \cdot k^{-1/2}}
  \sum_{m \in I_4} \frac{1}{\sqrt{|2k-p_m|}}   \end{split} \edm
  \begin{equation} \label{add8} \le \left\{\begin{array}{ll}
  ck^{-1+2\beta +12s_1+5\delta },&\mbox{if  } \beta \geq 1/6;\\ck^{-5/6+\beta +12s_1+5\delta },&\mbox{if  } \beta
  <1/6.\end{array}\right.
  \end{equation}
 Now we show \eqref{InE:pi_m}. We consider two cases:   $|2k-p_m |\leq k^{1-8s_1-4\delta }$ and $2k-p_m <-k^{1-8s_1-4\delta
 }$.
 We start with the former. Suppose, (\ref{InE:pi_m}) is not true, i.e.,  $\Pi_m
 \leq
 \frac{1}{4}k^{1-9s_1-4\delta}$. Then, there is a $q$ such that $0<p_q<k^{s_1}$
 and
 $$\bigl| \la \vec k(\vphi_m^\pm) +\vec p_m, \vec p_{q} \ra_*
 \bigr|\leq \frac{1}{4}k^{1-9s_1-4\delta}$$
 for $\vphi_m^+$ or/and $\vphi_m^-$. Note that $\vphi_m^\pm $ has a
 non-zero imaginary part when $p_m>2k$.
  Denoting  $\vec k(\vphi_m^\pm)+\vec p_m$ by $\vec
 k_m^\pm$, $\vec
 k_m^\pm \in \C^2$, we get \eqref{add1} for both $\vec
 k_m^{\pm}$.
 It follows from \eqref{add1} that $\vec k _m^{\pm
 }=\vec k(\varphi _{-m}^{\pm
 })$, we mean here that the pairs of vectors are the same.
 Therefore, there is at least one $\varphi _{-m}^{\pm}$, such that
 $\bigl| \la \vec k(\vphi_{-m}^\pm) , \vec p_{q} \ra_*
 \bigr|\leq \frac{1}{4}k^{1-9s_1-4\delta}$. We denote it by  $\varphi _{-m}^{+}$. The last inequality means
that $\varphi _{-m}^{+}$ is in
 $\mathcalO_q$, $\mathcalO_q=\mathcalO^{+}_q\cup \mathcalO^{-}_q $, and even relatively close to
 the center of a disc $\mathcalO^{\pm }_q $: its distance to the centrum is less than $r_qk^{-2s_1}$, the radius $r_q$ of
 $\mathcalO_q^{\pm}$ being given by $r_q=k^{-3s_1-\delta}{p_q}^{-1}$, see \eqref{r_m}.
 The distance between centers of $\mathcalO_q^{+}$ and
 $\mathcalO_q^{-}$ is $\pi +O(p_qk^{-1})$, It
 is easy to see also that $\varphi _{-m}^{\pm}=\varphi _{m}^{\pm}+\pi $ modulo $2\pi $, we mean here that the
 pairs of angles are the same.
 Considering the last two relations,  we obtain that $\varphi
 _{m}^{+}$ or/and $\varphi
 _{m}^{-}$ is also in $\mathcal O_q $.  Using
 $|2k-p_m|<k^{1-8s_1-4\delta }$, we get from
 \eqref{E:sin-cos}:
 $|\varphi _m^+-\varphi _m^-|<2k^{-4s_1-2\delta }$. It is much less
 than $r_q$.
  Hence,
 both $\varphi _m^{+}$ and $\varphi _m^{-}$ are in either  $\mathcal O_q^{+}$ or $\mathcal
 O_q^{-}$, relatively close to a center.  It is easy to show that  $r_m =o(k^{\beta -1})=o(r_q)$.
  This means $\mathcalO_m \subset \mathcalO_q$.
   We don't include include such $r_m$ in the sum
 \eqref{InE:sum-r_m}.
 Thus,
 \eqref{InE:pi_m} is proven for the case $|2k-p_m |\leq k^{1-8s_1-4\delta }$.
Let $2k-p_m <-k^{1-8s_1-4\delta
 }$.
By (\ref{E:sin-cos}), $$|\Im \varphi _{m}^{\pm
 }|=|\Im \varphi _{-m}^{\pm
 }|\approx { \small \sqrt{\frac{p_m^2}{4k^2}-1}}\geq \frac{1}{2}k^{-4s_1-2\delta
 }.$$
It is easy to see that from the definition of $\vphi_m^\pm$ that
$\vec k(\vphi_m^\pm) +\vec p_m= \vec k(\vphi_{-m}^\pm)$. Hence,
 $$\bigl| \la \vec k(\vphi_m^\pm) +\vec p_m, \vec p_q \ra_* \bigr|=
 \bigl| \la \vec k(\vphi_{-m}^\pm), \vec p_q \ra_* \bigr|=kp_q|\cos
 (\vphi_{-m}^\pm-\varphi _q)|>\frac{1}{4}kp_qk^{-8s_1-4\delta }\geq
 \frac{1}{4}k^{1-9s_1-4\delta }.$$ Therefore, $\Pi _m>\frac{1}{4}k^{1-9s_1-4\delta }$.
  Thus, we proved \eqref{InE:pi_m}.

 To show \eqref{InE:sum-2k-pm}, we split $I_4$ into three subsets $I_4=I_4'+I_4''+I_4'''$:
$$I_4'=\{m\in I_4:\ |2k - p_m |< k^{-1/3}\},\ \ \ I_4''=\{m\in I_4:\
k^{-1/3} < |2k - p_m |< k^\delta  \},$$ $$I_4'''=\{m\in I_4:\
|2k-p_m | \geq k^\delta  \}. $$ Correspondingly, $\omega =\omega
'+\omega ''+\omega '''$. Let us estimate $\omega '$.
 We use a well known estimate (see e.g. \cite{Lattice}) for the number
$N(k)$ of a rectangular lattice points in the
   circle
  of radius $k$: $N(k)=\pi v_{-1} k^2 +O(k^{2/3})$, where $v$ is the area of the elementary cell of the
  lattice and the implicit constant depend on the periods.
Considering that our lattice has a size of order $k^{-s_1}$, we
  rescale the last estimate, so it becomes: $N(k)=\pi  v_d^{-1}k^{2+2s_1}
  +O(k^{2/3+2s_1/3})$, $v_d=4\pi^2d_1^{-1}d_2^{-1}$. This means that
    $I_4'$ contains less than
  $ck^{2/3+2s_1}$ points. Note also that $|2k-p_m|> \frac{1}{4}k^{2\beta-2-s_1-\delta}$ follows from
$|1-\frac{p_m^2}{4k^2}| \ge \frac{k^{2\beta-2}}{4\Pi_m}$ and $\Pi_m
\le k^{1+s_1}$. Hence,
  $$\omega '<
  cv_d^{-1}k^{2/3+2s_1}\cdot k^{-\beta +1 +s_1/2+\delta /2}
  =o(k^{-\beta +5/3 +3s_1+\delta }).$$
  Next, considering that the region $I_4''$ contains no more than $cv_d^{-1}k^{1+\delta +2s_1}$ of points and
  $|2k-p_m|\geq k^{-1/3}$, we
  obtain
$$\omega ''<cv_d^{-1}k^{1+\delta +2s_1}k^{1/6}<cv_d^{-1}k^{7/6+2s_1+\delta}.$$
Further, we estimate $\omega '''$ by an integral:
  \begin{align*}
\omega '''
  < ck^{2s_1}v_d^{-1} \int _0^{2\pi }\cdot \int_{|2k-r|<2k} \frac{1}{\sqrt{|2k
  -r|}} r
  \,dr d\varphi
  \le cv_d^{-1} k^{\frac{3}{2}}
\end{align*} Combining the last three estimates, we obtain
\eqref{InE:sum-2k-pm}.
%%%%%%% $$\sum_{m \in
%%%%%%I_4}\frac{1}{\sqrt{|2k-p_m|}}\leq \left\{\begin{array}{ll}
%%%%%%ck^{3/2+2s_1+\delta }& \mbox{if }\beta >1/6\\ ck^{-\beta +5/3
%%%%%%+2s_1+\delta }& \mbox{if }\beta <1/6 .\end{array}\right. $$

 \noindent \itemthm As in the previous case, we have $\Pi_m > \frac{1}{4}k^{1-9s_1-4\delta}$.
 The number of points in $I_5$ admits the estimate:
 $\#(I_5) < cv_d^{-1} k^{\frac{2}{3}+2s_1}$. Therefore,
  \begin{equation*}
  \Sigma_5 = \sum_{m \in I_5}
  \frac{4k^{\beta}}{p_m \sqrt{\Pi_m}}
  \le \frac{8k^{\beta}}{k \cdot k^{(1-9s_1-4\delta)/2}}
  \sum_{m \in I_5} 1.
%%%%%%%%\label{add9}
   \end{equation*}
Hence, $\Sigma_5 < cv_d^{-1}k^{\beta-\frac{5}{6}+7s_1 +2\delta}.$
Adding (\ref{add6}),  (\ref{add7}),  (\ref{add8}), and the last
estimate, we
  obtain  (\ref{InE:sum-r_m}).
 \epf
\subsection{Proof of Geometric Lemma \ref{hope}. \label{5.2}} The set $\mathcalO ^{(1)}(\vec b)$ is defined by
formula \eqref{O^1(b)}, where $\mathcalO _m^{\pm }(\vec b)$ are
  open disks around $\vphi_m^\pm(\vec b)$, the points $\vphi_m^\pm(\vec b)$
being the two zeros in $\C$ of $|\vec k(\vphi)+\vec p_m(\vec
b)|_*^2=k^2$. Radius
 $r_m(\vec b)$ of $\mathcalO_m^\pm
(\vec b)$ is defined in Section   \ref{5.2.1}. Lemma \ref{hope} is
the combination of Lemma \ref{Lem:lowerbounds*} proven in Section
\ref{5.2.2} and Lemma \ref{Lem:calO+} proven in Section \ref{5.2.3}.
\subsubsection{Definition of the set $\mathcalO ^{(1)}(\vec b)$
\label{5.2.1}} Let us recall that $r_m$ are the radii of the discs
of the first non-resonant set  $\mathcalO ^{(1)}$. They are defined
by formula \eqref{r_m}. The radii $r_m(\vec b
 )$ of the discs in $\mathcalO ^{(1)}(\vec b)$ are, roughly speaking, defined as follows:
$r_m(\vec b
 )=k^{-2+2\beta _1+22s_1+15\delta}$, when $m=0$; $r_m(\vec b)\approx r_m$, when
 $p_m(\vec b)\geq 3k^{s_1},  |2k-p_m(\vec b)|>k^{1-8s_1-4\delta }$;  $r_m(\vec b)<<r_m $, when
  $|2k-p_m(\vec b)|\leq k^{1-8s_1-4\delta }$ or $p_m(\vec b)<3k^{s_1}$. We have to reduce the $r_m(b)$ for smaller
   $p_m(\vec b)$ and for $p_m(\vec b)$ close to $2k$ in order to
 ensure that each component of $\mathcal O ^{(1)}(\vec b)$ is
 sufficiently small (by \eqref{r_m}, $r_m$   becomes bigger, when $p_m$ is small or tends to $2k
 $).
 \begin{definition}\label{Def:r_m(b)}
  The radius $r_m(\vec b)$ of the two open disks $\mathcalO_m^\pm(\vec b)$ in $\C$ centered at $\vphi_m^\pm(\vec b)$ are defined by
   \beq \label{r_m*}
   r_m (\vec b):= \left\{
    \begin{array}{ll}
    k^{-2+2\beta _1+22s_1+15\delta } & \text{if } m=0,
      \\ [3em]
      \dfrac{k^{-2+2\beta _1+3s_1+2\delta }}{p_m(\vec b ) } &
      \text{if } m\neq 0, p_m (\vec b )< k^{1-8s_1-4\delta },\\ [1.5em]
     \dfrac{k^{-1+2\beta _1}}{p_m(\vec b )  \Pi_m (\vec b )\sqrt{\left|1-\frac{p_m^2(\vec b )}{4k^2}\right|}} &
      \text{if } p_m (\vec b)\geq  k^{1-8s_1-4\delta },\  |2k-p_m(\vec b)|>k^{1-8s_1-4\delta },\\ [3em]
      \dfrac{k^{-4+2\beta _1+10s_1+4\delta}}{\sqrt{\left|1-\frac{p_m^2(\vec b )}{4k^2}\right|}} &
      \text{if }|2k-p_m(\vec b)|\leq k^{1-8s_1-4\delta },\left|1-\frac{p_m^2(\vec b)}{4k^2}\right|\ge
       \frac{1}{4}k^{-4+2\beta _1+10s_1+4\delta},\\ [3em]
      %%%%%%%  \text{ and }
     %%%%%% 2k-p_m>k^{\delta }%%%5\left|1-\frac{p_m^2}{4k^2}\right|\ge \frac{k^{2\beta-2}}{4\Pi_m},\\
     %%%%%%555\dfrac{k^{1-3s_1-\delta}}{p_m k} &
     %%%%%%%% \text{if } 0< p_m (\vec b )\le k^{s_1},\\ [1.5em]
     4 k^{-2+\beta _1+5s_1+2\delta  }&
      \text{if } \left|1-\frac{p_m^2(\vec b )}{4k^2}\right| < \frac{1}{4}k^{-4+2\beta _1+10s_1+4\delta},
    \end{array}
   \right.
   \eeq
   where $\Pi _m(\vec b)$ is defined by (\ref{E:pi_m}),
   $\vec p_m(\vec b)$ being substituted instead of $\vec p_m$.
\end{definition}
 Considering as in Lemma \ref{Lem:r_m^2}, we easily show $r_m(\vec b)=o(1)$ as $k \to \infty$.
\subsubsection{Lemma \ref{Lem:lowerbounds*}.\label{5.2.2}}
 \begin{lem}\label{Lem:lowerbounds*} Let $100s_1<\beta
_1<1/12-28s_1-14\delta $.
 Suppose $\varphi \in  \hat \varPhi _0'\setminus \mathcal O^{(1)}(\vec b) $. Then
%%%%%%%%% \beq\label{InE:2nd-lowerbound*}
%%%%%%%%%   \bigl| |\vec k(\vphi)+\vec p_q(\vec b)|_*^2-k^2\bigr| >
%%%%%%%%%   k^{1-2s_1-\delta},\ \ \text{when } 0<p_{q}(\vec b)<k^{s_1},\ q\neq
%%%%%%%%%   0.
%%%%%%%%%   \eeq
%%%%%%%%% Furthermore,
   for every $m\in \Z^2$ such that
   \begin{equation}
   \bigl| |\vec k(\vphi)+\vec
 p_m(\vec b)|_*^2-k^2\bigr|<k^{\beta _1} \label{<beta _1}
 \end{equation}
  the following inequalities hold:
   \begin{equation} \min
_{0<p_q<k^{s_1}}\big||\vec k(\varphi )+\vec p_{m+q}(\vec
b)|_*^2-k^2\big| >k^{\beta _1}\label{dop1*} \end{equation}
  \beq\label{InE:1st-lowerbound*}
  \bigl| |\vec k(\vphi)+\vec p_m(\vec b)|_*^2-k^2\bigr|
  \bigl| |\vec k(\vphi)+\vec p_{m+q_1}(\vec b)|_*^2-k^2\bigr|\bigl| |\vec k(\vphi)+\vec p_{m+q_2}(\vec b)|_*^2-k^2\bigr| >k^{2\beta_1},
  \eeq
  when $0<p_{q_1}(0), p_{q_2}(0)<k^{s_1}$. This property
is preserved in the $k^{-4+2\beta _1-2s_1-\delta }$ neighborhood of
$\hat \varPhi _0'\setminus \mathcal O^{(1)}(\vec b)$.
%%%%  and
%%%%  \beq\label{InE:3rd-lowerbound}
%%%%  \bigl| |\vec k(\vphi)+\vec p_m|_*^2 - |\vec k(\vphi)+\vec p_{m+q}|_*^2 \bigr|
%%%%  > k^{2\beta} \bigl| |\vec k(\vphi)+\vec p_m|_*^2-k^2\bigr|^{-1},
%%%%  \eeq
 %%%%where $m \neq 0,\ m+q \neq 0,\ 0<p_q<k^{s_1}$.
 \end{lem}
%%%% \begin{cor} $\min _{m\in Z^2}\bigl| |\vec k(\vphi)+\vec p_m(\vec
%%%% b)|_*^2-k^2\bigr|>ck^{2\beta -2}.$ \label{cor}\end{cor}
%%%%Proof of (\ref{InE:2nd-lowerbound*}) is analogous to that of
%%%%(\ref{InE:2nd-lowerbound}).
 \bpf
 First, we consider the case $m=0$. It is easy to see that $\bigl||\vec k (\varphi ) +\vec b|_*^2-k^2\bigr|\geq
 kb_0r_0$, when $\varphi \not \in \mathcalO _0^{\pm }(\vec b)$. Using the estimate $b_0\geq k^{-1-16s_1-12\delta }$ and the formula for
 $r_0$, we get \begin{equation}
 \bigl||\vec k +b|_*^2-k^2\bigr|\geq k^{-2+2\beta
 _1 +6s_1+3\delta }.\label{+}
 \end{equation} Next,
  $$\bigl| |\vec k(\vphi)+\vec p_{q_i}(\vec b)|_*^2-k^2\bigr|=
  \bigl| |\vec k(\vphi)+\vec p_{q_i}(\vec b)|_*^2-|\vec k
  +b|_*^2\bigr|+O(k^{\beta _1})=$$
  $$2\left| \left<\vec k(\vphi),\vec p_{q_i}(\vec 0)\right>\right|+O(k^{\beta _1}),\ \ \ i=1,2.$$
   Considering that $\varphi \in \hat \varPhi _0'$, we obtain \beq \label{el5}2\left|\left<\vec k(\vphi),\vec p_{q_i}(\vec
  0)\right>\right|>k^{1-3s_1-\delta }.\end{equation}
  Hence,
  \begin{equation}\label{++}\bigl| |\vec k(\vphi)+\vec p_{q_i}(\vec
  b)|_*^2-k^2\bigr|>\frac{1}{4}k^{1-3s_1-\delta },\ \  i=1,2.\end{equation}
  Thus, \eqref{dop1*} holds for $m=0$.
  Multiplying  (\ref{+}) and (\ref{++}) for $i=1,2$,
   we arrive at (\ref{InE:1st-lowerbound*}) for
  $m=0$.

  Let $p_m(\vec b)<3k^{1-8s_1-4\delta}$, $m\neq 0$. It easily follows from the definition of
  $\mathcalO_m^{\pm }(\vec b)$,
  \beq\left||\vec k(\vphi)+\vec p_{m}(\vec
  b)|_*^2-k^2\right|\geq kp_m(\vec b)r_m\geq k^{-1+2\beta
  _1+3s_1+2\delta }. \label{el3}\eeq
Next,
  $$|\vec k(\vphi)+\vec p_{m+q_i}(\vec b)|_*^2-|\vec k(\vphi)+\vec p_{m}(\vec
  b)|_*^2=2\left|\left<\vec k(\vphi),\vec p_{q_i}(\vec 0)\right>\right|+O(k^{1-7s_1-4\delta}),\  i=1,2.$$
  Considering that $\varphi \in \hat \varPhi _0'$, we again obtain \eqref{el5}. Hence,
  $$\left||\vec k(\vphi)+\vec p_{m+q_i}(\vec b)|_*^2-|\vec k(\vphi)+\vec p_{m}(\vec
  b)|_*^2\right|>\frac{1}{2}k^{1-3s_1-\delta }.$$
  Taking into account \eqref{<beta _1}, we obtain
  \begin{equation}\left||\vec k(\vphi)+\vec p_{m+q_i}(\vec b)|_*^2-k^{2}\right|>\frac{1}{4}k^{1-3s_1-\delta
  }.\label{el4}
  \end{equation}
  Thus, \eqref{dop1*} holds for the case $p_m(\vec b)<3k^{1-8s_1-4\delta}$.
Multiplying \eqref{el3} and \eqref{el4} for i=1,2, we obtain
\eqref{InE:1st-lowerbound*}.

 If $p_m(\vec b)\geq k^{1-8s_1-4\delta}$, $|2k-p_m(\vec b)|>k^{1-8s_1-4\delta }$, we use the same considerations as
  in Lemma \ref{Lem:lowerbounds} to show that
  $$\Bigl| |\vec k(\vphi)+\vec p_m(\vec b)|_*^2-k^2\Bigr|
  \Bigl| |\vec k(\vphi)+\vec p_{m+q_i}(\vec
  b)|_*^2-k^2\Bigr|>k^{2\beta _1},\ \text{for } i=1,2.$$ The
  estimates \eqref{dop1*}, (\ref{InE:1st-lowerbound*}) easily follow.

Let $|2k-p_m(\vec b)|<k^{1-8s_1-4\delta }$.  Using
\eqref{E:first-factor} and considering as in Lemma
\ref{Lem:lowerbounds}, we obtain that \beq\label{InE:first-factor*}
  \bigl| |\vec k(\vphi)+\vec p_m(\vec b)|_*^2-k^2\bigr| >
  k^{-2+2\beta _1+10s_1+4\delta }
  \eeq
  when $\varphi \not \in O^{(1)}(\vec b)$.
  Suppose at least one of
$\varphi _m^{\pm}(\vec b)$  is in the $(k^{-4s_1-2\delta
})$-neighborhood of $\hat \varPhi _0'$.
 Then, $\Pi _m (\vec b)>\frac{1}{4}k^{1-9s_1-4\delta }$, see formula
 \eqref{InE:pi_m}  and the footnote there.
Using the  inequality
  \eqref{InE:p_m-p_m+q+}, see the footnote there, we obtain:
  \begin{equation}\label{i}
  \bigl| |\vec k(\vphi)+\vec p_{m+q_i}(\vec
  b)|_*^2-k^2\bigr|>\frac{1}{2}\Pi _m(\vec b),\ \
  i=1,2.\end{equation}
  Estimates \eqref{dop1*}, (\ref{InE:1st-lowerbound*}) easily
  follow. To finish the proof it remains to show that at least one  $\varphi
  _m^{\pm}(\vec b)$ is in the $k^{-4s_1-2\delta }$
neighborhood of $\hat \varPhi _0'$.
  If both $\varphi _m^{\pm}(\vec b)$ are not in the $k^{-4s_1-2\delta }$
neighborhood of $\hat \varPhi _0'$, then each is inside a disc
$O_q^{\pm}$, $0<p_q<k^{s_1}$ and further then $2k^{-4s_1-2\delta }$
from its boundary. Therefore, the distance of $\varphi $ to $\varphi
_m^{\pm}$ is greater than $k^{-4s_1-2\delta }$. However,
\eqref{<beta _1} and \eqref{r_m'}, yield $|\varphi -\varphi _m^{\pm
}|\leq 10k^{\beta _1/2-1}$. This contradiction proves that this case
is not possible. Since all estimates are stable with respect to the
perturbation of $\varphi $ of order $k^{-4+2\beta _1-2s_1-\delta }$,
the statement of the lemma holds in the $k^{-4+2\beta _1-2s_1-\delta
}$ neighborhood of $\hat \varPhi _0'\setminus \mathcal O^{(1)}(\vec
b)$ too.
  %%%%%%%%%%%%%%%%%%%%%%%%%%%%%%%
   %%%%%%%%%%%%%%%%%%%%%%%%%To obtain the above estimate, we first note that
   %%%%%%%%%%%%%%%%%%%%%%%%%$\left|\left<\vec k(\vphi),\vec p_{q_i}(\vec
   %%%%%%%%%%%%%%%%%%%%%%%%%0)\right>\right|>k^{1-3s_1 -\delta }$, because $\vphi \not \in
  %%%%%%%%%%%%%%%%%%%%%%%%% O_q$. Considering that
  %%%%%%%%%%%%%%%%%%%%%%%%% $\left|2\vec k(\vphi)+\vec p_m(\vec
  %%%%%%%%%%%%%%%%%%%%%%%%% b)\right|^2=4k^2-p_m^2<4k^{2-8s_1-4\delta }$, we obtain $\vec k(\vphi)+\vec p_m(\vec
 %%%%%%%%%%%%%%%%%%%%%%%%%  b)=-\vec k(\vphi)+O(k^{1-4s_1-2\delta })$. Hence,
  %%%%%%%%%%%%%%%%%%%%%%%%% $\left|\left<\vec k(\vphi)+\vec p_m(\vec b),\vec p_{q_i}(\vec
  %%%%%%%%%%%%%%%%%%%%%%%%% 0)\right>\right|>\frac{1}{2}k^{1-3\delta -\delta }$.
    %%%%%%%%%%%%%%%%%%%%%%%%%The estimate (\ref{i})
   %%%%%%%%%%%%%%%%%%%%%%%%%easily follows.
%%%%%%%%%%%%%%%%%%%%%%%%%   If $p_m>4k$, then \eqref{InE:1st-lowerbound*} is trivial,
 %%%%%%%%%%%%%%%%%%%%%%%%% since every multiplier on the left hand side is greater then $k^2$
 \epf
\subsubsection{Lemma \ref{Lem:calO+} \label{5.2.3}}
\begin{lem}\label{Lem:calO+} Suppose $100s_1<\beta
_1<1/12-28s_1-14\delta $ and ${\mathcal O}^{(1)}_c(\vec b)$ is a
connected component of ${\mathcal O}^{(1)}(\vec b)$. Then, the size
of ${\mathcal O}^{(1)}_c(\vec b)$
  does not exceed $ck^{-\gamma }$, $\gamma =11/6-\beta _1-12s_1-4\delta $.
  A
  component ${\mathcal O}^{(1)}_c(\vec b)$
  contains no more than $c_1k^{2/3 +2s_1}$ discs. The total size of
  ${\mathcal O}^{(1)}(\vec b)$ does not exceed $2\pi c_1k^{-5/6+\beta
  _1+12s_1+4\delta }
  $. The total number of discs in ${\mathcal O}^{(1)}(\vec b)$ is
  less than $ck^{2+2s_1}$.
  \end{lem}
  \bpf Let $\Delta _*$ be a rectangle in $\C$: $\Delta _*=\{\varphi : | \Re \varphi -\varphi _*|\leq k^{-1},
  \ |\Im \varphi |<k^{-\delta }\}$ for some $\varphi _*\in [0,2\pi )$.
  Clearly, $\varPhi _0$ is the union of such rectangles. Let
  $$I_*=\{m\in \Z^2,\ p_m(\vec b)<4k,\ \varphi _m^{\pm }(\vec b)\in \Delta
  _*  \},$$
  where $\varphi _m^{\pm }(\vec b)\in \Delta
  _*$ means that either $\varphi _m^+(\vec b)\in \Delta
  _*$ or $\varphi _m^{-}(\vec b)\in \Delta
  _*$. We will show that  the number of points in $I_*$ does not exceed
  $6c_0k^{1+2s_1}$ and
  \begin{equation}\Sigma _*:=\sum _{m\in I_*}r_m(\vec b)<ck^{-\gamma }
  \label{gamma} \end{equation}
  This means that the total size of ${\mathcal O}^{(1)}(\vec b)\cap \Delta
  _*$ is less than $k^{-\gamma }$.
  Considering that $\gamma >1$, we easily  obtain  that
 the size of each ${\mathcal O}^{(1)}_c(\vec b)$
   is less
  then $ck^{-\gamma }$. Since $\varPhi _0$ consists of $2\pi k$
  rectangles $\Delta _*$, the total size of
  ${\mathcal O}^{(1)}(\vec b)$ does not exceed $k^{-5/6+\beta
  _1+12s_1+4\delta }
  $. Let us prove \eqref{gamma}.
  In accordance with the Definition \ref{Def:r_m(b)} and by analogy with
  the proof of Lemma \ref{Lem:calO},
  we split $I_*$ into six sets:
  \begin{align*}
I_{*0} &:= \{m=0\},\ \ \ \   \
  I_{*1} := \{ m \in I_*\setminus\{0\} :p_m(\vec b) < k^{1-8s_1-4\delta}\}, \\
  I_{*2} &:= \{m \in I_* : p_m(\vec b) \geq k^{1-8s_1-4\delta},\ 2k-p_m(\vec b)
  > k^{1-8s_1-4\delta} \}, \\
  I_{*3} &:= \{m \in I_* : \ p_m(\vec b)-2k
  > k^{1-8s_1-4\delta}, \}, \\
  I_{*4} &:=\left\{m \in I_* : |2k-p_m(\vec b)| \leq k^{1-8s_1-4\delta},
  \left|1-\frac{p_m^2(\vec b)}{4k^2}\right| \ge \frac{1}{4}k^{-4+2\beta _1+10s_1+4\delta }\right\}, \\
  I_{*5} &:= \left\{m \in I_* : \left|1-\frac{p_m^2(\vec b)}{4k^2}\right| < \frac{1}{4}k^{-4+2\beta _1+10s_1+4\delta }
  \right\}.
  \end{align*}
 We define $\Sigma_{*j} := \sum_{m \in I_{*j}} r_m(\vec b),\ j=0,..,5$, and get an upper bound for each sum.
Formula \begin{equation}  \Sigma _{*0}= k^{-2+2\beta
_1+22s_1+15\delta } \label{Sigma_{*1}}\end{equation} immediately
follows from Definition \ref{Def:r_m(b)}.
%%%%%%%%%Considering as in Lemma \ref{Lem:calO}, we obtain:
%%%%%%%%% \begin{equation}
%%%%%%%%% \label{I_2 pi_m(b)}\Pi_m (\vec b)\ge \frac{1}{2}k^{1-3s_1-3\delta}, \ \ \ \ \ 1-\frac{p_m^2(\vec b)}{4k^2}=1+o(1)\
%%%%%%%%% \text{
%%%%%%%%% when  } m\in I_{*2}. \end{equation}
Obviously, \begin{equation}\label{Sigma _{*2}}
  \Sigma _{*1}=
  k^{-2+2\beta _1+3s_1+2\delta }\sum_{m \in I_{*1}}\frac{1}{p_m (\vec
  b)}.
  \end{equation}
  We estimate the sum in the right-hand side by an integral:
  %%%%%%%%Let
 %%%%%%%%L $$
 %%%%%%%%L  I_{*2}' := \{(r,\varphi ) : 1 < r \le k^{1-8s_1-4\delta},\ |\varphi _-\varphi _* |<
 %%%%%%%%L  r^{-1}\}.$$
  %%%%%%%%L It is easy to see that $I_{*2}\subset I_{*2}'$ and
$$\sum_{m \in
  I_{*1}}
  \frac{1}{p_m (\vec b)} \leq k^{2s_1}+k^{2s_1}\int _{I_{*1}'}\frac{1}{r} rdrd\varphi
  <k^{3s_1},$$
  where $I_{*1}'$ is a $(ck^{-s_1})$ neighborhood of $I_{*1}$.
  Substituting the  last estimate into \eqref{Sigma _{*2}}, we
  obtain:
  \begin{equation}\label{Sigma_{*2}}
  \Sigma _{*1}<ck^{-2+2\beta _1+6s_1+2\delta }.
  \end{equation}
Let us estimate $\Sigma _{*2}$.  First we note that
  %%%%%%%\beq\label{InE:sqrt*}
  $\sqrt{1-\frac{p_m^2(\vec b)}{4k^2}} >
  \frac{1}{4}k^{-4s_1-2\delta},$
  %%%%%%\eeq
 since $2k-p_m > k^{1-8s_1-4\delta}$. For the moment we assume
 that
  \beq\label{InE:sum-1/pi_m*}
  \sum_{m \in I_{*2}} \frac{1}{\Pi_m(\vec b)} \le ck^{7s_1+3\delta }.
  \eeq
 Then,
  \bdm
  \begin{split}
  \Sigma_{*2} = \sum_{m \in I_{*2}}
  \frac{k^{-1+2\beta _1}}{p_m \Pi_m (\vec b)\sqrt{1-\frac{p_m^2}{4k^2}}}
  \le \frac{ck^{-1+2\beta _1}}{k^{1-8s_1-4\delta}\cdot k^{-4s_1-2\delta}} \sum_{m \in I_{*3}} \frac{1}{\Pi_m(\vec b)}.
  %%%%%%% \\ \le ck^{-2+2\beta_1+12s_1+6\delta} \cdot k^{6s_1+2\delta }
  \end{split}
  \edm
  Thus,
  \begin{equation}
  \Sigma_{*2}<ck^{-2+2\beta _1+19s_1+9\delta}
 \label{Sigma_{*3}}.\end{equation}
 %%%%%where we used \eqref{InE:sqrt} in the first inequality and \eqref{InE:sum-1/pi_m} in the second.
 Let us prove \eqref{InE:sum-1/pi_m*}. Definition of $\varphi
 _m^{\pm}(\vec b)$ yields:
\begin{equation}\left|\vec k\left(\vphi_m^\pm(\vec b)\right)\right| = k,
  \  \ \ \left|\vec k\left(\vphi_m^\pm(\vec b)\right)+
 \vec p_m(\vec b)\right| =
 k.\label{again1}
 \end{equation}  Considering that $|\varphi _m^{\pm }(\vec b)-\varphi _*|<k^{-1}$, we
 obtain:
 $|\vec k(\vphi_*)| = k,$ $|\vec k(\vphi_*)+\vec p_m(\vec b)| =
 k +O(1),$  $|O(1)|<1.$
 This means that $\vec p_m(\vec b)$ belongs to the ring ${\mathcal R}$ of radii $k\pm 1$
 centered at $\vec k(\varphi _*)$. We split ${\mathcal R}$ into several
 components depending on the value of $\Pi _m(\vec b)$. Indeed, let
 $$\Omega _{*j}=\{m\in \Z^2:\ \vec p_m(\vec b)\in {\mathcal R},\ k^{2s_1+j\delta}\leq
 \Pi_m(\vec b)<k^{2s_1+(j+1)\delta}\},$$
 $j=0,\cdots,J,\ J=\lfloor \frac{1-s_1}{\delta} \rfloor.$
   Let us
  show that
  \begin{equation}
  \Sigma _{m\in \Omega _{*j}}\Pi _m^{-1}(\vec b)<4k^{7s_1+2\delta }.
  \label{Sigma_j*}
  \end{equation}
  We estimate the number of points in $\Omega _{*j}$. Indeed, by the definition of $\Pi _m$ for every
  $m\in \Omega _{*j}$ there is a $\vec p_{q_m}$,
  such that $0<p_{q_m}<k^{s_1}$ and
  $$\left|<\vec k\left(\varphi _m^{\pm }(\vec b)\right)+\vec p_m(\vec b),\vec p_{q_m}>\right|+k^{2s_1}<
  k^{2s_1+(j+1)\delta}.$$
  Considering that $|\varphi _m^{\pm }(\vec b)-\varphi _*|<k^{-1}$, we
  obtain that
  $$\left|<\vec k(\varphi _*)+\vec p_m(\vec b),\vec p_{q}>\right|<
  2k^{2s_1+(j+1)\delta} \mbox{ for a}\ q:\ 0<p_q<k^{s_1},\ \mbox{and } \vec p_m\in {\mathcal R}.$$
  The number of such points in ${\mathcal R}$, obviously, does not exceed
  $4k^{5s_1+(j+1)\delta}$ for a fixed $q$ and
  $4k^{9s_1+(j+1)\delta}$ for all $q$. Using now  the estimate
  $\Pi _m\geq k^{2s_1+j\delta}$, we arrive at \eqref{Sigma_j*}. The
  estimate \eqref{InE:sum-1/pi_m*} easily follows.

Let us estimate $\Sigma _{*3} $. Using \eqref{InE:pi_m}, we get
$r_m<2k^{-3+2\beta _1+13s_1+6\delta}$. By \eqref{E:sin-cos}, $\Re
\varphi _m ^{\pm }+\pi =\varphi _m$ (mod $2\pi $) when $m\in
I_{*3}$. Therefore, for any $m\in I_{*3}$, we have $|\varphi _m-\pi
-\varphi _*|<k^{-1}$ and $2k<p_m<4k$. Now it is obvious that the
number of points in $I_{*3}$ does not exceed $ck$. The estimate
\begin{equation}
\Sigma _{*3} < ck^{-2+2\beta _1+13s_1+6\delta} \label{Sigma+}
\end{equation}
easily follows. Next, we estimate $\Sigma _{*4} $.
%%%%%%%% We have proved in Lemma
%%%%%%%%\ref{Lem:calO} that
%%%%%%%%  \beq\label{InE:pi_m'}
%%%%%%%%  \Pi_m > \frac{1}{4}k^{1-5s_1-2\delta}
%%%%%%%%  \eeq
%%%%%%%% when $|2k-p_m|<k^{1-8s_1-4\delta}$.
 Suppose we have
 checked that
  \beq \label{InE:sum-2k-pm*}
  \sum_{m \in I_{*4}}\frac{1}{\sqrt{|2k-p_m(\vec b)|}} <
  ck^{5/3-\beta _1+2s_1}.
  \eeq
 Then,
  \bdm
  \begin{split}
  \Sigma_{*4}= \sum_{m \in I_{*4}}
  \frac{k^{-4+2\beta _1+10s_1+4\delta }}{\sqrt{\left|1-\frac{p_m^2}{4k^2}\right|}}
  \le ck^{-\frac{7}{2}+2\beta _1+10s_1+4\delta}
  \sum_{m \in I_{*4}} \frac{1}{\sqrt{|2k-p_m}|}.
  %%%%%%%%\\ \le ck^{-\frac{7}{2}+2\beta _1+10s_1+4\delta} \cdot k^{5/3-\beta_1+2s_1}.
  \end{split}
  \edm
  Therefore,
  \begin{equation} \label{Sigma_{*4}}
  \Sigma_{*4}< ck^{-11/6+\beta _1+12s_1+4\delta }.\end{equation}
%%%%%% where we use \eqref{InE:pi_m'} in the first inequality and \eqref{InE:sum-2k-pm}
%%%%%%  in the second.
  To show \eqref{InE:sum-2k-pm*}, we note that  $|2k-p_m|\geq
  \frac{1}{2}k^{-3+2\beta _1+10s_1+4\delta }$ follows
  from $1-\frac{p_m^2}{4k^2} \ge \frac{1}{4}k^{-4+2\beta _1
  +10s_1+4\delta }
  $.
   We split $I_{*4}$ into three regions:
  \begin{align*}
  \omega _{*1}&:=\{m\in I_{*4},\  k^{\delta } \leq |2k - p_m |\leq
  k^{1-8s_1-4\delta }\},\\
  \omega _{*2}&:=\{m\in I_{*4},\  k^{-1 } \leq  |2k - p_m |<
  k^{\delta }\},\\
  \omega _{*3}&:=\{m\in I_{*4},\  \dfrac{1}{2}k^{-3+2\beta _1+10s_1+4\delta } \leq |2k - p_m |<
  k^{-1}\}.\end{align*}
  The corresponding sums we denote as $\sigma _j$, $j=1,2,3$.
  It is
  easy to estimate $\sigma _1$ by an integral:
  \begin{align*}
\sigma _1:=\sum_{ k^\delta \le |2k-p_m| \leq k^{1-8s_1-4\delta}}
\frac{2}{\sqrt{|2k-p_m|}}
  < ck \int_{k^\delta}^{k^{1-8s_1-4\delta}} \int _0^{2\pi }\frac{1}{\sqrt{t}} \,d\varphi dt
  \le  ck^{\frac{3}{2}-4s_1-2\delta}.
\end{align*} Now we estimate $\sigma _2$.
It is easy to see that the number of points in $\omega _{*2}$ does
not exceed $c_0k^{1+\delta +2s_1}$. Using the estimate $2k - p_m
> k^{-1}$, we obtain: $$\sigma _2<ck^{3/2+2s_1+\delta
}.$$ Let us estimate $\sigma _3$ . Definition of $\vphi_m^\pm(\vec
b)$ yields $ \left|\vec k\left(\vphi_m^\pm(\vec b)\right)\right|^2_*
= k^2,
 \  \left|\vec k\left(\vphi_m^\pm(\vec b)\right)+
\vec p_m(\vec b)\right|_*^2 = k^2.$  By \eqref{E:sin-cos},  $|\Im
\varphi _m^{\pm }|<k^{-1}$ when $m\in \omega _{*3}$. Therefore,
$|\varphi _m^{\pm }-\varphi _*|<2k^{-1 }$. Hence, we
 obtain:
 $|\vec k(\vphi_*)| = k,$ $|\vec k(\vphi_*)+\vec p_m(\vec b)| =
 k +O(1),$ $|O(1)|<1$.
 This means that $\vec p_m(\vec b)$ belongs to the ring ${\mathcal R}$ of radii $k\pm
  1$
 centered at $\vec k(\varphi _*)$.
We also introduce the ring $\mathcalR_{*}=\{\vec k: \left||\vec
k|-2k\right|<k^{-1 }\}.$ Obviously $\omega_{*3}\subset {\mathcal
R_{*}}$. Thus, $\omega_{*3}\subset \mathcalR \cap \mathcalR _{*}$.
It is not difficult to show that the ring $\mathcal R$ cuts from
$\mathcalR_{*}$ an ``arc" of the length $O(k^{1/2})$. Using
well-known results from the theory of lattices, see \cite{Lattice},
we obtain that $\mathcalR \cap \mathcalR _{*}$ contains no more than
$ck^{1/6+s_1}$ points $\vec p_m(\vec b )$ for any $\vec b$.
Considering the estimate $|2k - p_m| \geq \frac{1}{2}k^{-3+2\beta
_1+10s_1+4\delta }$, we arrive at the inequality: $$\sigma
_3<ck^{5/3-\beta _1-4s_1-2\delta }.$$ Adding the estimates for
$\sigma _1$, $\sigma _2$ and $\sigma _3$ and considering that $\beta
_1<1/6-10s_1-11\delta $, we obtain \eqref{InE:sum-2k-pm*}.

 Let us estimate $\Sigma _{*5}$.
 %%%%%%%%As in the previous case,  $\Pi_m >
%%%%%%%%%% k^{1-5s_1-2\delta}$. Hence, $r_m(\vec b)<k^{-2+\beta _1+5s_1+2\delta }$.
 The number of points in $I_{*5}$ can be estimated the same way as the number of points in $\omega _{*3}$, i.e., it
 is less than $ck^{1/6+s_1}.$
 Hence,
  \begin{equation} \label{Sigma_{*5}}
  \Sigma_{*5}<
ck^{-11/6+\beta_1+6s_1+2\delta}.
  \end{equation}
  Adding the estimates \eqref{Sigma_{*1}}, \eqref{Sigma_{*2}},
  \eqref{Sigma+},
  \eqref{Sigma_{*3}}, \eqref{Sigma_{*4}} and \eqref{Sigma_{*5}} and considering that $\beta _1<1/6-10s_1-11\delta $, we
  obtain \eqref{gamma}.

  Let us show that the number of points in $I_*$ does not exceed
  $c_0k^{1+2s_1}$.  First, we count all real $\varphi _{m}^{\pm
  }(\vec b)$. Obviously, $|\varphi _{*}-\varphi _{m}^{\pm }(\vec
  b)|<k^{-1}$. Using \eqref{again1}, we obtain
  $|\vec k(\varphi _{*})+\vec p_m(\vec b)|=k+O(1)$, $|O(1)|<1$. The number of points
  $\vec p_m(\vec b)$ in this ring does not exceed
  $c_0k^{1+2s_1}$, $s_1$ appearing since the size of our lattice is of order
  $k^{-s_1}$. If $\Im \varphi _m^{\pm
  }(\vec b) \neq 0$, then, by \eqref{E:sin-cos}, $\varphi _m=\Re \varphi _{m}^{\pm
  }(\vec b)+\pi $ modulo $2\pi $. Hence, $|\varphi _m-\pi- \varphi
  _{*}|<k^{-1}$. Considering that $p_m(\vec b)<4k$, we obtain that
  the number of such points does not exceed $4c_0k^{1+2s_1}$. Hence,
  the total number of points in $I_*$ does not exceed $6c_0k^{1+2s_1}$.

  It remains to show that a component ${\mathcal O}^{(1)}_c(\vec b)$
  contains no more than $c_1k^{2/3 +s_1}$ discs. Indeed, the length of ${\mathcal O}^{(1)}_c(\vec b)$
   is less
  then $ck^{-\gamma }$. This means that all points $\varphi _m^{\pm
  }(\vec b)$ belonging to ${\mathcal O}^{(1)}_c(\vec b)$, are, in fact,
  in
  a square
   of the size $ck^{-\gamma }$ centered at a point
  $\varphi _{**}\in \Delta _*$.  Let us consider all real $\varphi _m^{\pm
  }(\vec b)$ in the component. Obviously $|\Re \varphi _{**}-\varphi _{m}^{\pm }(\vec b)|<ck^{-\gamma }$.
  Using \eqref{again1}, we obtain
  $|\vec k(\Re \varphi _{**})+\vec p_m(\vec b)|=k+O(k^{1-\gamma })$. By \cite{Lattice}, the number of points
  $\vec p_m(\vec b)$ in this ring does not exceed
  $c_1k^{2/3+s_1}$, $s_1$ appearing since the size of our lattice is of order
  $k^{-s_1}$. If $\Im \varphi _m^{\pm
  }(\vec b) \neq 0$, then, by \eqref{E:sin-cos}, $\varphi _m=\Re \varphi _{m}^{\pm
  }(\vec b)+\pi $ modulo $2\pi $. and $\cosh \Im \varphi _{m}^{\pm
  }(\vec b)=p_m/2k$. Hence, $|\varphi _m-\pi-\Re \varphi _{**}|<ck^{-\gamma }$
  and $|p_m-x_{**}<ck^{1-\gamma }$, $x_{**}=k\cosh \Im \varphi _{**}$.
  Obviously the number of points $\vec p_m(\vec b)$ is such a region
  is $O(1)$. Hence the total number of points $\varphi _{m}^{\pm}(\vec b)$ in ${\mathcal O}^{(1)}_c(\vec b)$
  is less than $c_1k^{2/3 +s_1}$. Considering that the number of
  points $\vec p_m(\vec b)$ satisfying the inequality $p_m(\vec
  b)<4k $ is less than $ck^{2+2s_1}$, we obtain that the total number of discs in ${\mathcal O}^{(1)}(\vec b)$ is
  less than $ck^{2+2s_1}$.
\epf

  \section{Appendices.} \label{appendices}

 \subsection*{Appendix 1. Proof of
Lemma \ref{L:3.7.1.1}}\label{A:5} By Lemma \ref{L:2.13}, Part 2, the
function  $\vec \varkappa _1(\varphi )$ is holomorphic in $\hat
\varPhi _1$ and  $ \lambda^{(1)}(\vec \varkappa _1(\varphi))=k^{2}.$
Hence, the equation (\ref{3.7.1.2}) is equivalent to
$\lambda^{(1)}(\vec y^{(1)}(\varphi))=\lambda ^{(1)}(\vec
y^{(1)}(\varphi)-\vec b)+\epsilon_0$. We use perturbation formula
(\ref{2.66}): $|\vec y^{(1)}(\varphi)|_*^{2}+f_1\bigl(\vec
y^{(1)}(\varphi)\bigr)
    =|\vec y^{(1)}(\varphi)-\vec b|_*^{2}+f_1\bigl(\vec y^{(1)}(\varphi)-\vec
    b\bigr)+\epsilon_0.$ This equation can be rewritten as
    \begin{equation}\label{3.7.1.5}
    2\langle \vec y^{(1)}(\varphi),\vec b\rangle _*-|\vec b|^2
+f_1\bigl(\vec y^{(1)}(\varphi)\bigr)
    -f_1\bigl(\vec y^{(1)}(\varphi)-\vec b\bigr)-\epsilon_0=0.
    \end{equation}
 Using the
notation $\vec b=b_0(\cos \varphi_b,\sin \varphi_b)$, dividing both
sides of the equation (\ref{3.7.1.5}) by $2b_0k$, and considering
that $\vec y^{(1)}(\varphi )=\vec \varkappa _1(\varphi)+\vec b=
    (k+h_1)\vec \nu +\vec b$, we obtain:
    \begin{equation}\label{3.7.1.6}
    \cos (\varphi-\varphi_b)-\epsilon_0g_1(\varphi)+g_2(\varphi)=0,
    \end{equation}
where $g_1(\varphi)
    =(2b_0k)^{-1}$
    and
    $$g_2(\varphi)=\dfrac{\langle \vec h_1(\varphi),\vec b\rangle
_*}{b_0k}-\dfrac{b_0}{2k} +\Bigl(f_1\bigl(\vec
y^{(1)}(\varphi)\bigr)
    -f_1\bigl(\vec y^{(1)}(\varphi)-\vec b\bigr)\Bigr)g_1(\varphi),\ \ \vec h_1(\varphi)=h_1(\varphi)\vec \nu.$$
Let us estimate $g_2(\varphi)$. Using the inequality (\ref{2.75})
for $h_1$, and considering that $b_0<k^{-1-16s_1-12\delta}$, we
easily obtain:
    $$
    \left|\frac{\langle \vec h_1(\varphi),\vec b\rangle
    _*}{b_0k}\right|\leq
    \frac{2h_1}{k}=O(k^{-4+30s_1+20\delta}),\ \ \
    \frac{b_0}{2k}\leq \frac{1}{2}k^{-2-16s_1-12\delta}.
    $$
 By (\ref{E:grad-f_1}),
    $ \left|f_1(\vec y^{(1)}(\varphi))-f_1(\vec y^{(1)}(\varphi)-\vec
    b)\right| \leq \sup |\nabla
    f_1|b_0=O(k^{-2+33s_1+22\delta }b_0),$
and therefore,
    $ \left|\bigl(f_1(\vec y^{(1)}(\varphi))-f_1(\vec y^{(1)}(\varphi)-\vec
    b)\bigr)g_1(\varphi)\right| =O(k^{-3+33s_1+22\delta }).$
Thus, we have
    $g_2(\varphi)=O(k^{-2-16s_1-12\delta}).$
Using $\epsilon_0<b_0k^{-1-16s_1-12\delta}$, we obtain
    $\epsilon_0g_1(\varphi)<k^{-2-16s_1-12\delta}.$ Thus,
    \begin{equation}
    g_2(\varphi)-\epsilon_0g_1(\varphi)=O(k^{-2-16s_1-12\delta}).\label{Nov14}
    \end{equation}
    Suppose that $\varphi_b\pm
\dfrac{\pi}{2}$ is in the
$\left(\frac{1}{8}k^{-2-16s_1-11\delta}\right)$-neighborhood of
$\hat \varPhi _1$. We draw two circles $C_{\pm}$ centered at
$\varphi_b\pm \dfrac{\pi}{2}$ with the radius
$\frac{1}{8}k^{-2-16s_1-11\delta}$. They are both inside the complex
$2k^{-2-16s_1-11\delta}$-neighborhood of $\varPhi _1$, the
perturbation series converging and the estimate (\ref{Nov14}) holds.
For any $\varphi$ on $C_{\pm}$, $|\varphi-(\varphi_b \pm
\pi/2)|=\frac{1}{8}k^{-2-16s_1-11\delta}$ and, therefore,
    $|\cos(\varphi-\varphi_b)|>\frac{1}{16}k^{-2-16s_1-11\delta}>|g_2(\varphi)-\epsilon_0g_1(\varphi)|$
    for any $\varphi \in C_{\pm}$.
By Rouch\'{e}'s Theorem, there is only one solution of the equation
(\ref{3.7.1.6}) inside each $C_{\pm}$. If $\varphi_b+\pi /2$ is not
in the $(\frac{1}{8}k^{-2-16s_1-11\delta})-$neighborhood of $\hat
\varPhi _1$, then
$|\cos(\varphi-\varphi_b)|>\frac{1}{16}k^{-2-16s_1-11\delta}$ in
$\hat \varPhi _1$ and, hence,  equation (\ref{3.7.1.6}) has no
solution.
 Thus, there are at most two solutions in $\hat \varPhi _1$ and
    $|\varphi^{\pm }_{\epsilon _0}-(\varphi_b \pm \pi /2)|<\frac{1}{8}k^{-2-16s_1-11\delta}.$

    \subsection*{Appendix 2. Proof of Lemma \ref{L:july5a}}\label{A:6}
Using the perturbation formula (\ref{2.66}), we obtain:
    \begin{align}\label{3.7.1.7}
    \lefteqn{
    \frac{\partial}{\partial \varphi}\lambda^{(1)}\bigl(\vec y^{(1)}(\varphi)\bigr)
    =\frac{\partial}{\partial \varphi}\left[\lambda^{(1)}\bigl(\vec
    y^{(1)}(\varphi)\bigr)-k^{2}\right]=
    \frac{\partial}{\partial \varphi}\left[\lambda^{(1)}\bigl(\vec
    y^{(1)}(\varphi)\bigr)-\lambda^{(1)}\bigl(\vec y^{(1)}(\varphi)-\vec b\bigr)\right]=
    }& \notag \\
    &\hspace{1cm}
    \left \langle \nabla_{\vec y}\lambda^{(1)}\bigl(\vec
    y^{(1)}(\varphi)\bigr)-\nabla_{\vec y}\lambda^{(1)}\bigl(\vec y^{(1)}(\varphi)-\vec
    b\bigr),\frac{\partial}{\partial \varphi}\vec y^{(1)}(\varphi)\right \rangle
    _*=
    \notag \\
    &\hspace{1cm}
    \left\langle \nabla \bigl|\vec y^{(1)}(\varphi)\bigr|_*^{2}-\nabla \bigl|\vec y^{(1)}(\varphi)-\vec
    b\bigr|_*^{2},(k+h_1)\vec{\mu}+h_1^{\prime}\vec{\nu}\right \rangle _* +\notag\\
    &\hspace{2cm}
    \left\langle \nabla f_1\bigl(\vec y^{(1)}(\varphi)\bigr)-\nabla f_1\bigl(\vec y^{(1)}(\varphi)-\vec
    b\bigr),(k+h_1)\vec{\mu}+h_1^{\prime}\vec{\nu}\right \rangle _*,
    \end{align}
where $\vec{\nu}=(\cos \varphi, \sin \varphi)$ and
$\vec{\mu}=\vec{\nu}^{\prime}=(-\sin \varphi,\cos \varphi)$. Note
that
    \begin{equation}\label{3.7.1.8}
    \nabla \bigl|\vec y^{(1)}(\varphi)\bigr|_*^{2}-\nabla \bigl|\vec y^{(1)}(\varphi)-\vec
    b\bigr|_*^{2}=\vec y^{(1)}(\varphi)-2\bigl(\vec y^{(1)}(\varphi)-\vec b\bigr)
    =2\vec b
    \end{equation}
Substituting (\ref{3.7.1.8}) into (\ref{3.7.1.7}), we get $
    \frac{\partial}{\partial \varphi}\lambda_j^{(1)}\bigl(\vec
    y^{(1)}(\varphi)\bigr)=T_1+T_2,$
   \begin{align*}
    T_1 &
    =2 \left\langle \vec
    b,(k+h_1)\vec{\mu}+h_1^{\prime} \vec{\nu} \right\rangle _* ,\\
    T_2  &=\left \langle \nabla f_1\bigl( \vec y^{(1)}(\varphi) \bigr) - \nabla
    f_1\bigl( \vec y^{(1)}(\varphi)-\vec
    b \bigr),(k+h_1)\vec{\mu}+h_1^{\prime}\vec{\nu}\right\rangle _*
   .
   \end{align*}
Considering that $\varphi$ is close to $\varphi_b \pm \pi /2$, we
readily obtain:
    $\langle \vec b,\vec{\nu}\rangle _*=o(b_0),$
 $\langle \vec
b,\vec{\mu}\rangle _*=\pm b_0(1+o(1))$.  Using also estimates
(\ref{2.75}) for $h_1$, we get $T_1=\pm 2b_0k(1+o(1))$.
 By (\ref{InE:second-derivative-lambda_j^1*}),
    $\Bigl|\nabla f_1\bigl(\vec y^{(1)}(\varphi)\bigr)-\nabla f_1\bigl(\vec y^{(1)}(\varphi)-\vec
    b\bigr)\Bigr|=O(b_0k^{-2+36s_1+24\delta }).$
Hence, $T_2=o\left(b_0k\right)$. Adding the estimates for $T_1,T_2$,
we get (\ref{3.7.1.6.1/2}).

\subsection*{Appendix 3. Proof of Corollary~\ref{C:3.3} }
\label{A:1}
 Let $\tau_0
\in \chi _2$. Taking into account the relation
$\lambda_j^{(1)}(\tau_0+2\pi p/N_1 a)=k^{2}$ and the definition of
$C_2$, we see that $|\lambda_j^{(1)}(\tau_0+2\pi p/N_1 a)-z|
=\epsilon_1/2$. Using  (\ref{3.6}) and the last equality, we easily
obtain:
   $|\lambda_n^{(1)}(\tau_0+2\pi \hat{p}/N_1a)-z| \geq \epsilon _1/2$
for $ \lambda_n^{(1)}(\tau_0+2\pi \hat{p}/N_1a) \neq
\lambda_j^{(1)}(\tau_0+2\pi p/N_1a)$. Therefore, for any $z \in
C_2$,
    \begin{equation} \label{Sept23}
    \|(\tilde{H}^{(1)}(\tau_0)-z)^{-1}\| \leq 2/\epsilon_1,
    \end{equation}
    i.e., (\ref{3.10}) is proved for $\tau _0\in \chi _2$.
Now we consider $\tau$ in the complex $(\epsilon_1
k^{-1-\delta})-$neighborhood of $\chi_2$. By  Hilbert relation,
    \begin{align*}(\tilde{H}^{(1)}(\tau)-z)^{-1}=(\tilde{H}^{(1)}(\tau_0)-z)^{-1}
    +T_1T_2
    (\tilde{H}^{(1)}(\tau)-z)^{-1},\ \ \ \ \ \ \ \ \ \ \ &\\   T_1=
    (\tilde{H}(\tau_0)-z)^{-1}(\tilde{H}_0(\tau_0)+k^{2})
,\ \ \
  T_2=(\tilde{H}_0(\tau_0)+k^{2})^{-1}(\tilde{H}_0(\tau_0)-\tilde{H}_0(\tau)).\end{align*}
    Suppose we have checked that $\|T_1T_2\|<k^{-\delta}$. Then,
    using (\ref{Sept23}),
    we easily arrive at (\ref{3.10}).
 The estimate $\|T_1\|<4k^{2}/\epsilon _1$,
easily follows from (\ref{Sept23}).   The estimate
$\|T_2\|<2\epsilon_1 k^{-2-\delta} $ easily follows from $|\tau
-\tau _0|<\epsilon_1 k^{-1-\delta}$.  Thus, $\|T_1T_2\|<8k^{-\delta
}$ and, hence, (\ref{3.10}) is proved.

\end{document}